%% file: main.tex
\input{alex_preamble}

\begin{document}

\title{Fast and fault-tolerant logical measurements: Auxiliary hypergraphs and transversal surgery}

\author{Alexander Cowtan}
\affiliation{Department of Computer Science, University of Oxford, Oxford OX1 3QD, UK}
\email{akcowtan@gmail.com}
\author{Zhiyang He}
\affiliation{Department of Mathematics, Massachusetts Institute of Technology, Cambridge, MA 02139, USA}
\author{Dominic J.~Williamson}
\affiliation{School of Physics, The University of Sydney, NSW 2006, Australia}
\affiliation{IBM Quantum, IBM Almaden Research Center, San Jose, CA 95120, USA}
\author{Theodore J.~Yoder}
\affiliation{IBM Quantum, IBM T.J. Watson Research Center, Yorktown Heights, NY 10598, USA}

\begin{abstract}
    Quantum code surgery is a promising technique to perform fault-tolerant computation on quantum low-density parity-check codes. Recent developments have significantly reduced the space overhead of surgery. However, generic surgery operations still require $O(d)$ rounds of repeated syndrome extraction to be made fault-tolerant. 
    In this work, we focus on reducing the time overhead of surgery. 
    We first present a general set of conditions that ensure fault-tolerant surgery operations can be performed with constant time overhead.
    This fast surgery necessarily makes use of an auxiliary complex described by a hypergraph rather than a graph.
    We then introduce a concrete scheme called block reading, which performs transversal surgery across multiple code blocks. 
    We further investigate surgery operations with intermediate time overhead, between $O(1)$ and $O(d)$, which apply to quantum locally testable codes.
    Finally, we establish a circuit equivalence between homomorphic measurement and hypergraph surgery and derive bounds on the time overhead of generic logical measurement schemes. 
    Overall, our results demonstrate that reducing the time cost of code surgery is not reliant on the quantum memory being single-shot. Instead it is chiefly the connectivity between a code and its measurement ancilla system that determines the achievable measurement time overhead.
\end{abstract}

\maketitle

\section{Introduction}
Quantum error correction~\cite{shor1995scheme,gottesman1997stabilizer,Kitaev1997qec,aharonov1997fault,shor1996fault} is an essential building block in any fault-tolerant quantum computer that is capable of running large-scale applications. 
The surface code and other topological codes~\cite{bravyi1998codes,Kitaev2003anyon,dennis2002memory,bombin2015gauge,eczoo_topological} have been extensively studied due to their rich mathematical structure and strong practical performance. 
Despite their many desirable properties, topological codes incur an appreciable space overhead~\cite{bravyi2010tradeoffs} when performing fault-tolerant quantum computation (FTQC).
This motivates the study of quantum low-density parity check (QLDPC) codes, which can achieve better parameters, notably constant encoding rate, in both the asymptotic and near-term regimes~\cite{eczoo_qldpc}. 
QLDPC codes possess the practically desirable LDPC property -- each check (or detector \cite{gidney2021stim}) involves a constant number of qubits and each qubit is in a constant number of checks.
Following exciting advancements in both theoretical code design~\cite{tillich2013quantum,kovalev2013quantum,panteleev2021degenerate,breuckmann2021balanced,lin2024quantum} and experimental qubit connectivity~\cite{Bravyi2022future,bluvstein2024logical,psiquantum2025manufacturable}, QLDPC codes are now competitive candidates to serve as low-overhead fault-tolerant quantum memories \cite{xu2024constant,bravyi2024highthreshold}. 

To realize FTQC in low space overhead with QLDPC codes, we need to design schemes for performing logical computation on their encoded qubits. 
This is a flourishing area of research with many recent works. 
Popular techniques include constant-depth unitary circuits~\cite{breuckmann2024foldtransversal,quintavalle2023partitioning,bravyi2024highthreshold,eberhardt2024operators,zhu2023gates,scruby2024quantum,breuckmann2024cups,hsin2024classifying,lin2024transversal,golowich2024quantum,malcolm2025computing,vuillot2022quantum,sayginel2024fault,berthusen2025automorphism}, 
resource state preparation and teleportation~\cite{gottesman2013fault,fawzi2020constant,tamiya2024polylog,nguyen2024quantum,he2025composable,huang2023homomorphic,xu2024fast,bergamaschi2024fault,Hong2025singleshot,li2025transversal}, 
and code switching~\cite{bombin2009quantum,breuckmann2017hyperbolic,Lavasani2018low,JochymOConnor2019faulttolerantgates,krishna2021fault,cohen2022low,cowtan2024css,cowtan2024ssip,cross2024improved, williamson2024low,swaroop2024universal, ide2024fault,zhang2024time,hillmann2024single, cowtan2025parallel,he2025extractors,yoder2025tour,poirson2025engineering,zheng2025high,xu2025batched,baspin2025fast,tan2025single,golowich2025constant}.
Here, we focus on the technique of QLDPC code surgery, which has seen rapid recent developments~\cite{cohen2022low,cowtan2024css,cowtan2024ssip,cross2024improved, williamson2024low,swaroop2024universal, ide2024fault,zhang2024time,hillmann2024single, cowtan2025parallel,he2025extractors,yoder2025tour,poirson2025engineering,zheng2025high,xu2025batched,baspin2025fast} due to its general applicability.
On a high level, code surgery is a technique for performing Pauli measurement on logical qubits encoded in a stabiliser code by introducing an ancilla system, consisting of new qubits and stabiliser generators, to deform the code in such a way that the chosen logical operators are decomposed into products of (newly introduced) stabilisers. 
The deformation procedure can be viewed as gauge fixing or weight reduction~\cite{vuillot2019code,hastings2016weight,hastings2021weight,sabo2024weight,Yuan2025unified}. 
Code surgery can be made fault-tolerant by repeatedly measuring all new stabiliser generators for a sufficient number of rounds (typically taken to be the code distance $d$) to prevent timelike errors.
After obtaining the logical measurement outcome, the ancillary qubits are individually measured out to return to the original code.

Unfortunately, measuring all stabiliser generators for $d$ rounds incurs a substantial time overhead when compared to protocols that use constant-depth gates.
For some computations this cost can be substantially ameliorated by using Pauli Based Computation~\cite{bravyi2016trading} to reduce the number of gates that must be implemented~\cite{litinski2019game,he2025extractors}. 
However, in the worst case a large time overhead remains. 
While a growing number of code families are known to be single-shot decodable~\cite{Bombin2015single,campbell2019theory,gu2024single,Kubica2022single,scruby2024high,quintavelle2021single,aasen2025topologically}, only high-dimensional topological codes are known to admit single-shot surgery~\cite{hillmann2024single}. 
Overall, prior to this work we lacked a rigorous understanding of the time overhead required for the fault-tolerance of surgery operations, except that $d$ rounds is typically sufficient.

In this work, we develop a theory of fast quantum code surgery, formulating the conditions needed for surgery operations to be fault-tolerant with only a constant number of rounds of error correction. 
Notably, the base memory code does \textit{not} need to be single-shot decodable.
As our primary example, we present a protocol called \textit{block reading} which enables \textit{transversal surgery} across identical CSS code blocks in constant space and time overhead. 
We discuss our main results in Section~\ref{sec:summary_results}, and discuss how our work relates to prior work in the literature, including independent works which appeared shortly before ours~\cite{baspin2025fast,xu2025batched,golowich2025constant,tan2025single}, in Section~\ref{sec:related_work}.

\subsection{Setting}

In this work, we aim to preserve the LDPC property of the codes and detectors throughout, and highlight where this is not guaranteed.
However, we ignore any further specific connectivity requirements of potential physical implementations. We assume a phenomenological Pauli and measurement noise model in our notion of fault tolerance. We focus primarily on asymptotics rather than specific constant factors. We do not concern ourselves with decoders for block reading, which depend greatly on the initial CSS codeblocks. We leave this important question to future work.

\subsection{Summary of main results}\label{sec:summary_results}

% Due to the length and intricacy of this work, we present a summary of our main results below.
To discuss our main results, we briefly review the usual setup of generalized code surgery, and refer readers to Section~\ref{sec:prelim} for a more thorough introduction. 
One starts with, in general, a collection of memory code blocks with encoded logical qubits, on which the aim is to perform fault-tolerant logical Pauli measurements. 
One constructs an ancilla system of qubits and stabilisers, which can be specified by a hypergraph~\cite{williamson2024low,ide2024fault}. 
One then connects this ancilla system to the memory blocks, in the process modifying the stabilisers of the ancilla system and the memory blocks according to the connections. 
The modified set of stabilisers is an abelian group, which specifies a stabiliser code (the deformed code).
Certain logical operators from the memory blocks become products of stabilisers in this deformed code, which means they are measured when we measure these new stabilisers. 
By measuring the deformed code stabilisers for a sufficient number of rounds (typically, in prior work, $d$ rounds, where $d$ is the distance of the memory codes), one can deduce the logical measurement results fault-tolerantly. 
Finally, one measures out the ancilla system to return to the original code. 
The operators measured depend largely on the ancilla system (equivalently the hypergraph) and its connections to the memory.
This formulation is general enough to capture all prior works on code surgery, see Section~3.2 of~\cite{he2025extractors} for a brief review.

In this work, we augment the above formulation by associating a 4-term chain complex with the hypergraph $H = (V, E)$. 
Specifically, let $C$ be a basis of cycles of $H$, where a cycle is a collections of hyperedges which include every vertex in $V$ an even number of times. 
Let $W$ be a set of $O(1)$-sized components of $H$, where a component is a collection of vertices which intersect every hyperedge on an even number of vertices.\footnote{Note here that $W$ is a set of our choice and is often not a basis of all components of $H$.} 
Then 
\[H^\bullet = \begin{tikzcd}\mathbb{F}_2^W & \mathbb{F}_2^V\arrow[l, "\delta_2"'] & \mathbb{F}_2^E\arrow[l, "\delta_1"'] & \mathbb{F}_2^C \arrow[l, "\delta_{0}"']\end{tikzcd}\]
is a 4-term cochain complex, where the coboundary maps $\delta_i$ are specified by the inclusion relations of $W, V, E, C$. 
Note that when $H$ is a simple connected graph (i.e., every hyperedge is a regular edge), the only component is the set of all vertices $V$. Typically when treating cell complexes the above coboundary maps would instead be boundary maps: we flip this convention to suit our conventions in the main body.

In prior works on surgery the focus was on the 3-term complex without $W$, which fully specifies the ancilla system. 
By including $W$, we can now consider the $1$-cosystolic distance of $H^\bullet$, which is 
\[
d^\bullet_1 = \min\{|u|: u\in \ker(\delta_2)\backslash\im(\delta_1)\}.
\]
Our main result states that $t\ge d$ hypergraph surgery operations, all satisfying certain distance conditions, can be performed sequentially with each operation taking a constant time, such that the overall protocol performs $t$ distinct logical operations fault-tolerantly in $\CO(t)$ time.

% Auxiliary complex surgery?

\begin{theorem}[Fast hypergraph surgery, informal statement of Theorem~\ref{thm:genhypersurg}]\label{thm:informal_hypergraph}
Let $Q$, $Q'$, $Q''...$ be a set of CSS LDPC codeblocks, each with distance at least $d$. 
Let $t\ge d$ and $\CH_{1}^\bullet, \cdots, \CH_{t}^\bullet$ be sparse chain complexes from hypergraphs which each define a $Z$-type surgery operation on the memory blocks, such that
\begin{itemize}
    \item All $\CH_{i}^\bullet$ have 1-cosystolic distances at least $d$, and
    \item The \textit{compacted code}, which is a CSS code defined by the $t$ surgery operations, has distance at least $d$.
\end{itemize}
Then the $t$ surgery operations can be performed sequentially with $O(1)$ time each. The phenomenological fault-distance of the protocol is at least $d$ and the LDPC property is preserved throughout.
\end{theorem}
We elaborate on the definition of the compacted code as well as other subtle conditions in the main text, but note here that there exist simpler sufficient conditions to require of each surgery operation individually or of the original code that together ensure the distance of the compacted code is large.

% We make an important clarifying remarks about this theorem. 
% First and foremost, 
Importantly,
we consider the fault-tolerance of performing $t\ge d$ surgery operations in time $\CO(t)$, instead of performing one surgery operation in time $O(1)$. 
This is because, strictly speaking, performing one surgery operation in time $O(1)$ is not fault-tolerant. 
We did not assume that the memory code blocks are single-shot decodable, and the deformed codes are not necessarily single-shot decodable. 
Therefore, the syndrome information from $O(1)$ rounds of noisy measurements is insufficient for fault-tolerance.

In contrast, performing $t$ operations in $\CO(t)$ time generates enough syndrome information to ensure the whole protocol has phenomenological fault-distance at least $d$.
In this work, we refer to this as \textit{amortised} constant time. %guarantee as constant time overhead \textit{in amortisation}.

This theorem formulates general conditions on surgery operations which enable constant-time-overhead logical gates. Hypergraph surgeries that measure stabilisers and logical operators of a code are forced to incorporate its local structure, up to changing the choice of generators. In the remaining parts of this work, we study a concrete fast surgery operation which we call \textit{block reading}. 
In its simplest form, given two identical code blocks of a CSS QLDPC code $Q$, block reading acts transversally on all pairs of data and check qubits of the two blocks and measure all pairs of logical qubits in the $Z\otimes Z$ (or similarly $X\otimes X$) basis. 
It does so by taking the auxiliary complex (or hypergraph) to be the code $Q$ itself, which by assumption has $1$-cosystolic distance at least $d$.
This example can be easily generalized to act on more than two code blocks and to measure more sophisticated Pauli product operators, including the checks of any CSS QLDPC code, transversally. 
We term this protocol full block reading.

% block reading acts transversally 

% The condition on sparsity of the logical operators being measured is defined in the main text.

\begin{theorem}[Full block reading, informal statement of Theorem~\ref{thm:full_block_amortised}]\label{thm:informal_fbr}
    Let $Q$, $Q'$, $Q''$... be a set of $c$ identical $\llbracket n, k, d \rrbracket$ CSS LDPC codeblocks. Then a set of sparse logical $Z$ Pauli operators across $Q$, $Q'$, $Q''$... can be measured in parallel using $\CO(cn)$ additional qubits and amortised $\CO(1)$ time overhead. Every measurement acts transversally on all $k$ logical qubits in a given codeblock. The phenomenological fault-distance of the protocol is at least $d$, and the LDPC property is preserved throughout.
\end{theorem}

% We generalise full block reading to hypergraph reading, where there is a hypergraph satisfying certain coherence conditions attached to subcodes of the original codeblocks, measuring the logicals across subcodes. This allows for better addressability on logical qubits while maintaining the constant time overhead, with a potential penalty in space cost.

Two important remarks are in order. First and foremost, parallels can be drawn between block reading and Steane-style error correction, or more generally logical measurements performed via ancillary code blocks and transversal CNOT gates.
In terms of logical action, full block reading has the same effect as measurements via transversal gates, except with a code-deformation procedure in contrast to a unitary circuit. 
Notably, however, the latter protocol usually assumes the ancillary code block is fault-tolerantly prepared, which in general requires $\CO(d)$ time overhead. 
The more subtle result of algorithmic fault-tolerance~\cite{zhou2024algorithmic} showed that this need not be the case, and an ancillary code block prepared in $O(1)$ time can still be used for measurements via transversal CNOT, such that the whole computation is fault-tolerant even though this individual operation is not. 
Our result on full block reading is similar to algorithmic fault-tolerance, in that we prove $d$ surgery operations can be performed fault-tolerantly in $\CO(d)$ time despite the fact that a single operation performed in constant time is not necessarily fault-tolerant. 
Notably we introduce new surgery gadgets and develop a very different proof to Ref.~\cite{zhou2024algorithmic}. This establishes a curious connection between transversal surgery and transversal gates.

Secondly, it is instructive to consider the simplest case of full block reading, where there is a single memory block and we measure all logical qubits in the $Z$ basis. This case mirrors Steane error correction, and showcases why a single fast hypergraph surgery operation is not necessarily fault-tolerant: if it were, we would have constant space-time overhead state preparation of arbitrary CSS QLDPC codes (see Remark~\ref{rem:state_prep})! 
Of course, full block reading with constant syndrome rounds {applied to a non-single-shot code} generates a frame error akin to that of Steane error correction with an ancillary code block (under-)prepared with constant syndrome rounds. 
% More precisely, 
This frame error is only eliminated in the subsequent $\CO(d)$ time steps.
Therefore, {this is one of the reasons that} performing either scheme in constant time is not strictly speaking fault-tolerant.

With full block reading as a motivating example, we study more flexible fast hypergraph surgery operations by taking subcodes of the memory code and thickening them appropriately. 
We call this \textit{partial block reading}, which acts transversally on all physical and logical qubits contained in the subcodes.

\begin{theorem}[Partial block reading, informal statement of Theorem~\ref{thm:fault_distance_many_subcode_1} and Corollary~\ref{coro:partial_BR_reduced_space_soundness}]\label{thm:informal_pbr}
Let $Q$, $Q'$, $Q''$... be a set of CSS LDPC codeblocks with distance at least $d$. Let $A$ be a subcode of each of the codeblocks $Q$, $Q'$, $Q''$... with length $n_A$ and distance at least $d$. Then a suitable hypergraph $\CH$ can be constructed using $\CO(n_A d)$ additional qubits. When $A$ has soundness $\rho_A$ this bound can be reduced to $\CO(\frac{n_A}{\rho_A})$, assuming the existence of a sparse cycle basis.
\end{theorem}

The precise notions of subcode, soundness and sparse cycle basis are defined in the main text.
As usual in the study of code surgery, the factor $\CO(d)$ space overhead is a worst case upper bound. 
When the subcode has suitable expansion properties, this space overhead can be significantly reduced, {all the way to $\CO(1)$ in some cases}. 
We expect the space overhead of partial block reading to be highly optimizable in practice, a problem we leave to future work.

Beyond constant time overhead surgery, we further study the case where the conditions of Theorem~\ref{thm:informal_hypergraph} are not satisfied, yet surgery can still be performed in lower than $\CO(d)$ time. 
We show that when the memory codes have soundness properties, then partial block reading can be performed with less than $O(d)$ syndrome rounds (in amortisation) when the subcode $A$ has distance lower than $d$.
\begin{theorem}[Intermediate time hypergraph surgery, informal statement of Theorem~\ref{thm:fault_distance_hypergraph_together}]\label{thm:informal_intermediate_time}
    Let $Q$, $Q'$, $Q''...$ be a set of CSS LDPC codeblocks with distance at least $d$, and Let $t\ge d$ and $\CH_{1}^\bullet, \cdots, \CH_{t}^\bullet$ be sparse chain complexes from hypergraphs which each define a surgery operation on the memory blocks, such that

    \begin{itemize}
        \item Each auxiliary complex $\CH_{1}^\bullet, \cdots, \CH_{t}^\bullet$ has $1$-cosystolic distance $d_i \geq \frac{1}{\alpha_i}d$,
        \item The deformed code each has distance at least $d$, and 
        \item The original codes have constant soundness.
    \end{itemize}
    Then the $t$ surgery operations can be performed sequentially with $\lceil\alpha_i\rceil$ time each. The phenomenological fault-distance of the protocol is at least $d$ and the LDPC property is preserved throughout.
\end{theorem}
 
Our hypergraph surgery constructions rely on homomorphic chain maps between the auxiliary complex and the memory. 
This fact draws strong parallel to the technique of homomorphic measurement~\cite{huang2023homomorphic}, which is a logical measurement scheme that employs transversal gates and generalizes Steane error correction.
We formalize this connection by proving a general circuit equivalence between surgery and homomorphic measurement.
% We then prove a general circuit equivalence between surgery and homomorphic measurement, another logical measurement scheme which uses transversal gates.
\begin{theorem}[Informal statement of results in Section~\ref{sec:hom_equivalence}]\label{thm:informal_hom_equivalence}
    Let $f$ define a homomorphic measurement protocol. Then the circuit for $f$ can be rewritten via ZX calculus to an equivalent hypergraph surgery protocol and vice versa.
\end{theorem}

Nevertheless, there are several subtleties, and this equivalence does not generally preserve the fault distance of the measurement protocols.
At a high level, the data (check) qubits in an ancillary block for homomorphic measurement correspond to new check (data) qubits in the corresponding surgery protocol respectively. As a consequence, timelike errors in surgery protocols can be resolved by measuring stabilisers for multiple rounds. However, the corresponding spacelike errors on a homomorphic measurement ancilla code are more difficult to resolve in a similar manner. This is why surgery protocols tend to be more flexible in addressing logical qubits: one can create a highly targeted ancilla system with low measurement fault distance, and then boost that measurement fault distance by measuring for multiple rounds.

Finally, we complement our constructions with a structural upper bound on the fault-distance of any logical measurement scheme, including surgery, homomorphic measurement and other similar protocols. 
The idea for this bound is based on quantifying the smallest undetectable fault that flips the logical representatives being measured while avoiding any ancilla system, see Proposition~\ref{prop:aux_fault_bound}. This logical fault corresponds to a spacetime stabiliser in the fault-tolerant memory on the code without any logical measurement, that becomes a nontrivial logical fault when the code is measured.
Here, we present a simplified statement of this general bound.

\begin{restatable}{theorem}{ConnectivityRequired}\label{thm:proof_connectivity}
    Any logical measurement protocol performed in $o(d)$ rounds on a quantum LDPC code by an auxiliary system requires connections to more than one logical representative to maintain phenomenological fault distance $\Omega(d)$, unless the correct logical measurement outcome is known in advance.
\end{restatable}

In other words, reducing the time cost of code surgery is not reliant on the quantum memory being single-shot, nor does fast logical measurement require transversal gates between codeblocks. Instead it is chiefly the degree of connectivity between a code and its measurement ancilla system that determines the achievable measurement time overhead.

\subsection{Related work}\label{sec:related_work}

Code surgery is an active topic of research, and our work has many connections to the literature. 
Notably, Ref.~\cite{baspin2025fast} appeared on arXiv shortly before this paper. 
Our works were developed independently, but share important results in common. We briefly discuss the areas of overlap and difference.

Most importantly, Theorem~5 of Ref.~\cite{baspin2025fast} states that a surgery operation defined by a sparse chain complex and a sparse chain homomorphism can be performed in constant time overhead fault-tolerantly if the auxiliary complex has suitable expansion properties. 
This is similar to our Theorem~\ref{thm:informal_hypergraph} (formally Theorem~\ref{thm:genhypersurg}), with a few points of distinction:
\begin{enumerate}
    \item We study the fault-tolerance of performing $t\ge d$ suitable surgery operations in $O(t)$ time, and claim that the time overhead of each logical measurement is constant in amortisation. In contrast, Ref.~\cite{baspin2025fast} study the fault-tolerance of performing one suitable surgery operation in constant time. 
    As we argued, this operation is not fault-tolerant by itself, and Ref.~\cite{baspin2025fast} remedies this by assuming that the syndrome information on the memory is noiseless before and after the constant time surgery operation (see their Section~IIIC). 
    This is a strict assumption in the context of single-shot operations -- for generic quantum codes, producing reliable syndrome information requires $O(d)$ rounds of syndrome measurement. 
    In this work, we consider a protocol where we start and end with $O(d)$ rounds of syndrome measurement on the memory, and perform $t\ge d$ hypergraph surgery operations in between. 
    We demonstrate that each hypergraph surgery operation generates a round of syndrome information on the memory blocks, and accumulating these rounds guarantees fault-tolerance of the full protocol.
    
    \item Ref.~\cite{baspin2025fast} asks for the auxiliary complex to have suitable expansion properties, which is a sufficient but not necessary condition for the deformed code to have high distance. From this perspective our requirements are less stringent. 
    Notably, full block reading for generic codes does not satisfy the expansion conditions in Ref.~\cite{baspin2025fast}.
    On the other hand, for partial block reading we prove that a similar expansion condition can satisfy the second condition of Theorem~\ref{thm:informal_hypergraph}.

    Both Ref.~\cite{baspin2025fast} and our work show that the desired expansion can be achieved by taking a tensor product with a repetition code. 
    The idea of using expansion to improve deformed code distance, and the idea of boosting expansion with a repetition code are both standard in the study of code surgery and weight reduction~\cite{hastings2021weight,cohen2022low, cross2024improved,williamson2024low,ide2024fault}.
\end{enumerate}

Our works are largely distinct and complimentary otherwise. 
As our primary examples we study full and partial block reading (Theorems~\ref{thm:informal_fbr} and \ref{thm:informal_pbr}), which perform transversal measurement of all logical operators contained in a (sub)code. 
Ref.~\cite{baspin2025fast} provides constructive, addressable surgery operations, capable of measuring one or more logical operators, using techniques including brute-force branching and auxiliary graph surgery~\cite{zhang2024time,cowtan2025parallel}.
Ref.~\cite{baspin2025fast} considered concrete examples on abelian multi-cycle codes and demonstrated promising numerical simulation results. 
We further study the case of hypergraph surgery with intermediate time overhead (Theorem~\ref{thm:informal_intermediate_time}), where the conditions of Theorem~\ref{thm:informal_hypergraph} are not satisfied yet soundness of the memory codes enables fault-tolerant surgery in less than $O(d)$ time.\footnote{Ref.~\cite{baspin2025fast} has a brief remark with a similar observation.}
Our results on circuit-equivalence between surgery and homomorphic measurement, Theorem~\ref{thm:informal_hom_equivalence}, and on the connectivity requirements for fast logical measurement, Theorem~\ref{thm:proof_connectivity}, have no counterparts in Ref.~\cite{baspin2025fast}.

% This is observed as a brief remark in 

There are several other recent independent works which share some overlap with our results. 
Ref.~\cite{zheng2025high} also considered the use of hypergraphs for code surgery.
They explored randomized constructions demonstrating promising practical performance.
While they did not explore the time overhead of these surgery operations, they stated a similar condition to boost the distance of the deformed code with expansion properties of the hypergraph (see their Lemma~2). 

In parallel, Ref.~\cite{tan2025single} and~\cite{golowich2025constant} developed single-shot dimension jumping techniques\footnote{Dimension jump is a special form of code switching, where one switch from (multiple copies of) lower dimensional codes to (fewer copies) of higher dimensional codes and vice versa.} for three-dimensional homological product codes. 
Relatedly, Ref.~\cite{xu2025batched} developed single-shot code switching techniques for more general homological product codes.
In its general form, full block reading can be seen as measuring, for $\CO(1)$ rounds, the stabilisers of a homological product code: that of the memory code and the pattern matrix.\footnote{More generally the pattern complex, if we use full block reading for both $X$ and $Z$ basis transversal logical measurements.} 
Our results show that this protocol can be combined with further hypergraph surgery protocols to be made fault-tolerant. 
In comparison, Refs.~\cite{tan2025single,golowich2025constant,xu2025batched} all proved stronger fault-tolerance guarantees in their specific settings. 
They also developed techniques to switch from the resultant homological product code to other codes, which enable a range of interesting logical computation gadgets. 
We remark that the aforementioned works~\cite{baspin2025fast,zheng2025high,tan2025single,golowich2025constant,xu2025batched} all have other contributions not mentioned in our brief discussion above.

Connecting to prior literature, 
while conventional quantum LDPC code surgery takes as input the support of a logical operator to be measured \cite{cohen2022low,cowtan2024css,cowtan2024ssip,cross2024improved,williamson2024low,ide2024fault},
hypergraph surgery must incorporate the local structure of the code on the region being measured. 
We focus on the simplest choice of hypergraph corresponding to the subcode itself, namely block reading.
This is similar to the CSS code surgery described in Ref.~\cite{poirson2025engineering}, as well as the finely-devised sticking in Ref.~\cite{zhang2024time}.
Our work is also related to Ref.~\cite{hillmann2024single}, in which single-shot surgery with 3D and 4D toric codes was studied using the formalism of fault complexes. Our proofs in Appendix~\ref{app:proof_fault_distance} use this formalism, and extend it to include deformations on spacetime volumes.

We are aware of upcoming work in which the time overhead of lattice surgery with surface codes is reduced by using the complementary gap to estimate confidence in a surgery protocol and stop early if the confidence is high~\cite{shutty2025early}. Our work is grounded in a different setting, concerned with phenomenological fault-distance and general CSS LDPC codes, and focuses on reducing the time overhead asymptotically.
% rather than by constant factors.

\subsection{Organization}
This work is lengthy and our results involve many subtleties. For the sake of the reader, we have deferred the majority of technical proofs to the appendices and aimed to build up complicated results from simpler ones. 
In Section~\ref{sec:prelim}, we review important background and definitions. 
In Section~\ref{sec:block_reading}, we introduce full block reading as the motivating example, and demonstrate fault-tolerance guarantees with amortised constant time overhead surgery operations. 
In Section~\ref{sec:partial_block_reading}, we study partial block reading. 
In Section~\ref{sec:hyper-surgery}, we generalize this to hypergraph surgery. 
For readers familiar with the techniques of code surgery, we recommend skimming Sections~\ref{sec:block_reading},~\ref{sec:partial_block_reading} and~\ref{sec:hyper-surgery} on an initial reading. 

In Section~\ref{sec:block_ltcs}, we study hypergraph surgery operations with intermediate time overheads, between $\CO(1)$ and $\CO(d)$. This type of surgery is enabled by quantum locally testable codes with constant soundness. 
In Section~\ref{sec:mod_expansion}, we define a novel generalisation of the notions of both soundness and relative expansion from Ref.~\cite{swaroop2024universal} which we call modular expansion. We show that hypergraphs with modular expansion can be used to perform hypergraph surgery in lower space overhead. 
In Section~\ref{sec:examples} we demonstrate our block reading schemes on a variety of examples. We show that our formalism captures fast surgery with topological codes such as 4D toric codes, and we describe block reading with 2D topological codes and bivariate bicycle codes \cite{bravyi2024high}.
Finally, in Section~\ref{sec:hom_equivalence} we prove the circuit equivalence between homomorphic measurement and surgery, and prove an upper bound on the phenomenological fault distance for generic logical measurement protocol on quantum LDPC codes.

\section{Preliminaries}\label{sec:prelim}
\subsection{Stabiliser codes and Tanner graphs}

\begin{definition}
Let $\CP^n$ be the Pauli group over $n$ qubits. A qubit stabiliser code $Q$ is specified by an Abelian subgroup $- \mathds{1} \notin \CS \subset \CP^n$ such that the codespace $\CH$ is the mutual $+1$ eigenspace of $\CS$, i.e.
\[U \ket{\psi} = \ket{\psi}\quad \forall\ U \in \CS,\ \ket{\psi}\in \CH.\]
We say $\CS$ is the stabiliser group for $Q$.
\end{definition}

A qubit stabiliser code $Q$ can be specified by a stabiliser check matrix
\[H =[H_X | H_Z ] \in \F_2^{r\times 2n},\ \textrm{such that}\ H\begin{pmatrix}0 & 1 \\ 1 & 0\end{pmatrix}H^\intercal = 0,\]
where a row $[u | v]$ corresponds to a generator of the stabiliser group, and therefore check on $Q$, $i^{u\cdot v}X(u)Z(v)$, for $u, v \in \F_2^n$.

\begin{definition}
A qubit CSS code $Q$ is a qubit stabiliser code where the generators of $\CS$ can be split into two sets $\CS_X$ and $\CS_Z$. $\CS_X$ contains Pauli products with terms drawn from $\{X,I\}$ and $\CS_Z$ terms drawn from $\{Z,I\}$.
\end{definition}

Thus there is a stabiliser check matrix $H$ for $Q$ such that
\[H = \begin{pmatrix}H_X & 0 \\ 0 & H_Z\end{pmatrix},\qquad H_X H_Z^\intercal = 0.\]

\begin{definition}
A CSS low-density parity check (LDPC) code family is a family of CSS codes such that the column and row weights of $H_X$ and $H_Z$ are bounded above by some constant $\sigma$.
\end{definition}

Logical operators in a stabiliser code are Pauli products that commute with all the checks of the code.
As the stabiliser group is Abelian, all stabilisers are equivalent to the trivial logical operator.
The code distance $d$ of $Q$ is the minimum Pauli weight of all the nontrivial logical operator representatives.

% \begin{figure}[t]\centering
% \[\tikzfig{tikz_files/shor_tanner}\]
% \caption{A Tanner graph for the $\llbracket 9,1,3 \rrbracket$ Shor code.}
% \label{fig:ShorTanner}
% \end{figure}

\begin{definition}
A Tanner graph is a bipartite graph with a vertex for each qubit and each check. A qubit is connected to each check in which it participates
with an edge labeled $[1|0]$, $[0|1]$, or $[1|1]$ depending on whether the check acts on the qubit as $X$, $Z$, or $Y$.
\end{definition}

In our convention, data qubits are black circles and checks are boxes. 
If a check only measures $Z$ or $X$ terms then we omit the edge label, and instead label the box with a $Z$ or $X$. 
Therefore for a CSS code we have no edge labels, and all boxes contain either $Z$ or $X$ labels. 
In this case, the condition that all stabilisers commute is the same as saying that the number of paths between every $Z$ and $X$ check is even.

Depicting large Tanner graphs directly becomes unwieldy, so we use \textit{scalable} notation\footnote{Terminology borrowed from Ref.~\cite{carette2019szx}.} to make them more compact. Qubits are gathered into disjoint named sets $\CQ_0, \CQ_1, \CQ_2, ...,$ and the same for checks $\CC_0, \CC_1, \CC_2, ...\,$. An edge between $\CQ_i$ and $\CC_j$ is then labeled with the stabiliser check matrix of $Q$ restricted to $\CQ_i$ and lifted to $\CC_j$, so we have
\[ [C_X|C_Z] \in \F_2^{|\CC_j|\times 2|\CQ_i|}.\]
If that check matrix is all-zeros then omit the edge entirely. If the checks are all of the same type then we omit the all-zeros half of the check matrix; for example if $\CC_j$ contains only $Z$ checks then we label the check box $Z$ and the edge
\[ [C_Z] \in \F_2^{|\CC_j| \times |\CQ_i|}.\]

\begin{figure}[tb]
    \hfill
     \begin{subfigure}[t]
     {0.36\textwidth}
         \centering
         \input{Figures-Tikz/tanner_eg1.tikz}
         \caption{A generic stabiliser code,}
     \end{subfigure}
     \begin{subfigure}[t]{0.6\textwidth}
     \centering
         \input{Figures-Tikz/tanner_eg2.tikz}
             \caption{A CSS code,}
     \end{subfigure}
     \hfill
     \vspace{5mm}
     \begin{subfigure}[t]{\textwidth}
     \centering
     \input{Figures-Tikz/hypergraph_prod.tikz}
     \caption{A hypergraph product code.}
     \end{subfigure}
     \caption{Examples of scalable Tanner graphs.}
         \label{fig:example_tanner_graphs}
\end{figure}
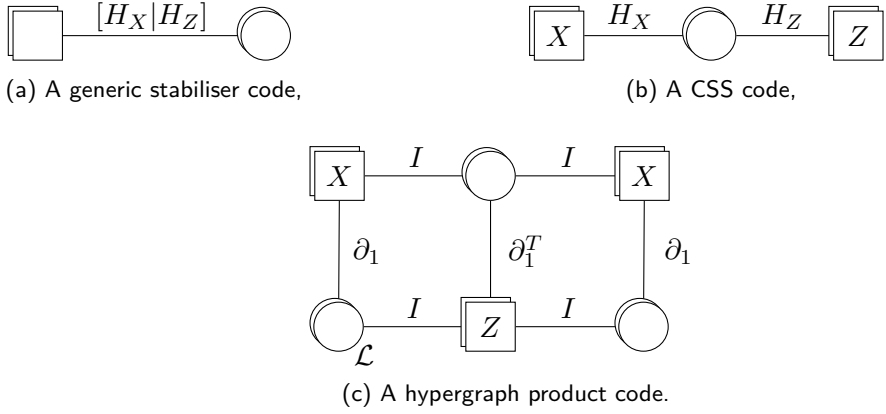

For example, in Figure~\ref{fig:example_tanner_graphs} we show some basic scalable Tanner graphs with qubit sets drawn as circles and check sets drawn as boxes.
We have (a) a generic stabiliser code, (b) a CSS code, and (c) the hypergraph product~\cite{tillich2013quantum} of a classical code $C$, with bits $\mathcal{L}$ and check matrix $\del_1$, by a repetition code $R$ with blocksize 2 and check matrix $\begin{pmatrix}
    1 & 1
\end{pmatrix}$.

We use scalable Tanner graphs to describe our procedures.

\subsection{CSS codes and chain complexes}

The well-known CSS code-homology correspondence relates quantum CSS codes to chain complexes, allowing for the study of various properties using the techniques of homological algebra~\cite{breuckmann2021ldpc,breuckmann2021balanced,breuckmann2024cups,panteleev2022good,cowtan2024css,ide2024fault,poirson2025engineering}. As we employ this formalism to study block reading, we give a brief summary of this correspondence here.

A based chain complex over $\F_2$,
\[C_\bullet = \begin{tikzcd} \cdots \arrow[r, "\del_{l+1}"] & C_l \arrow[r, "\del_l"] & C_{l-1} \arrow[r, "\del_{l-1}"] & C_{l-2} \arrow[r, "\del_{l-2}"] & \cdots \end{tikzcd}\]
is a $\Z$-graded vector space $\bigoplus_{i\in \Z} C_i$ over $\F_2$ equipped with a basis at each degree $i$, and boundary operators $\del_i : C_{i} \rightarrow C_{i-1}$ such that $\del_{i-1}\circ\del_i = 0$, $\forall i\in \Z$. Equivalently, $\im\del_{i+1}\subseteq \ker\del_{i}$. The homology space at degree $i$ is $H_i(C_\bullet) = \ker \del_{i}/\im \del_{i+1}$.

A CSS code must satisfy the property that $H_XH_Z^\intercal = 0$, and therefore we can view an $\llbracket n,k,d\rrbracket$ CSS code as a chain complex with 3 nonzero terms:
\[\begin{tikzcd}C_\bullet = C_{2}\arrow[r, "\del_2"]& C_1\arrow[r, "\del_{1}"]& C_{0}\end{tikzcd}\]
where explicitly at each degree we have
\[\begin{tikzcd}C_\bullet = \F_2^{m_Z}\arrow[r, "H_Z^\intercal"]& \F_2^n\arrow[r, "H_X"]& \F_2^{m_X}\end{tikzcd}\]
An undetectable $Z$ operator must have support on a vector $v \in \ker H_X$, and so every such operator is of the form $Z(v)$. $Z$ stabilisers are elements of $\im H_Z^\intercal$. Thus equivalence classes of $\overline{Z}$ logical operators are elements of $H_1(C_\bullet) = \ker H_X/\im H_Z^\intercal$, the homology space of $C_\bullet$ at degree 1. Hence $k = \dim H_1(C_\bullet)$, as the number of logical qubits must be equal to the number of independent equivalence classes of $\overline{Z}$ logical operators. The $Z$-distance $d_Z$ of a CSS code is the weight of the smallest nontrivial $\overline{Z}$ operator, so $d_Z = \min\limits_{v \in \ker H_X\backslash\im H_Z^\intercal}|v|$.

Dual to the notion of chain complex is a cochain complex,
\[C^\bullet = \begin{tikzcd} \cdots \arrow[r, "\delta^{l-1}"] & C^l \arrow[r, "\delta^l"] & C^{l+1} \arrow[r, "\delta^{l+1}"] & C^{l+2} \arrow[r, "\delta^{l+2}"] & \cdots \end{tikzcd}\]
which is defined similarly to a chain complex but with boundary operators transposed to become coboundary operators $\delta^i$, over $\F_2$. The cohomology space is $H^i(C^\bullet) = \ker\delta^i/\im\delta^{i-1}$, and enjoys the isomorphism $H^i(C^\bullet) \cong H_i(C_\bullet)$.

As $H_XH_Z^\intercal = 0$, we also have $H_ZH_X^\intercal = 0$, hence we can equally view a CSS code as a cochain complex with 3 terms:
\[\begin{tikzcd}C^\bullet = \F_2^{m_X}\arrow[r, "H_X^\intercal"]& \F_2^n\arrow[r, "H_Z"]& \F_2^{m_Z}\end{tikzcd}\]
This view prioritises $X$ operators, rather than $Z$ operators. We then have $k = \dim H^1(C^\bullet)$ and $d_X = \min\limits_{u \in \ker H_Z\backslash\im H_X^\intercal}|u|$. Furthermore, $d = \min(d_X,d_Z)$ as the error types are detected independently.

% \begin{remark}
%     The Tanner graph of a CSS code is the Levi graph of the corresponding (co)chain complex, and so in this case the perspectives of Tanner graphs and homology are largely equivalent.
% \end{remark}

A chain map, or homomorphism between chain complexes, is a linear map:
\[f_\bullet: C_\bullet \rightarrow D_\bullet\]
which has a component $f_i : C_i \rightarrow D_i$ at each degree $i$, such that the following diagram commutes,
\begin{equation}\begin{tikzcd}\cdots \arrow[r] & C_{l+1}\arrow[r, "\del^{C}_{l+1}"]\arrow[d, "f_{l+1}"] & C_{l}\arrow[r, "\del^{C}_{l}"]\arrow[d, "f_{l}"] & C_{l-1}\arrow[r]\arrow[d,"f_{l-1}"] & \cdots\\
\cdots \arrow[r] & D_{l+1}\arrow[r, "\del^{D}_{l+1}"] & D_{l}\arrow[r, "\del^{D}_{l}"] & D_{l-1}\arrow[r] & \cdots\end{tikzcd}
\label{eq:cd_chain_map}
\end{equation}
i.e. $\del_{i+1}^Df_{i+1} = f_i\del_{i+1}^C$ $\forall i \in \Z$.

A dual definition applies to a cochain map $f^\bullet$, which can be found by taking the transpose of each map in the diagram.

\begin{definition}[Subcode]\label{def:subcode}
    Let a subcode of the CSS code defined by a chain complex $C_\bullet$ be a code defined by the chain complex $A_\bullet$ with an injective chain map $f_\bullet : A_\bullet \rightarrow C_\bullet$. We stipulate further that $f_\bullet$ must take each data qubit in $A_\bullet$ to a single data qubit in $C_\bullet$, and the same for any $Z$ or $X$ checks in $A_\bullet$.\footnote{In Ref.~\cite{cowtan2024css} this property of $f_\bullet$ is called \textit{basis-preservation}.} 
\end{definition}

The additional stipulation is made to maintain control over weights, as this is essential for our analysis; if $C_\bullet$ is LDPC, so is $A_\bullet$. Each matrix in the chain map $f_\bullet$ is therefore a permutation matrix, up to some all-zero rows.

We can equally take the subcode in a dual manner, viewing the original CSS code via its cochain complex $C^\bullet$ and stipulating that there is a similar cochain map $g^\bullet: A^\bullet \rightarrow C^\bullet$. In this work, it is always made clear from context whether the subcode is taken in the chain ($Z$) or cochain ($X$) sense.

\subsubsection{Tensor products}
The tensor product of (co)chain complexes is inherited from the tensor product of $\Z$-graded vector spaces. Namely,
\begin{definition}\label{def:tensor}\cite[Sec.~2.7]{Weib1994}
Let $C_\bullet,D_\bullet$ be chain complexes over $\F_2$. Define $(C\tens D)_{\bullet}$ with components
\[(C\tens D)_l = \bigoplus_{i+j=l}C_i \tens D_j \]
where the latter tensor product is the normal tensor product of vector spaces. Differentials between components are given for $\del^{(C\tens D)}_{l}$ by matrices
\[\begin{pmatrix}\del^{C}_i \tens I \\ I \tens \del^{D}_j \end{pmatrix}\]
for a given $i, j$, then stacked horizontally for each term $i, j\ |\ i + j = l$, and vertically for each $i', j'\ |\ i' + j' = l-1$. One can check that $\del^{(C\tens D)}_{l-1}\circ\del^{(C\tens D)}_{l}=0 \pmod 2$, as desired.\end{definition}

\begin{lemma}[K{\"u}nneth formula] \label{lem:hom_factorA}\cite[Thm~3.6.3]{Weib1994}
\[H_l(C\tens D) \cong \bigoplus_{i+j=l} H_i(C)\tens H_j(D)\]
\end{lemma}
That is, the homology spaces factor through the tensor product conveniently. Both the tensor product and K{\"u}nneth formula can be applied dually to cochain complexes.

By virtue of the fact that we are working with finite-dimensional vector spaces with explicit bases, we can strengthen the K{\"u}nneth formula:
\begin{lemma}\cite[Prop.~1.13]{audoux2019ontensor} \label{lem:hom_factorB}
\[H_l(C\tens D) = \bigoplus_{i+j=l} H_i(C)\tens H_j(D)\footnote{Strictly speaking this is still an isomorphism, but the isomorphism is canonical.}\]
\end{lemma}
\proof
We first observe that any basis element of $H_i(C)\tens H_j(D)$ for $i+j = l$, which is an equivalence class of elements in $C_i \tens D_j$, is an element of $H_l(C\tens D)$, and that linearly independent elements remain independent. Then by a counting argument, $H_l(C\tens D) = \bigoplus_{i+j=l} H_i(C)\tens H_j(D)$.
\endproof

As CSS codes can be viewed as (co)chain complexes, the tensor product can be used to construct new codes \cite{audoux2019ontensor,tillich2013quantum}.

\subsubsection{Mapping cones}\label{sec:mapping_cones}
Given a chain map $f_\bullet: A_\bullet \rightarrow C_\bullet$ over $\F_2$, the \textit{mapping cone} of $f_\bullet$ is defined to be the chain complex with spaces $\cone(f)_i = C_i \oplus A_{i-1}$ and maps $\partial_i^{\cone(f)}: \cone(f)_i \to \cone(f)_{i - 1}$ given by
\[\label{eq:cone}
    \begin{tikzcd}
        {A_{i-1}} & {A_{i-2}}  \\
        {C_i} & {C_{i-1}}
        \arrow["{f_{i-1}}", from=1-1, to=2-2]
        \arrow["{\partial^{C}_i}"', from=2-1, to=2-2]
        \arrow["{\partial^{A}_{i-1}}", from=1-1, to=1-2]
        \arrow["\oplus", phantom, from=1-1, to=2-1]
        \arrow["\oplus", phantom, from=1-2, to=2-2]
    \end{tikzcd}
\]
\[
\partial_i^{\cone(f)} = \, \begin{pmatrix}
        
        \partial^{C}_i & f_{i-1} \\
        0 & \partial^{A}_{i-1}
    \end{pmatrix}.
\]

The homology of mapping cones can be studied using exact sequences and the Snake Lemma~\cite[Lem.~1.3.2]{Weib1994}, see App.~\ref{app:proof_fault_distance}.

Using chain maps and mapping cones, one can construct new quantum codes. In particular, $Z$ logical measurement of a CSS code $C_\bullet$ can be seen as constructing a particular auxiliary system $A_\bullet$ and then using the chain map $f_\bullet$ to build the deformed code $\cone(f_\bullet)$ \cite{ide2024fault}; $X$ logical measurements are then mapping cones from cochain maps. This perspective originates from the quantum weight reduction of Hastings \cite{hastings2016weight,hastings2021weight}.

\subsection{Metachecks and soundness}
Typically, quantum codes require $\CO(d)$ rounds of error correction for every logical timestep in order to maintain fault-tolerance. This is because in addition to errors occurring on data qubits, measurement errors can occur on checks. As checks are not typically protected against measurement errors, one must measure repeatedly in order to diagnose such errors successfully. This can be seen as constructing a spacetime code whose timelike component is a repetition code \cite{hillmann2024single}.

There are, however, quantum codes that do have protection against measurement errors \cite{Bombin2015single,campbell2019theory,gu2024single,Kubica2022single,scruby2024high,quintavelle2021single}. Although there are multiple forms of such protection, here we focus on the approach that uses metachecks \cite{campbell2019theory}. A metacheck is a set of stabilisers that has a product of outcomes known to be $+1$ in advance, even in the presence of faults on data qubits. Hence a metacheck is a detector~\cite{McEwen2023relaxing} on the spacetime code, which detects solely measurement errors.

It is possible to construct codes with sufficient metacheck structures, such that the minimum number of faulty measurements in a single error-correction round required to cause a logical fault is the same as the code distance $d$. Such codes can be single-shot codes, which in principle require only a single error-correction round against adversarial noise \cite{campbell2019theory}. Single-shot LDPC codes with metachecks require redundancy in their stabiliser generators and so can be expensive to construct, as having enough redundant checks to yield sufficient detectors for protection against up to $d$ measurement errors can entail a larger space overhead.

Given a scalable Tanner graph for a CSS code, we can incorporate metachecks with new sets of vertices and edges:
\[\input{Figures-Tikz/CSS_code_metachecks.tikz}\]

where blue (red) diamonds represent metachecks on $Z$ ($X$) checks respectively, and $\mathfrak{M}_Z$ ($\mathfrak{M}_X$) is the $\F_2$-linear map taking a syndrome outcome to its corresponding set of metachecks.

Equivalently, metachecks can be viewed as additional terms in the (co)chain complex of a code:
\[C_\bullet = \begin{tikzcd}C_3 \arrow[r, "\del_3"] & C_{2}\arrow[r, "\del_2"]& C_1\arrow[r, "\del_{1}"]& C_{0} \arrow[r, "\del_0"] & C_{-1}\end{tikzcd}\]
\[C^\bullet = \begin{tikzcd}C^3 & \arrow[l, "\delta^2"'] C^{2}& \arrow[l, "\delta^1"'] C_1& \arrow[l, "\delta^0"']C^{0} & \arrow[l, "\delta^{-1}"'] C^{-1}\end{tikzcd}\]
where $C_3=C^3$ represents metachecks on $Z$ checks and $C_{-1}=C^{-1}$ represents metachecks on $X$ checks.

In scalable Tanner graph notation, maps extend outwards from data qubits, to checks and then metachecks. In (co)chain complexes, maps begin at the lowest or highest degree and point towards the higher or lower degree respectively.

\begin{definition}\cite{campbell2019theory}
    Let $Q$ be a CSS code. The $Z$-single shot distance is
    \[d^{ss}_Z=\min\limits_{v \in \ker\mathfrak{M}_Z\backslash \im H_Z} |v| = \min\limits_{v \in \ker\delta^2\backslash \im\delta^1}|v|\]
\end{definition}
and is the weight of the smallest $Z$-measurement fault that is undetectable by $Z$-metachecks and is not equivalent to an $X$ fault on data qubits. Similarly, 
\[d^{ss}_X = \min\limits_{u \in \ker\mathfrak{M}_X\backslash \im H_X} |u|= \min\limits_{u \in \ker\del_{0}\backslash \im\del_{1}}|u|\]
The single-shot distance of $Q$ is therefore $d^{ss}=\min(d^{ss}_X,d^{ss}_Z)$, as the $Z$ and $X$ checks are independent. If $\ker\del_{0}\backslash \im\del_{1} =\emptyset$ then by convention we say that $d^{ss}_X = \infty$, and similar for $d^{ss}_Z$.

Like logical operators, measurement faults can be partitioned into equivalence classes, in $H^2(C)$ and $H_0(C)$ for $Z$ measurement errors and $X$ measurement errors respectively.
Even if the single-shot distance is at least the distance of the code, i.e. $d^{ss}\geq d$, it is still possible for a low-weight measurement fault to prevent fault-tolerant single-shot error correction. This is because there could be a low-weight measurement fault which \textit{is} equivalent to a fault on data qubits, but that fault is very large. Hence the measurement fault may be interpreted by the decoder as a large fault on data qubits, and a large, erroneous recovery operator applied. As a consequence, the next logical cycle would require only a small fault on data qubits to cause a logical error.

This motivates the study of LDPC codes with good soundness \cite{aharonov2015quantum,campbell2019theory,panteleev2022asymptotically}, where every valid small set of syndrome outcomes has a sufficiently small set of data qubit errors which produces them; thus a minimum distance decoder will not misinterpret a small measurement fault as a large data qubit fault.

There are several different definitions of soundness for both classical and quantum codes. For classical codes, we adopt a simple combinatorial definition of local-testability from Ref.~\cite[Def.~11]{leverrier2022towards}.
\begin{definition}[Local testability]\label{def:classical_soundness}
    A binary linear code $\CC$ is $(\omega,\rho)$-locally testable if it has a parity-check matrix $H : \F_2^n \rightarrow \F_2^m$ with rows of weight at most $\omega$ such that for any vector $v \in \F_2^n$,
    \[\frac{1}{m}|H v| \geq \frac{\rho}{n}d(v,\CC)\]
    where $d(v, \CC) = \min_{x \in \CC}(|v+x|)$ and $|\cdot |$ is the Hamming weight. The values $\omega$ and $\rho$ are the locality and soundness of the code respectively.
\end{definition}

For quantum stabiliser codes, we follow Ref.~\cite{eldar2015local}. For an $n$-qubit quantum code $Q$, let
\[Q_t = {\rm Span}\{(A_1\otimes\cdots\otimes A_n)\ket{\psi} : \ket{\psi} \in Q, \#\{i\in [n], A_i\neq I\} \leq t\}\]
be the \textit{t-fattening} of $Q$, which is the set of states with distance at most $t$ from the codespace. Then define
\[D_Q = \sum_{t \geq 1}t(\Pi_{Q_t}-\Pi_{Q_{t-1}}).\]
where $\Pi_{Q_t}$ is the projector onto the space $Q_t$.
\begin{definition}\label{def:quantum_soundness}
    A qubit stabiliser code $Q$ with generators $S_1,\cdots,S_m$ is locally testable with locality $\omega$ and soundness $\rho$ if all generators have weight at most $\omega$ and
    \[\frac{1}{m}\sum\limits_{i=1}^m\frac{1}{2}(I-S_i) \succeq \frac{\rho}{n}D_Q\]
    where $A \succeq B$ means the operator $A-B$ is positive semidefinite.
\end{definition}

\begin{lemma}\cite[Fact~17]{eldar2015local}\label{lem:fact_17}
    A quantum CSS code $\textrm{CSS}(H_X,H_Z)$ with parity-check matrices $H_X$, $H_Z$ is quantum locally testable with soundness $\rho$ if $\ker H_X$, $\ker H_Z$ are classical locally testable codes with soundness $\rho$. Conversely, if $\textrm{CSS}(H_X,H_Z)$ has soundness $\rho$ then $\ker H_X$, $\ker H_Z$ are classical locally testable codes with soundness at least $\rho/2$.
\end{lemma}

\begin{remark}
Note that Def.~\ref{def:quantum_soundness} is subtly different from the version of soundness given in Ref.~\cite{campbell2019theory}, and is closer to the definition of \textit{confinement} in Ref.~\cite{quintavelle2021single}. Intuitively, the version in Ref.~\cite{campbell2019theory} stipulates that every valid small set of syndrome outcomes has a sufficiently small set of data qubit errors which produces them. Def.~\ref{def:quantum_soundness} and confinement instead stipulate that every small set of data qubit errors produces a sufficiently large set of syndrome outcomes. See Ref.~\cite{quintavelle2021single} for more discussion about the distinction between these definitions. Def.~\ref{def:quantum_soundness} aligns more closely with the definition of soundness for classical codes, so is preferable for our purposes.
\end{remark}

It is well-known that a quantum code with both  sufficiently high single-shot distance and sufficient soundness, of a certain type, can be decoded in a single-shot manner against adversarial noise using a minimum-weight decoder \cite{campbell2019theory}, and similar results apply too to confinement \cite{quintavelle2021single}.
We remark that the development of quantum locally testable codes with good parameters is an ongoing line of research~\cite{Cross2024quantumlocally,wills2024tradeoff,dinur2024expansion,kalachev2025maximally}.

In summary, a metachecked code being single-shot is reliant on the decoder and noise model, in addition to other code properties than just the single-shot distance. For phenomenological adversarial noise, a minimum weight decoder is sufficient when the codes have sufficient properties related to soundness~\cite{campbell2019theory,quintavelle2021single}. In this work we do not require the quantum memory to be single-shot, or even have high single-shot distance, but we make use of the general framework of metachecks, soundness and their relation to homology.

\subsection{Quantum code surgery}

Generalised lattice surgery, or quantum code surgery, is a method of performing computation by code deformation, see Section~3.2 of Ref.~\cite{he2025extractors} for a brief review. In this work, we presume that the codes are all stabiliser codes, although one can perform surgery with non-Abelian codes \cite{cowtan2022algebraic,Bravyi2022,Lyons2024,Ren2025,davydova2025universal}. We also focus on logical measurements by the introduction of an auxiliary system, while surgery can be performed in other ways to implement a larger class of operations \cite{deBeaudrap2020zx,cowtan2024css,poirson2025engineering,aasen2025geometrically}.

Those exceptions aside, all prior works in quantum LDPC code surgery are known to be unified by the construction of a \textit{measurement hypergraph} \cite{williamson2024low,ide2024fault,he2025extractors}. In brief, for a given initial quantum memory $Q$ and logical measurement to be performed, we construct a hypergraph $\CH(\CV,\CE)$, consisting of vertices $\CV$, edges $\CE$ and cycles.\footnote{A cycle in a hypergraph is a collection of edges that contains every vertex an even number of times.} Edges are associated to new data qubits, and there are two types of new stabiliser checks: vertex checks,  whose measurement outcomes are used to infer the logical measurement outcome, and cycle checks, which are present to gauge-fix and thus prevent any new logical qubits from appearing in the deformed code.\footnote{These new logical qubits could have low weight logical operators, and lower the dressed weights of other operators acting on logical qubits in the code \cite{cohen2022low}, which would lower the fault-distance.} The measurement hypergraph is connected to the initial code $Q$. Specifically, a subset of vertices (new checks) are connected to qubits in $Q$ in order to perform the logical measurement, while a subset of edges (new data qubits) are connected to checks in $Q$ such that the deformed code commutes as a stabiliser code.

\begin{definition}
    Let $\CH$ be a hypergraph with edge-vertex incidence matrix $G: \F_2\CE \rightarrow \F_2\CV$. A cycle basis is a full-rank matrix $N: \F_2\CF \rightarrow \F_2\CE$ such that $\im N = \ker G$. $\CF$ is identified as a set of faces, or cycles, in $\CH$. 
    We say that the cycle basis is sparse if $N$ is a sparse matrix.
\end{definition}

Given a cycle basis of a measurement hypergraph, each cycle in the basis can be associated to a new check, which fixes the gauge of the logical which would otherwise have support on the cycle. If the cycle basis is sparse then this can be done while maintaining the LDPC property of the deformed code.

There are several methods of constructing such measurement hypergraphs \cite{cohen2022low,cowtan2024css,cowtan2024ssip,cross2024improved,zhang2024time,williamson2024low,ide2024fault,swaroop2024universal,cowtan2025parallel,he2025extractors}, which are described in more detail in Ref.~\cite[Sec.~3]{he2025extractors}. The most asymptotically efficient method currently known to measure a generic set of Pauli product operators on disjoint logical qubits is by brute-force branching and gauging logical measurements with universal adapters \cite{cowtan2025parallel}. For an arbitrary collection of logically disjoint Pauli product measurements supported on $t$ logical qubits, this scheme uses $\mathcal{O}\big(t \omega (\log t + \log^3\omega)\big)$ ancilla qubits, where $\omega \geq d$ is the maximum weight of the single logical Pauli representatives involved in the measurements, and $d$ is the code distance.
This is all done in time $\mathcal{O}(d)$ independent of $t$, while preserving the phenomenological fault distance and LDPC property.

This is a vast improvement to the original work on surgery with quantum LDPC codes~\cite{cohen2022low}, which required $\CO(d\omega)$ ancilla qubits to measure a single weight $\omega$ logical operator representative\footnote{The representative was also required to be irreducible, meaning that it did not contain any other logical representatives, a problem which was fully resolved in Refs~\cite{williamson2024low,ide2024fault}.} and could not generally perform parallel measurements without losing the LDPC property. However, despite these advances, all contemporary methods require $\CO(d)$ rounds of error correction per logical cycle for generic codes in order to maintain fault-tolerance\footnote{This assumes a single representative is measured; Ref.~\cite{williamson2024low} described a method to decrease the time overhead to $O(\frac{d}{m})$ via parallel surgery measurement of $2m-1$ equivalent logical operators.}. 
% This is in stark contrast to transversal gates \cite{dennis2002memory,zhou2024algorithmic} which can be performed in constant time,
% and homomorphic measurements \cite{xu2024fast}, which can in principle leverage transversal gates to achieve similar time overheads. 
It also means that the study of LDPC code surgery does not, prior to this work, cover the well-known cases of single-shot surgery with 3D and 4D topological codes \cite{hillmann2024single}.

The primary reason for this stark contrast is that, in LDPC code surgery, the measurement outcomes of new checks on vertices in the measurement hypergraph $\CH$ are used to infer logical measurement outcomes; and in prior works on LDPC code surgery these new checks did not have any protection against measurement errors. When measuring a single logical operator using gauging logical measurements or homological measurements \cite{williamson2024low,ide2024fault}, for example, the product of vertex check outcomes is interpreted as the logical measurement outcome. Hence, were they measured for only one round, just a single measurement error would flip the product, and constitute a logical error. As such, the checks must be measured for $\CO(d)$ rounds in order to infer the correct logical measurement outcome.

In this work, we focus on reducing the number of rounds required by introducing metachecks which protect against measurement errors in $\CH$. All of our protocols yield CSS-type surgery.

\begin{definition}\label{def:CSS-type}
    A CSS-type code surgery is a surgery protocol which begins with a CSS code memory, then performs code deformation such that the code remains CSS throughout.
\end{definition}
CSS-type code surgery is a limited subset of more general code surgery, as one cannot measure e.g. $\overline{Y}$ terms or products of the form $\overline{X}\otimes\overline{Z}$ while preserving the CSS property. Thus all product measurements of CSS-type each contain terms drawn from either $\{\overline{Z},\overline{I}\}$ or $\{\overline{X},\overline{I}\}$. We remark that CSS surgery is easily extended to measure more general logical Pauli operators on codes that are self-dual. 
% A proof of this elementary fact follows from e.g. Ref.~\cite[Prop.~4.11]{cowtan2024css}.

\begin{lemma}\label{lem:X_dist_preserved}
    Given any CSS code $Q$, measurement of $\overline{Z}$ logicals by a measurement hypergraph results in a deformed code with $X$-distance at least as high as the $X$-distance of $Q$, so long as all cycles are gauge-fixed or otherwise stabilisers.
\end{lemma}
This is an immediate corollary of Ref.~\cite[Thm.~1]{ide2024fault}. A dual result follows for the $Z$-distance when performing measurement of $\overline{X}$ logicals.

\section{Full block reading}\label{sec:block_reading}

To understand and reduce the time overhead of surgery operations, we first consider \textit{full block reading}, which is a surgery operation that acts transversally on all the physical qubits, and thereby logical qubits, of identical code blocks. 
% Block reading begins with a set of codeblocks and performs efficient logical measurements between them. 
The fact that one can perform block reading with surface codes is a widely known folklore result, as these are in a sense `transversal' measurements between patches. 
Our results extend substantially beyond this simple case.
% In \textit{full block reading}, every block is identical and each logical qubit in a block is involved in a logical measurement. Full block reading is analogous to Steane measurement, in a manner which we formalize in Section~\ref{sec:hom_equivalence}.
We start with the simple example of two blocks.

\subsection{Full block reading on two blocks}\label{sec:two_blocks}

Given two identical $\llbracket n,k,d \rrbracket$ quantum memories, which are CSS LDPC codeblocks labelled $Q$ and $Q'$,
\[\input{Figures-Tikz/two_blocks.tikz}\]
we can initialise a hypergraph $\CH(\CV,\CE)$ between the two
\begin{equation}
\input{Figures-Tikz/two_blocks_2.tikz}
\label{eq:two_blocks}\end{equation}
such that new data qubits are edges, new $Z$ checks are vertices and $\CH$ has the edge-vertex incidence matrix $G \in \F_2^{\CV\times\CE}$. $\CH$ has 1-to-1 maps from edges to $X$ checks in $Q$ and to $X$ checks in $Q'$, and similar for vertices. Thus in order for the checks of the deformed code to commute, we have $G = H_X^\intercal$. Let $D$ be this deformed code.

\begin{lemma}
    Measuring the checks of $D$ performs a $\overline{Z}\otimes\overline{Z}$ measurement on each pair of logical qubits $(\overline{q}_i,\overline{q}'_i)$ for $i \in [k]$, where $\overline{q}_i$ is a logical qubit in $Q$ and $\overline{q}'_i$ the same logical qubit in $Q'$.
\end{lemma}
\proof
Consider an arbitrary $\overline{Z}$ logical operator representative acting on $\overline{q}_i$, and call it $\overline{Z}_i = Z(v_i)$. Let $\overline{Z}_i'$ be the identical copy of $\overline{Z}_i$ in $Q'$.

By definition, $v_i \in \ker(H_X)$. Multiplying $\overline{Z}_i$ by the bijective $Z$ checks $P$ in $\CV$ does not leave any support on $\CE$, as $Pv_i\in \ker(G^\intercal)=\ker(H_X)$, but does clean $\overline{Z}_i$ into $Q'$. The operator which it is cleaned to is precisely $\overline{Z}_i'$, hence the two logicals are now stabiliser equivalent and so we have performed a measurement of $\overline{Z}_i\otimes\overline{Z}_i'$.

Now because $\ker(G^\intercal)=\ker(H_X)$, the only operators which can be cleaned are of this form and so we have performed a $\overline{Z}\otimes\overline{Z}$ measurement on each pair of logical qubits between the two codes.
\endproof

Specifically, each set of $Z$ checks in $\CV$ in bijection with a logical $\overline{Z}$ operator in $Q$ and $Q'$ measures the product of those operators. The logical measurement outcome is inferred as the product of the syndrome outcomes in that set.

\begin{lemma}\label{lem:two_blocks_k}
    $D$ has no additional logical qubits.
\end{lemma}
\proof
New logical qubits can only appear when performing measurement hypergraph surgery if there are cycles in the hypergraph \cite{williamson2024low,ide2024fault}, which do not have corresponding faces, i.e. checks, to gauge-fix. In general if $\ker(G) \neq 0$ then there can exist a new $\overline{X}$ logical operator with support on $u \in \ker(G)$.

This hypergraph may have cycles, but all logical operators on these cycles are also products of $X$ stabilisers from $Q$ (or, alternatively and symmetrically, from $Q'$). In particular, given any element $u \in \ker(G)$, multiply $X(u)$ by the bijective $X$ checks $S$ in $Q$. Because $u \in \ker(G)$, $Su \in \ker(H_X^\intercal)$ and so the cleaned operator maps to 0 on data qubits in $Q$, hence is a stabiliser.
\endproof

We shall presently prove that the code distance is preserved by the deformation, but in order to do so it is easier to use the language of cohomology.

\begin{lemma}\label{lem:prod_complex_two_blocks}
    Let $C^\bullet$ be the initial codeblock $Q$ viewed as a cochain complex. Then the deformed code  $D$ has the cochain complex
    \[D^\bullet = (C \otimes R)^\bullet\]
    where
    \[R^\bullet = \begin{tikzcd}R^0 \arrow[r, "\mathbb{P}"] & R^1\end{tikzcd},\]
    \[R^0 = \F_2^2,\qquad R^1 = \F_2,\qquad \mathbb{P} = \begin{pmatrix}1 & 1\end{pmatrix}.\]
\end{lemma}
\proof
\[(C \otimes R)^\bullet =\begin{tikzcd}C^0\otimes \F_2^2 \arrow[r] & C^1\otimes \F_2^2 \oplus C^0\otimes \F_2 \arrow[r] & C^2 \otimes \F_2^2 \oplus C^1\otimes \F_2 \arrow[r] & C^2 \otimes \F_2 \end{tikzcd}\]
In words,
\begin{itemize}
    \item at the level of $X$ checks we have two copies of $C^0$, the $X$ checks of $Q$,
    \item at the level of data qubits we have two copies of $C^1$ the data qubits of $Q$, and an additional set of data qubits in bijection with the $X$ checks of $Q$,
    \item at the level of $Z$ checks we have two copies of $C^2$ and an additional set of $Z$ checks in bijection with the data qubits of $Q$,
    \item there is another term, which we ignore for now and consider only the first 3 terms.
\end{itemize}
We now check that the coboundary maps are correct.
\[\delta^0_{C\otimes R} = \begin{pmatrix}
        H_X^\intercal \otimes I\\
        I\otimes \mathbb{P}
    \end{pmatrix},\qquad \delta^1_{C\otimes R} = \begin{pmatrix}
        H_Z \otimes I & 0\\
        I \otimes \mathbb{P} & H_X^\intercal\otimes I
    \end{pmatrix}\]
which we recognise as the check matrices, with appropriate transposes, from Eq.~\ref{eq:two_blocks}.
\endproof

Using the perspective of cochains it is easy to verify the logical dimension using the K{\"u}nneth formula. Moreover, by Lemma~\ref{lem:hom_factorB} we have the equation:
\[H^1(C\otimes R) = H^1(C)\otimes H^0(R) \oplus H^0(C)\otimes H^1(R)\]
hence the set of $\overline{X}$ logical operators in $D$ is given by representatives of
\[H^1(C)\otimes H^0(R) \oplus H^0(C)\otimes H^1(R)\]
which explicitly is
\[\ker H_Z/\im H_X^\intercal \otimes \ker\mathbb{P} \oplus C^2/\im H_Z\otimes 0 = \ker H_Z/\im H_X^\intercal \otimes \ker\mathbb{P}\]
so every $\overline{X}$ logical operator belongs to a stabiliser equivalence class $[u_i,u_i']$ of the identical logical operators $u_i$, $u_i'$ in $Q$ and $Q'$, or a nontrivial sum thereof.

\begin{lemma}\label{lem:two_block_distance}
    The distance of $D$ is at least $d$.
\end{lemma}
\proof
As there are no new logical qubits, to prove the distance bound it is sufficient to ensure that cleaning $\overline{Z}$ logicals cannot reduce their weight below $d$. The $X$-distance is guaranteed by Lemma~\ref{lem:X_dist_preserved}.

For $\overline{Z}$ logicals, observe that multiplying any $\overline{Z}$ logical in $Q$ by $Z$ checks in $\CV$ adds support to the corresponding data qubits in $Q'$ by a 1-to-1 map, and potentially adds support on $\CE$ as well. Any $\overline{Z}$ logical which has support on both $Q$ and $Q'$, say $v$ and $v'$, can in the worst case be cleaned to have weight $|v+v'|$ on $Q$ and $Q'$, considering both $v$ and $v'$ to be in $\F_2^n$, and $v+v'$ is a valid logical of $Q$ so must have weight at least $d$.
\endproof

As a consequence of Lemmas~\ref{lem:two_blocks_k} and~\ref{lem:two_block_distance} it is guaranteed by the arguments in Ref.~\cite{williamson2024low} that the block reading can be performed with fault distance $d$ in $d$ rounds of error correction, as the procedure can be viewed as a parallel gauging logical measurement. All new data qubits are initialised in $\ket{+}$, and all checks are measured for $d$ rounds, before measuring out the new data qubits in the $X$ basis. We reduce this from $d$ to $\CO(1)$ in Section~\ref{sec:time_overhead}.

\subsection{Full block reading of $Z$ type}\label{sec:fbr_Z_type}
We can extend block reading straightforwardly to acting on many codeblocks simultaneously.
We start with a set of $c$ identical $\llbracket n, k, d \rrbracket$ LDPC codeblocks.

\begin{definition}[Pattern Matrix]\label{def:pattern_matrix}
    Let $\mathbb{P}:\F_2^c\rightarrow \F_2^\phi$ be a full-rank matrix, which defines the pattern of logical measurements to be performed between codeblocks by deforming to a new code $D$. We call $\mathbb{P}$ the pattern matrix.

    Each row $r$ of $\mathbb{P}$ corresponds to a set of parallel measurements to be performed. Each entry of 1 in $r$ denotes that a codeblock is involved in that set of measurements, so if $r^j = 1$ for column $j$ then codeblock $j$ is involved, and if $r^j=0$ then it is not. Then $r$ specifies a measurement of $P_i^1\otimes P_i^2\otimes P_i^3\cdots$ for every logical qubit $i \in [k]$ in codeblocks $Q^1$, $Q^2$, $Q^3$... in the basis $P^j \in \{\overline{Z},\overline{I}\}$ depending on whether $r^j = 1$ or $0$.
\end{definition}

If $\mathbb{P}$ is not full rank, and has some linearly dependent rows, then those rows correspond to logical measurements which are already performed by other rows and so are redundant. Additionally, if any row has weight 1, with the 1 in column $j$, that measurement is equivalent to measuring out codeblock $j$ in the $Z$ basis. Hence for simplicity we assume that $\mathbb{P}$ is full rank, with no linearly dependent rows, and that there are no rows with weight 1.

\begin{definition}[$Z$-type full block reading]\label{def:Z_full_block}
    A full block reading of $Z$ type takes as input $c$ CSS LDPC codes, each with cochain complex $C^\bullet$, and a pattern matrix $\mathbb{P} : \F_2^c \rightarrow \F_2^\phi$.
    
    The cochain complex for the deformed code $D$ of the full block reading denoted by $\mathbb{P}$ is given by $(C\otimes R)^\bullet$ where
    \[R^\bullet = \begin{tikzcd}\F_2^c \arrow[r, "\mathbb{P}"] & \F_2^\phi\end{tikzcd},\]
\end{definition}
First we check that the definition above is consistent with the definition of a pattern matrix. We have
\[(C\otimes R)^\bullet = \begin{tikzcd}C^0\otimes \F_2^c \arrow[r] & C^1\otimes \F_2^c \oplus C^0 \otimes \F_2^\phi \arrow[r] & C^2 \otimes \F_2^c \oplus C^1 \otimes \F_2^\phi \arrow[r] & C^2 \otimes \F_2^\phi\end{tikzcd}\]
where we currently ignore the last $C^2\otimes \F_2^\phi$ term. We can pick out
\[C^0\otimes \F_2^c\rightarrow C^1\otimes \F_2^c \rightarrow C^2\otimes \F_2^c\]
as the $c$ original codeblocks. Then,
\[\delta^0_{C\otimes R} = \begin{pmatrix}
        H_X^\intercal \otimes I\\
        I\otimes \mathbb{P}
    \end{pmatrix},\qquad \delta^1_{C\otimes R} = \begin{pmatrix}
        H_Z \otimes I & 0\\
        I \otimes \mathbb{P} & H_X^\intercal\otimes I
    \end{pmatrix}\]
as for the example in Eq.~\ref{eq:two_blocks} but where $\mathbb{P}$ is now a more general matrix. The $\overline{Z}$ logical operators in $D$ are then
\[H_1(C\otimes R) = H_1(C)\otimes H_0(R)\oplus H_0(C)\otimes H_1(R) = \ker H_X/\im H_Z^\intercal \otimes \F_2^c/\im \mathbb{P}^\intercal \oplus C^0/\im H_X \otimes \ker\mathbb{P}^\intercal\]
where $\ker\mathbb{P}^\intercal = 0$ as $\mathbb{P}$ has no surplus rows, so
\[H_1(C\otimes R) = \ker H_X/\im H_Z^\intercal \otimes \F_2^c/\im \mathbb{P}^\intercal .\]
Given that our initial code had $\overline{Z}$ logical operators given by $\ker H_X/\im H_Z^\intercal \otimes \F_2^c$, we see that those in $\im \mathbb{P}^\intercal$ are now stabilisers, and so block reading has measured precisely the rows of $\mathbb{P}$ on each logical qubit. Furthermore, no new equivalence classes are present, so there are no new logical qubits.

\begin{remark}
    Let $\mathbb{P}$ be LDPC and each identical codeblock be LDPC. Then the deformed code $(C \otimes R)^\bullet$ during measurement is LDPC, and the procedure uses $\CO(cn)$ ancilla qubits.
\end{remark}
This is immediate from Definition~\ref{def:Z_full_block}.

\begin{lemma}\label{lem:fullblock_distance}
    The deformed code has code distance at least $d$.
\end{lemma}
\proof
The $X$-distance follows from Lemma~\ref{lem:X_dist_preserved}.

The $\overline{Z}$ logicals are given by $H_1(C\otimes R) = \ker H_X/\im H_Z^\intercal \otimes \F_2^c/\im \mathbb{P}^\intercal$. We start with a $\overline{Z}$ logical $\overline{\Lambda}_Z$ composed of different $Z$ logicals $v, v',v'',...$ in each codeblock, at least one of which must have weight at least $d$ and be a nontrivial $\overline{Z}$ logical. Applying a new $Z$ check in $\CV$ only cleans support from the codeblocks if the incident data qubits all have support in $\overline{\Lambda}_Z$; otherwise the check applies at least 1 new $Z$ to a data qubit in a codeblock, and the weight of $\overline{\Lambda}_Z$ cannot be reduced below $d$ in this way.

If we apply new $Z$ checks in $\CV$ to clean support from $c_{\rm cl}$ codeblocks, in the worst case we have support $|v| + |v'| + |v''| + \cdots - c_{\rm cl}|v \cap v' \cap v''\cap \cdots|$ left in the codeblocks. But 
\begin{align*}
    |v| + |v'| + |v''| + \cdots - c_{\rm cl}|v \cap v' \cap v''\cap \cdots| &= |v \cup v' \cup v'' \cup \cdots| - |v\cap v' \cap v'' \cap \cdots| \\
    &\geq |v\Delta v'\Delta v''\Delta \cdots|\\
   & = |v + v' + v'' + \cdots| \geq d
\end{align*}
where $\Delta$ is the symmetric difference of sets.
\endproof

Should the row weights of $\mathbb{P}$ be too high, meaning that the weights of these logical measurements are high and so the degree of the Tanner graph is high, then one can introduce ancilla blocks to break those measurements up, making $\mathbb{P}$ sparse. This corresponds to adding new bits, i.e., columns to the check matrix $\mathbb{P}$, such that a row can be split up into multiple rows.
The ancilla codeblocks used to reduce its weight can be initialised at the same time as the code deformation procedure begins, and can be considered part of the measurement hypergraph, where data qubits in ancilla blocks are edges, $Z$ checks are vertices and $X$ checks are cycles, as normal. Consequently, the data qubits are initialised in $\ket{+}$. This can be thought of as thickening the hypergraph, see Figure~\ref{fig:thickened_partial_reading_metachecks} later on.

\subsection{Time overhead of full block reading}\label{sec:time_overhead}

We now return to our original example, Eq.~\ref{eq:two_blocks} from Section~\ref{sec:two_blocks}. When computing the cochain complex in Lemma~\ref{lem:prod_complex_two_blocks} we had an additional term $(C\otimes R)^3=C^2\otimes \F_2 = C^2$, which corresponds to a collection of metachecks on the $Z$-checks of the deformed code. Thus the full scalable Tanner graph for the logical measurement is
\[\input{Figures-Tikz/two_blocks_3.tikz}\]
where $\mathfrak{G} = H_Z$ and
$\mathfrak{M}_Z = \delta^2_{C\otimes R} = \begin{pmatrix}
    I \otimes \mathbb{P} & H_Z\otimes I
\end{pmatrix} = \begin{pmatrix}
    I \otimes \mathbb{P} & H_Z 
\end{pmatrix}$.

Consider first the case where measurement faults can occur only on checks in $\CV$. As $\mathfrak{G} = H_Z$, if a measurement fault $v \notin \ker(H_Z)$ then it is detected by the metachecks. If $v\in \ker(H_Z)$ then either (a) it is equivalent to an $X$ data qubit error on $\CE$ or (b) it has weight on checks at least $d$. (b) follows from the fact that $G = H_X^\intercal$, and the lowest weight vector in $\ker H_Z\backslash H_X^\intercal$ has by definition weight at least $d$. For (a), as all the data qubits in $\CE$ are initialised in the $\ket{+}$ basis, $X$ errors leave the state invariant so the logical measurement outcome is unaffected. Therefore to cause an incorrect logical measurement outcome, the measurement fault weight on $\CV$ must be at least $d$, in the case where measurement faults cannot happen elsewhere. Note further that any vector $v \in \ker(\mathfrak{G})$ representing an undetectable measurement fault on $\CE$, regardless of weight, is equivalent to an undetectable $X$ data qubit fault on $Q$ (or $Q'$), as multiplying by $X$ faults on bijective qubits leaves $0$ on the $Z$ checks in $Q$ (or $Q'$).

To study the more general case it is fruitful to use the cohomology of $(C\otimes R)^\bullet$.
\[H^2(C\otimes R) = H^2(C)\otimes H^0(R)\oplus H^1(C)\otimes H^1(R),\]
which is explicitly
\[C^2/\im H_Z\otimes \ker\mathbb{P}\oplus \ker H_Z/\im H_X^\intercal \otimes 0 = C^2/\im H_Z\otimes \ker\mathbb{P}.\]
Thus any $Z$ measurement fault which is neither detected by metachecks nor equivalent to a data qubit fault must be a linear combination of pairs of identical $Z$ check faults in $Q$ and $Q'$, which are \textit{not} equivalent to $X$ data qubit faults, and $Z$ check faults which \textit{are} equivalent to $X$ data qubit faults. Note that the $Z$ single shot distance $d^{ss}_Z$ of $(C\otimes R)^\bullet$ can be very low even in the more general case, as $C^2/\im H_Z$ can have low weight representatives. Nevertheless, any measurement fault which includes $Z$ checks in $\CV$ must be able to be cleaned, by applying some subset of $X$ errors to $Q$ and $Q'$, so that it has support only on the $Z$ checks of $Q$ and $Q'$. Only if the set of $X$ errors which it is cleaned to is large does the measurement fault cause a logical measurement fault.

In our analysis of the time overhead, we have first a weak result:
\begin{restatable}{proposition}{FDfullblock}\label{prop:fault_distance_fullblock}
    Let $Q$, $Q'$, $Q'',$... be a set of identical CSS LDPC codeblocks with distance $d$, and let $\mathbb{P}$ be a pattern matrix determining a full block reading.
    
    Assume that there are $d$ rounds of syndrome measurements on $Q$, $Q'$, $Q''$... before and after the full block reading procedure. Then full block reading performs logical measurement in $\CO(1)$ rounds of syndrome measurements with fault-distance $d$.
\end{restatable}
\proof
See Appendix~\ref{app:proof_fault_distance}.
\endproof

The intuition is that in the timestep where the full block reading is performed, any measurement fault on the check qubits in $\CV$ with weight $< d$ which is undetectable by metachecks is either (a) in $\im G$ and so does not affect the logical measurement outcome or (b) has a paired measurement fault on $Z$ checks outside of $\CV$, in the original code, such that the measurement faults on the original code must extend to future and previous timesteps to avoid detection.

In order to maintain the fault-distance of the entire procedure we must assume $d$ rounds of syndrome measurements on the original code before and after block reading. This is because there may be low weight faults on data qubits, $Z$ checks and $X$ checks outside of $\CH$ which are not detected within $1$ timestep, and so could lead to logical errors by connecting to timelike errors extending from the top or bottom boundaries. This is equivalent to $d$ measurement rounds of `padding' before and after block reading \cite{beverland2024fault}. 

\begin{remark}[State preparation]\label{rem:state_prep}
Proposition~\ref{prop:fault_distance_fullblock} does not allow one to generically use full block reading on a single codeblock to prepare a logical $\ket{\overline{+}}^k$ or $\ket{\overline{0}}^k$ state in constant time, because of the padding required to provably preserve fault-distance.
\end{remark}

\begin{remark}[Frame change]
    At the end of the block reading protocol, the ancilla data qubits are measured out in the $X$-basis, and the $X$ checks in the original code are un-deformed back to their original incidence. These measurements do not commute with the preceding $Z$ checks, and so have random outcomes, although products of these measurement outcomes which correspond to cycles in the hypergraph must be +1 in the absence of errors, as these are precisely the sets of $X$ measurements which do commute with the preceding $Z$ checks. Because each new data qubit is in 1-to-1 correspondence with an $X$ check in each incident codeblock, the anticipated outcomes of $X$-checks in the original codeblocks are updated after the block reading protocol by taking a product with the outcomes of those data qubit $X$ measurements, inducing a frame change.

    When the $X$ measurements on ancilla data qubits undergo faults, the frame is updated incorrectly. These measurement errors are therefore in bijection with $X$ check errors on the original code, detected by preceding and subsequent rounds of measurement.
    
    The reason why preceding rounds of measurement can diagnose such frame errors is because in the measurement round where ancilla data qubits are measured out the product of such a measurement and its corresponding $X$-check in an original codeblock commutes with the $Z$ checks in the deformed code, and so must yield a +1 outcome in the absence of errors. Thus having a consistent history of checks before the frame change can detect frame errors.
\end{remark}

Proposition~\ref{prop:fault_distance_fullblock} is at first not particularly helpful, because the overall time cost of the procedure is still $\CO(d)$. However, the proof extends to performing full block readings sequentially \textit{without padding} between logical measurement rounds, with each logical measurement taking $\CO(1)$ rounds.
\begin{restatable}{theorem}{algoft}\label{thm:full_block_amortised}
    Let $\Xi$ be a set of $\eta$ full block reading procedures  of $Z$-type such that no block reading procedure in $\Xi$ is a product of any others.
    
    Then all $\eta$ logical measurement rounds of full block reading can be performed sequentially using $\CO(1)$ rounds of syndrome measurement each, and using $d$ rounds of padding before and after the full set of procedures. The procedure has phenomenological fault-distance $d$, for any $\eta \in \N$.
\end{restatable}
\proof
See Appendix~\ref{app:proof_fault_distance}.
\endproof

\begin{remark}
    In the above statement, the assumption that the pattern matrix is full rank is for technical reasons related to our proof technique. We expect that this requirement can be relaxed. 
\end{remark}

\subsection{Full block reading of CSS-type}

Full block reading can be generalised to measure sets of commuting operators of the form $\overline{Z}\otimes\overline{Z}\otimes\cdots\otimes\overline{Z}$ and $\overline{X}\otimes\overline{X}\otimes\cdots\otimes\overline{X}$ simultaneously. 
%This allows for fast code switching from layers of an initial code to layers of a new code, along with logical measurements corresponding to the stabilisers of the initial code. 
This allows for fast code switching into a transversally concatenated code, similar to a procedure in Ref.~\cite{xu2025batched} and the basic operation in the protocol of Ref.~\cite{Litinski2025Blocklet}. 

To see this, let $\mathbb{P}_Z$ be the pattern matrix for $Z$-type logical measurements. Recall that the rows of $\mathbb{P}_Z$ each specify a set of parallel $Z$-type measurements as per Definition~\ref{def:pattern_matrix}. In order for us to perform $X$-type block reading at the same time, the $X$-type logical operators to measure must commute with the $Z$-type logical operators. Thus each row $s$ of a $\mathbb{P}_X$ pattern matrix, specifying a set of parallel $X$-type logical measurements, must satisfy $r\cdot s$ for each row $r$ of $\mathbb{P}_Z$, where ${}\cdot{}$ is the usual dot product on vectors over $\F_2$. Therefore, $\mathbb{P}_Z\mathbb{P}_X^\intercal = 0$.

Our pattern matrix is now upgraded to a \textit{pattern complex}.

\begin{definition}[CSS-type full block reading]
    A full block reading of CSS-type takes as input $c$ CSS LDPC codes, each with cochain complex $C^\bullet$, and a pattern cochain complex $R^\bullet$.
    
    The cochain complex for the deformed code of the full block reading denoted by $\mathbb{P}$ is given by $(C\otimes R)^\bullet$ where
    \[R^\bullet = \begin{tikzcd}\F_2^\theta \arrow[r,"\mathbb{P}_X^\intercal"] &  \F_2^c \arrow[r, "\mathbb{P}_Z"] & \F_2^\phi\end{tikzcd}\]
    with $R^{-1}=\F_2^\theta$, $R^0 = \F_2^c$ and $R^1 = \F_2^\phi$, and we assume that both $\mathbb{P}_Z$ and $\mathbb{P}_X$ are full rank with no redundant rows, and that all rows have weight greater than 1.
\end{definition}

As the tensor product of cochain complexes is also a cochain complex, the checks in the deformed code commute. We have
\begin{align*}&(C\otimes R)^{-1} = 
C^0\otimes R^{-1} \\ 
&(C\otimes R)^{0} = C^1\otimes R^{-1}\oplus C^0 \otimes R^0 \\
&(C\otimes R)^{1} = C^2\otimes R^{-1}\oplus C^1\otimes R^0 \oplus C^0\otimes R^1 \\
&(C\otimes R)^{2} = C^2\otimes R^{0} \oplus C^1\otimes R^{1}\\
&(C\otimes R)^{3} = C^2\otimes R^1\end{align*}
and
\[H^1(C\otimes R) = C^2/\im H_Z\otimes \ker\mathbb{P}_X^\intercal \oplus \ker H_Z/\im H_X^\intercal\otimes \ker\mathbb{P}_Z/\im\mathbb{P}_X^\intercal \oplus C^0 \otimes R^1/\im \mathbb{P}_Z  \]
which given that $\ker \mathbb{P}_X^\intercal = 0$ and $R^1/\im \mathbb{P}_Z = 0$ reduces to 
\[H^1(C\otimes R) = \ker H_Z/\im H_X^\intercal\otimes \ker\mathbb{P}_Z/\im\mathbb{P}_X^\intercal.\]
A similar calculation gives
\[H_1(C\otimes R) = \ker H_X/\im H_Z^\intercal\otimes \ker\mathbb{P}_X/\im\mathbb{P}_Z^\intercal,\]
so we see that logical $\overline{Z}$ operators in $\im \mathbb{P}_Z^\intercal$, i.e. the row space of $\mathbb{P}_Z$, are now stabilisers, and the same for logical $\overline{X}$ operators in the row space of $\mathbb{P}_X$, for every logical qubit in the codeblocks.

That the distance of the CSS-type block reading deformed code is at least $d$ is immediate by applying the arguments of Lemma~\ref{lem:fullblock_distance} twice, first for the $Z$-type auxiliary system and then for the $X$-type.

In the CSS-type full block reading measurement protocol, all new data qubits included in the $Z$-type ($X$-type) measurements are initialised in $\ket{+}$ ($\ket{0}$) states respectively. All syndromes are measured for $\CO(1)$ rounds, and then all new data qubits are measured out in the $X$ ($Z$) bases respectively.

\begin{remark}
    Let $R^\bullet$ be LDPC and each identical codeblock $Q, Q', Q"...$ be LDPC. Then the deformed code $(C\otimes R)^\bullet$ is LDPC.
\end{remark}

\begin{restatable}{lemma}{CSSfbreading}\label{lem:CSS_fullblock_reading}
Let $\Xi$ be a set of $\eta$ full block reading procedures of CSS-type such that no block reading procedure in $\Xi$ is a product of any others, and each of the measurements commute.
    
Then all $\eta$ logical measurement rounds of full block reading can be performed sequentially using $\CO(1)$ rounds of syndrome measurement each, and using $d$ rounds of padding before and after the full set of procedures. The procedure has phenomenological fault-distance $d$, for any $\eta \in \N$.
\end{restatable}
\proof
See Appendix~\ref{app:proof_fault_distance}.
\endproof

In Appendix~\ref{app:nonCSS_block} we also show that full block reading can be extended to perform non-CSS measurements, namely those with $\overline{X}\otimes\overline{Z}$ and $\overline{Y}$ terms, on CSS codes, but that our constructions for these require specific structure on the codes, such as transversal $S$ gates, so they add no extra computational power to full block reading with CSS-type measurements.

Full block reading gives no addressability on logical qubits within a codeblock, as the measurements are on all logical qubits, so the utility of full block reading is limited in the same manner as 1-to-1 transversal CNOT gates between CSS codes. The logical computational power of a full block reading is identical to initialising an ancilla block, then applying transversal CNOTs, then measuring out the ancilla block.

Consequently, it is unclear how useful full block reading would be for practical fault-tolerant computation. Compared to measurement using transversal gates and ancilla codeblocks, full block reading requires fewer qubits overall, but full block reading also requires more connections between the initial codeblocks and the auxiliary system.

Nevertheless, our analysis of full block reading sets the stage for the more sophisticated operations in the next sections.
We first consider the case of reading a subcode, then present our general result on fast hypergraph surgery.

\section{Partial block reading}\label{sec:partial_block_reading}

Returning again to the example in Eq.~\ref{eq:two_blocks}, consider a different generalisation. We start with two codeblocks $Q$ and $Q'$, which may be different,
\[\input{Figures-Tikz/two_blocks_partial1.tikz}\]
and construct the measurement hypergraph $\CH$ to connect only to a \textit{subcode} of each, but where the subcodes are identical.
\[\input{Figures-Tikz/two_blocks_partial2.tikz}\]
That is, up to permutation and the addition of all-zero rows, the matrices satisfy
\[M_l = M_r; \qquad F_l = F_r.\]
Given an appropriate choice of subcode and measurement hypergraph, the measurements are only between those logical qubits whose logical representatives have connections to the hypergraph, and so are contained in the subcode. This is called a \textit{partial block reading}.

The construction no longer yields a tensor product of cochain complexes, so we instead use \textit{mapping cones} from Section~\ref{sec:mapping_cones} to describe partial block reading. As the constructions become quite complicated, we also refrain from studying multiple partial block readings occurring simultaneously, unlike in the full block reading case. All partial block readings are assumed to be performed sequentially, albeit with only a constant number of syndrome rounds between them.

\begin{definition}\label{def:partial_block_reading}
    Let $A$ be a subcode of $Q$, $Q'$, $Q''...$, each of which have chain complexes $C_\bullet, C_\bullet', C_\bullet''...$, such that there is a collection of injective chain maps $f_\bullet$, $f_\bullet'$, $f_\bullet''$... of the forms \[f_\bullet: A_\bullet \rightarrow C_\bullet; \quad f_\bullet': A_\bullet\rightarrow C_\bullet'; \quad f_\bullet'': A_\bullet \rightarrow C_\bullet''...\]

    Then the partial block reading of $Z$ type defined by $\{f_\bullet, f_\bullet',f_\bullet''...\}$ is
    \[D_\bullet = \cone(h_\bullet); \quad h_\bullet = f_\bullet + f_\bullet'+f_\bullet''+...\]
\end{definition}

Explicitly, $D_\bullet$ is the chain complex
\[\begin{tikzcd}
    A_1 & A_0 & 0 \\
    C_2 & C_1 & C_0\\
    C_2' & C_1' & C_0'\\
    C_2'' & C_1'' & C_0''\\
    \vdots & \vdots & \vdots\\
    \arrow[from=1-1, to=2-2]
    \arrow[from=1-2, to=2-3]
    \arrow[from=1-1, to=3-2]
    \arrow[from=1-2, to=3-3]
    \arrow[from=1-1, to=4-2]
    \arrow[from=1-2, to=4-3]
    \arrow["{\partial^{A}_{1}}", from=1-1, to=1-2]
    \arrow["0", from=1-2, to=1-3]
    \arrow[from=2-1, to=2-2]
    \arrow[from=2-2, to=2-3]
    \arrow[from=3-1, to=3-2]
    \arrow[from=3-2, to=3-3]
    \arrow[from=4-1, to=4-2]
    \arrow[from=4-2, to=4-3]
    \arrow["\oplus", phantom, from=1-1, to=2-1]
    \arrow["\oplus", phantom, from=1-2, to=2-2]
    \arrow["\oplus", phantom, from=2-1, to=3-1]
    \arrow["\oplus", phantom, from=2-2, to=3-2]
    \arrow["\oplus", phantom, from=3-1, to=4-1]
    \arrow["\oplus", phantom, from=3-2, to=4-2]
    \arrow["\oplus", phantom, from=1-3, to=2-3]
    \arrow["\oplus", phantom, from=2-3, to=3-3]
    \arrow["\oplus", phantom, from=3-3, to=4-3]
    \arrow["\oplus", phantom, from=4-3, to=5-3]
    \arrow["\oplus", phantom, from=4-1, to=5-1]
    \arrow["\oplus", phantom, from=4-2, to=5-2]
\end{tikzcd}
\]
where labels for the differentials in each codeblock, and labels for the chain maps, have been suppressed for visibility. Note that also $A_\bullet$ generally possesses a nonzero $A_2$ component. This component corresponds to metachecks, so we ignore that term for now.

It is trivial to compute, using the Snake Lemma~\cite[Lem.~1.3.2]{Weib1994}, that the measured logical qubits are those in each code which have $\overline{Z}$ logical representatives contained in the subcode $A_\bullet$.

As a Tanner graph, we have
\begin{equation}
    \input{Figures-Tikz/three_blocks_partial.tikz}
    \label{eq:three_blocks_partial}
\end{equation}
for a partial block reading between three codeblocks. Observe that the chain maps $f_\bullet, f_\bullet', f_\bullet''$ determine the connectivity between the blocks and the auxiliary hypergraph, and that $\CE = A_0$, $\CV = A_1$.

Unlike when performing full block reading, the code distance of the deformed code $D_\bullet$ in partial block reading is not generically preserved, and can drop below $d$. Elementary examples of this phenomenon include surface codes with defects and toric codes, which we detail in Section~\ref{sec:examples}.

This drop in distance can come from two causes. First, it is possible to multiply existing $\overline{Z}$ logicals in the codeblocks by new $Z$ checks in order to lower the weights below $d$. Second, viewing the ancilla system defined by the mapping cone as a hypergraph $\CH$, there may be cycles in $\CH$ which are nontrivial $\overline{X}$ logicals, meaning that they do not correspond to stabilisers of the deformed code. When unmeasured, the cycles have symplectically paired $\overline{Z}$ logicals, which when multiplied by other logicals in the original code can result in weights lower than $d$ \cite{cohen2022low}.

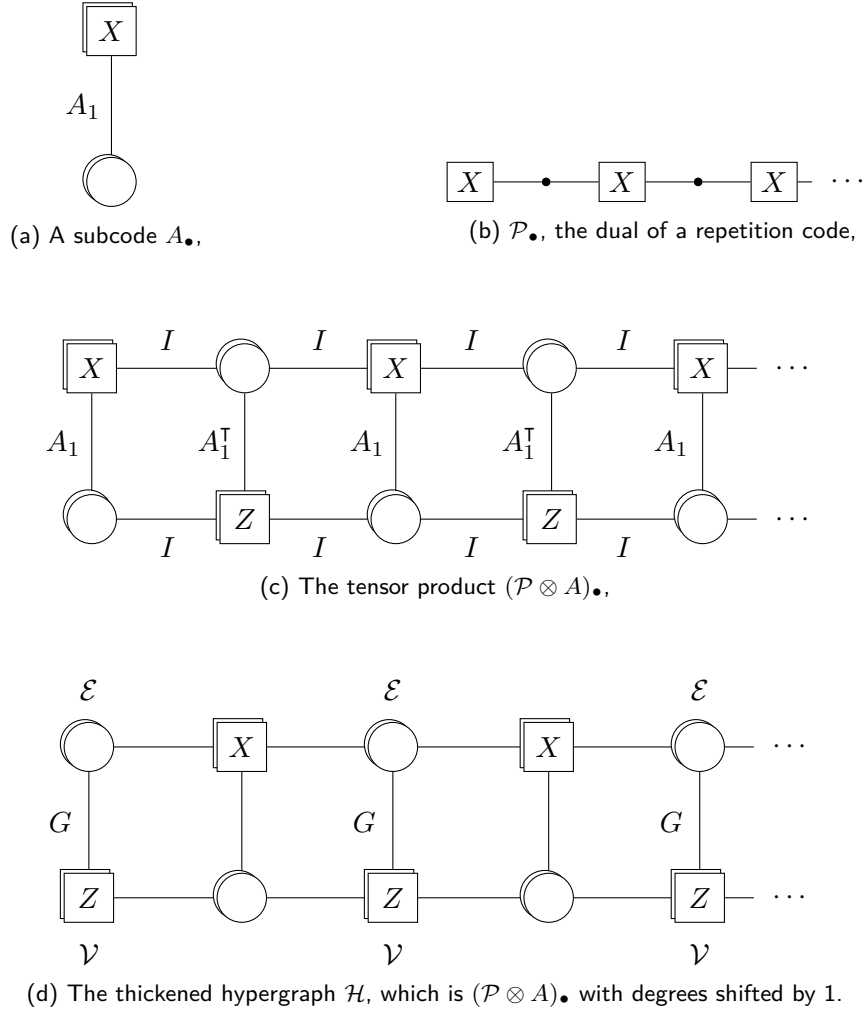
\begin{figure}[tb]
    \hfill
     \begin{subfigure}[t]
     {0.36\textwidth}
         \centering
         \input{Figures-Tikz/hypergraph_diag.tikz}
         \caption{A subcode $A_\bullet$,}
     \end{subfigure}
     \begin{subfigure}[t]{0.6\textwidth}
     \centering
         \input{Figures-Tikz/dual_path_graph.tikz}
             \caption{$\CP_\bullet$, the dual of a repetition code,}
     \end{subfigure}

     \vspace{10mm}
     \begin{subfigure}[t]{\textwidth}
     \centering
     \input{Figures-Tikz/tensored_subcode.tikz}
     \caption{The tensor product $(\CP\otimes A)_\bullet$,}
     \end{subfigure}
     
     \vspace{10mm}
     \begin{subfigure}[t]{\textwidth}
     \centering
     \input{Figures-Tikz/thickened_hypergraph.tikz}
     \caption{The thickened hypergraph $\CH$, which is $(\CP\otimes A)_\bullet$ with degrees shifted by 1.}
     \end{subfigure}
     
     \caption{Constructing a thickened hypergraph. Note that (b) is a conventional Tanner graph, not a scalable Tanner graph.}
         \label{fig:thickened_hypergraph}
\end{figure}

\begin{figure}[tb]
     \centering
     \input{Figures-Tikz/thickened_partial_reading.tikz}
     \caption{Example of partial block reading with a thickened hypergraph. The logical measurement is the same as in Eq.~\ref{eq:three_blocks_partial}, but the hypergraph has been thickened to prevent the weight of $\overline{Z}$ operators dropping below $d$.}
     \label{fig:thickened_partial_reading}
\end{figure}
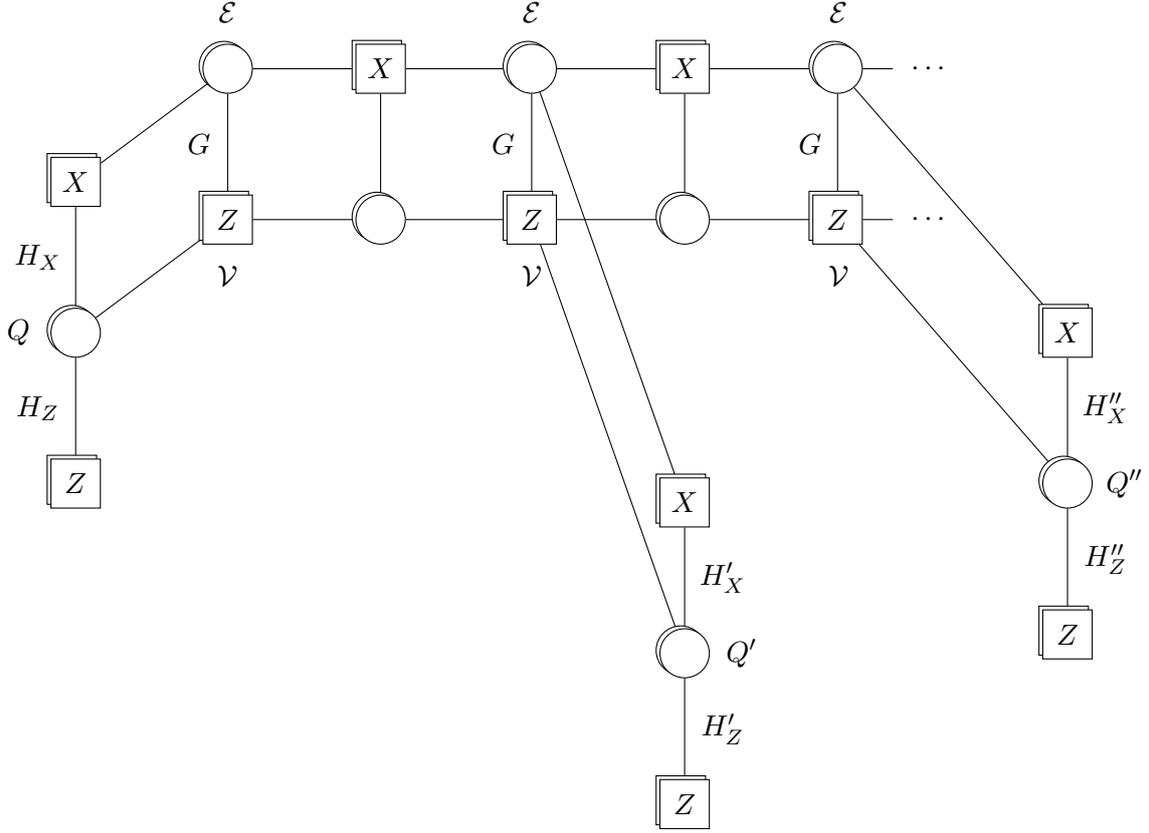

In this section, we focus on partial block reading where the subcode has distance $d$, and the measurements are assumed to be in the $Z$-basis. Dualising appropriately yields the same results for the $X$-basis.
For the first cause of the deformed code distance dropping, we can thicken the hypergraph. This is equivalent to taking a tensor product of $A_\bullet$ and the dual of a repetition code $\CP_\bullet$ before defining the mapping cone, as shown in Figure~\ref{fig:thickened_hypergraph}. We can define chain maps from any copy of $A_\bullet$ in the tensor product to $Q$, and the same for $Q'$, $Q''...$, and this choice does not need to be the same for each. An example is shown in Figure~\ref{fig:thickened_partial_reading}

\begin{lemma}\label{lem:deformed_dist_d}
    Let $(\CP\otimes A)_\bullet$ be the thickened subcode, and let $f_\bullet, f_\bullet',f_\bullet''...$ be chain maps from any copies of $A_\bullet$ in $(\CP\otimes A)_\bullet$ into $Q, Q', Q''...$, with $h_\bullet = f_\bullet + f_\bullet' + f_\bullet''\cdots$.
    
    Then if $(\CP\otimes A)_\bullet$ has been thickened at least $d$ times the deformed code $D_\bullet$ has distance at least $d$.
\end{lemma}
\proof
Immediate from Corollary~\ref{cor:thickening}, Lemma~\ref{lem:multiblock_mod_exp_dist} and Remark~\ref{rem:subsystem_CKBB} further down, so we defer the proof until then.
\endproof

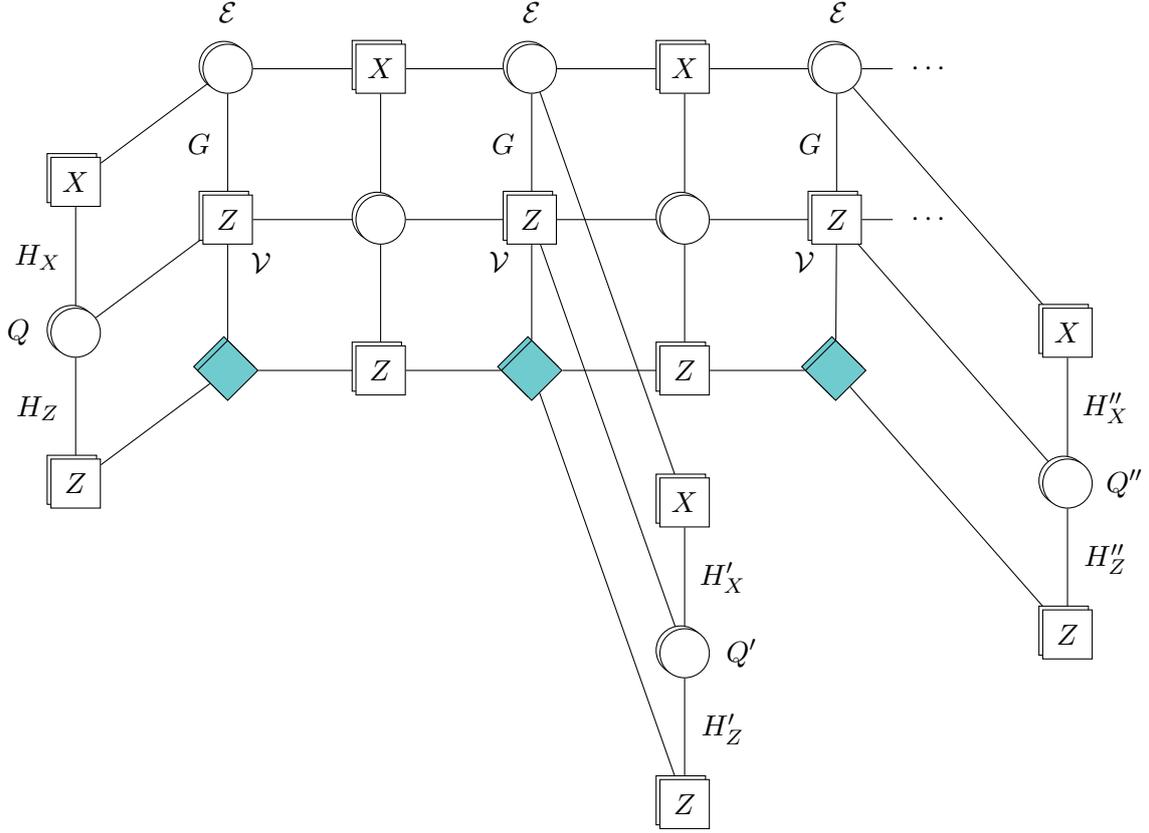
\begin{figure}[tb]
     \centering
     \input{Figures-Tikz/thickened_partial_reading_metachecks.tikz}
     \caption{The same partial block reading as in Figure~\ref{fig:thickened_partial_reading} but with additional redundant checks and metachecks.}
    \label{fig:thickened_partial_reading_metachecks}
\end{figure}

This is similar to devised sticking \cite{zhang2024time} and the original CKBB scheme \cite{cohen2022low} in that the auxiliary system is thickened $d$ times, but in this case we are measuring many representatives of the same equivalence classes of operators simultaneously. 
As in the CKBB scheme, thickening $d$ times is generally overkill \cite{cowtan2024ssip}, and distance preservation commonly requires a smaller ancilla system than this upper bound. Indeed, in Section~\ref{sec:block_ltcs} we demonstrate that soundness properties of the subcode can be used to reduce the required thickening while provably preserving the distance, and even that gives a sufficient but not generally necessary amount of thickening. If thickening $d$ times were truly necessary then our time overhead of $d$ would essentially be replaced by a space overhead factor of $d$, at least when $A_\bullet$ is large, which is interesting but less useful than achieving a genuine reduction in spacetime overhead.

For the second cause of the deformed code distance dropping, that there can be cycles in $\CH$ which correspond to $\overline{X}$ logicals, we use the following fact. As the subcode has distance $d$, the logical measurement protocol can also be performed in $1$ round of error correction, similar to full block reading,  by introducing redundant $Z$ checks between each level of the hypergraph as shown in Figure~\ref{fig:thickened_partial_reading_metachecks}.

This fact helps us to gauge-fix the cycles,  because each data qubit in each cycle is initialised in $\ket{+}$ immediately before and measured out in the $X$ basis immediately after the single round of error correction. The results of those $X$ measurements can be used to infer the outcome of $X$ stabiliser measurements which would fix the gauges of the cycles, and so we do not need to explicitly add checks for those cycles \cite{williamson2024low}. However, if the set of cycles does not have a basis of low weight cycles, this can result in detectors with high weight in the space direction, which impinge on the fault-tolerance against stochastic noise.

\begin{restatable}{proposition}{FDSubcode}\label{prop:fault_distance_subcode_1}
    Let $Q$, $Q'$, $Q''...$ be a set of CSS LDPC codeblocks with distance $d$, and let $A_\bullet$ be a subcode with distance $d$, with chain maps from $A_\bullet$, or a thickened version thereof, to $Q, Q', Q''...$. Let $D_\bullet = \cone(h_\bullet)$ be the deformed code, which has code distance at least $d$.
    
    Assume that there are $d$ rounds of syndrome measurements on $Q$, $Q'$, $Q''$... before and after the partial block reading procedure. Then partial block reading performs logical measurement in $\CO(1)$ rounds of syndrome measurements with fault-distance $d$.
\end{restatable}
\proof
See Appendix~\ref{app:proof_fault_distance}.
\endproof
The key here is that when thickening, each $Z$ check error in level $a$ is equivalent to a $Z$ check error in level $a-1$ along with an $X$ error on the data qubit between them. In this way, we form a chain of metachecks up the thickened hypergraph. These allow all stabilisers in the subcode to be matched to sets of checks in each level of the hypergraph.
Evidently, if the subcode $A_\bullet$ has $n_A$ qubits, and the amount of thickening required is $L$, then the space cost of a partial block reading is $\CO(n_AL)$.

The above proposition considers performing one partial block reading with $d$ rounds of padding before and after the procedure. 
Once again, we would like to generalize this to performing $t\ge d$ operations in $O(t)$ time, thereby achieving constant time overhead in amortisation. 
This introduces an extra complication. 
Instead of the deformed code, we consider the \textit{compacted code}, defined as follows.

\begin{definition}[Compacted code]\label{def:compacted_code}
    Let $C_\bullet$ be a complex representing the memory CSS code blocks, and let $A_{1,\bullet}, \cdots, A_{t,\bullet}$ be chain complexes each with a chain map $f_{i,\bullet}: A_{i,\bullet}\rightarrow C_\bullet$. The compacted code is the code taken by applying all the mapping cones to the original code, i.e., 
    \[\mathtt{CC}_\bullet = \cone(\sum_i (f_{i,\bullet}).\]
    % i.e. the code taken by applying all the mapping cones to the original code.
\end{definition}
While this code is most likely not LDPC and potentially massive, it is just a proof construct and is never explicitly used in our surgery operations. 

There is an intuitive reason behind defining the compacted code. 
Ideally, we would like to construct partial block readings assuming just that every deformed code has distance $d$. 
This assumption, however, may not be sufficient: if we perform two partial block readings one after another with only $O(1)$ syndrome rounds between them, we must carefully consider how the weight of an unmeasured logical may be reduced by multiplication with the new deformed code checks. 
While, by assumption, both the first and second partial block readings would individually guarantee any weight-reduced operator remains weight $d$, there would be no such guarantee when both weight reductions occur together.
This may seem irrelevant -- after all, we are not performing the partial block readings simultaneously -- but it happens that a sequence of check measurement errors in the $O(1)$ syndrome rounds between the block readings can effectively reproduce this simultaneous scenario using potentially just $O(1)$ additional faults (since we are not assuming the original code is sound).\footnote{We note that this simultaneous weight reduction issue, a scenario not considered in detail by prior code surgery work, is also a concern when doing auxiliary graph surgeries \cite{williamson2024low,ide2024fault} separated by $O(1)$ syndrome rounds if one does not assume expansion of the auxiliary graphs and instead just assumes large deformed code distances. In practical constructions, e.g.~Ref.~\cite{cross2024improved}, where one indeed tends not to guarantee expansion but just deformed distance, this is an important issue to be aware of.}
%While the first weight reduction guarantees that the reduced operator still have weight at least $d$, the second weight reduction may not (even though the second deformed code has distance $d$).
Therefore, to ensure the overall surgery protocol is fault-tolerant, we impose the condition that the compacted code, which can be seen as the memory blocks deformed by all the surgery operations applied simultaneously, has distance at least $d$.

\begin{restatable}{theorem}{FDmanysubcode}\label{thm:fault_distance_many_subcode_1}
    Let $Q$, $Q'$, $Q''...$ be a set of CSS LDPC codeblocks with distance $d$, and let $\Xi$ be a set of $Z$-type partial block readings $\{\mathbf{Partial_1}, \mathbf{Partial_2}, \mathbf{Partial_3},...,\mathbf{Partial_\eta}\}$ where each subcode has distance $d$ and the compacted code has distance $d$.

    Then all $\eta$ logical measurement rounds of partial block reading can be performed sequentially using $\CO(1)$ rounds of syndrome measurement each, and using $d$ rounds of padding before and after the set of logical measurements. The procedure has phenomenological fault-distance $d$, for any $\eta \in \N$.
\end{restatable}
\proof
See Appendix~\ref{app:proof_fault_distance}.
\endproof

While it may seem daunting to study the compacted code and satisfy the distance condition, we show in  Section~\ref{sec:mod_expansion} that standard techniques in code surgery, notably thickening and expansion, can be applied to satisfy this condition. 
We further remark that in the case of full block reading with a full rank pattern matrix $\mathbb{P}$, the resultant compacted code is a homological product code $(C\otimes R)^\bullet$, which has distance at least $d$.

\begin{remark}
    In Section~\ref{sec:hom_equivalence} we study the relationship between surgery and homomorphic measurement \cite{huang2023homomorphic}. For now, observe that given a subcode of distance $d$ performing homomorphic measurement is straightforward, as one initialises an ancilla system that is a copy of that subcode and performs transversal CNOTs and single-qubit measurements. This initialisation can take up to $\CO(d)$ rounds to preserve fault-tolerance. 
    Note that the method of algorithmic fault-tolerance~\cite{zhou2024algorithmic} does not apply to generic homomorphic measurements.
    Performing partial block reading with a distance $d$ subcode requires additional conditions on the distance of the deformed code to maintain fault-tolerance, which can be resolved at some additional space cost by the thickening described above.
\end{remark}

\section{Fast hypergraph surgery}\label{sec:hyper-surgery}

% There are ways to design a hypergraph with high modular expansion other than just taking a subcode $A_\bullet$ and thickening it appropriately. As thickening is expensive in terms of space, these alternatives are promising for near-term codes. For instance, given a small subcode one can numerically add hyperedges to $A_\bullet$ in order to boost the modular expansion while retaining a degree of sparsity. More generally, one could construct a bespoke hypergraph system for a given subcode with enough brute-force. 

In this section we generalise our partial block reading results to the case where $A_\bullet$ is not a subcode but some other complex with a suitable chain map into the original code.
% This homomorphic chain map requires that the auxiliary complex has structural symmetry with the memory code. 
% In particular, the space $A_2$ is mapped to $C_2$, which is the space of $Z$ stabilisers of the memory. 
% As $A_2$ becomes metacheck in the surgery operation, this means the space of metacheck 

\[\begin{tikzcd}
    A_2 \arrow[r]\arrow[d, "f_2"] & A_1 \arrow[r]\arrow[d, "f_1"] & A_0 \arrow[r]\arrow[d, "f_0"] & A_{-1}\arrow[d, "0"] \\
    C_2 \arrow[r] & C_1 \arrow[r] & C_0 \arrow[r, "0"] & 0
\end{tikzcd}\]

\begin{definition}\label{def:inj_on_verts}
    We say that the chain map $f_\bullet: A_\bullet 
    \rightarrow C_\bullet$ is non-overlapping on vertices if each row in the matrix $f_1: A_1 \rightarrow C_1$ contains at most a single non-zero entry.
\end{definition}
The degree 1 of $A_\bullet$ is shifted by the mapping cone construction to be in $\cone(f)_2$, i.e. the set of vertices in the hypergraph is the set of basis elements of $A_1$.

We can directly transport Definition~\ref{def:partial_block_reading} to this more general setting.
\begin{definition}
    Let $A_\bullet$ be an auxiliary chain complex and let $Q$, $Q'$, $Q''...$ be CSS LDPC codes, each of which have chain complexes $C_\bullet, C_\bullet', C_\bullet''...$, such that there is a collection of sparse chain maps $f_\bullet$, $f_\bullet'$, $f_\bullet''$... of the forms \[f_\bullet: A_\bullet \rightarrow C_\bullet; \quad f_\bullet': A_\bullet\rightarrow C_\bullet'; \quad f_\bullet'': A_\bullet \rightarrow C_\bullet''...\]
    which are non-overlapping on vertices.

    Then the hypergraph surgery of $Z$ type defined by $\{f_\bullet, f_\bullet',f_\bullet''...\}$ is
    \[D_\bullet = \cone(h_\bullet); \quad h_\bullet = f_\bullet + f_\bullet'+f_\bullet''+...\]
\end{definition}

As with partial block reading it is trivial to compute which logical operators are measured using the Snake Lemma~\cite[Lem.~1.3.2]{Weib1994}.

\begin{restatable}{theorem}{GenHyperSurg}\label{thm:genhypersurg}
    Let $Q$, $Q'$, $Q''...$ be a set of CSS LDPC codeblocks with distance at least $d$, and let $\Xi$ be a set of $Z$-type hypergraph surgeries $\{\mathbf{Hyper_1}, \mathbf{Hyper_2}, \mathbf{Hyper_3},...,\mathbf{Hyper_\eta}\}$ such that:
    
    \begin{itemize}
        \item Each auxiliary complex $(A_\bullet)_1, (A_\bullet)_2, (A_\bullet)_3, ..., (A_\bullet)_\eta$ has 1-cosystolic distance $d$, with sparse cycle bases.
        \item The compacted code has distance at least $d$.
    \end{itemize}

    Then all $\eta$ logical measurement rounds of generalised hypergraph reading can be performed sequentially using $\CO(1)$ rounds of syndrome measurement each, and using $d$ rounds of padding before and after the set of logical measurements. The procedure has phenomenological fault-distance $d$, for any $\eta \in \N$.
\end{restatable}
\proof
See Appendix~\ref{app:proof_fault_distance}.
\endproof

% Past works on surgery have focused on ensuring the deformed code has distance $d$, which provides fault-tolerance guarantees for $\CO(d)$ rounds measurements. 
% This result establishes that if further the hypergraph complexes have 1-cosystolic distance at least $d$, then their respective surgery operations can be performed in amotised constant time. 
In the mapping cone that describes the surgery operation, the 1-cosystolic distance captures the metacheck distance.
To satisfy this distance condition, it is often easier to start with a homomorphic, high distance quantum code, as we did in full block reading. 

Returning to the second condition, in Section~\ref{sec:mod_expansion} we define a notion of hypergraph expansion called modular expansion and show that it is a sufficient condition for the compacted code to have distance $d$.
We then show that a generic hypergraph can always be thickened (by up to $d$ layers) to a graph with the desired expansion properties. 
This is in analogy with prior surgery works studying graphs, where it is known that thickening can be used to boost a property called relative expansion~\cite{swaroop2024universal}. 
We remark that for both our work and prior works, expansion is a sufficient but not necessary condition, in practice one can often find good constructions heuristically, such as adding random hyperedges.

% past work has developed a full set of tools for designing simple expander graphs which ensures the deformed code has high distance and is LDPC.

\begin{remark}
In a generic hypergraph surgery, constant-weight local operators are measured alongside high-weight global operators. 
If we require that the original codespace is restored after the hypergraph surgery, the constant-weight local operators must all be in the stabiliser group, and the high-weight operators must be logical operators. 
For a $Z$-type measurement on a single code block, with an injective $f_1$ function, this requires the group generated by $\mathbb{F}_2$-linear combinations of the hyperedges to include the truncation of the $X$-type stabiliser group to the qubits addressed by the measurement. 
In the previous sections we have chosen the $X$-type checks of the subcode, but more generally we can pick a different, potentially overcomplete, generating set.
\end{remark}

\section{Intermediate time overhead with locally testable codes}\label{sec:block_ltcs}

% When the codeblocks or subcodes prior to surgery already have properties related to soundness, we can improve our bounds on the overheads of surgery, and perform more elaborate surgeries.

We can consider (a) block reading with a distance $d$ subcode and (b) conventional LDPC code surgery, which uses a distance 1 subcode, to each sit at opposite ends of an axis determining the time cost of logical measurements. In this section we consider interpolating along this axis, to find surgeries which use between $\CO(1)$ and $\CO(d)$ time each, while remaining fault-tolerant.

\subsection{Partial block reading with a low distance subcode}

\begin{restatable}{proposition}{FDSubcodeAlpha}
\label{prop:fault_distance_subcode}
    Let $Q$, $Q'$, $Q''...$ be a set of CSS LDPC codeblocks with distance $d$, and let $A_\bullet$ be a subcode with distance $d_A = \frac{1}{\alpha}d$, with chain maps from $A_\bullet$, or a thickened version thereof, to $Q, Q', Q''...$. Let $D_\bullet = \cone(h_\bullet)$ be the deformed code with code distance $d$ and let $\CH$ have a sparse cycle basis.
    
    Assume that there are $d$ rounds of syndrome measurements on $Q$, $Q'$, $Q''$... before and after the partial block reading procedure. First, $\lceil\alpha\rceil$ rounds of syndrome measurements are necessary (but not sufficient) to perform partial block reading with fault-distance $d$.
    
    Then if either: (a) $\alpha \leq 2$ or (b) the initial codeblocks are each locally testable codes with soundness $\rho \geq \frac{2n}{m}$, where $n$ is the codeblock length and $m$ the number of checks, then partial block reading performs logical measurement in $\lceil\alpha\rceil$ rounds of syndrome measurements with fault-distance $d$, maintaining the LDPC property throughout.
\end{restatable}
\proof
See Appendix~\ref{app:proof_fault_distance}.
\endproof

\begin{remark}[Cycle checks]
When the block reading takes longer than $\CO(1)$ round we genuinely require a sparse cycle basis for $\CH$ to maintain the LDPC property of the deformed code, as the gauge checks at each timestep cannot easily be inferred from the state preparation and measurement of data qubits in the auxiliary system. In general, it is not guaranteed that $\CH$ possesses a sparse cycle basis. In practice we expect to find sufficiently sparse cycle bases numerically for small codes.

Additionally, if a subcode has low weight $X$ metachecks which have support on checks in the subcode, after degree shifting these metachecks become low weight cycle checks for a $Z$ basis partial block reading.

\end{remark}

Proposition~\ref{prop:fault_distance_subcode} is interesting primarily because there are codes that are not expected to have distance $d$ subcodes addressing targeted subsets of logical qubits. Indeed only topological codes and certain product codes are known to allow for addressing any logical qubit in this manner \cite{huang2023homomorphic,xu2024fast}. Instead, one can find subcodes with distance between 1 and $d$ which still permit fast measurement, but not necessarily in constant time. The distance of $A_\bullet$ is the minimum number of check errors on the auxiliary system which are required to cause an undetectable logical measurement error in a single round;  this is not the same as the single-shot distance of the entire deformed code, because there can be checks in the bulk which have far less protection against measurement errors.

However, the errors which can occur in the bulk do not prevent us from performing many of these partial block readings sequentially without $d$ rounds of padding in between, implying that interpolated partial block reading can indeed speed up computation substantially.

\begin{restatable}{theorem}{FDSubcodeTogether}
\label{thm:fault_distance_subcode_together}
    Let $Q$, $Q'$, $Q''...$ be a set of CSS LDPC codeblocks with distance  at least $d$, and let $\Xi$ be a set of partial block readings $\{\mathbf{Partial_1}, \mathbf{Partial_2}, \mathbf{Partial_3},...,\mathbf{Partial_\eta}\}$ where each subcode has distance $d_i \geq \frac{1}{\alpha_i}d$ for $i \in [\eta]$ and the deformed codes each have code distance $d$, with sparse cycle bases. 
    % Let either (a) $d_i \geq d/2$ $\forall i\in [\eta]$ or (b) the original codes have soundness $\rho \geq \frac{2n}{m}$.
    Suppose the original codes have soundness $\rho \geq \frac{2n}{m}$.

    Then all $\eta$ logical measurement rounds of partial block reading can be performed sequentially using $\lceil\alpha_i\rceil$ rounds of syndrome measurement each, and using $d$ rounds of padding before and after the logical measurements. The procedure has phenomenological fault-distance $d$, for any $\eta \in \N$.
\end{restatable}
\proof
See Appendix~\ref{app:proof_fault_distance}.
\endproof

We remark that when the base codes are locally testable codes with constant soundness, we only require the weaker condition that each deformed code has distance $d$, instead of the stronger condition that the compacted code has distance $d$. 
Intuitively, this is because when considering the potential weight reduction of a logical fault that is deformed into two different surgery ancilla systems, the support of that logical fault is necessarily shifted onto timelike check errors, which are proportional to the reduced space support as the codes have constant soundness.

\begin{remark}[Interpolated surgery with lower soundness]\label{rem:interp_lower}
    Proposition~\ref{prop:fault_distance_subcode} and Theorem~\ref{thm:fault_distance_subcode_together} demand soundness $\rho \geq \frac{2n}{m}$, which is asymptotically constant as $m \in \Theta(n)$ for LDPC codes. If the soundness is lower than constant, which one may expect for practical codes, it is sufficient to use asymptotically more rounds of syndrome measurement per logical measurement, proportional to $1/\rho$, to maintain phenomenological fault-distance $d$. Furthermore, one cannot guarantee that the surgeries can be performed within $\CO(1)$ time of each other, without considering the compacted code (see Def.~\ref{def:compacted_code}). Scaling the time between logical measurements to $\CO(1/\rho)$ is then sufficient to preserve fault-distance.

    If the soundness is vanishing then there is still a reduction in time overhead from measuring multiple representatives, but the reduction is additive rather than multiplicative, see Proposition~\ref{prop:aux_fault_bound}, and one is forced to consider the compacted code to ensure that the logical measurements can be performed close together in time.
\end{remark}

\subsection{Intermediate time hypergraph surgery}

Similar to the previous section on partial block reading, we can generalize Theorem~\ref{thm:fault_distance_subcode_together} to the case of hypergraph surgeries. 

\begin{restatable}{theorem}{FDHyperTogether}
\label{thm:fault_distance_hypergraph_together}
    Let $Q$, $Q'$, $Q''...$ be a set of CSS LDPC codeblocks with distance at least $d$, and let $\Xi$ be a set of hypergraph surgeries $\{\mathbf{Hyper_1}, \mathbf{Hyper_2}, \mathbf{Hyper_3},...,\mathbf{Hyper_\eta}\}$ such that:

    \begin{itemize}
        \item Each auxiliary complex $(A_\bullet)_1, (A_\bullet)_2, (A_\bullet)_3, ..., (A_\bullet)_\eta$ has $1$-cosystolic distance $d_i \geq \frac{1}{\alpha_i}d$ for $i \in [\eta]$, with sparse cycle bases.
        \item Each deformed code has distance at least $d$.
        \item The original codes have soundness $\rho \geq \frac{2n}{m}$.
        
    \end{itemize}
    
    Then all $\eta$ logical measurement rounds of hypergraph surgery can be performed sequentially using $\lceil\alpha_i\rceil$ rounds of syndrome measurement each, and using $d$ rounds of padding before and after the logical measurements. The procedure has phenomenological fault-distance $d$, for any $\eta \in \N$.
\end{restatable}
\proof
This follows by combining the proofs of Theorem~\ref{thm:genhypersurg} and Theorem~\ref{thm:fault_distance_subcode_together}.
\endproof

Gauging logical measurement~\cite{williamson2024low} can be seen as an instance of hypergraph surgery where the subcode $A_\bullet$ has distance 1, and therefore requires $\CO(d)$ rounds of syndrome measurement to maintain fault-distance $d$.

\section{Modular expansion}\label{sec:mod_expansion}

In previous sections, we studied hypergraph surgery operations where the compacted code or the deformed codes has distance $d$. 
This is critical for fault-tolerance.
In this section, we discuss sufficient expansion properties on the hypergraph for these conditions to hold. 
In context of past surgery works, when $A_\bullet$ contains only one logical operator (so $\alpha = d$ and $d_A = 1$), partial block reading recovers the CKBB scheme \cite{cohen2022low}. 
It is known that there are substantially more efficient schemes for measuring individual logical operators in $\CO(d)$ time by surgery \cite{williamson2024low,ide2024fault}, using expander graphs. 
This section is essentially an extension of these results to hypergraphs, which can measure multiple logical representatives simultaneously and quickly.

\begin{definition}
    Let $\CH(\CV, \CE)$ be a hypergraph, with edge-vertex incidence $\F_2$-matrix $G$. Let $I$ be an index set of basis elements $v_i$ of a subspace $W\subset \F_2 \CV$. Then we can index element in $W$ as $\kappa_\lambda$ for $\lambda \in 2^I$.
\end{definition}

We conflate $\kappa_\lambda$ as a vector and as a set of vertices in $\CV$. To each basis element $v_i$ we associate its set of vertices $V_i = \supp(v_i)$, and define the vertex set $\CU = \bigcup\limits_i V_i$.

\begin{definition}[Modular expansion]\label{def:modular_exp}
Let $\CH(\CV,\CE)$ be a hypergraph with incidence matrix $G: \F_2\CE \rightarrow \F_2\CV$ and specified subspace $W \subset \F_2\CV$ containing elements $\kappa_\lambda$. Let $t \in \R_+$ be a positive real number. The modular expansion $\CM_t$ of $\CH$ is the largest real number such that, for all $v \subseteq \CV$,
\[|G^\intercal v| \geq  \CM_t \min(t, |\kappa_\lambda| - |v \cap \kappa_\lambda| + |v \cap \CU \backslash \kappa_\lambda|\ :\ \kappa_\lambda \in W)\]
\end{definition}

As this definition is quite complicated, we give some orienting remarks. Modular expansion is a generalisation of both relative expansion from Ref.~\cite{swaroop2024universal} and the soundness of a classical locally testable code, and is therefore also as generalisation of the edge-expansion of a graph. It is `modular' in the sense that it contains  distinguished subsets $\kappa_\lambda$ such that, when they all exist in the same hypergraph, the notion of expansion for each of them combines to give a complete notion of expansion for the whole hypergraph.

In detail, recall that edge-expansion (henceforth called global expansion) can be defined for hypergraphs as follows.
\begin{definition}[Global expansion]
    Let $\CH(\CV,\CE)$ be a hypergraph with incidence matrix $G: \F_2\CE \rightarrow \F_2\CV$. The global expansion of $\CH$ is the largest real number $\beta$ such that
    \[|G^\intercal v| \geq \beta\min(|v|,|\CV\backslash v|).\]
\end{definition}
When $G$ is a graph this reduces to traditional edge-expansion and $\beta$ is the Cheeger constant.

Ref.~\cite{swaroop2024universal} defined relative expansion, a generalisation of global expansion.
\begin{definition}[Relative expansion]
    Let $\CH(\CV,\CE)$ be a hypergraph with incidence matrix $G: \F_2\CE \rightarrow \F_2\CV$, distinguished subset $\CU \subset \CV$ and chosen parameter $t \in \R_+$. The relative expansion of $\CH$ is the largest real number $\beta_t$ such that
    \[|G^\intercal v| \geq \beta_t\min(t,|v \cap \CU|,|\CU| - |v\cap \CU|).\]
\end{definition}
When $t \geq |\CV|$ and $\CU = \CV$ this reduces to the definition of global expansion. Similarly, when $|I| = 1$ modular expansion reduces to relative expansion, observing that there are only two $\kappa_\lambda$ vectors, one being the empty set and the other $\CU$.

A different generalisation of global expansion is to the soundness of a locally testable code. Recall Definition~\ref{def:classical_soundness}, with variables replaced to be suggestive of our hypergraph setting.
\begin{definition}
    A binary linear code $\CC \subset \F_2\CV$ is $(\omega,\rho)$-locally testable if it has a parity-check matrix $G^\intercal$ with rows of weight at most $\omega$ such that for any vector $v \in \F_2\CV$,
    \[\frac{1}{m}|G^\intercal v| \geq \frac{\rho}{n}d(v,\CC)\]
    where $m = |\CE|$, $n = |\CV|$ and $d(v, \CC) = \min_{x \in \CC}(|v+x|)$. The values $\omega$ and $\rho$ are the locality and soundness of the code respectively.
\end{definition}

In particular, if we rearrange the equation defining soundness $\rho$ then we get
$|G^\intercal v| \geq \beta d(v,\CC)$ for $\beta = \frac{\rho m}{n}$. We observe that in the case where the only codewords are $0$ and $\CV$, $\beta$ coincides with the definition of global expansion. For any sparse hypergraph $\CH$, if it has global expansion it is also a locally testable code, albeit a rather trivial one.
In this way the soundness of a code can be thought of as a generalisation of expansion to the case where a hypergraph has $\ker G^\intercal$ with dimension greater than 1. As $\dim G^\intercal$ is equal to the number of logical representatives which are measured simultaneously by the auxiliary hypergraph, we require a more general definition than relative expansion in order to perform block reading.

Modular expansion coincides with soundness when $\CU = \CV$ and $t \geq |\CV|$ under the rearrangement $\CM_t = \frac{\rho m}{n}$, and when $W = \ker G^\intercal$, so each element $\kappa_\lambda$ is an element of the kernel. We thus have the informal commuting diagram,
\[\input{Figures-Tikz/expansion_commuting_diag.tikz}\]
where $k = \dim G^\intercal$, $n = |\CV|$, and each $\kappa_\lambda$ is assumed for simplicity of the diagram to correspond to an element of $\ker G^\intercal$.

\begin{remark}
    For a simply connected graph, the edge-expansion can be related to spectral properties of the adjacency matrix or Laplacian of the graph. We do not know of a similar relation between relative expansion, soundness or modular expansion and spectral properties of the Laplacian of the hypergraph.
\end{remark}

To explain our results on hypergraphs with modular expansion we use the terminology of port functions, similar to Ref.~\cite{swaroop2024universal}. In short, given a CSS code $Q$ and a hypergraph $\CH(\CV, \CE)$, a port function $f_p$ is a map from a subset of data qubits in $Q$ to a set of vertices in $\CV$. The deformed code is then defined as $Q\leftrightarrow \CH$, with connections introduced between $Q$ and $\CH$ given by the port function $f_p$. This definition of the deformed code is similar to the definition of the mapping cone by a chain map, although the port function does not specify the connectivity of deformed checks into the auxiliary hypergraph, and the port function is in the opposite direction to $f_1$ in a chain map. Moreover, the port function definition can be applied to codes which are not CSS.

\begin{restatable}[Distance preservation]{theorem}{DistPreservation}\label{thm:distance_preservation}
    Let $\{\overline{Z}_i\}_{i\in I}$ be a set of logical representatives (trivial or nontrivial) on a distance $d$ CSS code $Q$ with supports $\xi_i = \supp(\overline{Z}_i)$.
    Let $\CH(\CV,\CE)$ be a hypergraph with vertices in $\CV$ being $Z$ checks, hyperedges in $\CE$ data qubits and $X$ gauge checks forming a cycle basis, such that $\CH$ measures all the logical representatives $\{\overline{Z}_i\}_{i\in I}$. Let $f_p:\bigcup_i \xi_i\rightarrow \CV$ be an injective port function such that $\CU = \im f_p \subset \CV$, and $f_p(\xi_i) = V_i \subset \CV$, $\forall i \in I$. Let $\CM_d(\CH) \geq 1$, where $W$ is generated by $f_p(\xi_i)$. 
    Then the distance of the deformed code $Q \leftrightarrow \CH$ is at least $d$.
\end{restatable}
\proof
See Appendix~\ref{app:mod_exp_proofs}.
\endproof

As the above proof demands the port function $f_p$ from the original code to the hypergraph be injective, it does not capture the most general form of hypergraph surgeries. 
In that case, scaling the modular expansion proportional to the check incidence to the original code is sufficient to preserve the distance, see Remark~\ref{rem:hyper_incidence_boost}.

\begin{restatable}[Thickening]{theorem}{Thickening}\label{thm:thickening}
    Let $\CH(\CV, \CE)$ be a hypergraph with modular expansion $\CM_t(\CH)$. Let $\CJ_L$ be the path graph with length $L \geq \frac{1}{\CM_t(\CH)}$, i.e. $L$ vertices and $L-1$ edges. Let  $\CH_L := \CH \square \CJ_L$ be the hypergraph of $\CH$ thickened $L$ times. Then $\CH_L$ has modular expansion $\CM_t(\CH_L) \geq 1$ for $\CU^\ell = \bigcup\limits_i V^\ell_i$ at any level $\ell \in \{1,2,\cdots,L\}$, where $V^\ell_i$ is the copy of $V_i$ in the $\ell$th level of the thickened hypergraph.
\end{restatable}
\proof
See Appendix~\ref{app:mod_exp_proofs}.
\endproof

\begin{lemma}\label{lem:min_expansion}
    Let $\CH(\CV,\CE)$ be a hypergraph such that each element in $\ker(G^\intercal)$ has the supporting set $\kappa_\lambda$ for some $\lambda \in 2^I$. Then $\CM_t(\CH) \geq \frac{1}{t}$.
\end{lemma}
\proof
Consider an arbitrary vertex set $v$. We consider two cases: first, where $|G^\intercal v| =0$, and then where $|G^\intercal v| \geq 1$.

In the first case, we know that $v \in \ker(G^\intercal)$ and therefore $v = \kappa_m$ for some $m \in 2^I$. Therefore $|\kappa_m| - |v\cap \kappa_m| + |v\cap \CU\backslash \kappa_m| = 0$, and so
$|G^\intercal v| \geq \CM_t(\CH) \min(t, |\kappa_\lambda| - |v\cap \kappa_\lambda| + |v\cap \CU\backslash \kappa_\lambda| : \lambda\in 2^I)$ for any $\CM_t(\CH)$.

In the second case, $|G^\intercal v| \geq 1$. We now consider the two values of the minimum. If $\min(\cdot) = t$ then $|G^\intercal v| \geq \frac{1}{t}t$. If $\min(\cdot) = |\kappa_\lambda| - |v\cap \kappa_\lambda| + |v\cap \CU\backslash \kappa_\lambda|$ for some $\lambda\in 2^I$ then $t \geq |\kappa_\lambda| - |v\cap \kappa_\lambda| + |v\cap \CU\backslash \kappa_\lambda|$, and so $|G^\intercal v| \geq 1 \geq \frac{1}{t}(|\kappa_\lambda| - |v\cap \kappa_\lambda| + |v\cap \CU\backslash \kappa_\lambda|)$.
\endproof

\begin{corollary}\label{cor:thickening}
    Let $\CH(\CV,\CE)$ be any hypergraph such that each element in $\ker(G^\intercal)$ has the supporting set $\kappa_\lambda$ for some $\lambda \in 2^I$. Let $\CH_t := \CH \square\CJ_t$ be the hypergraph of $\CH$ thickened $t$ times. Then $\CH_t$ has modular expansion at least $\CM_t(\CH_t) \geq 1$ for any level $\ell$.
\end{corollary}
\proof
As $\CM_t(\CH) \geq \frac{1}{t}$ and $\CJ_L$ in Theorem~\ref{thm:thickening} has $L = t$ then $L \geq \frac{1}{\CM_t(\CH)}$.
\endproof

Corollary~\ref{cor:thickening} post-hoc justifies \textit{devised sticking} from Ref.~\cite{zhang2024time} by setting $t = d$, as there the hypergraph is thickened $d$ times to preserve fault-tolerance, and the CKBB scheme \cite{cohen2022low} is a special case of this.

\begin{remark}\label{rem:subsystem_CKBB}
    The expansion-based argument does not account for the cycles in the hypergraph, however. For these see Ref.~\cite[Thm.~1]{cohen2022low}, which shows that when the hypergraph is thickened $d$ times the cycles can be considered gauge logicals and the deformed code still has distance at least $d$ as a subsystem code.
\end{remark}

Theorem~\ref{thm:thickening} shows that it is not necessary to thicken $d$ times, if the modular expansion before thickening is greater than $\frac{1}{d}$. This has previously been proved for the simpler case of measuring only one logical operator \cite{cross2024improved, swaroop2024universal}.

\begin{lemma}[Cross-block measurements]\label{lem:multiblock_mod_exp_dist}
    Let $\CH$ be a hypergraph with modular expansion $\CM_d(\CH)\geq \frac{1}{L}$. Let $\CH$ correspond to the degree-shifted subcode $A_\bullet$ with distance at least $d$. Vertices in $\CH$ are $Z$ checks and edges are data qubits. Let $f_\bullet, f_\bullet', f_\bullet''...$ be the inclusion chain maps of $A_\bullet$ into $C_\bullet, C_\bullet',C_\bullet''...$, each of which has distance at least $d$.

    Let $\CH_L$ be the the hypergraph thickened $L$ times, with modular expansion $\CM_d(\CH_L) \geq 1$ at each level $\ell \in \{1, 2, \cdots, L\}$. Then the deformed code $D_\bullet$ given by modifying each chain map $f_\bullet, f_\bullet', f_\bullet''...$ to have their preimages on any arbitrary levels of the hypergraph and taking the mapping cone has distance $d$ if all cycles are gauge-fixed.
\end{lemma}
\proof
As all cycles are gauge-fixed, there are no new logical qubits, and no $\overline{X}$ logical can have weight lower than $d$. We must therefore ensure that no $\overline{Z}$ logical which is unmeasured can have its weight reduced lower than $d$ by cleaning.

We prove the lemma by contradiction. If any $\overline{Z}$ logical extending over all initial codeblocks $C_\bullet, C_\bullet',C_\bullet''...$, but not the hypergraph, could have its weight reduced below $d$ by cleaning into the hypergraph, then a single logical in one of $C_\bullet, C_\bullet', C_\bullet''...$ must also have its weight reduced below $d$ by cleaning. This fact is because removing the chain maps to all other codeblocks would leave the same set of $Z$ checks which could clean $\overline{\Lambda}_Z$ to the same data qubits $\sigma$, with the exception of data qubits in the other codeblocks. This set of data qubits must satisfy $|\sigma| < d$ for this lemma to be false. However, we know that $\CM_d(\CH_L) \geq 1$ at each level of the hypergraph, therefore $|\sigma| \geq d$ and the lemma holds.
\endproof

By a combination of Corollary~\ref{cor:thickening} and Lemma~\ref{lem:multiblock_mod_exp_dist}, observe that the earlier Lemma~\ref{lem:deformed_dist_d} from Section~\ref{sec:partial_block_reading} holds.

To summarise this section, if a subcode $A_\bullet$ for partial block reading has sufficient soundness then it must always also have modular expansion, which can then be boosted by thickening an appropriate amount to guarantee the deformed code's distance is preserved. 

Recall the definition of soundness for quantum codes in Def.~\ref{def:quantum_soundness}.
\begin{corollary}\label{coro:partial_BR_reduced_space_soundness}
    Let $A_\bullet$ be the subcode used in a partial block reading. Let $A_\bullet$ have soundness $\rho_A$. Then, viewed as a hypergraph $\CH$, thickening $\frac{2n_A}{\rho_A m_A}$ times preserves the distance of the deformed code when all cycles are gauge-fixed.
\end{corollary}
\proof
By Lemma~\ref{lem:fact_17}, $\CH$ has soundness $\frac{\rho_A m_A}{2n_A}$ as a classical code with parity-check matrix $G^\intercal$. Setting $W = \ker G^\intercal$, $\CH$ also has modular expansion $\CM_d \geq \frac{\rho_A m_A}{2n_A}$. By Lemma~\ref{lem:multiblock_mod_exp_dist}, set $L = \frac{2n_A}{\rho_A m_A}$ and the thickened hypergraph $\CH_L$ has modular expansion $\CM_d(\CH_L) \geq 1$.
\endproof

\begin{remark}
    As $\frac{m_A}{2n_A} \in \CO(1)$ for LDPC codes, the asymptotic space cost of partial block reading when all cycles are gauge-fixed can be reduced to $\CO(\frac{n_A}{\rho_A})$.
\end{remark}

Assuming the modular expansion $\CM_d(\CH)$ before thickening is greater than $\frac{1}{d}$, which is its minimum by Lemma~\ref{lem:min_expansion}, this thickening requires fewer than the $d$ layers required by the CKBB scheme and similar \cite{cohen2022low, zhang2024time}. Note that as the soundness of a subcode is not generally inherited from the soundness of the full code, finding subcodes with high soundness is generally difficult.

Regardless, having hypergraphs equipped with modular expansion is extremely useful to perform fast surgery.

\begin{lemma}
    If each hypergraph $\CH_i$ associated to a partial block reading $\mathbf{Partial_i}$ for $i \in [n]$ has modular expansion $\CM_d(\CH_i) \geq 1$ then the compacted code $\mathtt{CC}_\bullet$ has distance $d$.
\end{lemma}
\proof
By Theorem~\ref{thm:distance_preservation}, attaching $\CH_1$ to the original code via the chain map $(h_\bullet)_1$, to yield $\cone((h_\bullet)_1)$, must always preserve the distance of the code. Applying the other mapping cones iteratively will then each preserve the distance, so $\mathtt{CC}_\bullet$ has distance $d$.
\endproof

As a consequence, the conditions of Theorem~\ref{thm:fault_distance_many_subcode_1} can be satisfied constructively, by choosing subcodes of distance $d$ and then thickening such that each hypergraph has modular expansion.

\begin{remark}\label{rem:hyper_incidence_boost}
A similar approach applies to Theorem~\ref{thm:genhypersurg}, where instead it is sufficient for each hypergraph to have modular expansion $\CM_d(\CH_i) \geq w_i$, where $w_i$ is the maximal column weight of $(f_1)_i, (f_1')_i, (f_1'')_i,\cdots$. 
\end{remark}

\section{Examples}\label{sec:examples}

In this section we illustrate block reading with a series of examples. We focus first on 2D surface codes, upon which traditional lattice surgery is extremely well-understood \cite{Horsman2012LatticeSurgery,Litinski2018latticesurgery,deBeaudrap2020zx}. 

\subsection{Topological codes}
\subsubsection{2D surface codes}
We demonstrate on unrotated surface codes, as the connection to block reading, and LDPC code surgery in general, is more apparent in this case.

We start with two square surface code patches with distance $d$, with $n = d^2 + (d-1)^2$ data qubits each,

\[\input{Figures-Tikz/SC_fullblock.tikz}\]

In these diagrams qubits are edges, $X$ and $Z$ checks are vertices and faces respectively.

The deformed code for the full block reading performing a $\overline{Z}\otimes\overline{Z}$ is then given by

\[\input{Figures-Tikz/SC_fullblock2.tikz}\]

where the new edges are dashed grey lines. There are new vertically-oriented faces where there are new squares, and at the boundaries.

Evidently the connectivity requirements are close to the requirements for transversal CNOTs.

The only distance $d$ subcode (as defined in Def.~\ref{def:subcode}) of the square surface code is itself, and so the only partial block reading with a distance $d$ subcode is full block reading.
A partial block reading using a subcode $A_\bullet$ with distance $d_A$ less than $d$ can be constructed by choosing a strip of the surface codes and then connecting these together.

\[\input{Figures-Tikz/SC_partialblock.tikz}\]

In the simple case of surface code patches as above the distance is always preserved by partial block reading, because the only logicals which can have reduced weight are those on $\overline{Z}\otimes \overline{Z}$ or $\overline{X}\otimes\overline{X}$, which are now stabilisers. If the surface code patches also have defects, or are toric codes instead, partial block reading does not generically preserve code distance and increasing modular expansion in the auxiliary hypergraph is necessary, by thickening or otherwise.

Recall that surface codes have asymptotically vanishing soundness by ~\cite[Cor.~1]{campbell2019theory} so using a partial block reading with $d_A < d$ on the surface code is not particularly useful for reducing the time overhead, unless $d_A \geq \frac{d}{2}$ by Proposition~\ref{prop:fault_distance_subcode}.

In the case where $d_A = 1$ and the chosen subcode is along the boundary of the patches, we recover traditional lattice surgery with unrotated surface codes:

\[\input{Figures-Tikz/SC_latticesurgery.tikz}\]

\subsubsection{2D toric codes}

Partial block reading with toric codes is illustrative, as it showcases how the distance of the deformed code can drop without thickening.

First, consider the following $d=5$ toric codeblocks, shown by tessellating their fundamental polygons. Blue and red edge pairs of the same lattice are identified to form loops.

\[\input{Figures-Tikz/toric_codes1.tikz}\]

A partial block reading between them can be performed with a subcode corresponding to a cylinder around the torus. Measuring only one representative from each torus recovers traditional surgery with 2D toric codes,

\[\input{Figures-Tikz/toric_codes2.tikz}\]

which measures $\overline{Z}\otimes\overline{Z}$ on one logical qubit from each codeblock, but not the other. 

Growing the width of the cylinder such that the distance of $A_\bullet$ is 4, we acquire the deformed code:

\[\input{Figures-Tikz/toric_codes3.tikz}\]

Depending on the chosen subcode, the distance may drop in the deformed code, as there can be a stringlike $\overline{Z}$ operator which extends from one torus to the other and then back, skipping the part of the torus with the new connections and making a short noncontractible closed curve. In our example the deformed code distance is still $5$, as these curves take the form shown by green edges,

\[\input{Figures-Tikz/toric_codes4.tikz}\]

which have length at least $6$. However, were we to start with $d= 7$ toric codes, with the width of the cylinder being 6, the distance would drop as a consequence of these same curves. Increasing the length of this closed curve is then possible by thickening the subcode, or boosting the modular expansion in some other manner. If we were to perform a full block reading, these closed curves would become contractible due to the additional stabilisers, so the deformed code distance would be preserved regardless.

\subsubsection{4D toric codes}
4D toric codes are known to be decodable in a single-shot manner, as they have some soundness and metachecks of both $X$ and $Z$-type \cite{campbell2019theory,quintavelle2021single,aasen2025topologically, aasen2025geometrically}.

A 4D toric code can be constructed by a 4-fold tensor product of the chain complex
\[R_\bullet = \begin{tikzcd}
    R_1 \arrow[r, "\del_1"] & R_0
\end{tikzcd}\]
where $R_1 = R_0 = \F_2^m$ and the differential $\del_1$ is the incidence matrix of the cycle graph with $m$ vertices and edges, so $\dim \ker(\del_1) = 1$ and $\dim \F_2^m/\im(\del_1) = 1$. 

The 4-fold tensor product yields a chain complex with 5 nonzero components. Let the 4D toric code be 
\[C_\bullet = (R\otimes R\otimes R\otimes R)_\bullet = \begin{tikzcd}C_4 \arrow[r] & C_3 \arrow[r] & C_2 \arrow[r] & C_1 \arrow[r] & C_0\end{tikzcd}\] 
and the qubit component is fixed to be $C_2$.\footnote{This shift in qubit degree compared to convention is for convenience, so that there are no negative indices and basis elements in $C_2$ can be identified with faces in a cell complex.} More explicitly we have
\[C_2 = \bigoplus_{i+j+k+l = 2} R_i\otimes R_j \otimes R_k \otimes R_l,\]
and the entire chain complex is given by
\[\begin{tikzcd}
    & & R_1 R_1 R_0 R_0 \arrow[dr]\arrow[ddr] & & \\
    & R_1 R_1 R_1 R_0 \arrow[ur]\arrow[r]\arrow[dddr] & R_1 R_0 R_1 R_0 \arrow[r]\arrow[dddr] & R_1 R_0 R_0 R_0 \arrow[ddr] & \\
    & R_1 R_1 R_0 R_1 \arrow[uur]\arrow[r]\arrow[dddr] & R_1 R_0 R_0 R_1 \arrow[ur]\arrow[dddr] & R_0 R_1 R_0 R_0 \arrow[dr] & \\
    R_1 R_1 R_1 R_1 \arrow[uur]\arrow[ur]\arrow[dr]\arrow[ddr] & & & & R_0 R_0 R_0 R_0\\
    & R_1 R_0 R_1 R_1 \arrow[uuur]\arrow[uur]\arrow[ddr]& R_0 R_1 R_1 R_0 \arrow[uur]\arrow[r]& R_0 R_0 R_1 R_0 \arrow[ur]& \\
    & R_0 R_1 R_1 R_1 \arrow[ur]\arrow[r]\arrow[dr]& R_0 R_1 R_0 R_1 \arrow[uuur]\arrow[r]& R_0 R_0 R_0 R_1 \arrow[uur]& \\
    & & R_0 R_0 R_1 R_1 \arrow[uur]\arrow[ur] & &
\end{tikzcd}\]
where tensor products are suppressed for visibility and columns of terms are taken in a direct sum.

The K{\"u}nneth formula implies that $\dim H_2(C) = 6$. The distance of $C_\bullet$ is $d = m^2$, which can be computed using Lemma~\ref{lem:hom_factorB}. The intuition is that all nontrivial logical operators in a 4D toric code form membranes through the complex, as opposed to strings in 2D toric codes. For a given summand in $C_2$, say $R_1\otimes R_1\otimes R_0 \otimes R_0$, one choice of corresponding nontrivial $\overline{Z}$ logical operator representative is then $\underline{1}\otimes\underline{1}\otimes (1,0,\cdots,0)\otimes (1,0,\cdots,0)$, where $\underline{1} = (1,1,\cdots,1)$.

To give a treatment of partial block reading with 4D toric codes we specify a subcode $A_\bullet$. There are evidently many such choices. One choice is to measure just a single logical representative, which by Thm.~\ref{thm:proof_connectivity} must take $\Theta(d)$ syndrome measurement rounds to maintain phenomenological fault distance.

Another choice of $A_\bullet$ is to specify a set of $Z$-checks in $C_3$, and to take the summands in $C_2$, in the image of that set of $Z$-checks and so on. This is trivially guaranteed to be a subcode. For example, set $A_\bullet$ to be
\[
\begin{tikzcd}
    & R_1R_1R_0 \F_2 \arrow[r]\arrow[dr] & R_1R_0R_0\F_2 \arrow[dr] & \\
    R_1R_1R_1\F_2 \arrow[ur]\arrow[r]\arrow[dr] & R_1R_0R_1\F_2 \arrow[ur]\arrow[dr] & R_0R_1R_0\F_2 \arrow[r] & R_0R_0R_0\F_2 \\
    & R_0R_1R_1\F_2 \arrow[ur]\arrow[r] & R_0R_0R_1\F_2 \arrow[ur] &
\end{tikzcd}
\]
where the chain map component $f_3$ maps $R_1R_1R_1\F_2$ in $A_3$ into $R_1R_1R_1R_0$ in $C_3$ and so on, choosing a consistent basis element of $R_0$ to map $\F_2$ into.
This choice measures 3 $\overline{Z}$ logical operators, and many representatives of each. This is the choice taken in Ref.~\cite{aasen2025geometrically}, and can be seen as setting $A_\bullet$ to be a hyperplane of the lattice. While each $\overline{Z}$ logical representative contained in $A_\bullet$ must have weight at least $m^2$, the $X$-distance of $A_\bullet$ is only $m$, as this is a 3D toric code by identifying that $R_1R_1R_1\F_2 \cong R_1R_1R_1$ etc, with the qubit component at $A_2$ rather than the customary $A_1$.

In Ref.~\cite{hillmann2024single} it was observed that while one can perform single-shot lattice surgery using a subcode given by a hyperplane with 4D surface and toric codes, the performance was reduced unless $\CO(\sqrt{d}) = \CO(m)$ syndrome rounds were used. 
By Proposition~\ref{prop:fault_distance_subcode} we in fact see that $\Omega(\sqrt{d})$ rounds are required to maintain fault-distance, as $\alpha = \sqrt{d}$ in this case. Thus one can perform single-shot lattice surgery, or lattice surgery using a constant number of syndrome rounds, with 4D toric codes, but with only $\CO(\sqrt{d})$ fault-distance, given that choice of subcode.

\begin{remark}
    Proposition~\ref{prop:fault_distance_subcode} is not quite sufficient to justify $\Theta(\sqrt{d})$ rounds of syndrome measurement to maintain fault-distance, only $\Omega(\sqrt{d})$, as 4D toric codes do not have constant soundness \cite{campbell2019theory,quintavelle2021single}, see Remark~\ref{rem:interp_lower}.
\end{remark}

\begin{remark}
    Lattice surgery with `geometrically enhanced' 4D toric codes was studied in Ref.~\cite[Sec.~V]{aasen2025geometrically}. There, the $X$-distance of $A_\bullet$ was called the `boundary distance', and similar arguments about the number of rounds required for fault tolerance were employed. The lattice surgery there is performed in an `ancilla-free' manner, where no new data qubits are used; only new checks. This is a quotient rather than a mapping cone, see Refs.~\cite{cowtan2024css,cowtan2024ssip,poirson2025engineering}.
\end{remark}

\subsection{Bivariate bicycle codes}

Bivariate bicycle (BB) codes~\cite{bravyi2024high} are a class of Abelian 2-block group-algebra (2BGA) codes~\cite{lin2024quantum,wang2023abelian}, and as such are lifted product and balanced product codes~\cite{panteleev2021degenerate,panteleev2021quantum,breuckmann2021balanced}. They are of substantial practical interest, as some bivariate bicycle codes are known to exhibit high pseudo-thresholds and low logical error rates under circuit noise and suitable decoders, while having a higher rate than surface codes~\cite{bravyi2024high,yoder2025tour}, albeit requiring more connectivity. BB codes can also be seen as toric codes boosted to have limited non-local connectivity~\cite{liang2025generalized}.

It is immediate that full block reading can be used to perform parallel $\overline{Z}\otimes\overline{Z}$ or  $\overline{X}\otimes\overline{X}$ measurements between two BB codes using $n + m_Z = n+m_X = \frac{3}{2}n$ new qubits, including check qubits, while maintaining the distance and LDPC property. For instance, the $\llbracket 144, 12, 12 \rrbracket$ gross code, which uses 288 total qubits as a quantum memory when including check qubits, requires 216 additional qubits in total for a full block reading. This is lower than for the same set of parallel measurements in Ref.~\cite[Sec.~3.3]{cowtan2024ssip}, which used  360 total new qubits (216 new data qubits and 144 new check qubits), increased the check weight of the deformed code to 12, and required $\CO(d)$ rounds of syndrome measurement. Thus even a naive application of block reading can reduce the spacetime overhead of logical measurements by an order of magnitude compared to conventional surgery, where the connectivity allows for it.

We suspect that one can perform partial block readings using subcodes of the gross code, and similar BB codes, in order to address desired logical qubits and reduce space overheads while maintaining fast and parallel logical measurement at high fault distance, and retaining a degree of addressability. As this requires substantial numerics we defer this to future work.

\section{Equivalence between surgery and homomorphic measurement}\label{sec:hom_equivalence}

The abstract relationship between (a) surgery by an auxiliary system and (b) homomorphic measurement \cite{huang2023homomorphic} was folklore prior to this work, and indeed is implied by Ref.~\cite{ide2024fault}. That is, given a chain map one can always construct a mapping cone and vice versa. The relationship we show in this section is more concrete: given the circuit for homomorphic measurement on a CSS code, it can always be rewritten into a logical measurement by surgery using a mapping cone (called homological measurement in Ref.~\cite{ide2024fault}) and vice versa.

\begin{restatable}{theorem}{SZXHomToSurgery}
    Let $C_\bullet$ be a CSS code with a homomorphic measurement procotol specified by a chain map $f_\bullet : A_\bullet \rightarrow C_\bullet$. Then the circuit corresponding to the homomorphic measurement can be rewritten to a surgery specified by $\cone(f_\bullet)$.
\end{restatable}
\proof
See Appendix~\ref{app:proof_circuit_equiv}.
\endproof

\begin{restatable}{corollary}{SZXSurgeryToHom}
    Let $C_\bullet$ be a CSS code with a surgery protocol specified by $\cone(f_\bullet)$ for a chain map $f_\bullet : A_\bullet \rightarrow C_\bullet$. Then the circuit corresponding to the surgery protocol can be rewritten to a homomomorphic measurement specified by $f_\bullet$.
\end{restatable}
\proof
See Appendix~\ref{app:proof_circuit_equiv}.
\endproof

These proofs are performed using the ZX-calculus \cite{coecke2011interacting}, a formal calculus for rewriting tensor networks on qubits. This is a convenient formalism, as the rewriting requires converting checks and data qubits in the homomorphic measurements into data qubits and checks respectively in the surgery procedure. In essence, the time and space axes of the ancilla circuit are inverted.

However, the rewriting does not generally preserve fault-distance, unlike the rewrites in Ref.~\cite{rodatz2025fault,Rusch2025Completeness}. The reason why converting a surgery to a homomorphic measurement does not generally preserve fault-distance is because having a low timelike distance in a surgery protocol can be resolved by measuring for multiple rounds. However, the equivalent homomorphic measurement has low space distance, which is fatal. We expect that one can prepare many of those ancilla states and measure repeatedly, in the style of a Shor measurement \cite{shor1996fault}. However, this is no longer a homomorphic measurement in the sense defined in Ref.~\cite{huang2023homomorphic}.

Conversely, we do not know whether rewriting any homomorphic measurement to a hypergraph surgery preserves fault-distance.

Our last results focus on general bounds on fast measurement by an auxiliary system. The results are insensitive to the nature of the auxiliary system: whether it is surgery, homomorphic measurement or any other scheme.

\begin{restatable}{proposition}{AuxiliaryFaultBound}\label{prop:aux_fault_bound}
    Let $Q$ be a quantum LDPC code. Let $S$ be a set of logical operator representatives measured at the same time by one or more auxiliary system for $T$ syndrome rounds. Assume that the correct logical measurement outcomes of $S$ are not known in advance. Let $f$ be the lowest weight data qubit fault in $Q$ such that:
    \begin{itemize}
        \item $f$ flips at least one logical representative in $S$.
        \item If $f$ flips a logical representative in $S$ then it flips all other representatives in the same equivalence class in $S$.
    \end{itemize}
    Let $w$ be the weight of $f$, and $s$ be the syndrome weight of $f$ in $Q$. Then the phenomenological fault-distance of the measurement protocol is upper bounded by $2w + Ts$.
\end{restatable}
\proof
See Appendix~\ref{app:proof_connectivity}.
\endproof

When $Q$ is CSS and the set $S$ forms a $Z$-type subcode $A_\bullet$, $w$ is the $X$-distance of $A_\bullet$.

% Our last result proves that block reading and similar protocols are essentially the only way to reduce the number of rounds required for surgery asymptotically. The result is sufficiently general to also cover other methods of measurement by an auxiliary system such as homomorphic measurements.

\ConnectivityRequired*
\proof
See Appendix~\ref{app:proof_connectivity}.
\endproof

Importantly, for fast surgery it is not enough for the deformed code in the idling operation to be a single-shot code. Because the correct outcomes of the new checks are not generally known in advance, many connections between representatives of the original code and the auxiliary hypergraph are essential to correct for measurement errors on these new checks. Indeed, our earlier results on full and partial block reading with distance $d$ subcodes imply that this is the only thing that is required, assuming that the auxiliary hypergraph is sensibly constructed and that sufficent rounds of padding are used.

\section{Discussion}

In this work we have introduced a variety of new, general LDPC code surgeries which reduce the time overhead of logical measurement below $\CO(d)$ rounds and allow the measurement of logical operators in parallel. We generalised prior work~\cite{cross2024improved,williamson2024low,ide2024fault,swaroop2024universal}, which used expansion properties of hypergraphs to efficiently measure individual logical operators, to use soundness instead, enabling measurement of many logical operators in parallel. We elucidated the connection between homomorphic measurement and surgery, and gave general bounds on the fault distance for any logical measurement by an auxiliary system.

To prove fault distances for fast surgery in Appendix~\ref{app:proof_fault_distance} we developed an extension of the fault complexes formalism~\cite{hillmann2024single} by taking mapping cones on idling spacetime volumes, which may be of independent interest.

Our work raises many further questions. First, when is it possible to efficiently devise a sparse auxiliary hypergraph $\CH$ with high modular expansion even if the original subcodes have low soundness? In the case of individual logical operator representatives this is straightforward by constructing an expander graph, but for more complicated parallel measurements involving many representatives the numerical construction of suitable hypergraphs is challenging for large codes, and brute-force branching techniques~\cite{zhang2024time,cowtan2025parallel} have formidable constant factor overheads.

Similarly, when $\CH$ is a graph one can appeal to the Decongestion Lemma~\cite[App.~A]{freedman2021building}. When $\CH$ is a small hypergraph one can typically devise a sparse cycle basis for $\CH$ by numerical brute-force~\cite{cross2024improved}, but for large codes such a method becomes impractical. Are there families of LDPC codes for which such bases are efficient to construct, or unnecessary entirely? We expect this to be true of any code that is geometrically local in a fixed spatial dimension. In fact, constant-time surgery operations raise the intriguing possibility of avoiding the need for a sparse cycle basis at all. This is because cycle checks do not need to be explicitly measured as they can be inferred from the initialization and readout steps, forming detectors that have a constant extent in the time direction. These detectors must provide fault-tolerance, but need not be strictly local. For instance, see the nonlocal hierarchical fault-tolerant detector complexes derived from local measurements in Ref.~\cite{Litinski2025Blocklet}. 

An issue that arises when using hypergraphs is that the method of universal adapters~\cite{swaroop2024universal} to connect heterogeneous subcodes between codeblocks does not generally apply. This is because the SkipTree algorithm assumes that the auxiliary system is a graph, rather than a hypergraph. We do not know whether generalisations of the algorithm can be invented to connect heterogeneous hypergraphs for product measurements, either in full generality or for specific cases.

{Here, we have focused on hypergraph measurements that only measure local checks and global logicals, i.e. they do not locally deform the code after the measurement. It is possible to relax these requirements, which leads to a setting of dynamical local codes. Could this lead to more addressable fast surgery without sacrificing performance?}

Our proofs concerning interpolated partial block reading in Sec.~\ref{sec:block_ltcs} used a strict definition of soundness, see Def.~\ref{def:quantum_soundness}, which we believe can be loosened to a `small set soundness', with a meaning similar to small set expansion or modular expansion. 
Our proofs regarding fault-distance require $d$ rounds of syndrome measurement (`padding') before and after the surgery protocols in order to provably protect against timelike faults extending from one timelike boundary to the other. We expect that these are typically unnecessary given sensible fault-tolerant operations performed before and after the surgery protocol. However, we have not fully characterized when these rounds of padding can be reduced in number or removed.

It is known that high-distance subcodes which address specific logical qubits exist for topological and tensor product codes \cite{huang2023homomorphic,xu2024fast}, but for other families of LDPC codes little is known. It would be interesting to consider algebraic methods to find high-distance subcodes for lifted or balanced product codes.

All our surgery methods used auxiliary hypergraphs, and there are other approaches to surgery which are `ancilla-free', meaning that they use no new data qubits~\cite{aasen2025geometrically,aasen2025topologically}. Such approaches fall under the umbrella of quotients of codes, see Refs.~\cite{cowtan2024css,poirson2025engineering}, rather than mapping cones~\cite{ide2024fault}. The arguments concerning time overhead appear to translate between the two approaches, but discerning a rigorous connection is left to future work. We suggest that one could take a pushout or similar quotient directly on the spacetime volumes described by fault complexes~\cite{hillmann2024single} in order to study such surgeries.

We do not know whether it is possible to take an arbitrary quantum LDPC code and provably measure an arbitrary set of commuting logical Pauli products in constant time overhead and linear (up to polylog) space overhead while maintaining fault-distance and the LDPC property. Such a proof is possible for the case of $\CO(d)$ time overhead \cite{cowtan2025parallel},
but the difficulties of finding appropriate subcodes and cycle bases require new methods. Conversely, it may be possible to show that such a generic protocol is impossible, see Ref.~\cite{Krishna2025Tradeoffs} for a related result in the classical setting.

We have developed a formalism for studying the phenomenological faults in generic CSS LDPC code surgery operations using fault complexes. We expect it is possible to prove thresholds of surgery on spacetime volumes in this formalism using combinatorial methods on the relevant fault complexes~\cite{he2025composable}.

Block reading entails metachecks, which means the addition of many more detectors to the decoding graph for a single round of measurement. We do not know whether there exist decoders which can efficiently decode such graphs quickly enough to be suitable for quantum hardware. In Ref.~\cite{aasen2025topologically} a powerset decoder was used to decode the a geometrically enhanced 4D toric code memory, which also has metachecks. However, this was far too slow for realistic hardware, and we do not know how well the powerset decoder would handle surgery operations. More realistic decoders for high-dimensional topological and tensor product codes were studied in Refs.~\cite{hillmann2024single,higgott2023improved}.

Finally, to bring the theory of block reading to application it will be necessary to perform many simulations and stability experiments~\cite{gidney2022stability}, and numerically determine pseudo-thresholds and logical error rates of the surgery operations.

\section{Acknowledgements}
We are grateful to Katie Chang and Louis Golowich for insightful discussions and for spotting an error in an earlier version of this paper. 
A.C.~is employed by Xanadu and would like to thank Ben Ide, Eric Sabo, Timo Hillmann, Michael Vasmer and Guillaume Dauphinais for helpful discussions.
Z.H.~is supported by MIT Department of Mathematics and thanks Harry Zhou, Tomas Jochym-O'Connor and Guanyu Zhu for helpful discussions.
D.J.W.~is supported by the Australian Research Council Discovery Early Career Research Award (DE220100625). 
T.J.Y.~thanks Qian Xu for sharing his perspective that homomorphic measurements and code surgery should be in some sense equivalent.
We used TikZit~\cite{TkZ} to generate the Tanner graph figures in this paper.

\bibliography{main}
\input{appendix}
\end{document}

%% file: alex_preamble.tex
\documentclass[a4paper, onecolumn, 11pt, unpublished]{quantumarticle}
\pdfoutput=1
\usepackage{breakurl}
\usepackage{stmaryrd}
\usepackage{amsthm,amstext,amssymb,amsmath}
\usepackage{epsfig}
\usepackage{slashed}
\usepackage{fancyhdr}
\usepackage{graphicx}
\usepackage{pdfpages}
\usepackage{float}
\usepackage{simplewick}
\usepackage{graphics,psfrag}
\usepackage{pstricks}
\usepackage{latexsym}
\usepackage{changepage}
\usepackage{caption}
\usepackage{paralist}
\usepackage{tikzit}
\input{basic.tikzstyles}
\input{quantum.tikzstyles}
\input{zx.tikzstyles}
\usepackage{graphicx}
\usepackage{array}
\usepackage{hyperref}
\usepackage{tikz-cd}
\usepackage{mathrsfs}
\usepackage{braket}
\usepackage{soul}

\usepackage{subcaption}

\usepackage{hhline}
\usepackage{algorithm}
\usepackage{algpseudocode}

\usepackage{scalefnt}
\usepackage[all]{xypic}
\usepackage{slashed}
\usepackage{extarrows}
\usepackage{lettrine}
\usepackage{booktabs}
\usepackage{paralist}
\usepackage{wrapfig}
\usepackage{xcolor}
\usepackage{dsfont}

\usepackage{thmtools}

\DeclareMathAlphabet{\mathcal}{OMS}{cmsy}{m}{n}
\newtheorem{lemma}{Lemma}[section] 

\newtheorem{corollary}[lemma]{Corollary}
\newtheorem{example}[lemma]{Example}
\newtheorem{theorem}[lemma]{Theorem}

\newtheorem{definition}[lemma]{Definition}
\newtheorem{remark}[lemma]{Remark}

\renewcommand{\imath}{\mathrm{i}}

\newcommand{\CC}{\hbox{{$\mathcal C$}}}

\newcommand{\CF}{\hbox{{$\mathcal F$}}}
\newcommand{\CG}{\hbox{{$\mathcal G$}}}

\newcommand{\CV}{\hbox{{$\mathcal V$}}}
\newcommand{\CH}{\hbox{{$\mathcal H$}}}
\newcommand{\CS}{\hbox{{$\mathcal S$}}}
\newcommand{\CM}{\hbox{{$\mathcal M$}}}

\newcommand{\CP}{\hbox{{$\mathcal P$}}}
\newcommand{\CQ}{\hbox{{$\mathcal Q$}}}
\newcommand{\CR}{\hbox{{$\mathcal R$}}}
\newcommand{\CU}{\hbox{{$\mathcal U$}}}
\newcommand{\CE}{\hbox{{$\mathcal E$}}}
\newcommand{\CO}{\hbox{{$\mathcal O$}}}
\newcommand{\CJ}{\hbox{{$\mathcal J$}}}

\newcommand{\C}{\mathbb{C}}
\newcommand{\R}{\mathbb{R}}
\newcommand{\F}{\mathbb{F}}
\newcommand{\Z}{\mathbb{Z}}
\newcommand{\N}{\mathbb{N}}

\newcommand{\del}{\partial}

\newcommand{\tens}{\mathop{{\otimes}}}

\newcommand{\im}{\mathrm{im}}
\newcommand{\cone}{\mathrm{cone}}

\newcommand{\supp}{{\rm supp}}

\newsavebox{\pullback}
\sbox\pullback{%
\begin{tikzpicture}%
\draw (0,0) -- (2ex,0ex);%
\draw (2ex,0ex) -- (2ex,2ex);%
\end{tikzpicture}}

\newsavebox{\pushout}
\sbox\pushout{%
\begin{tikzpicture}%
\draw (0,0) -- (0,2ex);%
\draw (0,2ex) -- (2ex,2ex);%
\end{tikzpicture}}

\setstcolor{red}

\usepackage{cite}

\newcommand{\sunny}[1]{\textcolor{blue}{}}
\newcommand{\ted}[1]{\textcolor{teal}{}}
\newcommand{\alex}[1]{\textcolor{magenta}{}}
\newcommand{\dom}[1]{\textcolor{brown}{}}

\newcommand{\textnode}[1]{\parbox{2.8cm}{\centering #1}}

\usepackage[numbers, sort, compress]{natbib}

%% file: basic.tikzstyles
\tikzstyle{boundary vertex}=[inner sep=0mm, minimum size=1mm, shape=circle, draw=black, fill=black]
\tikzstyle{grey_dot}=[fill={rgb,255: red,191; green,191; blue,191}, draw={rgb,255: red,191; green,191; blue,191}, shape=circle, minimum size=1mm, inner sep=0mm]
\tikzstyle{blue_dot}=[fill={rgb,255: red,202; green,251; blue,255}, draw=black, shape=circle, minimum size=1.5mm, inner sep=0mm]
\tikzstyle{white_dot}=[fill=white, draw=black, shape=circle, minimum size=1.5mm, inner sep=0mm]
\tikzstyle{red_dot}=[fill=red, draw=black, shape=circle, minimum size=1.5mm, inner sep=0mm]
\tikzstyle{box}=[fill=white, draw=black, shape=rectangle]
\tikzstyle{scalable_dot}=[fill=white, draw=black, shape=circle, minimum size=6.5mm, inner sep=0mm]
\tikzstyle{scalable_box}=[fill=white, draw=black, shape=rectangle, minimum height=6.5mm, minimum width=6.5mm]
\tikzstyle{scalable_meta_X}=[fill={rgb,255: red,255; green,170; blue,162}, draw=black, shape=diamond, minimum height=8mm, minimum width=8mm]
\tikzstyle{scalable_meta_Z}=[fill={rgb,255: red,112; green,203; blue,207}, draw=black, shape=diamond, minimum height=8mm, minimum width=8mm]
\tikzstyle{X_detector}=[fill={rgb,255: red,191; green,0; blue,64}, draw=black, shape=rectangle]
\tikzstyle{Z_fault}=[fill={rgb,255: red,112; green,203; blue,207}, draw=black, shape=circle, minimum size=1mm, inner sep=0.75mm]
\tikzstyle{X_fault}=[fill={rgb,255: red,210; green,203; blue,0}, draw=black, shape=circle, minimum size=1mm, inner sep=0.75mm]
\tikzstyle{Z_detector}=[fill={rgb,255: red,105; green,105; blue,255}, draw=black, shape=rectangle]
\tikzstyle{scalable_meta_mixed}=[fill={rgb,255: red,255; green,105; blue,255}, draw=black, shape=diamond, minimum height=8mm, minimum width=8mm]
\tikzstyle{green_fault}=[fill={rgb,255: red,4; green,220; blue,0}, draw=black, shape=circle, minimum size=1mm, inner sep=0.75mm]

\tikzstyle{arrow}=[->]
\tikzstyle{red_arrow}=[->, draw=red]
\tikzstyle{cyan_arrow}=[->, draw=cyan]
\tikzstyle{red_dash}=[-, dashed, draw=red]
\tikzstyle{grey dash}=[-, fill=none, draw={rgb,255: red,191; green,191; blue,191}, dashed]
\tikzstyle{dashed arrow}=[->, dashed]
\tikzstyle{blue_edge}=[-, draw=cyan]
\tikzstyle{red_edge}=[-, draw=red]
\tikzstyle{green_edge}=[-, fill=none, draw=green]
\tikzstyle{green_dash}=[-, draw=green, dashed]
\tikzstyle{X_edge}=[-, dashed, draw={rgb,255: red,191; green,0; blue,64}]
\tikzstyle{Z_edge}=[dashed, draw={rgb,255: red,105; green,105; blue,255}, -]

%% file: quantum.tikzstyles
\tikzstyle{gate}=[shape=rectangle, text height=1.5ex, text depth=0.25ex, yshift=0.5mm, fill=white, draw=black, minimum height=5mm, yshift=-0.5mm, minimum width=5mm, font={\small}, tikzit category=circuit]
\tikzstyle{big gate}=[shape=rectangle, text height=1.5ex, text depth=0.25ex, yshift=0.5mm, fill=white, draw=black, minimum height=18mm, yshift=-0.5mm, minimum width=5mm, font={\small}, tikzit category=circuit]
\tikzstyle{Z dot}=[inner sep=0mm, minimum size=2mm, shape=circle, draw=black, fill={rgb,255: red,221; green,255; blue,221}, tikzit category=zx]
\tikzstyle{Z phase dot}=[minimum size=5mm, font={\footnotesize\boldmath}, shape=rectangle, rounded corners=2mm, inner sep=0.2mm, outer sep=-2mm, scale=0.8, tikzit shape=circle, draw=black, fill={rgb,255: red,221; green,255; blue,221}, tikzit draw=blue, tikzit category=zx]
\tikzstyle{X dot}=[Z dot, shape=circle, draw=black, fill={rgb,255: red,255; green,136; blue,136}, tikzit category=zx]
\tikzstyle{X phase dot}=[Z phase dot, tikzit shape=circle, tikzit draw=blue, fill={rgb,255: red,255; green,136; blue,136}, font={\footnotesize\boldmath}, tikzit category=zx]
\tikzstyle{hadamard}=[fill=yellow, draw=black, shape=rectangle, inner sep=0.6mm, minimum height=1.5mm, minimum width=1.5mm, tikzit category=zx]
\tikzstyle{paulibox}=[fill={rgb,255: red,221; green,221; blue,255}, draw=black, shape=rectangle, inner sep=0.6mm, minimum height=5mm, minimum width=5mm, font={\footnotesize}, text height=1.5ex, text depth=0.25ex, tikzit category=zx]
\tikzstyle{vertex}=[inner sep=0mm, minimum size=1mm, shape=circle, draw=black, fill=black, tikzit category=misc]
\tikzstyle{vertex set}=[inner sep=0mm, minimum size=1mm, shape=circle, draw=black, fill=white, font={\footnotesize\boldmath}, tikzit category=misc]
\tikzstyle{small black dot}=[fill=black, draw=black, shape=circle, inner sep=0pt, minimum width=1.2mm, tikzit category=circuit]
\tikzstyle{cnot ctrl}=[fill=black, draw=black, shape=circle, inner sep=0pt, minimum width=1.2mm, tikzit category=circuit]
\tikzstyle{cnot targ}=[fill=white, draw=white, shape=circle, tikzit category=circuit, label={center:$\oplus$}, inner sep=0pt, minimum width=2.1mm, tikzit fill={rgb,255: red,102; green,204; blue,255}, tikzit draw=black]
\tikzstyle{ket}=[fill=white, draw=black, shape=regular polygon, regular polygon sides=3, regular polygon rotate=-30, scale=0.7, inner sep=1pt, tikzit category=circuit, tikzit shape=rectangle, tikzit fill=green]
\tikzstyle{bra}=[fill=white, draw=black, shape=regular polygon, regular polygon sides=3, regular polygon rotate=30, scale=0.7, inner sep=1pt, tikzit category=circuit, tikzit shape=rectangle, tikzit fill=red]
\tikzstyle{scalar}=[shape=rectangle, text height=1.5ex, text depth=0.25ex, yshift=0.5mm, fill=white, draw=black, minimum height=5mm, yshift=-0.5mm, minimum width=5mm, font={\small}]
\tikzstyle{clabel}=[fill=white, draw=none, shape=rectangle, tikzit fill={rgb,255: red,56; green,255; blue,242}, font={\footnotesize}, inner sep=1pt, tikzit category=labels]
\tikzstyle{empty diagram}=[draw={gray!40!white}, dashed, shape=rectangle, minimum width=1cm, minimum height=1cm, tikzit category=misc]
\tikzstyle{bigger gate}=[fill=white, draw=black, shape=rectangle, minimum height=28mm, minimum width=5mm, font={\small}, tikzit category=circuit]
\tikzstyle{wider gate}=[fill=white, draw=black, shape=rectangle, minimum height=28mm, minimum width=9mm, font={\small}]
\tikzstyle{biggest gate}=[fill=white, draw=black, shape=rectangle, minimum height=36mm, minimum width=9mm, font={\small}]

\tikzstyle{hadamard edge}=[-, dashed, dash pattern=on 2pt off 0.5pt, thick, draw={rgb,255: red,68; green,136; blue,255}]
\tikzstyle{box edge}=[-, dashed, dash pattern=on 2pt off 0.5pt, thick, draw={rgb,255: red,203; green,192; blue,225}]
\tikzstyle{brace edge}=[-, tikzit draw=blue, decorate, decoration={brace,amplitude=1mm,raise=-1mm}]
\tikzstyle{diredge}=[->]
\tikzstyle{double edge}=[-, double, shorten <=-1mm, shorten >=-1mm, double distance=2pt]
\tikzstyle{gray edge}=[-, {gray!60!white}]
\tikzstyle{pointer edge}=[->, very thick, gray]
\tikzstyle{boldedge}=[-, line width=1.6pt, shorten <=-0.17mm, shorten >=-0.17mm]

%% file: zx.tikzstyles
\tikzstyle{boundary vertex}=[inner sep=0mm, minimum size=1mm, shape=circle, draw=black, fill=black]
\tikzstyle{grey_dot}=[fill={rgb,255: red,191; green,191; blue,191}, draw={rgb,255: red,191; green,191; blue,191}, shape=circle, minimum size=1mm, inner sep=0mm]
\tikzstyle{blue_dot}=[fill={rgb,255: red,202; green,251; blue,255}, draw=black, shape=circle, minimum size=1mm, inner sep=0mm]
\tikzstyle{white_dot}=[fill=white, draw=black, shape=circle, minimum size=1.5mm, inner sep=0mm]
\tikzstyle{gate}=[shape=rectangle, text height=1.5ex, text depth=0.25ex, yshift=0.5mm, fill=white, draw=black, minimum height=5mm, yshift=-0.5mm, minimum width=5mm, font={\small}, tikzit category=circuit]
\tikzstyle{big gate}=[shape=rectangle, text height=1.5ex, text depth=0.25ex, yshift=0.5mm, fill=white, draw=black, minimum height=18mm, yshift=-0.5mm, minimum width=5mm, font={\small}, tikzit category=circuit]
\tikzstyle{Z dot}=[inner sep=0mm, minimum size=2mm, shape=circle, draw=black, fill={rgb,255: red,221; green,255; blue,221}, tikzit category=zx]
\tikzstyle{Z phase dot}=[minimum size=5mm, font={\footnotesize\boldmath}, shape=rectangle, rounded corners=2mm, inner sep=0.2mm, outer sep=-2mm, scale=0.8, tikzit shape=circle, draw=black, fill={rgb,255: red,221; green,255; blue,221}, tikzit draw=blue, tikzit category=zx]
\tikzstyle{X dot}=[Z dot, shape=circle, draw=black, fill={rgb,255: red,255; green,136; blue,136}, tikzit category=zx]
\tikzstyle{X phase dot}=[Z phase dot, tikzit shape=circle, tikzit draw=blue, fill={rgb,255: red,255; green,136; blue,136}, font={\footnotesize\boldmath}, tikzit category=zx]
\tikzstyle{hadamard}=[fill=yellow, draw=black, shape=rectangle, inner sep=0.6mm, minimum height=1.5mm, minimum width=1.5mm, tikzit category=zx]
\tikzstyle{paulibox}=[fill={rgb,255: red,221; green,221; blue,255}, draw=black, shape=rectangle, inner sep=0.6mm, minimum height=5mm, minimum width=5mm, font={\footnotesize}, text height=1.5ex, text depth=0.25ex, tikzit category=zx]
\tikzstyle{vertex}=[inner sep=0mm, minimum size=1mm, shape=circle, draw=black, fill=black, tikzit category=misc]
\tikzstyle{vertex set}=[inner sep=0mm, minimum size=1mm, shape=circle, draw=black, fill=white, font={\footnotesize\boldmath}, tikzit category=misc]
\tikzstyle{small black dot}=[fill=black, draw=black, shape=circle, inner sep=0pt, minimum width=1.2mm, tikzit category=circuit]
\tikzstyle{cnot ctrl}=[fill=black, draw=black, shape=circle, inner sep=0pt, minimum width=1.2mm, tikzit category=circuit]
\tikzstyle{cnot targ}=[fill=white, draw=white, shape=circle, tikzit category=circuit, label={center:$\oplus$}, inner sep=0pt, minimum width=2.1mm, tikzit fill={rgb,255: red,102; green,204; blue,255}, tikzit draw=black]
\tikzstyle{ket}=[fill=white, draw=black, shape=regular polygon, regular polygon sides=3, regular polygon rotate=-30, scale=0.7, inner sep=1pt, tikzit category=circuit, tikzit shape=rectangle, tikzit fill=green]
\tikzstyle{bra}=[fill=white, draw=black, shape=regular polygon, regular polygon sides=3, regular polygon rotate=30, scale=0.7, inner sep=1pt, tikzit category=circuit, tikzit shape=rectangle, tikzit fill=red]
\tikzstyle{scalar}=[shape=rectangle, text height=1.5ex, text depth=0.25ex, yshift=0.5mm, fill=white, draw=black, minimum height=5mm, yshift=-0.5mm, minimum width=5mm, font={\small}]
\tikzstyle{clabel}=[fill=white, draw=none, shape=rectangle, tikzit fill={rgb,255: red,56; green,255; blue,242}, font={\footnotesize}, inner sep=1pt, tikzit category=labels]
\tikzstyle{empty diagram}=[draw={gray!40!white}, dashed, shape=rectangle, minimum width=1cm, minimum height=1cm, tikzit category=misc]
\tikzstyle{rmat}=[draw, signal, fill={gray!30}, signal to=east, signal from=west, inner sep=1.5pt, minimum height=9pt]
\tikzstyle{lmat}=[draw, signal, fill={gray!30}, signal to=west, signal from=east, inner sep=1.5pt, minimum height=9pt]
\tikzstyle{ggreen}=[fill=green, draw=black, shape=circle, tikzit category=SZX, tikzit fill=green, tikzit draw=black, line width=1pt, inner sep=2pt]
\tikzstyle{gred}=[fill=red, draw=black, shape=circle, rounded corners=2mm, tikzit category=SZX, inner sep=2pt, tikzit fill=red, line width=1pt]
\tikzstyle{divide}=[regular polygon, regular polygon sides=3, shape border rotate=90, draw=black, fill={gray!30}, inner sep=1.6pt, tikzit category=scal, rounded corners=0.8mm]
\tikzstyle{gather}=[fill={gray!30}, draw=black, tikzit category=scal, rounded corners=0.8mm, regular polygon, regular polygon sides=3, shape border rotate=-90, inner sep=1.6pt]
\tikzstyle{A}=[fill=white, shape=circle, tikzit category=scal, inner sep=1pt]
\tikzstyle{umat}=[draw, signal, fill={gray!30}, signal to=north, signal from=south, inner sep=1.5pt, minimum width=9pt]
\tikzstyle{dmat}=[draw, signal, fill={gray!30}, signal to=south, signal from=north, inner sep=1.5pt, minimum width=9pt]

\tikzstyle{arrow}=[->]
\tikzstyle{red_arrow}=[->, draw=red]
\tikzstyle{cyan_arrow}=[->, draw=cyan]
\tikzstyle{red_dash}=[-, dashed, draw=red]
\tikzstyle{grey dash}=[-, fill=none, draw={rgb,255: red,191; green,191; blue,191}, dashed]
\tikzstyle{dashed arrow}=[->, dashed]
\tikzstyle{blue_edge}=[-, draw={rgb,255: red,46; green,126; blue,255}]
\tikzstyle{red_edge}=[-, draw=red]
\tikzstyle{hadamard edge}=[-, dashed, dash pattern=on 2pt off 0.5pt, thick, draw={rgb,255: red,68; green,136; blue,255}]
\tikzstyle{box edge}=[-, dashed, dash pattern=on 2pt off 0.5pt, thick, draw={rgb,255: red,203; green,192; blue,225}]
\tikzstyle{brace edge}=[-, tikzit draw=blue, decorate, decoration={brace,amplitude=1mm,raise=-1mm}]
\tikzstyle{diredge}=[->]
\tikzstyle{double edge}=[-, double, shorten <=-1mm, shorten >=-1mm, double distance=2pt]
\tikzstyle{gray edge}=[-, {gray!60!white}]
\tikzstyle{pointer edge}=[->, very thick, gray]
\tikzstyle{boldedge}=[-, line width=1.6pt, shorten <=-0.17mm, shorten >=-0.17mm]
\tikzstyle{very thick}=[-, line width=1pt, tikzit draw=red]

%% file: Figures-Tikz/tanner_eg1.tikz
\begin{tikzpicture}
	\begin{pgfonlayer}{nodelayer}
		\node [style={scalable_box}] (5) at (-3.125, 0.125) {};
		\node [style={scalable_dot}] (3) at (2.9, 0.1) {};
		\node [style={scalable_box}] (0) at (-3, 0) {};
		\node [style={scalable_dot}] (1) at (3, 0) {};
		\node [style=none] (2) at (0, 0.5) {$[H_X|H_Z]$};
	\end{pgfonlayer}
	\begin{pgfonlayer}{edgelayer}
		\draw (0) to (1);
	\end{pgfonlayer}
\end{tikzpicture}

%% file: Figures-Tikz/tanner_eg2.tikz
\begin{tikzpicture}
	\begin{pgfonlayer}{nodelayer}
		\node [style={scalable_box}] (0) at (-4, 0.125) {};
		\node [style={scalable_dot}] (1) at (0.025, 0.1) {};
		\node [style={scalable_box}] (2) at (-3.875, 0) {$X$};
		\node [style={scalable_dot}] (3) at (0.125, 0) {};
		\node [style={scalable_box}] (4) at (3.875, 0.125) {};
		\node [style={scalable_box}] (5) at (4, 0) {$Z$};
		\node [style=none] (6) at (-2, 0.5) {$H_X$};
		\node [style=none] (7) at (2, 0.5) {$H_Z$};
	\end{pgfonlayer}
	\begin{pgfonlayer}{edgelayer}
		\draw (2) to (3);
		\draw (3) to (5);
	\end{pgfonlayer}
\end{tikzpicture}

%% file: Figures-Tikz/hypergraph_prod.tikz
\begin{tikzpicture}
	\begin{pgfonlayer}{nodelayer}
		\node [style={scalable_box}] (0) at (-2.1, 2.125) {};
		\node [style={scalable_box}] (1) at (-1.975, 2) {$X$};
		\node [style={scalable_dot}] (2) at (-2.1, -1.9) {};
		\node [style={scalable_dot}] (3) at (-2, -2) {};
		\node [style={scalable_box}] (8) at (1.875, -1.875) {};
		\node [style={scalable_box}] (9) at (2, -2) {$Z$};
		\node [style={scalable_dot}] (10) at (1.9, 2.1) {};
		\node [style={scalable_dot}] (11) at (2, 2) {};
		\node [style={scalable_box}] (12) at (5.925, 2.125) {};
		\node [style={scalable_box}] (13) at (6.05, 2) {$X$};
		\node [style={scalable_dot}] (14) at (5.925, -1.9) {};
		\node [style={scalable_dot}] (15) at (6.025, -2) {};
		\node [style=none] (16) at (-1.325, -2.75) {$\mathcal{L}$};
		\node [style=none] (17) at (-1.25, 0) {$\del_1$};
		\node [style=none] (18) at (2.925, 0) {$\del_1^T$};
		\node [style=none] (19) at (6.95, 0) {$\del_1$};
		\node [style=none] (20) at (0, -1.5) {$I$};
		\node [style=none] (21) at (0.05, 2.5) {$I$};
		\node [style=none] (22) at (4.05, 2.5) {$I$};
		\node [style=none] (23) at (4.05, -1.5) {$I$};
	\end{pgfonlayer}
	\begin{pgfonlayer}{edgelayer}
		\draw (1) to (3);
		\draw (1) to (11);
		\draw (3) to (9);
		\draw (9) to (11);
		\draw (13) to (15);
		\draw (11) to (13);
		\draw (9) to (15);
	\end{pgfonlayer}
\end{tikzpicture}

%% file: Figures-Tikz/CSS_code_metachecks.tikz
\begin{tikzpicture}
	\begin{pgfonlayer}{nodelayer}
		\node [style={scalable_meta_X}] (16) at (7.825, 0.025) {};
		\node [style={scalable_box}] (0) at (3.9, 0.1) {};
		\node [style={scalable_dot}] (1) at (-0.125, 0.075) {};
		\node [style={scalable_dot}] (2) at (0, 0) {};
		\node [style={scalable_box}] (3) at (4, 0) {$X$};
		\node [style={scalable_box}] (4) at (-4.1, 0.1) {};
		\node [style={scalable_box}] (5) at (-4, 0) {$Z$};
		\node [style=none] (6) at (2, 0.5) {$H_X$};
		\node [style=none] (7) at (-2, 0.5) {$H_Z$};
		\node [style={scalable_meta_Z}] (11) at (-8.2, 0.025) {};
		\node [style={scalable_meta_Z}] (12) at (-8.025, 0) {};
		\node [style={scalable_meta_X}] (15) at (8, 0) {};
		\node [style=none] (17) at (-6, 0.5) {$\mathfrak{M}_Z$};
		\node [style=none] (18) at (6, 0.5) {$\mathfrak{M}_X$};
	\end{pgfonlayer}
	\begin{pgfonlayer}{edgelayer}
		\draw (2) to (5);
		\draw (2) to (3);
		\draw (5) to (12);
		\draw (15) to (3);
	\end{pgfonlayer}
\end{tikzpicture}

%% file: Figures-Tikz/two_blocks.tikz
\begin{tikzpicture}
	\begin{pgfonlayer}{nodelayer}
		\node [style={scalable_box}] (0) at (-4.1, 4.1) {};
		\node [style={scalable_dot}] (1) at (-4.1, 0.075) {};
		\node [style={scalable_dot}] (2) at (-4, 0) {};
		\node [style={scalable_box}] (3) at (-4, 4) {$X$};
		\node [style={scalable_box}] (4) at (-4.1, -3.9) {};
		\node [style={scalable_box}] (5) at (-4, -4) {$Z$};
		\node [style=none] (6) at (-5, 2) {$H_X$};
		\node [style=none] (7) at (-5, -2) {$H_Z$};
		\node [style=none] (8) at (-5.5, 0) {$Q$};
		\node [style={scalable_box}] (9) at (3.9, 4.1) {};
		\node [style={scalable_dot}] (10) at (3.9, 0.075) {};
		\node [style={scalable_dot}] (11) at (4, 0) {};
		\node [style={scalable_box}] (12) at (4, 4) {$X$};
		\node [style={scalable_box}] (13) at (3.9, -3.9) {};
		\node [style={scalable_box}] (14) at (4, -4) {$Z$};
		\node [style=none] (15) at (5, 2) {$H_X$};
		\node [style=none] (16) at (5, -2) {$H_Z$};
		\node [style=none] (17) at (5.5, 0) {$Q'$};
	\end{pgfonlayer}
	\begin{pgfonlayer}{edgelayer}
		\draw (2) to (5);
		\draw (2) to (3);
		\draw (11) to (14);
		\draw (11) to (12);
	\end{pgfonlayer}
\end{tikzpicture}

%% file: Figures-Tikz/two_blocks_2.tikz
\begin{tikzpicture}
	\begin{pgfonlayer}{nodelayer}
		\node [style={scalable_box}] (0) at (-4.1, 4.1) {};
		\node [style={scalable_dot}] (1) at (-4.1, 0.075) {};
		\node [style={scalable_dot}] (2) at (-4, 0) {};
		\node [style={scalable_box}] (3) at (-4, 4) {$X$};
		\node [style={scalable_box}] (4) at (-4.1, -3.9) {};
		\node [style={scalable_box}] (5) at (-4, -4) {$Z$};
		\node [style=none] (6) at (-5, 2) {$H_X$};
		\node [style=none] (7) at (-5, -2) {$H_Z$};
		\node [style=none] (8) at (-5.5, 0) {$Q$};
		\node [style={scalable_box}] (9) at (3.9, 4.1) {};
		\node [style={scalable_dot}] (10) at (3.9, 0.075) {};
		\node [style={scalable_dot}] (11) at (4, 0) {};
		\node [style={scalable_box}] (12) at (4, 4) {$X$};
		\node [style={scalable_box}] (13) at (3.9, -3.9) {};
		\node [style={scalable_box}] (14) at (4, -4) {$Z$};
		\node [style=none] (15) at (5, 2) {$H_X$};
		\node [style=none] (16) at (5, -2) {$H_Z$};
		\node [style=none] (17) at (5.5, 0) {$Q'$};
		\node [style={scalable_box}] (18) at (-0.1, 0.1) {};
		\node [style={scalable_box}] (19) at (0, 0) {$Z$};
		\node [style={scalable_dot}] (20) at (-0.1, 4.075) {};
		\node [style={scalable_dot}] (21) at (0, 4) {};
		\node [style=none] (22) at (-2, 4.5) {$I$};
		\node [style=none] (23) at (2, 4.5) {$I$};
		\node [style=none] (24) at (-2, -0.75) {$I$};
		\node [style=none] (25) at (2, -0.75) {$I$};
		\node [style=none] (26) at (0.75, 2) {$G$};
		\node [style=none] (27) at (0, 5.5) {$\CE$};
		\node [style=none] (28) at (0, -1.5) {$\CV$};
	\end{pgfonlayer}
	\begin{pgfonlayer}{edgelayer}
		\draw (2) to (5);
		\draw (2) to (3);
		\draw (11) to (14);
		\draw (11) to (12);
		\draw (2) to (19);
		\draw (19) to (11);
		\draw (21) to (12);
		\draw (21) to (3);
		\draw (21) to (19);
	\end{pgfonlayer}
\end{tikzpicture}

%% file: Figures-Tikz/two_blocks_3.tikz
\begin{tikzpicture}
	\begin{pgfonlayer}{nodelayer}
		\node [style={scalable_box}] (0) at (-4.1, 4.1) {};
		\node [style={scalable_dot}] (1) at (-4.1, 0.075) {};
		\node [style={scalable_dot}] (2) at (-4, 0) {};
		\node [style={scalable_box}] (3) at (-4, 4) {$X$};
		\node [style={scalable_box}] (4) at (-4.1, -3.9) {};
		\node [style={scalable_box}] (5) at (-4, -4) {$Z$};
		\node [style=none] (6) at (-5, 2) {$H_X$};
		\node [style=none] (7) at (-5, -2) {$H_Z$};
		\node [style=none] (8) at (-5.5, 0) {$Q$};
		\node [style={scalable_box}] (9) at (3.9, 4.1) {};
		\node [style={scalable_dot}] (10) at (3.9, 0.075) {};
		\node [style={scalable_dot}] (11) at (4, 0) {};
		\node [style={scalable_box}] (12) at (4, 4) {$X$};
		\node [style={scalable_box}] (13) at (3.9, -3.9) {};
		\node [style={scalable_box}] (14) at (4, -4) {$Z$};
		\node [style=none] (15) at (5, 2) {$H_X$};
		\node [style=none] (16) at (5, -2) {$H_Z$};
		\node [style=none] (17) at (5.5, 0) {$Q'$};
		\node [style={scalable_box}] (18) at (-0.1, 0.1) {};
		\node [style={scalable_box}] (19) at (0, 0) {$Z$};
		\node [style={scalable_dot}] (20) at (-0.1, 4.075) {};
		\node [style={scalable_dot}] (21) at (0, 4) {};
		\node [style=none] (22) at (-2, 4.5) {$I$};
		\node [style=none] (23) at (2, 4.5) {$I$};
		\node [style=none] (24) at (-2, -0.75) {$I$};
		\node [style=none] (25) at (2, -0.75) {$I$};
		\node [style=none] (26) at (0.75, 2) {$G$};
		\node [style=none] (27) at (0, 5.5) {$\CE$};
		\node [style={scalable_meta_Z}] (32) at (-0.175, -3.975) {};
		\node [style={scalable_meta_Z}] (33) at (0, -4) {};
		\node [style=none] (34) at (2, -4.75) {$I$};
		\node [style=none] (35) at (-2, -4.75) {$I$};
		\node [style=none] (36) at (0.75, -2) {$\mathfrak{G}$};
		\node [style=none] (37) at (-1, 1) {$\CV$};
	\end{pgfonlayer}
	\begin{pgfonlayer}{edgelayer}
		\draw (2) to (5);
		\draw (2) to (3);
		\draw (11) to (14);
		\draw (11) to (12);
		\draw (2) to (19);
		\draw (19) to (11);
		\draw (21) to (12);
		\draw (21) to (3);
		\draw (21) to (19);
		\draw (19) to (33);
		\draw (33) to (14);
		\draw (33) to (5);
	\end{pgfonlayer}
\end{tikzpicture}

%% file: Figures-Tikz/two_blocks_partial1.tikz
\begin{tikzpicture}
	\begin{pgfonlayer}{nodelayer}
		\node [style={scalable_box}] (0) at (-4.1, 4.1) {};
		\node [style={scalable_dot}] (1) at (-4.1, 0.075) {};
		\node [style={scalable_dot}] (2) at (-4, 0) {};
		\node [style={scalable_box}] (3) at (-4, 4) {$X$};
		\node [style={scalable_box}] (4) at (-4.1, -3.9) {};
		\node [style={scalable_box}] (5) at (-4, -4) {$Z$};
		\node [style=none] (6) at (-5, 2) {$H_X$};
		\node [style=none] (7) at (-5, -2) {$H_Z$};
		\node [style=none] (8) at (-5.5, 0) {$Q$};
		\node [style={scalable_box}] (9) at (3.9, 4.1) {};
		\node [style={scalable_dot}] (10) at (3.9, 0.075) {};
		\node [style={scalable_dot}] (11) at (4, 0) {};
		\node [style={scalable_box}] (12) at (4, 4) {$X$};
		\node [style={scalable_box}] (13) at (3.9, -3.9) {};
		\node [style={scalable_box}] (14) at (4, -4) {$Z$};
		\node [style=none] (15) at (5, 2) {$H_X'$};
		\node [style=none] (16) at (5, -2) {$H_Z'$};
		\node [style=none] (17) at (5.5, 0) {$Q'$};
	\end{pgfonlayer}
	\begin{pgfonlayer}{edgelayer}
		\draw (2) to (5);
		\draw (2) to (3);
		\draw (11) to (14);
		\draw (11) to (12);
	\end{pgfonlayer}
\end{tikzpicture}

%% file: Figures-Tikz/two_blocks_partial2.tikz
\begin{tikzpicture}
	\begin{pgfonlayer}{nodelayer}
		\node [style={scalable_box}] (0) at (-4.1, 4.1) {};
		\node [style={scalable_dot}] (1) at (-4.1, 0.075) {};
		\node [style={scalable_dot}] (2) at (-4, 0) {};
		\node [style={scalable_box}] (3) at (-4, 4) {$X$};
		\node [style={scalable_box}] (4) at (-4.1, -3.9) {};
		\node [style={scalable_box}] (5) at (-4, -4) {$Z$};
		\node [style=none] (6) at (-5, 2) {$H_X$};
		\node [style=none] (7) at (-5, -2) {$H_Z$};
		\node [style=none] (8) at (-5.5, 0) {$Q$};
		\node [style={scalable_box}] (9) at (3.9, 4.1) {};
		\node [style={scalable_dot}] (10) at (3.9, 0.075) {};
		\node [style={scalable_dot}] (11) at (4, 0) {};
		\node [style={scalable_box}] (12) at (4, 4) {$X$};
		\node [style={scalable_box}] (13) at (3.9, -3.9) {};
		\node [style={scalable_box}] (14) at (4, -4) {$Z$};
		\node [style=none] (15) at (5, 2) {$H_X'$};
		\node [style=none] (16) at (5, -2) {$H_Z'$};
		\node [style=none] (17) at (5.5, 0) {$Q'$};
		\node [style={scalable_box}] (18) at (-0.1, 0.1) {};
		\node [style={scalable_box}] (19) at (0, 0) {$Z$};
		\node [style={scalable_dot}] (20) at (-0.1, 4.075) {};
		\node [style={scalable_dot}] (21) at (0, 4) {};
		\node [style=none] (22) at (-2, 4.5) {$M_l$};
		\node [style=none] (23) at (2, 4.5) {$M_r$};
		\node [style=none] (24) at (-2, -0.75) {$F_l$};
		\node [style=none] (25) at (2, -0.75) {$F_r$};
		\node [style=none] (26) at (0.75, 2) {$G$};
		\node [style=none] (27) at (0, 5.5) {$\CE$};
		\node [style=none] (28) at (0, -1.5) {$\CV$};
	\end{pgfonlayer}
	\begin{pgfonlayer}{edgelayer}
		\draw (2) to (5);
		\draw (2) to (3);
		\draw (11) to (14);
		\draw (11) to (12);
		\draw (2) to (19);
		\draw (19) to (11);
		\draw (21) to (12);
		\draw (21) to (3);
		\draw (21) to (19);
	\end{pgfonlayer}
\end{tikzpicture}

%% file: Figures-Tikz/three_blocks_partial.tikz
\begin{tikzpicture}
	\begin{pgfonlayer}{nodelayer}
		\node [style={scalable_box}] (0) at (-5.1, 1.1) {};
		\node [style={scalable_dot}] (1) at (-5.1, -2.925) {};
		\node [style={scalable_dot}] (2) at (-5, -3) {};
		\node [style={scalable_box}] (3) at (-5, 1) {$X$};
		\node [style={scalable_box}] (4) at (-5.1, -6.9) {};
		\node [style={scalable_box}] (5) at (-5, -7) {$Z$};
		\node [style=none] (6) at (-6, -1) {$H_X$};
		\node [style=none] (7) at (-6, -5) {$H_Z$};
		\node [style=none] (8) at (-6.5, -3) {$Q$};
		\node [style={scalable_box}] (9) at (6.9, 8.1) {};
		\node [style={scalable_dot}] (10) at (6.9, 4.075) {};
		\node [style={scalable_dot}] (11) at (7, 4) {};
		\node [style={scalable_box}] (12) at (7, 8) {$X$};
		\node [style={scalable_box}] (13) at (6.9, 0.1) {};
		\node [style={scalable_box}] (14) at (7, 0) {$Z$};
		\node [style=none] (15) at (8, 6) {$H_X'$};
		\node [style=none] (16) at (8, 2) {$H_Z'$};
		\node [style=none] (17) at (8.5, 4) {$Q'$};
		\node [style={scalable_box}] (18) at (-0.1, 0.1) {};
		\node [style={scalable_box}] (19) at (0, 0) {$Z$};
		\node [style={scalable_dot}] (20) at (-0.1, 4.075) {};
		\node [style={scalable_dot}] (21) at (0, 4) {};
		\node [style=none] (22) at (-2.75, 3) {$f_0$};
		\node [style=none] (24) at (-2.5, -2.25) {$f_1^\intercal$};
		\node [style=none] (26) at (-0.75, 2) {$G$};
		\node [style={scalable_box}] (29) at (3.9, -1.9) {};
		\node [style={scalable_dot}] (30) at (3.9, -5.925) {};
		\node [style={scalable_dot}] (31) at (4, -6) {};
		\node [style={scalable_box}] (32) at (4, -2) {$X$};
		\node [style={scalable_box}] (33) at (3.9, -9.9) {};
		\node [style={scalable_box}] (34) at (4, -10) {$Z$};
		\node [style=none] (35) at (5, -4) {$H_X''$};
		\node [style=none] (36) at (5, -8) {$H_Z''$};
		\node [style=none] (37) at (5.5, -6) {$Q''$};
		\node [style=none] (40) at (4, 7) {$f_0'$};
		\node [style=none] (41) at (4.25, 1.75) {${f_1'}^\intercal$};
		\node [style=none] (42) at (2.125, -0.5) {$f_0''$};
		\node [style=none] (43) at (2.125, -4.75) {${f_1''}^\intercal$};
		\node [style=none] (44) at (0, 5.5) {$\CE$};
		\node [style=none] (45) at (0, -1.5) {$\CV$};
	\end{pgfonlayer}
	\begin{pgfonlayer}{edgelayer}
		\draw (2) to (5);
		\draw (2) to (3);
		\draw (11) to (14);
		\draw (11) to (12);
		\draw (2) to (19);
		\draw (19) to (11);
		\draw (21) to (12);
		\draw (21) to (3);
		\draw (21) to (19);
		\draw (31) to (34);
		\draw (31) to (32);
		\draw (19) to (31);
		\draw (32) to (21);
	\end{pgfonlayer}
\end{tikzpicture}

%% file: Figures-Tikz/hypergraph_diag.tikz
\begin{tikzpicture}
	\begin{pgfonlayer}{nodelayer}
		\node [style={scalable_box}] (0) at (-0.1, 4.1) {};
		\node [style={scalable_box}] (1) at (0, 4) {$X$};
		\node [style={scalable_dot}] (2) at (-0.1, 0.075) {};
		\node [style={scalable_dot}] (3) at (0, 0) {};
		\node [style=none] (4) at (-0.75, 2) {$A_1$};
	\end{pgfonlayer}
	\begin{pgfonlayer}{edgelayer}
		\draw (3) to (1);
	\end{pgfonlayer}
\end{tikzpicture}

%% file: Figures-Tikz/dual_path_graph.tikz
\begin{tikzpicture}
	\begin{pgfonlayer}{nodelayer}
		\node [style=box] (0) at (-3, 0) {$X$};
		\node [style=boundary vertex] (1) at (-1, 0) {};
		\node [style=box] (2) at (1, 0) {$X$};
		\node [style=boundary vertex] (3) at (3, 0) {};
		\node [style=none] (5) at (6, 0) {};
		\node [style=none] (6) at (7, 0) {$\cdots$};
		\node [style=box] (7) at (5, 0) {$X$};
	\end{pgfonlayer}
	\begin{pgfonlayer}{edgelayer}
		\draw (0) to (1);
		\draw (1) to (2);
		\draw (2) to (3);
		\draw (3) to (7);
		\draw (7) to (5.center);
	\end{pgfonlayer}
\end{tikzpicture}

%% file: Figures-Tikz/tensored_subcode.tikz
\begin{tikzpicture}
	\begin{pgfonlayer}{nodelayer}
		\node [style={scalable_box}] (0) at (-8.1, 2.1) {};
		\node [style={scalable_box}] (1) at (-8, 2) {$X$};
		\node [style={scalable_dot}] (2) at (-8.1, -1.875) {};
		\node [style={scalable_dot}] (3) at (-8, -2) {};
		\node [style={scalable_box}] (7) at (-0.1, 2.1) {};
		\node [style={scalable_box}] (8) at (0, 2) {$X$};
		\node [style={scalable_dot}] (9) at (-0.1, -1.875) {};
		\node [style={scalable_dot}] (10) at (0, -2) {};
		\node [style={scalable_box}] (14) at (-4.075, -1.9) {};
		\node [style={scalable_box}] (15) at (-3.975, -2) {$Z$};
		\node [style={scalable_dot}] (16) at (-4.075, 2.125) {};
		\node [style={scalable_dot}] (17) at (-3.975, 2) {};
		\node [style={scalable_box}] (18) at (7.975, 2.1) {};
		\node [style={scalable_box}] (19) at (8.075, 2) {$X$};
		\node [style={scalable_dot}] (20) at (7.975, -1.875) {};
		\node [style={scalable_dot}] (21) at (8.075, -2) {};
		\node [style={scalable_box}] (25) at (4, -1.9) {};
		\node [style={scalable_box}] (26) at (4.1, -2) {$Z$};
		\node [style={scalable_dot}] (27) at (4, 2.125) {};
		\node [style={scalable_dot}] (28) at (4.1, 2) {};
		\node [style=none] (29) at (9.5, -2) {};
		\node [style=none] (30) at (9.5, 2) {};
		\node [style=none] (31) at (10.5, -2) {$\cdots$};
		\node [style=none] (32) at (10.5, 2) {$\cdots$};
		\node [style=none] (33) at (-8.75, 0) {$A_1$};
		\node [style=none] (34) at (-0.75, 0) {$A_1$};
		\node [style=none] (35) at (7.25, 0) {$A_1$};
		\node [style=none] (36) at (-4.75, 0) {$A_1^\intercal$};
		\node [style=none] (37) at (3.25, 0) {$A_1^\intercal$};
		\node [style=none] (38) at (-6, 2.75) {$I$};
		\node [style=none] (39) at (-2, 2.75) {$I$};
		\node [style=none] (40) at (2, 2.75) {$I$};
		\node [style=none] (41) at (6, 2.75) {$I$};
		\node [style=none] (42) at (-6, -2.75) {$I$};
		\node [style=none] (43) at (-2, -2.75) {$I$};
		\node [style=none] (44) at (2, -2.75) {$I$};
		\node [style=none] (45) at (6, -2.75) {$I$};
	\end{pgfonlayer}
	\begin{pgfonlayer}{edgelayer}
		\draw (3) to (1);
		\draw (10) to (8);
		\draw (3) to (15);
		\draw (15) to (10);
		\draw (15) to (17);
		\draw (17) to (1);
		\draw (17) to (8);
		\draw (21) to (19);
		\draw (26) to (21);
		\draw (26) to (28);
		\draw (28) to (19);
		\draw (10) to (26);
		\draw (8) to (28);
		\draw (19) to (30.center);
		\draw (21) to (29.center);
	\end{pgfonlayer}
\end{tikzpicture}

%% file: Figures-Tikz/thickened_hypergraph.tikz
\begin{tikzpicture}
	\begin{pgfonlayer}{nodelayer}
		\node [style={scalable_box}] (0) at (-0.1, 0.1) {};
		\node [style={scalable_box}] (1) at (0, 0) {$Z$};
		\node [style={scalable_dot}] (2) at (-0.1, 4.075) {};
		\node [style={scalable_dot}] (3) at (0, 4) {};
		\node [style=none] (4) at (-0.75, 2) {$G$};
		\node [style=none] (5) at (0, 5.5) {$\CE$};
		\node [style=none] (6) at (0, -1.5) {$\CV$};
		\node [style={scalable_box}] (7) at (7.9, 0.1) {};
		\node [style={scalable_box}] (8) at (8, 0) {$Z$};
		\node [style={scalable_dot}] (9) at (7.9, 4.075) {};
		\node [style={scalable_dot}] (10) at (8, 4) {};
		\node [style=none] (11) at (7.25, 2) {$G$};
		\node [style=none] (12) at (8, 5.5) {$\CE$};
		\node [style=none] (13) at (8, -1.5) {$\CV$};
		\node [style={scalable_box}] (28) at (3.925, 4.1) {};
		\node [style={scalable_box}] (29) at (4.025, 4) {$X$};
		\node [style={scalable_dot}] (30) at (3.925, 0.075) {};
		\node [style={scalable_dot}] (31) at (4.025, 0) {};
		\node [style={scalable_box}] (32) at (15.975, 0.1) {};
		\node [style={scalable_box}] (33) at (16.075, 0) {$Z$};
		\node [style={scalable_dot}] (34) at (15.975, 4.075) {};
		\node [style={scalable_dot}] (35) at (16.075, 4) {};
		\node [style=none] (36) at (15.325, 2) {$G$};
		\node [style=none] (37) at (16.075, 5.5) {$\CE$};
		\node [style=none] (38) at (16.075, -1.5) {$\CV$};
		\node [style={scalable_box}] (39) at (12, 4.1) {};
		\node [style={scalable_box}] (40) at (12.1, 4) {$X$};
		\node [style={scalable_dot}] (41) at (12, 0.075) {};
		\node [style={scalable_dot}] (42) at (12.1, 0) {};
		\node [style=none] (43) at (17.5, 4) {};
		\node [style=none] (44) at (17.5, 0) {};
		\node [style=none] (45) at (18.5, 4) {$\cdots$};
		\node [style=none] (46) at (18.5, 0) {$\cdots$};
	\end{pgfonlayer}
	\begin{pgfonlayer}{edgelayer}
		\draw (3) to (1);
		\draw (10) to (8);
		\draw (3) to (29);
		\draw (29) to (10);
		\draw (29) to (31);
		\draw (31) to (1);
		\draw (31) to (8);
		\draw (35) to (33);
		\draw (40) to (35);
		\draw (40) to (42);
		\draw (42) to (33);
		\draw (10) to (40);
		\draw (8) to (42);
		\draw (33) to (44.center);
		\draw (35) to (43.center);
	\end{pgfonlayer}
\end{tikzpicture}

%% file: Figures-Tikz/thickened_partial_reading.tikz
\begin{tikzpicture}
	\begin{pgfonlayer}{nodelayer}
		\node [style={scalable_box}] (0) at (-8.1, -5.4) {};
		\node [style={scalable_box}] (1) at (-8, -5.5) {$Z$};
		\node [style={scalable_dot}] (2) at (-8.1, -1.425) {};
		\node [style={scalable_dot}] (3) at (-8, -1.5) {};
		\node [style=none] (4) at (-8.75, -3.5) {$G$};
		\node [style=none] (5) at (-8, 0) {$\CE$};
		\node [style=none] (6) at (-8, -7) {$\CV$};
		\node [style={scalable_box}] (7) at (-0.1, -5.4) {};
		\node [style={scalable_box}] (8) at (0, -5.5) {$Z$};
		\node [style={scalable_dot}] (9) at (-0.1, -1.425) {};
		\node [style={scalable_dot}] (10) at (0, -1.5) {};
		\node [style=none] (11) at (-0.75, -3.5) {$G$};
		\node [style=none] (12) at (0, 0) {$\CE$};
		\node [style=none] (13) at (0, -7) {$\CV$};
		\node [style={scalable_box}] (28) at (-4.075, -1.4) {};
		\node [style={scalable_box}] (29) at (-3.975, -1.5) {$X$};
		\node [style={scalable_dot}] (30) at (-4.075, -5.425) {};
		\node [style={scalable_dot}] (31) at (-3.975, -5.5) {};
		\node [style={scalable_box}] (32) at (7.95, -5.4) {};
		\node [style={scalable_box}] (33) at (8.05, -5.5) {$Z$};
		\node [style={scalable_dot}] (34) at (7.95, -1.425) {};
		\node [style={scalable_dot}] (35) at (8.05, -1.5) {};
		\node [style=none] (36) at (7.325, -3.5) {$G$};
		\node [style=none] (37) at (8.075, 0) {$\CE$};
		\node [style=none] (38) at (8.075, -7) {$\CV$};
		\node [style={scalable_box}] (39) at (3.925, -1.4) {};
		\node [style={scalable_box}] (40) at (4.025, -1.5) {$X$};
		\node [style={scalable_dot}] (41) at (3.925, -5.425) {};
		\node [style={scalable_dot}] (42) at (4.025, -5.5) {};
		\node [style=none] (43) at (9.5, -1.5) {};
		\node [style=none] (44) at (9.5, -5.5) {};
		\node [style=none] (45) at (10.5, -1.5) {$\cdots$};
		\node [style=none] (46) at (10.5, -5.5) {$\cdots$};
		\node [style={scalable_box}] (47) at (-12.1, -4.4) {};
		\node [style={scalable_dot}] (48) at (-12.1, -8.425) {};
		\node [style={scalable_dot}] (49) at (-12, -8.5) {};
		\node [style={scalable_box}] (50) at (-12, -4.5) {$X$};
		\node [style={scalable_box}] (51) at (-12.1, -12.4) {};
		\node [style={scalable_box}] (52) at (-12, -12.5) {$Z$};
		\node [style=none] (53) at (-13, -6.5) {$H_X$};
		\node [style=none] (54) at (-13, -10.5) {$H_Z$};
		\node [style=none] (55) at (-13.5, -8.5) {$Q$};
		\node [style={scalable_box}] (56) at (3.925, -12.9) {};
		\node [style={scalable_dot}] (57) at (3.925, -16.925) {};
		\node [style={scalable_dot}] (58) at (4.025, -17) {};
		\node [style={scalable_box}] (59) at (4.025, -13) {$X$};
		\node [style={scalable_box}] (60) at (3.925, -20.9) {};
		\node [style={scalable_box}] (61) at (4.025, -21) {$Z$};
		\node [style=none] (62) at (5.025, -15) {$H_X'$};
		\node [style=none] (63) at (5.025, -19) {$H_Z'$};
		\node [style=none] (64) at (5.525, -17) {$Q'$};
		\node [style={scalable_box}] (65) at (14, -8.4) {};
		\node [style={scalable_dot}] (66) at (14, -12.425) {};
		\node [style={scalable_dot}] (67) at (14.1, -12.5) {};
		\node [style={scalable_box}] (68) at (14.1, -8.5) {$X$};
		\node [style={scalable_box}] (69) at (14, -16.4) {};
		\node [style={scalable_box}] (70) at (14.1, -16.5) {$Z$};
		\node [style=none] (71) at (15.1, -10.5) {$H_X''$};
		\node [style=none] (72) at (15.1, -14.5) {$H_Z''$};
		\node [style=none] (73) at (15.6, -12.5) {$Q''$};
	\end{pgfonlayer}
	\begin{pgfonlayer}{edgelayer}
		\draw (3) to (1);
		\draw (10) to (8);
		\draw (3) to (29);
		\draw (29) to (10);
		\draw (29) to (31);
		\draw (31) to (1);
		\draw (31) to (8);
		\draw (35) to (33);
		\draw (40) to (35);
		\draw (40) to (42);
		\draw (42) to (33);
		\draw (10) to (40);
		\draw (8) to (42);
		\draw (33) to (44.center);
		\draw (35) to (43.center);
		\draw (49) to (52);
		\draw (49) to (50);
		\draw (49) to (1);
		\draw (50) to (3);
		\draw (58) to (61);
		\draw (58) to (59);
		\draw (8) to (58);
		\draw (59) to (10);
		\draw (67) to (70);
		\draw (67) to (68);
		\draw (33) to (67);
		\draw (68) to (35);
	\end{pgfonlayer}
\end{tikzpicture}

%% file: Figures-Tikz/thickened_partial_reading_metachecks.tikz
\begin{tikzpicture}
	\begin{pgfonlayer}{nodelayer}
		\node [style={scalable_meta_Z}] (61) at (-8.1, -9.4) {};
		\node [style={scalable_box}] (0) at (-8.1, -5.4) {};
		\node [style={scalable_box}] (1) at (-8, -5.5) {$Z$};
		\node [style={scalable_dot}] (2) at (-8.1, -1.425) {};
		\node [style={scalable_dot}] (3) at (-8, -1.5) {};
		\node [style=none] (4) at (-8.75, -3.5) {$G$};
		\node [style=none] (5) at (-8, 0) {$\CE$};
		\node [style=none] (6) at (-7.125, -6.65) {$\CV$};
		\node [style={scalable_box}] (7) at (-0.1, -5.4) {};
		\node [style={scalable_box}] (8) at (0, -5.5) {$Z$};
		\node [style={scalable_dot}] (9) at (-0.1, -1.425) {};
		\node [style={scalable_dot}] (10) at (0, -1.5) {};
		\node [style=none] (11) at (-0.75, -3.5) {$G$};
		\node [style=none] (12) at (0, 0) {$\CE$};
		\node [style=none] (13) at (-0.85, -6.625) {$\CV$};
		\node [style={scalable_box}] (14) at (-4.075, -1.4) {};
		\node [style={scalable_box}] (15) at (-3.975, -1.5) {$X$};
		\node [style={scalable_dot}] (16) at (-4.075, -5.425) {};
		\node [style={scalable_dot}] (17) at (-3.975, -5.5) {};
		\node [style={scalable_box}] (18) at (7.925, -5.4) {};
		\node [style={scalable_box}] (19) at (8.025, -5.5) {$Z$};
		\node [style={scalable_dot}] (20) at (7.925, -1.425) {};
		\node [style={scalable_dot}] (21) at (8.025, -1.5) {};
		\node [style=none] (22) at (7.325, -3.5) {$G$};
		\node [style=none] (23) at (8.075, 0) {$\CE$};
		\node [style=none] (24) at (7.175, -6.625) {$\CV$};
		\node [style={scalable_box}] (25) at (3.925, -1.4) {};
		\node [style={scalable_box}] (26) at (4.025, -1.5) {$X$};
		\node [style={scalable_dot}] (27) at (3.925, -5.425) {};
		\node [style={scalable_dot}] (28) at (4.025, -5.5) {};
		\node [style=none] (29) at (9.5, -1.5) {};
		\node [style=none] (30) at (9.5, -5.5) {};
		\node [style=none] (31) at (10.5, -1.5) {$\cdots$};
		\node [style=none] (32) at (10.5, -5.5) {$\cdots$};
		\node [style={scalable_box}] (33) at (-12.1, -4.4) {};
		\node [style={scalable_dot}] (34) at (-12.1, -8.425) {};
		\node [style={scalable_dot}] (35) at (-12, -8.5) {};
		\node [style={scalable_box}] (36) at (-12, -4.5) {$X$};
		\node [style={scalable_box}] (37) at (-12.1, -12.4) {};
		\node [style={scalable_box}] (38) at (-12, -12.5) {$Z$};
		\node [style=none] (39) at (-13, -6.5) {$H_X$};
		\node [style=none] (40) at (-13, -10.5) {$H_Z$};
		\node [style=none] (41) at (-13.5, -8.5) {$Q$};
		\node [style={scalable_box}] (42) at (3.925, -12.9) {};
		\node [style={scalable_dot}] (43) at (3.925, -16.925) {};
		\node [style={scalable_dot}] (44) at (4.025, -17) {};
		\node [style={scalable_box}] (45) at (4.025, -13) {$X$};
		\node [style={scalable_box}] (46) at (3.925, -20.9) {};
		\node [style={scalable_box}] (47) at (4.025, -21) {$Z$};
		\node [style=none] (48) at (5.025, -15) {$H_X'$};
		\node [style=none] (49) at (5.025, -19) {$H_Z'$};
		\node [style=none] (50) at (5.525, -17) {$Q'$};
		\node [style={scalable_box}] (51) at (14, -8.4) {};
		\node [style={scalable_dot}] (52) at (14, -12.425) {};
		\node [style={scalable_dot}] (53) at (14.1, -12.5) {};
		\node [style={scalable_box}] (54) at (14.1, -8.5) {$X$};
		\node [style={scalable_box}] (55) at (14, -16.4) {};
		\node [style={scalable_box}] (56) at (14.1, -16.5) {$Z$};
		\node [style=none] (57) at (15.1, -10.5) {$H_X''$};
		\node [style=none] (58) at (15.1, -14.5) {$H_Z''$};
		\node [style=none] (59) at (15.6, -12.5) {$Q''$};
		\node [style={scalable_meta_Z}] (60) at (-8, -9.5) {};
		\node [style={scalable_box}] (62) at (-4.075, -9.4) {};
		\node [style={scalable_box}] (63) at (-3.975, -9.5) {$Z$};
		\node [style={scalable_meta_Z}] (64) at (-0.1, -9.4) {};
		\node [style={scalable_meta_Z}] (65) at (0, -9.5) {};
		\node [style={scalable_box}] (66) at (3.925, -9.4) {};
		\node [style={scalable_box}] (67) at (4.025, -9.5) {$Z$};
		\node [style={scalable_meta_Z}] (68) at (7.9, -9.4) {};
		\node [style={scalable_meta_Z}] (69) at (8, -9.5) {};
	\end{pgfonlayer}
	\begin{pgfonlayer}{edgelayer}
		\draw (3) to (1);
		\draw (10) to (8);
		\draw (3) to (15);
		\draw (15) to (10);
		\draw (15) to (17);
		\draw (17) to (1);
		\draw (17) to (8);
		\draw (21) to (19);
		\draw (26) to (21);
		\draw (26) to (28);
		\draw (28) to (19);
		\draw (10) to (26);
		\draw (8) to (28);
		\draw (19) to (30.center);
		\draw (21) to (29.center);
		\draw (35) to (38);
		\draw (35) to (36);
		\draw (35) to (1);
		\draw (36) to (3);
		\draw (44) to (47);
		\draw (44) to (45);
		\draw (8) to (44);
		\draw (45) to (10);
		\draw (53) to (56);
		\draw (53) to (54);
		\draw (19) to (53);
		\draw (54) to (21);
		\draw (1) to (60);
		\draw (17) to (63);
		\draw (60) to (63);
		\draw (63) to (65);
		\draw (8) to (65);
		\draw (67) to (69);
		\draw (65) to (67);
		\draw (67) to (28);
		\draw (19) to (69);
		\draw (38) to (60);
		\draw (69) to (56);
		\draw (65) to (47);
	\end{pgfonlayer}
\end{tikzpicture}

%% file: Figures-Tikz/expansion_commuting_diag.tikz
\begin{tikzpicture}
	\begin{pgfonlayer}{nodelayer}
		\node [style=none] (0) at (-6, 3) {Modular expansion};
		\node [style=none] (1) at (-2, 3) {};
		\node [style=none] (2) at (2, 3) {};
		\node [style=none] (3) at (6, 3) {Relative expansion};
		\node [style=none] (4) at (0, 4) {$k = 1$};
		\node [style=none] (5) at (-6, 2) {};
		\node [style=none] (6) at (-6, -2) {};
		\node [style=none] (7) at (-6, -3) {Soundness of an LTC};
		\node [style=none] (8) at (-8, 0.5) {$t \geq n$};
		\node [style=none] (9) at (-2, -3) {};
		\node [style=none] (10) at (2, -3) {};
		\node [style=none] (11) at (6, -3) {Global expansion};
		\node [style=none] (12) at (6, 2) {};
		\node [style=none] (13) at (6, -2) {};
		\node [style=none] (14) at (0, -4) {$k = 1$};
		\node [style=none] (15) at (-8, -0.5) {$\mathcal{U} = \mathcal{V}$};
		\node [style=none] (16) at (8, 0.5) {$t \geq n$};
		\node [style=none] (17) at (8, -0.5) {$\mathcal{U} = \mathcal{V}$};
	\end{pgfonlayer}
	\begin{pgfonlayer}{edgelayer}
		\draw [style=arrow] (1.center) to (2.center);
		\draw [style=arrow] (5.center) to (6.center);
		\draw [style=arrow] (9.center) to (10.center);
		\draw [style=arrow] (12.center) to (13.center);
	\end{pgfonlayer}
\end{tikzpicture}

%% file: Figures-Tikz/SC_fullblock.tikz
\begin{tikzpicture}
	\begin{pgfonlayer}{nodelayer}
		\node [style=boundary vertex] (15) at (-6, 7) {};
		\node [style=boundary vertex] (16) at (-4, 8) {};
		\node [style=boundary vertex] (17) at (-2, 9) {};
		\node [style=boundary vertex] (18) at (-4, 6) {};
		\node [style=boundary vertex] (19) at (-2, 7) {};
		\node [style=boundary vertex] (20) at (0, 8) {};
		\node [style=none] (21) at (-8, 8) {};
		\node [style=none] (22) at (-6, 9) {};
		\node [style=none] (23) at (-4, 10) {};
		\node [style=none] (24) at (6, 5) {};
		\node [style=none] (25) at (4, 4) {};
		\node [style=none] (26) at (2, 3) {};
		\node [style=boundary vertex] (27) at (0, 10) {};
		\node [style=boundary vertex] (30) at (2, 9) {};
		\node [style=none] (33) at (-2, 11) {};
		\node [style=none] (38) at (8, 6) {};
		\node [style=boundary vertex] (43) at (-2, 5) {};
		\node [style=boundary vertex] (44) at (0, 6) {};
		\node [style=boundary vertex] (45) at (2, 7) {};
		\node [style=boundary vertex] (46) at (4, 8) {};
		\node [style=boundary vertex] (67) at (2, 11) {};
		\node [style=boundary vertex] (68) at (4, 10) {};
		\node [style=none] (69) at (0, 12) {};
		\node [style=none] (70) at (10, 7) {};
		\node [style=boundary vertex] (71) at (6, 9) {};
		\node [style=boundary vertex] (72) at (0, 4) {};
		\node [style=boundary vertex] (73) at (2, 5) {};
		\node [style=boundary vertex] (74) at (4, 6) {};
		\node [style=boundary vertex] (75) at (6, 7) {};
		\node [style=boundary vertex] (76) at (8, 8) {};
		\node [style=boundary vertex] (77) at (-8, -3) {};
		\node [style=boundary vertex] (78) at (-6, -2) {};
		\node [style=boundary vertex] (79) at (-4, -1) {};
		\node [style=boundary vertex] (80) at (-6, -4) {};
		\node [style=boundary vertex] (81) at (-4, -3) {};
		\node [style=boundary vertex] (82) at (-2, -2) {};
		\node [style=none] (83) at (-10, -2) {};
		\node [style=none] (84) at (-8, -1) {};
		\node [style=none] (85) at (-6, 0) {};
		\node [style=none] (86) at (4, -5) {};
		\node [style=none] (87) at (2, -6) {};
		\node [style=none] (88) at (0, -7) {};
		\node [style=boundary vertex] (89) at (-2, 0) {};
		\node [style=boundary vertex] (90) at (0, -1) {};
		\node [style=none] (91) at (-4, 1) {};
		\node [style=none] (92) at (6, -4) {};
		\node [style=boundary vertex] (93) at (-4, -5) {};
		\node [style=boundary vertex] (94) at (-2, -4) {};
		\node [style=boundary vertex] (95) at (0, -3) {};
		\node [style=boundary vertex] (96) at (2, -2) {};
		\node [style=boundary vertex] (97) at (0, 1) {};
		\node [style=boundary vertex] (98) at (2, 0) {};
		\node [style=none] (99) at (-2, 2) {};
		\node [style=none] (100) at (8, -3) {};
		\node [style=boundary vertex] (101) at (4, -1) {};
		\node [style=boundary vertex] (102) at (-2, -6) {};
		\node [style=boundary vertex] (103) at (0, -5) {};
		\node [style=boundary vertex] (104) at (2, -4) {};
		\node [style=boundary vertex] (105) at (4, -3) {};
		\node [style=boundary vertex] (106) at (6, -2) {};
	\end{pgfonlayer}
	\begin{pgfonlayer}{edgelayer}
		\draw (15) to (17);
		\draw (18) to (20);
		\draw (21.center) to (26.center);
		\draw (25.center) to (22.center);
		\draw (23.center) to (24.center);
		\draw (33.center) to (38.center);
		\draw (17) to (27);
		\draw (20) to (30);
		\draw (43) to (45);
		\draw (45) to (46);
		\draw (69.center) to (70.center);
		\draw (30) to (68);
		\draw (46) to (71);
		\draw (67) to (27);
		\draw (72) to (74);
		\draw (74) to (75);
		\draw (75) to (76);
		\draw (77) to (79);
		\draw (80) to (82);
		\draw (83.center) to (88.center);
		\draw (87.center) to (84.center);
		\draw (85.center) to (86.center);
		\draw (91.center) to (92.center);
		\draw (79) to (89);
		\draw (82) to (90);
		\draw (93) to (95);
		\draw (95) to (96);
		\draw (99.center) to (100.center);
		\draw (90) to (98);
		\draw (96) to (101);
		\draw (97) to (89);
		\draw (102) to (104);
		\draw (104) to (105);
		\draw (105) to (106);
	\end{pgfonlayer}
\end{tikzpicture}

%% file: Figures-Tikz/SC_fullblock2.tikz
\begin{tikzpicture}
	\begin{pgfonlayer}{nodelayer}
		\node [style=boundary vertex] (15) at (-6, 7) {};
		\node [style=boundary vertex] (16) at (-4, 8) {};
		\node [style=boundary vertex] (17) at (-2, 9) {};
		\node [style=boundary vertex] (18) at (-4, 6) {};
		\node [style=boundary vertex] (19) at (-2, 7) {};
		\node [style=boundary vertex] (20) at (0, 8) {};
		\node [style=none] (21) at (-8, 8) {};
		\node [style=none] (22) at (-6, 9) {};
		\node [style=none] (23) at (-4, 10) {};
		\node [style=none] (24) at (6, 5) {};
		\node [style=none] (25) at (4, 4) {};
		\node [style=none] (26) at (2, 3) {};
		\node [style=boundary vertex] (27) at (0, 10) {};
		\node [style=boundary vertex] (30) at (2, 9) {};
		\node [style=none] (33) at (-2, 11) {};
		\node [style=none] (38) at (8, 6) {};
		\node [style=boundary vertex] (43) at (-2, 5) {};
		\node [style=boundary vertex] (44) at (0, 6) {};
		\node [style=boundary vertex] (45) at (2, 7) {};
		\node [style=boundary vertex] (46) at (4, 8) {};
		\node [style=boundary vertex] (67) at (2, 11) {};
		\node [style=boundary vertex] (68) at (4, 10) {};
		\node [style=none] (69) at (0, 12) {};
		\node [style=none] (70) at (10, 7) {};
		\node [style=boundary vertex] (71) at (6, 9) {};
		\node [style=boundary vertex] (72) at (0, 4) {};
		\node [style=boundary vertex] (73) at (2, 5) {};
		\node [style=boundary vertex] (74) at (4, 6) {};
		\node [style=boundary vertex] (75) at (6, 7) {};
		\node [style=boundary vertex] (76) at (8, 8) {};
		\node [style=boundary vertex] (77) at (-8, -3) {};
		\node [style=boundary vertex] (78) at (-6, -2) {};
		\node [style=boundary vertex] (79) at (-4, -1) {};
		\node [style=boundary vertex] (80) at (-6, -4) {};
		\node [style=boundary vertex] (81) at (-4, -3) {};
		\node [style=boundary vertex] (82) at (-2, -2) {};
		\node [style=none] (83) at (-10, -2) {};
		\node [style=none] (84) at (-8, -1) {};
		\node [style=none] (85) at (-6, 0) {};
		\node [style=none] (86) at (4, -5) {};
		\node [style=none] (87) at (2, -6) {};
		\node [style=none] (88) at (0, -7) {};
		\node [style=boundary vertex] (89) at (-2, 0) {};
		\node [style=boundary vertex] (90) at (0, -1) {};
		\node [style=none] (91) at (-4, 1) {};
		\node [style=none] (92) at (6, -4) {};
		\node [style=boundary vertex] (93) at (-4, -5) {};
		\node [style=boundary vertex] (94) at (-2, -4) {};
		\node [style=boundary vertex] (95) at (0, -3) {};
		\node [style=boundary vertex] (96) at (2, -2) {};
		\node [style=boundary vertex] (97) at (0, 1) {};
		\node [style=boundary vertex] (98) at (2, 0) {};
		\node [style=none] (99) at (-2, 2) {};
		\node [style=none] (100) at (8, -3) {};
		\node [style=boundary vertex] (101) at (4, -1) {};
		\node [style=boundary vertex] (102) at (-2, -6) {};
		\node [style=boundary vertex] (103) at (0, -5) {};
		\node [style=boundary vertex] (104) at (2, -4) {};
		\node [style=boundary vertex] (105) at (4, -3) {};
		\node [style=boundary vertex] (106) at (6, -2) {};
	\end{pgfonlayer}
	\begin{pgfonlayer}{edgelayer}
		\draw (15) to (17);
		\draw (18) to (20);
		\draw (21.center) to (26.center);
		\draw (25.center) to (22.center);
		\draw (23.center) to (24.center);
		\draw (33.center) to (38.center);
		\draw (17) to (27);
		\draw (20) to (30);
		\draw (43) to (45);
		\draw (45) to (46);
		\draw (69.center) to (70.center);
		\draw (30) to (68);
		\draw (46) to (71);
		\draw (67) to (27);
		\draw (72) to (74);
		\draw (74) to (75);
		\draw (75) to (76);
		\draw (77) to (79);
		\draw (80) to (82);
		\draw (83.center) to (88.center);
		\draw (87.center) to (84.center);
		\draw (85.center) to (86.center);
		\draw (91.center) to (92.center);
		\draw (79) to (89);
		\draw (82) to (90);
		\draw (93) to (95);
		\draw (95) to (96);
		\draw (99.center) to (100.center);
		\draw (90) to (98);
		\draw (96) to (101);
		\draw (97) to (89);
		\draw (102) to (104);
		\draw (104) to (105);
		\draw (105) to (106);
		\draw [style=grey dash] (72) to (102);
		\draw [style=grey dash] (103) to (73);
		\draw [style=grey dash] (74) to (104);
		\draw [style=grey dash] (105) to (75);
		\draw [style=grey dash] (76) to (106);
		\draw [style=grey dash] (101) to (71);
		\draw [style=grey dash] (46) to (96);
		\draw [style=grey dash] (45) to (95);
		\draw [style=grey dash] (44) to (94);
		\draw [style=grey dash] (43) to (93);
		\draw [style=grey dash] (80) to (18);
		\draw [style=grey dash] (19) to (81);
		\draw [style=grey dash] (82) to (20);
		\draw [style=grey dash] (30) to (90);
		\draw [style=grey dash] (98) to (68);
		\draw [style=grey dash] (67) to (97);
		\draw [style=grey dash] (27) to (89);
		\draw [style=grey dash] (17) to (79);
		\draw [style=grey dash] (16) to (78);
		\draw [style=grey dash] (15) to (77);
	\end{pgfonlayer}
\end{tikzpicture}

%% file: Figures-Tikz/SC_partialblock.tikz
\begin{tikzpicture}
	\begin{pgfonlayer}{nodelayer}
		\node [style=boundary vertex] (15) at (-6, 7) {};
		\node [style=boundary vertex] (16) at (-4, 8) {};
		\node [style=boundary vertex] (17) at (-2, 9) {};
		\node [style=boundary vertex] (18) at (-4, 6) {};
		\node [style=boundary vertex] (19) at (-2, 7) {};
		\node [style=boundary vertex] (20) at (0, 8) {};
		\node [style=none] (21) at (-8, 8) {};
		\node [style=none] (22) at (-6, 9) {};
		\node [style=none] (23) at (-4, 10) {};
		\node [style=none] (24) at (6, 5) {};
		\node [style=none] (25) at (4, 4) {};
		\node [style=none] (26) at (2, 3) {};
		\node [style=boundary vertex] (27) at (0, 10) {};
		\node [style=boundary vertex] (30) at (2, 9) {};
		\node [style=none] (33) at (-2, 11) {};
		\node [style=none] (38) at (8, 6) {};
		\node [style=boundary vertex] (43) at (-2, 5) {};
		\node [style=boundary vertex] (44) at (0, 6) {};
		\node [style=boundary vertex] (45) at (2, 7) {};
		\node [style=boundary vertex] (46) at (4, 8) {};
		\node [style=boundary vertex] (67) at (2, 11) {};
		\node [style=boundary vertex] (68) at (4, 10) {};
		\node [style=none] (69) at (0, 12) {};
		\node [style=none] (70) at (10, 7) {};
		\node [style=boundary vertex] (71) at (6, 9) {};
		\node [style=boundary vertex] (72) at (0, 4) {};
		\node [style=boundary vertex] (73) at (2, 5) {};
		\node [style=boundary vertex] (74) at (4, 6) {};
		\node [style=boundary vertex] (75) at (6, 7) {};
		\node [style=boundary vertex] (76) at (8, 8) {};
		\node [style=boundary vertex] (77) at (-8, -3) {};
		\node [style=boundary vertex] (78) at (-6, -2) {};
		\node [style=boundary vertex] (79) at (-4, -1) {};
		\node [style=boundary vertex] (80) at (-6, -4) {};
		\node [style=boundary vertex] (81) at (-4, -3) {};
		\node [style=boundary vertex] (82) at (-2, -2) {};
		\node [style=none] (83) at (-10, -2) {};
		\node [style=none] (84) at (-8, -1) {};
		\node [style=none] (85) at (-6, 0) {};
		\node [style=none] (86) at (4, -5) {};
		\node [style=none] (87) at (2, -6) {};
		\node [style=none] (88) at (0, -7) {};
		\node [style=boundary vertex] (89) at (-2, 0) {};
		\node [style=boundary vertex] (90) at (0, -1) {};
		\node [style=none] (91) at (-4, 1) {};
		\node [style=none] (92) at (6, -4) {};
		\node [style=boundary vertex] (93) at (-4, -5) {};
		\node [style=boundary vertex] (94) at (-2, -4) {};
		\node [style=boundary vertex] (95) at (0, -3) {};
		\node [style=boundary vertex] (96) at (2, -2) {};
		\node [style=boundary vertex] (97) at (0, 1) {};
		\node [style=boundary vertex] (98) at (2, 0) {};
		\node [style=none] (99) at (-2, 2) {};
		\node [style=none] (100) at (8, -3) {};
		\node [style=boundary vertex] (101) at (4, -1) {};
		\node [style=boundary vertex] (102) at (-2, -6) {};
		\node [style=boundary vertex] (103) at (0, -5) {};
		\node [style=boundary vertex] (104) at (2, -4) {};
		\node [style=boundary vertex] (105) at (4, -3) {};
		\node [style=boundary vertex] (106) at (6, -2) {};
	\end{pgfonlayer}
	\begin{pgfonlayer}{edgelayer}
		\draw (15) to (17);
		\draw (18) to (20);
		\draw (21.center) to (26.center);
		\draw (25.center) to (22.center);
		\draw (23.center) to (24.center);
		\draw (33.center) to (38.center);
		\draw (17) to (27);
		\draw (20) to (30);
		\draw (43) to (45);
		\draw (45) to (46);
		\draw (69.center) to (70.center);
		\draw (30) to (68);
		\draw (46) to (71);
		\draw (67) to (27);
		\draw (72) to (74);
		\draw (74) to (75);
		\draw (75) to (76);
		\draw (77) to (79);
		\draw (80) to (82);
		\draw (83.center) to (88.center);
		\draw (87.center) to (84.center);
		\draw (85.center) to (86.center);
		\draw (91.center) to (92.center);
		\draw (79) to (89);
		\draw (82) to (90);
		\draw (93) to (95);
		\draw (95) to (96);
		\draw (99.center) to (100.center);
		\draw (90) to (98);
		\draw (96) to (101);
		\draw (97) to (89);
		\draw (102) to (104);
		\draw (104) to (105);
		\draw (105) to (106);
		\draw [style=grey dash] (74) to (104);
		\draw [style=grey dash] (105) to (75);
		\draw [style=grey dash] (76) to (106);
		\draw [style=grey dash] (101) to (71);
		\draw [style=grey dash] (46) to (96);
		\draw [style=grey dash] (45) to (95);
		\draw [style=grey dash] (82) to (20);
		\draw [style=grey dash] (30) to (90);
		\draw [style=grey dash] (98) to (68);
		\draw [style=grey dash] (67) to (97);
		\draw [style=grey dash] (27) to (89);
		\draw [style=grey dash] (17) to (79);
	\end{pgfonlayer}
\end{tikzpicture}

%% file: Figures-Tikz/SC_latticesurgery.tikz
\begin{tikzpicture}
	\begin{pgfonlayer}{nodelayer}
		\node [style=boundary vertex] (0) at (-6, 7) {};
		\node [style=boundary vertex] (1) at (-4, 8) {};
		\node [style=boundary vertex] (2) at (-2, 9) {};
		\node [style=boundary vertex] (3) at (-4, 6) {};
		\node [style=boundary vertex] (4) at (-2, 7) {};
		\node [style=boundary vertex] (5) at (0, 8) {};
		\node [style=none] (6) at (-8, 8) {};
		\node [style=none] (7) at (-6, 9) {};
		\node [style=none] (8) at (-4, 10) {};
		\node [style=none] (9) at (6, 5) {};
		\node [style=none] (10) at (4, 4) {};
		\node [style=none] (11) at (2, 3) {};
		\node [style=boundary vertex] (12) at (0, 10) {};
		\node [style=boundary vertex] (13) at (2, 9) {};
		\node [style=none] (14) at (-2, 11) {};
		\node [style=none] (15) at (8, 6) {};
		\node [style=boundary vertex] (16) at (-2, 5) {};
		\node [style=boundary vertex] (17) at (0, 6) {};
		\node [style=boundary vertex] (18) at (2, 7) {};
		\node [style=boundary vertex] (19) at (4, 8) {};
		\node [style=boundary vertex] (20) at (2, 11) {};
		\node [style=boundary vertex] (21) at (4, 10) {};
		\node [style=none] (22) at (0, 12) {};
		\node [style=none] (23) at (10, 7) {};
		\node [style=boundary vertex] (24) at (6, 9) {};
		\node [style=boundary vertex] (25) at (0, 4) {};
		\node [style=boundary vertex] (26) at (2, 5) {};
		\node [style=boundary vertex] (27) at (4, 6) {};
		\node [style=boundary vertex] (28) at (6, 7) {};
		\node [style=boundary vertex] (29) at (8, 8) {};
		\node [style=boundary vertex] (30) at (-8, -3) {};
		\node [style=boundary vertex] (31) at (-6, -2) {};
		\node [style=boundary vertex] (32) at (-4, -1) {};
		\node [style=boundary vertex] (33) at (-6, -4) {};
		\node [style=boundary vertex] (34) at (-4, -3) {};
		\node [style=boundary vertex] (35) at (-2, -2) {};
		\node [style=none] (36) at (-10, -2) {};
		\node [style=none] (37) at (-8, -1) {};
		\node [style=none] (38) at (-6, 0) {};
		\node [style=none] (39) at (4, -5) {};
		\node [style=none] (40) at (2, -6) {};
		\node [style=none] (41) at (0, -7) {};
		\node [style=boundary vertex] (42) at (-2, 0) {};
		\node [style=boundary vertex] (43) at (0, -1) {};
		\node [style=none] (44) at (-4, 1) {};
		\node [style=none] (45) at (6, -4) {};
		\node [style=boundary vertex] (46) at (-4, -5) {};
		\node [style=boundary vertex] (47) at (-2, -4) {};
		\node [style=boundary vertex] (48) at (0, -3) {};
		\node [style=boundary vertex] (49) at (2, -2) {};
		\node [style=boundary vertex] (50) at (0, 1) {};
		\node [style=boundary vertex] (51) at (2, 0) {};
		\node [style=none] (52) at (-2, 2) {};
		\node [style=none] (53) at (8, -3) {};
		\node [style=boundary vertex] (54) at (4, -1) {};
		\node [style=boundary vertex] (55) at (-2, -6) {};
		\node [style=boundary vertex] (56) at (0, -5) {};
		\node [style=boundary vertex] (57) at (2, -4) {};
		\node [style=boundary vertex] (58) at (4, -3) {};
		\node [style=boundary vertex] (59) at (6, -2) {};
	\end{pgfonlayer}
	\begin{pgfonlayer}{edgelayer}
		\draw (0) to (2);
		\draw (3) to (5);
		\draw (6.center) to (11.center);
		\draw (10.center) to (7.center);
		\draw (8.center) to (9.center);
		\draw (14.center) to (15.center);
		\draw (2) to (12);
		\draw (5) to (13);
		\draw (16) to (18);
		\draw (18) to (19);
		\draw (22.center) to (23.center);
		\draw (13) to (21);
		\draw (19) to (24);
		\draw (20) to (12);
		\draw (25) to (27);
		\draw (27) to (28);
		\draw (28) to (29);
		\draw (30) to (32);
		\draw (33) to (35);
		\draw (36.center) to (41.center);
		\draw (40.center) to (37.center);
		\draw (38.center) to (39.center);
		\draw (44.center) to (45.center);
		\draw (32) to (42);
		\draw (35) to (43);
		\draw (46) to (48);
		\draw (48) to (49);
		\draw (52.center) to (53.center);
		\draw (43) to (51);
		\draw (49) to (54);
		\draw (50) to (42);
		\draw (55) to (57);
		\draw (57) to (58);
		\draw (58) to (59);
		\draw [style=grey dash] (29) to (59);
		\draw [style=grey dash] (54) to (24);
		\draw [style=grey dash] (51) to (21);
		\draw [style=grey dash] (20) to (50);
	\end{pgfonlayer}
\end{tikzpicture}

%% file: Figures-Tikz/toric_codes1.tikz
\begin{tikzpicture}
	\begin{pgfonlayer}{nodelayer}
		\node [style=boundary vertex] (30) at (-8, -6) {};
		\node [style=boundary vertex] (31) at (-6, -5) {};
		\node [style=boundary vertex] (32) at (-4, -4) {};
		\node [style=boundary vertex] (33) at (-6, -7) {};
		\node [style=boundary vertex] (34) at (-4, -6) {};
		\node [style=boundary vertex] (35) at (-2, -5) {};
		\node [style=none] (36) at (-9.5, -5.25) {};
		\node [style=none] (37) at (-7.5, -4.25) {};
		\node [style=none] (38) at (-5.5, -3.25) {};
		\node [style=none] (39) at (5.5, -8.75) {};
		\node [style=none] (40) at (3.5, -9.75) {};
		\node [style=none] (41) at (1.5, -10.75) {};
		\node [style=boundary vertex] (42) at (-2, -3) {};
		\node [style=boundary vertex] (43) at (0, -4) {};
		\node [style=none] (44) at (-3.5, -2.25) {};
		\node [style=none] (45) at (7.5, -7.75) {};
		\node [style=boundary vertex] (46) at (-4, -8) {};
		\node [style=boundary vertex] (47) at (-2, -7) {};
		\node [style=boundary vertex] (48) at (0, -6) {};
		\node [style=boundary vertex] (49) at (2, -5) {};
		\node [style=boundary vertex] (50) at (0, -2) {};
		\node [style=boundary vertex] (51) at (2, -3) {};
		\node [style=none] (52) at (-1.5, -1.25) {};
		\node [style=none] (53) at (9.5, -6.75) {};
		\node [style=boundary vertex] (54) at (4, -4) {};
		\node [style=boundary vertex] (55) at (-2, -9) {};
		\node [style=boundary vertex] (56) at (0, -8) {};
		\node [style=boundary vertex] (57) at (2, -7) {};
		\node [style=boundary vertex] (58) at (4, -6) {};
		\node [style=boundary vertex] (59) at (6, -5) {};
		\node [style=none] (60) at (1, -1.5) {};
		\node [style=none] (61) at (3, -2.5) {};
		\node [style=none] (62) at (5, -3.5) {};
		\node [style=none] (63) at (7, -4.5) {};
		\node [style=none] (64) at (-0.5, -0.75) {};
		\node [style=none] (65) at (10.5, -6.25) {};
		\node [style=none] (66) at (-3, -9.5) {};
		\node [style=none] (67) at (-5, -8.5) {};
		\node [style=none] (68) at (0.5, -11.25) {};
		\node [style=none] (69) at (-7, -7.5) {};
		\node [style=none] (70) at (-9, -6.5) {};
		\node [style=none] (71) at (-10.5, -5.75) {};
		\node [style=boundary vertex] (72) at (0, -10) {};
		\node [style=boundary vertex] (73) at (2, -9) {};
		\node [style=boundary vertex] (74) at (4, -8) {};
		\node [style=boundary vertex] (75) at (6, -7) {};
		\node [style=boundary vertex] (76) at (8, -6) {};
		\node [style=none] (77) at (9, -5.5) {};
		\node [style=none] (78) at (-1, -10.5) {};
		\node [style=boundary vertex] (79) at (-6, 9) {};
		\node [style=boundary vertex] (80) at (-4, 10) {};
		\node [style=boundary vertex] (81) at (-2, 11) {};
		\node [style=boundary vertex] (82) at (-4, 8) {};
		\node [style=boundary vertex] (83) at (-2, 9) {};
		\node [style=boundary vertex] (84) at (0, 10) {};
		\node [style=none] (85) at (-7.5, 9.75) {};
		\node [style=none] (86) at (-5.5, 10.75) {};
		\node [style=none] (87) at (-3.5, 11.75) {};
		\node [style=none] (88) at (7.5, 6.25) {};
		\node [style=none] (89) at (5.5, 5.25) {};
		\node [style=none] (90) at (3.5, 4.25) {};
		\node [style=boundary vertex] (91) at (0, 12) {};
		\node [style=boundary vertex] (92) at (2, 11) {};
		\node [style=none] (93) at (-1.5, 12.75) {};
		\node [style=none] (94) at (9.5, 7.25) {};
		\node [style=boundary vertex] (95) at (-2, 7) {};
		\node [style=boundary vertex] (96) at (0, 8) {};
		\node [style=boundary vertex] (97) at (2, 9) {};
		\node [style=boundary vertex] (98) at (4, 10) {};
		\node [style=boundary vertex] (99) at (2, 13) {};
		\node [style=boundary vertex] (100) at (4, 12) {};
		\node [style=none] (101) at (0.5, 13.75) {};
		\node [style=none] (102) at (11.5, 8.25) {};
		\node [style=boundary vertex] (103) at (6, 11) {};
		\node [style=boundary vertex] (104) at (0, 6) {};
		\node [style=boundary vertex] (105) at (2, 7) {};
		\node [style=boundary vertex] (106) at (4, 8) {};
		\node [style=boundary vertex] (107) at (6, 9) {};
		\node [style=boundary vertex] (108) at (8, 10) {};
		\node [style=none] (109) at (3, 13.5) {};
		\node [style=none] (110) at (5, 12.5) {};
		\node [style=none] (111) at (7, 11.5) {};
		\node [style=none] (112) at (9, 10.5) {};
		\node [style=none] (113) at (1.5, 14.25) {};
		\node [style=none] (114) at (12.5, 8.75) {};
		\node [style=none] (115) at (-1, 5.5) {};
		\node [style=none] (116) at (-3, 6.5) {};
		\node [style=none] (117) at (2.5, 3.75) {};
		\node [style=none] (118) at (-5, 7.5) {};
		\node [style=none] (119) at (-7, 8.5) {};
		\node [style=none] (120) at (-8.5, 9.25) {};
		\node [style=boundary vertex] (121) at (2, 5) {};
		\node [style=boundary vertex] (122) at (4, 6) {};
		\node [style=boundary vertex] (123) at (6, 7) {};
		\node [style=boundary vertex] (124) at (8, 8) {};
		\node [style=boundary vertex] (125) at (10, 9) {};
		\node [style=none] (126) at (11, 9.5) {};
		\node [style=none] (127) at (1, 4.5) {};
	\end{pgfonlayer}
	\begin{pgfonlayer}{edgelayer}
		\draw (30) to (32);
		\draw (33) to (35);
		\draw (36.center) to (41.center);
		\draw (40.center) to (37.center);
		\draw (38.center) to (39.center);
		\draw (44.center) to (45.center);
		\draw (32) to (42);
		\draw (35) to (43);
		\draw (46) to (48);
		\draw (48) to (49);
		\draw (52.center) to (53.center);
		\draw (43) to (51);
		\draw (49) to (54);
		\draw (50) to (42);
		\draw (55) to (57);
		\draw (57) to (58);
		\draw (58) to (59);
		\draw (66.center) to (63.center);
		\draw (62.center) to (67.center);
		\draw (69.center) to (61.center);
		\draw (60.center) to (70.center);
		\draw [style={red_arrow}] (68.center) to (39.center);
		\draw [style={red_arrow}] (71.center) to (38.center);
		\draw (78.center) to (77.center);
		\draw [style={cyan_arrow}] (65.center) to (62.center);
		\draw [style={cyan_arrow}] (68.center) to (67.center);
		\draw [style={blue_edge}] (67.center) to (71.center);
		\draw [style={blue_edge}] (64.center) to (62.center);
		\draw [style={red_edge}] (39.center) to (65.center);
		\draw [style={red_edge}] (38.center) to (64.center);
		\draw (79) to (81);
		\draw (82) to (84);
		\draw (85.center) to (90.center);
		\draw (89.center) to (86.center);
		\draw (87.center) to (88.center);
		\draw (93.center) to (94.center);
		\draw (81) to (91);
		\draw (84) to (92);
		\draw (95) to (97);
		\draw (97) to (98);
		\draw (101.center) to (102.center);
		\draw (92) to (100);
		\draw (98) to (103);
		\draw (99) to (91);
		\draw (104) to (106);
		\draw (106) to (107);
		\draw (107) to (108);
		\draw (115.center) to (112.center);
		\draw (111.center) to (116.center);
		\draw (118.center) to (110.center);
		\draw (109.center) to (119.center);
		\draw [style={red_arrow}] (117.center) to (88.center);
		\draw [style={red_arrow}] (120.center) to (87.center);
		\draw (127.center) to (126.center);
		\draw [style={cyan_arrow}] (114.center) to (111.center);
		\draw [style={cyan_arrow}] (117.center) to (116.center);
		\draw [style={blue_edge}] (116.center) to (120.center);
		\draw [style={blue_edge}] (113.center) to (111.center);
		\draw [style={red_edge}] (88.center) to (114.center);
		\draw [style={red_edge}] (87.center) to (113.center);
	\end{pgfonlayer}
\end{tikzpicture}

%% file: Figures-Tikz/toric_codes2.tikz
\begin{tikzpicture}
	\begin{pgfonlayer}{nodelayer}
		\node [style=boundary vertex] (30) at (-8, -6) {};
		\node [style=boundary vertex] (31) at (-6, -5) {};
		\node [style=boundary vertex] (32) at (-4, -4) {};
		\node [style=boundary vertex] (33) at (-6, -7) {};
		\node [style=boundary vertex] (34) at (-4, -6) {};
		\node [style=boundary vertex] (35) at (-2, -5) {};
		\node [style=none] (36) at (-9.5, -5.25) {};
		\node [style=none] (37) at (-7.5, -4.25) {};
		\node [style=none] (38) at (-5.5, -3.25) {};
		\node [style=none] (39) at (5.5, -8.75) {};
		\node [style=none] (40) at (3.5, -9.75) {};
		\node [style=none] (41) at (1.5, -10.75) {};
		\node [style=boundary vertex] (42) at (-2, -3) {};
		\node [style=boundary vertex] (43) at (0, -4) {};
		\node [style=none] (44) at (-3.5, -2.25) {};
		\node [style=none] (45) at (7.5, -7.75) {};
		\node [style=boundary vertex] (46) at (-4, -8) {};
		\node [style=boundary vertex] (47) at (-2, -7) {};
		\node [style=boundary vertex] (48) at (0, -6) {};
		\node [style=boundary vertex] (49) at (2, -5) {};
		\node [style=boundary vertex] (50) at (0, -2) {};
		\node [style=boundary vertex] (51) at (2, -3) {};
		\node [style=none] (52) at (-1.5, -1.25) {};
		\node [style=none] (53) at (9.5, -6.75) {};
		\node [style=boundary vertex] (54) at (4, -4) {};
		\node [style=boundary vertex] (55) at (-2, -9) {};
		\node [style=boundary vertex] (56) at (0, -8) {};
		\node [style=boundary vertex] (57) at (2, -7) {};
		\node [style=boundary vertex] (58) at (4, -6) {};
		\node [style=boundary vertex] (59) at (6, -5) {};
		\node [style=none] (60) at (1, -1.5) {};
		\node [style=none] (61) at (3, -2.5) {};
		\node [style=none] (62) at (5, -3.5) {};
		\node [style=none] (63) at (7, -4.5) {};
		\node [style=none] (64) at (-0.5, -0.75) {};
		\node [style=none] (65) at (10.5, -6.25) {};
		\node [style=none] (66) at (-3, -9.5) {};
		\node [style=none] (67) at (-5, -8.5) {};
		\node [style=none] (68) at (0.5, -11.25) {};
		\node [style=none] (69) at (-7, -7.5) {};
		\node [style=none] (70) at (-9, -6.5) {};
		\node [style=none] (71) at (-10.5, -5.75) {};
		\node [style=boundary vertex] (72) at (0, -10) {};
		\node [style=boundary vertex] (73) at (2, -9) {};
		\node [style=boundary vertex] (74) at (4, -8) {};
		\node [style=boundary vertex] (75) at (6, -7) {};
		\node [style=boundary vertex] (76) at (8, -6) {};
		\node [style=none] (77) at (9, -5.5) {};
		\node [style=none] (78) at (-1, -10.5) {};
		\node [style=boundary vertex] (79) at (-6, 9) {};
		\node [style=boundary vertex] (80) at (-4, 10) {};
		\node [style=boundary vertex] (81) at (-2, 11) {};
		\node [style=boundary vertex] (82) at (-4, 8) {};
		\node [style=boundary vertex] (83) at (-2, 9) {};
		\node [style=boundary vertex] (84) at (0, 10) {};
		\node [style=none] (85) at (-7.5, 9.75) {};
		\node [style=none] (86) at (-5.5, 10.75) {};
		\node [style=none] (87) at (-3.5, 11.75) {};
		\node [style=none] (88) at (7.5, 6.25) {};
		\node [style=none] (89) at (5.5, 5.25) {};
		\node [style=none] (90) at (3.5, 4.25) {};
		\node [style=boundary vertex] (91) at (0, 12) {};
		\node [style=boundary vertex] (92) at (2, 11) {};
		\node [style=none] (93) at (-1.5, 12.75) {};
		\node [style=none] (94) at (9.5, 7.25) {};
		\node [style=boundary vertex] (95) at (-2, 7) {};
		\node [style=boundary vertex] (96) at (0, 8) {};
		\node [style=boundary vertex] (97) at (2, 9) {};
		\node [style=boundary vertex] (98) at (4, 10) {};
		\node [style=boundary vertex] (99) at (2, 13) {};
		\node [style=boundary vertex] (100) at (4, 12) {};
		\node [style=none] (101) at (0.5, 13.75) {};
		\node [style=none] (102) at (11.5, 8.25) {};
		\node [style=boundary vertex] (103) at (6, 11) {};
		\node [style=boundary vertex] (104) at (0, 6) {};
		\node [style=boundary vertex] (105) at (2, 7) {};
		\node [style=boundary vertex] (106) at (4, 8) {};
		\node [style=boundary vertex] (107) at (6, 9) {};
		\node [style=boundary vertex] (108) at (8, 10) {};
		\node [style=none] (109) at (3, 13.5) {};
		\node [style=none] (110) at (5, 12.5) {};
		\node [style=none] (111) at (7, 11.5) {};
		\node [style=none] (112) at (9, 10.5) {};
		\node [style=none] (113) at (1.5, 14.25) {};
		\node [style=none] (114) at (12.5, 8.75) {};
		\node [style=none] (115) at (-1, 5.5) {};
		\node [style=none] (116) at (-3, 6.5) {};
		\node [style=none] (117) at (2.5, 3.75) {};
		\node [style=none] (118) at (-5, 7.5) {};
		\node [style=none] (119) at (-7, 8.5) {};
		\node [style=none] (120) at (-8.5, 9.25) {};
		\node [style=boundary vertex] (121) at (2, 5) {};
		\node [style=boundary vertex] (122) at (4, 6) {};
		\node [style=boundary vertex] (123) at (6, 7) {};
		\node [style=boundary vertex] (124) at (8, 8) {};
		\node [style=boundary vertex] (125) at (10, 9) {};
		\node [style=none] (126) at (11, 9.5) {};
		\node [style=none] (127) at (1, 4.5) {};
	\end{pgfonlayer}
	\begin{pgfonlayer}{edgelayer}
		\draw (30) to (32);
		\draw (33) to (35);
		\draw (36.center) to (41.center);
		\draw (40.center) to (37.center);
		\draw (38.center) to (39.center);
		\draw (44.center) to (45.center);
		\draw (32) to (42);
		\draw (35) to (43);
		\draw (46) to (48);
		\draw (48) to (49);
		\draw (52.center) to (53.center);
		\draw (43) to (51);
		\draw (49) to (54);
		\draw (50) to (42);
		\draw (55) to (57);
		\draw (57) to (58);
		\draw (58) to (59);
		\draw (66.center) to (63.center);
		\draw (62.center) to (67.center);
		\draw (69.center) to (61.center);
		\draw (60.center) to (70.center);
		\draw [style={red_arrow}] (68.center) to (39.center);
		\draw [style={red_arrow}] (71.center) to (38.center);
		\draw (78.center) to (77.center);
		\draw [style={cyan_arrow}] (65.center) to (62.center);
		\draw [style={cyan_arrow}] (68.center) to (67.center);
		\draw [style={blue_edge}] (67.center) to (71.center);
		\draw [style={blue_edge}] (64.center) to (62.center);
		\draw [style={red_edge}] (39.center) to (65.center);
		\draw [style={red_edge}] (38.center) to (64.center);
		\draw (79) to (81);
		\draw (82) to (84);
		\draw (85.center) to (90.center);
		\draw (89.center) to (86.center);
		\draw (87.center) to (88.center);
		\draw (93.center) to (94.center);
		\draw (81) to (91);
		\draw (84) to (92);
		\draw (95) to (97);
		\draw (97) to (98);
		\draw (101.center) to (102.center);
		\draw (92) to (100);
		\draw (98) to (103);
		\draw (99) to (91);
		\draw (104) to (106);
		\draw (106) to (107);
		\draw (107) to (108);
		\draw (115.center) to (112.center);
		\draw (111.center) to (116.center);
		\draw (118.center) to (110.center);
		\draw (109.center) to (119.center);
		\draw [style={red_arrow}] (117.center) to (88.center);
		\draw [style={red_arrow}] (120.center) to (87.center);
		\draw (127.center) to (126.center);
		\draw [style={cyan_arrow}] (114.center) to (111.center);
		\draw [style={cyan_arrow}] (117.center) to (116.center);
		\draw [style={blue_edge}] (116.center) to (120.center);
		\draw [style={blue_edge}] (113.center) to (111.center);
		\draw [style={red_edge}] (88.center) to (114.center);
		\draw [style={red_edge}] (87.center) to (113.center);
		\draw [style=grey dash] (99) to (50);
		\draw [style=grey dash] (100) to (51);
		\draw [style=grey dash] (103) to (54);
		\draw [style=grey dash] (59) to (108);
		\draw [style=grey dash] (125) to (76);
	\end{pgfonlayer}
\end{tikzpicture}

%% file: Figures-Tikz/toric_codes3.tikz
\begin{tikzpicture}
	\begin{pgfonlayer}{nodelayer}
		\node [style=boundary vertex] (30) at (-8, -6) {};
		\node [style=boundary vertex] (31) at (-6, -5) {};
		\node [style=boundary vertex] (32) at (-4, -4) {};
		\node [style=boundary vertex] (33) at (-6, -7) {};
		\node [style=boundary vertex] (34) at (-4, -6) {};
		\node [style=boundary vertex] (35) at (-2, -5) {};
		\node [style=none] (36) at (-9.5, -5.25) {};
		\node [style=none] (37) at (-7.5, -4.25) {};
		\node [style=none] (38) at (-5.5, -3.25) {};
		\node [style=none] (39) at (5.5, -8.75) {};
		\node [style=none] (40) at (3.5, -9.75) {};
		\node [style=none] (41) at (1.5, -10.75) {};
		\node [style=boundary vertex] (42) at (-2, -3) {};
		\node [style=boundary vertex] (43) at (0, -4) {};
		\node [style=none] (44) at (-3.5, -2.25) {};
		\node [style=none] (45) at (7.5, -7.75) {};
		\node [style=boundary vertex] (46) at (-4, -8) {};
		\node [style=boundary vertex] (47) at (-2, -7) {};
		\node [style=boundary vertex] (48) at (0, -6) {};
		\node [style=boundary vertex] (49) at (2, -5) {};
		\node [style=boundary vertex] (50) at (0, -2) {};
		\node [style=boundary vertex] (51) at (2, -3) {};
		\node [style=none] (52) at (-1.5, -1.25) {};
		\node [style=none] (53) at (9.5, -6.75) {};
		\node [style=boundary vertex] (54) at (4, -4) {};
		\node [style=boundary vertex] (55) at (-2, -9) {};
		\node [style=boundary vertex] (56) at (0, -8) {};
		\node [style=boundary vertex] (57) at (2, -7) {};
		\node [style=boundary vertex] (58) at (4, -6) {};
		\node [style=boundary vertex] (59) at (6, -5) {};
		\node [style=none] (60) at (1, -1.5) {};
		\node [style=none] (61) at (3, -2.5) {};
		\node [style=none] (62) at (5, -3.5) {};
		\node [style=none] (63) at (7, -4.5) {};
		\node [style=none] (64) at (-0.5, -0.75) {};
		\node [style=none] (65) at (10.5, -6.25) {};
		\node [style=none] (66) at (-3, -9.5) {};
		\node [style=none] (67) at (-5, -8.5) {};
		\node [style=none] (68) at (0.5, -11.25) {};
		\node [style=none] (69) at (-7, -7.5) {};
		\node [style=none] (70) at (-9, -6.5) {};
		\node [style=none] (71) at (-10.5, -5.75) {};
		\node [style=boundary vertex] (72) at (0, -10) {};
		\node [style=boundary vertex] (73) at (2, -9) {};
		\node [style=boundary vertex] (74) at (4, -8) {};
		\node [style=boundary vertex] (75) at (6, -7) {};
		\node [style=boundary vertex] (76) at (8, -6) {};
		\node [style=none] (77) at (9, -5.5) {};
		\node [style=none] (78) at (-1, -10.5) {};
		\node [style=boundary vertex] (79) at (-6, 9) {};
		\node [style=boundary vertex] (80) at (-4, 10) {};
		\node [style=boundary vertex] (81) at (-2, 11) {};
		\node [style=boundary vertex] (82) at (-4, 8) {};
		\node [style=boundary vertex] (83) at (-2, 9) {};
		\node [style=boundary vertex] (84) at (0, 10) {};
		\node [style=none] (85) at (-7.5, 9.75) {};
		\node [style=none] (86) at (-5.5, 10.75) {};
		\node [style=none] (87) at (-3.5, 11.75) {};
		\node [style=none] (88) at (7.5, 6.25) {};
		\node [style=none] (89) at (5.5, 5.25) {};
		\node [style=none] (90) at (3.5, 4.25) {};
		\node [style=boundary vertex] (91) at (0, 12) {};
		\node [style=boundary vertex] (92) at (2, 11) {};
		\node [style=none] (93) at (-1.5, 12.75) {};
		\node [style=none] (94) at (9.5, 7.25) {};
		\node [style=boundary vertex] (95) at (-2, 7) {};
		\node [style=boundary vertex] (96) at (0, 8) {};
		\node [style=boundary vertex] (97) at (2, 9) {};
		\node [style=boundary vertex] (98) at (4, 10) {};
		\node [style=boundary vertex] (99) at (2, 13) {};
		\node [style=boundary vertex] (100) at (4, 12) {};
		\node [style=none] (101) at (0.5, 13.75) {};
		\node [style=none] (102) at (11.5, 8.25) {};
		\node [style=boundary vertex] (103) at (6, 11) {};
		\node [style=boundary vertex] (104) at (0, 6) {};
		\node [style=boundary vertex] (105) at (2, 7) {};
		\node [style=boundary vertex] (106) at (4, 8) {};
		\node [style=boundary vertex] (107) at (6, 9) {};
		\node [style=boundary vertex] (108) at (8, 10) {};
		\node [style=none] (109) at (3, 13.5) {};
		\node [style=none] (110) at (5, 12.5) {};
		\node [style=none] (111) at (7, 11.5) {};
		\node [style=none] (112) at (9, 10.5) {};
		\node [style=none] (113) at (1.5, 14.25) {};
		\node [style=none] (114) at (12.5, 8.75) {};
		\node [style=none] (115) at (-1, 5.5) {};
		\node [style=none] (116) at (-3, 6.5) {};
		\node [style=none] (117) at (2.5, 3.75) {};
		\node [style=none] (118) at (-5, 7.5) {};
		\node [style=none] (119) at (-7, 8.5) {};
		\node [style=none] (120) at (-8.5, 9.25) {};
		\node [style=boundary vertex] (121) at (2, 5) {};
		\node [style=boundary vertex] (122) at (4, 6) {};
		\node [style=boundary vertex] (123) at (6, 7) {};
		\node [style=boundary vertex] (124) at (8, 8) {};
		\node [style=boundary vertex] (125) at (10, 9) {};
		\node [style=none] (126) at (11, 9.5) {};
		\node [style=none] (127) at (1, 4.5) {};
	\end{pgfonlayer}
	\begin{pgfonlayer}{edgelayer}
		\draw (30) to (32);
		\draw (33) to (35);
		\draw (36.center) to (41.center);
		\draw (40.center) to (37.center);
		\draw (38.center) to (39.center);
		\draw (44.center) to (45.center);
		\draw (32) to (42);
		\draw (35) to (43);
		\draw (46) to (48);
		\draw (48) to (49);
		\draw (52.center) to (53.center);
		\draw (43) to (51);
		\draw (49) to (54);
		\draw (50) to (42);
		\draw (55) to (57);
		\draw (57) to (58);
		\draw (58) to (59);
		\draw (66.center) to (63.center);
		\draw (62.center) to (67.center);
		\draw (69.center) to (61.center);
		\draw (60.center) to (70.center);
		\draw [style={red_arrow}] (68.center) to (39.center);
		\draw [style={red_arrow}] (71.center) to (38.center);
		\draw (78.center) to (77.center);
		\draw [style={cyan_arrow}] (65.center) to (62.center);
		\draw [style={cyan_arrow}] (68.center) to (67.center);
		\draw [style={blue_edge}] (67.center) to (71.center);
		\draw [style={blue_edge}] (64.center) to (62.center);
		\draw [style={red_edge}] (39.center) to (65.center);
		\draw [style={red_edge}] (38.center) to (64.center);
		\draw (79) to (81);
		\draw (82) to (84);
		\draw (85.center) to (90.center);
		\draw (89.center) to (86.center);
		\draw (87.center) to (88.center);
		\draw (93.center) to (94.center);
		\draw (81) to (91);
		\draw (84) to (92);
		\draw (95) to (97);
		\draw (97) to (98);
		\draw (101.center) to (102.center);
		\draw (92) to (100);
		\draw (98) to (103);
		\draw (99) to (91);
		\draw (104) to (106);
		\draw (106) to (107);
		\draw (107) to (108);
		\draw (115.center) to (112.center);
		\draw (111.center) to (116.center);
		\draw (118.center) to (110.center);
		\draw (109.center) to (119.center);
		\draw [style={red_arrow}] (117.center) to (88.center);
		\draw [style={red_arrow}] (120.center) to (87.center);
		\draw (127.center) to (126.center);
		\draw [style={cyan_arrow}] (114.center) to (111.center);
		\draw [style={cyan_arrow}] (117.center) to (116.center);
		\draw [style={blue_edge}] (116.center) to (120.center);
		\draw [style={blue_edge}] (113.center) to (111.center);
		\draw [style={red_edge}] (88.center) to (114.center);
		\draw [style={red_edge}] (87.center) to (113.center);
		\draw [style=grey dash] (99) to (50);
		\draw [style=grey dash] (100) to (51);
		\draw [style=grey dash] (103) to (54);
		\draw [style=grey dash] (59) to (108);
		\draw [style=grey dash] (125) to (76);
		\draw [style=grey dash] (91) to (42);
		\draw [style=grey dash] (92) to (43);
		\draw [style=grey dash] (98) to (49);
		\draw [style=grey dash] (58) to (107);
		\draw [style=grey dash] (124) to (75);
		\draw [style=grey dash] (123) to (74);
		\draw [style=grey dash] (57) to (106);
		\draw [style=grey dash] (97) to (48);
		\draw [style=grey dash] (35) to (84);
		\draw [style=grey dash] (81) to (32);
		\draw [style=grey dash] (80) to (31);
		\draw [style=grey dash] (83) to (34);
		\draw [style=grey dash] (96) to (47);
		\draw [style=grey dash] (105) to (56);
		\draw [style=grey dash] (122) to (73);
	\end{pgfonlayer}
\end{tikzpicture}

%% file: Figures-Tikz/toric_codes4.tikz
\begin{tikzpicture}
	\begin{pgfonlayer}{nodelayer}
		\node [style=boundary vertex] (30) at (-8, -6) {};
		\node [style=boundary vertex] (31) at (-6, -5) {};
		\node [style=boundary vertex] (32) at (-4, -4) {};
		\node [style=boundary vertex] (33) at (-6, -7) {};
		\node [style=boundary vertex] (34) at (-4, -6) {};
		\node [style=boundary vertex] (35) at (-2, -5) {};
		\node [style=none] (36) at (-9.5, -5.25) {};
		\node [style=none] (37) at (-7.5, -4.25) {};
		\node [style=none] (38) at (-5.5, -3.25) {};
		\node [style=none] (39) at (5.5, -8.75) {};
		\node [style=none] (40) at (3.5, -9.75) {};
		\node [style=none] (41) at (1.5, -10.75) {};
		\node [style=boundary vertex] (42) at (-2, -3) {};
		\node [style=boundary vertex] (43) at (0, -4) {};
		\node [style=none] (44) at (-3.5, -2.25) {};
		\node [style=none] (45) at (7.5, -7.75) {};
		\node [style=boundary vertex] (46) at (-4, -8) {};
		\node [style=boundary vertex] (47) at (-2, -7) {};
		\node [style=boundary vertex] (48) at (0, -6) {};
		\node [style=boundary vertex] (49) at (2, -5) {};
		\node [style=boundary vertex] (50) at (0, -2) {};
		\node [style=boundary vertex] (51) at (2, -3) {};
		\node [style=none] (52) at (-1.5, -1.25) {};
		\node [style=none] (53) at (9.5, -6.75) {};
		\node [style=boundary vertex] (54) at (4, -4) {};
		\node [style=boundary vertex] (55) at (-2, -9) {};
		\node [style=boundary vertex] (56) at (0, -8) {};
		\node [style=boundary vertex] (57) at (2, -7) {};
		\node [style=boundary vertex] (58) at (4, -6) {};
		\node [style=boundary vertex] (59) at (6, -5) {};
		\node [style=none] (60) at (1, -1.5) {};
		\node [style=none] (61) at (3, -2.5) {};
		\node [style=none] (62) at (5, -3.5) {};
		\node [style=none] (63) at (7, -4.5) {};
		\node [style=none] (64) at (-0.5, -0.75) {};
		\node [style=none] (65) at (10.5, -6.25) {};
		\node [style=none] (66) at (-3, -9.5) {};
		\node [style=none] (67) at (-5, -8.5) {};
		\node [style=none] (68) at (0.5, -11.25) {};
		\node [style=none] (69) at (-7, -7.5) {};
		\node [style=none] (70) at (-9, -6.5) {};
		\node [style=none] (71) at (-10.5, -5.75) {};
		\node [style=boundary vertex] (72) at (0, -10) {};
		\node [style=boundary vertex] (73) at (2, -9) {};
		\node [style=boundary vertex] (74) at (4, -8) {};
		\node [style=boundary vertex] (75) at (6, -7) {};
		\node [style=boundary vertex] (76) at (8, -6) {};
		\node [style=none] (77) at (9, -5.5) {};
		\node [style=none] (78) at (-1, -10.5) {};
		\node [style=boundary vertex] (79) at (-6, 9) {};
		\node [style=boundary vertex] (80) at (-4, 10) {};
		\node [style=boundary vertex] (81) at (-2, 11) {};
		\node [style=boundary vertex] (82) at (-4, 8) {};
		\node [style=boundary vertex] (83) at (-2, 9) {};
		\node [style=boundary vertex] (84) at (0, 10) {};
		\node [style=none] (85) at (-7.5, 9.75) {};
		\node [style=none] (86) at (-5.5, 10.75) {};
		\node [style=none] (87) at (-3.5, 11.75) {};
		\node [style=none] (88) at (7.5, 6.25) {};
		\node [style=none] (89) at (5.5, 5.25) {};
		\node [style=none] (90) at (3.5, 4.25) {};
		\node [style=boundary vertex] (91) at (0, 12) {};
		\node [style=boundary vertex] (92) at (2, 11) {};
		\node [style=none] (93) at (-1.5, 12.75) {};
		\node [style=none] (94) at (9.5, 7.25) {};
		\node [style=boundary vertex] (95) at (-2, 7) {};
		\node [style=boundary vertex] (96) at (0, 8) {};
		\node [style=boundary vertex] (97) at (2, 9) {};
		\node [style=boundary vertex] (98) at (4, 10) {};
		\node [style=boundary vertex] (99) at (2, 13) {};
		\node [style=boundary vertex] (100) at (4, 12) {};
		\node [style=none] (101) at (0.5, 13.75) {};
		\node [style=none] (102) at (11.5, 8.25) {};
		\node [style=boundary vertex] (103) at (6, 11) {};
		\node [style=boundary vertex] (104) at (0, 6) {};
		\node [style=boundary vertex] (105) at (2, 7) {};
		\node [style=boundary vertex] (106) at (4, 8) {};
		\node [style=boundary vertex] (107) at (6, 9) {};
		\node [style=boundary vertex] (108) at (8, 10) {};
		\node [style=none] (109) at (3, 13.5) {};
		\node [style=none] (110) at (5, 12.5) {};
		\node [style=none] (111) at (7, 11.5) {};
		\node [style=none] (112) at (9, 10.5) {};
		\node [style=none] (113) at (1.5, 14.25) {};
		\node [style=none] (114) at (12.5, 8.75) {};
		\node [style=none] (115) at (-1, 5.5) {};
		\node [style=none] (116) at (-3, 6.5) {};
		\node [style=none] (117) at (2.5, 3.75) {};
		\node [style=none] (118) at (-5, 7.5) {};
		\node [style=none] (119) at (-7, 8.5) {};
		\node [style=none] (120) at (-8.5, 9.25) {};
		\node [style=boundary vertex] (121) at (2, 5) {};
		\node [style=boundary vertex] (122) at (4, 6) {};
		\node [style=boundary vertex] (123) at (6, 7) {};
		\node [style=boundary vertex] (124) at (8, 8) {};
		\node [style=boundary vertex] (125) at (10, 9) {};
		\node [style=none] (126) at (11, 9.5) {};
		\node [style=none] (127) at (1, 4.5) {};
	\end{pgfonlayer}
	\begin{pgfonlayer}{edgelayer}
		\draw (33) to (35);
		\draw (36.center) to (41.center);
		\draw (40.center) to (37.center);
		\draw (38.center) to (39.center);
		\draw (44.center) to (45.center);
		\draw (32) to (42);
		\draw (35) to (43);
		\draw (46) to (48);
		\draw (48) to (49);
		\draw (52.center) to (53.center);
		\draw (43) to (51);
		\draw (49) to (54);
		\draw (50) to (42);
		\draw (55) to (57);
		\draw (57) to (58);
		\draw (58) to (59);
		\draw (66.center) to (63.center);
		\draw (62.center) to (67.center);
		\draw (69.center) to (61.center);
		\draw [style={red_arrow}] (68.center) to (39.center);
		\draw [style={red_arrow}] (71.center) to (38.center);
		\draw (78.center) to (77.center);
		\draw [style={cyan_arrow}] (65.center) to (62.center);
		\draw [style={cyan_arrow}] (68.center) to (67.center);
		\draw [style={blue_edge}] (67.center) to (71.center);
		\draw [style={blue_edge}] (64.center) to (62.center);
		\draw [style={red_edge}] (39.center) to (65.center);
		\draw [style={red_edge}] (38.center) to (64.center);
		\draw (82) to (84);
		\draw (85.center) to (90.center);
		\draw (89.center) to (86.center);
		\draw (87.center) to (88.center);
		\draw (93.center) to (94.center);
		\draw (81) to (91);
		\draw (84) to (92);
		\draw (95) to (97);
		\draw (97) to (98);
		\draw (101.center) to (102.center);
		\draw (92) to (100);
		\draw (98) to (103);
		\draw (99) to (91);
		\draw (104) to (106);
		\draw (106) to (107);
		\draw (107) to (108);
		\draw (115.center) to (112.center);
		\draw (111.center) to (116.center);
		\draw (118.center) to (110.center);
		\draw [style={red_arrow}] (117.center) to (88.center);
		\draw [style={red_arrow}] (120.center) to (87.center);
		\draw (127.center) to (126.center);
		\draw [style={cyan_arrow}] (114.center) to (111.center);
		\draw [style={cyan_arrow}] (117.center) to (116.center);
		\draw [style={blue_edge}] (116.center) to (120.center);
		\draw [style={blue_edge}] (113.center) to (111.center);
		\draw [style={red_edge}] (88.center) to (114.center);
		\draw [style={red_edge}] (87.center) to (113.center);
		\draw [style=grey dash] (100) to (51);
		\draw [style=grey dash] (103) to (54);
		\draw [style=grey dash] (59) to (108);
		\draw [style=grey dash] (125) to (76);
		\draw [style=grey dash] (91) to (42);
		\draw [style=grey dash] (92) to (43);
		\draw [style=grey dash] (98) to (49);
		\draw [style=grey dash] (58) to (107);
		\draw [style=grey dash] (124) to (75);
		\draw [style=grey dash] (123) to (74);
		\draw [style=grey dash] (57) to (106);
		\draw [style=grey dash] (97) to (48);
		\draw [style=grey dash] (35) to (84);
		\draw [style=grey dash] (81) to (32);
		\draw [style=grey dash] (83) to (34);
		\draw [style=grey dash] (96) to (47);
		\draw [style=grey dash] (105) to (56);
		\draw [style=grey dash] (122) to (73);
		\draw [style={green_edge}] (99) to (109.center);
		\draw [style={green_edge}] (50) to (60.center);
		\draw [style={green_edge}] (70.center) to (30);
		\draw [style={green_edge}] (119.center) to (79);
		\draw [style={green_edge}] (79) to (80);
		\draw (81) to (80);
		\draw (31) to (32);
		\draw [style={green_edge}] (30) to (31);
		\draw [style={green_edge}] (80) to (31);
		\draw [style={green_edge}] (50) to (99);
	\end{pgfonlayer}
\end{tikzpicture}

%% file: appendix.tex
\appendix

\section{Proofs concerning fault-distance}\label{app:proof_fault_distance}

In this appendix we provide detailed proofs of the phenomenological fault-distance lower bounds described in the main text. 
The standard approach to proving phenomenological fault-distance is to construct a measurement circuit, where Pauli error faults can occur at integer timesteps and check faults can occur at half-integer timesteps, with detectors \cite{McEwen2023relaxing, derks2024designing}, and then to show that any fault that does not trigger any detectors and is not a product of spacetime stabilisers must have at least a given weight $d$.

It is difficult to prove the fault-distances which we require in that setting, and so we use an alternative, almost equivalent one: that of fault complexes \cite{hillmann2024single}. This enables us to use homology to classify the logical faults and bound their corresponding weights. In this setting, Pauli error faults of $Z$ type and $Z$ check errors occur on even timesteps, while Pauli error faults of $X$ type and $X$ check errors occur on odd timesteps.

A sketch of the proofs in this section works as follows: we construct fault complexes for the idling operation, and then modify them using mapping cones to obtain fault complexes for various surgery protocols. These result in spacetime volumes where the code is initially in the idling operation for $d$ rounds, then undergoes a series of deformations to perform surgery with no padding between surgery rounds, and finally all auxiliary systems are measured out and the code returns to the idling operation for $d$ rounds. Because the spacetime volumes are defined in terms of chain complexes, and the deformations are described in terms of mapping cones, we can construct long exact sequences which relate the classes of logical faults in the spacetime volumes for surgery to the classes of logical faults in the idling spacetime volume. We then organise these classes of logical faults and show that they each must have weight at least $d$.

 Our techniques may be of independent interest to study other protocols in circuit-based or measurement-based quantum computing.

Before we move to discussing fault complexes, we give some elementary homological background which is required.
The Snake Lemma \cite[Lem.~1.3.2]{Weib1994} and rank-nullity together imply that for a mapping cone $\cone(f_\bullet)$ defined by a chain map $f_\bullet: A_\bullet \rightarrow C_\bullet$, we have $|H_i(\cone(f))| = |H_i(C)| + |H_{i-1}(A)| - |\im H_i(f)| - |\im H_{i-1}(f)|$, where $|V|$ is shorthand for $\dim V$ when $V$ is a vector space.

In more detail, the Snake Lemma dictates that for the mapping cone $\cone(f_\bullet)$ we have a long exact sequence \footnote{This means that the image of each map coincides with the kernel of the next map.}:
\[\begin{tikzcd}
    \cdots \arrow[r] & H_i(A) \arrow[r, "H_i(f)"] & H_i(C) \arrow[r, "\iota_*"] & H_i(\cone(f)) \arrow[r, "\pi_*"] & H_{i-1}(A) \arrow[r, "H_{i-1}(f)"] & H_{i-1}(C) \arrow[r] & \cdots
\end{tikzcd}\]
In this sequence the $H_i(f)$ are the maps between homology spaces $H_i(A) \rightarrow H_i(C)$, which are well-defined by the commutative diagram in Eq.~\ref{eq:cd_chain_map}.

$\iota_*$ is a map $H_i(C) \rightarrow H_i(\cone (f))$ generated by the Snake Lemma. In homological measurement~\cite{ide2024fault}, this map at $i=1$ takes logical operators in the original code to the logical operators in the deformed code when performing surgery. In that case, the map $\iota_*$ will be a surjection, as some of the logical qubits from the original code are measured out and no new logical qubits are added when the auxiliary hypergraph's cycles are gauge-fixed.

$\pi_*$ is a map $H_i(\cone (f)) \rightarrow H_{i-1}(A)$. In homological measurement~\cite{ide2024fault}, this map at $i=1$ picks out the new contributions to the logical space of the deformed code from $H_{i-1}(A)$; exactness at $H_i(\cone (f))$ shows that the only new logical qubits which can arise are from cycles in the hypergraph.

In our case we will take mapping cones on spacetime volumes, rather than just codes, but a similar intuition applies.
 
For any linear map $T: V \rightarrow W$ rank-nullity implies that $|V| = |\ker T| + |\im T|$, so we can compute
\begin{align}
    |H_i(\cone(f))| &= |\ker \pi_*| + |\im \pi_*| = |\im \iota_*| + |\im \pi_*| \nonumber \\
    &= |H_i(C)| - |\ker \iota_*| + |\ker H_{i-1}(f)| \nonumber \\ 
    &= |H_i(C)| - |\im H_i(f)| + |H_{i-1}(A)| - |\im H_{i-1}(f)|.
    \label{eq:exact_dims}
\end{align}

The premise of all of our proofs is that in a CSS-type code surgery, where the initial code and deformed codes are all CSS codes, all logical faults must be either of $\overline{Z}$ type in spacetime, composed of $Z$ Pauli errors and $X$ check errors, or $\overline{X}$ type, composed of $X$ Pauli errors and $Z$ check errors. As any logical fault composed of a product of these is at least as high weight as the corresponding logical fault of either $\overline{Z}$ or $\overline{X}$ type, we need only show that the weights of all $\overline{Z}$ and $\overline{X}$ logical faults are at least $d$ to prove that the fault-distance of a protocol is at least $d$.

We will frequently make use of the fact that when $H_i(f)$ is injective $|\im H_i(f)| = |H_i(A)|$, by definition.

\subsection{Fault complexes}

A fault complex is defined as a chain complex with 4 terms,
\[F_\bullet = \begin{tikzcd}
    F_3 \arrow[r] & F_2 \arrow[r] & F_1 \arrow[r] & F_0 
\end{tikzcd}\]
where by convention the $F_3$ and $F_0$ components correspond to $Z$ and $X$ detectors respectively. In fault-tolerant measurement-based quantum computing (MBQC), $F_2$ corresponds to a set of $X$ fault locations, i.e. locations in spacetime which can experience only $X$-type faults, while $F_1$ corresponds to a set of $Z$-type fault locations. The fault complex is therefore in bijection with a graph state.

When the graph state corresponds to a foliated CSS code, this can be reinterpreted in the circuit model, undergoing phenomenological noise. Now, $F_2$ corresponds to a combined set of spacetime locations at which data qubits can experience $X$ Pauli faults and also $Z$ checks, which can experience measurement errors.
$F_1$ is the same but for $Z$ Pauli faults and $X$ checks. In this case, we can expand out a fault complex to have the following diagram:
\[\begin{tikzcd}
    & F_{1,1} \arrow[r]\arrow[ddr] & F_{1,0} \arrow[dr] & \\
    F_{1,2} \arrow[ur] \arrow[dr] & & & F_{0,0} \\
    & F_{0,2} \arrow[r]& F_{0,1} \arrow[ur] &
\end{tikzcd}\]
where components in the overall fault complex are given by taking direct sums vertically in the diagram, so we have
\[\begin{tikzcd}F_{1,2}\arrow[r] & F_{1,1}\oplus F_{0,2} \arrow[r] & F_{1,0}\oplus F_{0,1} \arrow[r] & F_{0,0} \end{tikzcd}\]
Throughout this section every diagram drawn in the above form is assumed to have direct sums vertically in the diagram. \footnote{This is essentially passing to the ``total complex" \cite{Weib1994}.}

In the phenomenological circuit model:
\begin{itemize}
    \item $F_{0,0}$ corresponds to $X$ detectors.
    \item $F_{0,1}$ corresponds to locations for $Z$ Pauli faults on data qubits.
    \item $F_{1,0}$ corresponds to locations for $X$ check faults.
    \item $F_{0,2}$ corresponds to locations for $Z$ check faults.
    \item $F_{1,1}$ corresponds to locations for $X$ Pauli faults on data qubits.
    \item $F_{1,2}$ corresponds to $Z$ detectors.
\end{itemize}

Expressed as a commutative diagram the fault complex then has the structure:
\[\begin{tikzcd}
    & \textnode{$X$ Pauli faults} \arrow[r]\arrow[ddr] & \textnode{$X$ check faults} \arrow[dr] & \\
    \textnode{$Z$ detectors} \arrow[ur] \arrow[dr] & & & \textnode{$X$ detectors} \\
    & \textnode{$Z$ check faults} \arrow[r]& \textnode{$Z$ Pauli faults} \arrow[ur] &
\end{tikzcd}\]

\[\begin{tikzcd}
    & \textnode{$F_{1,1}$ \\ ($X$ Pauli faults)} \arrow[r]\arrow[ddr] & \textnode{$F_{1,0}$ \\ ($X$ check faults)} \arrow[dr] & \\
    \textnode{$F_{1,2}$ \\ ($Z$ detectors)} \arrow[ur] \arrow[dr] & & & \textnode{$F_{0,0}$ \\ ($X$ detectors)} \\
    & \textnode{$F_{0,2}$ \\ ($Z$ check faults)} \arrow[r]& \textnode{$F_{0,1}$ \\ ($Z$ Pauli faults)} \arrow[ur] &
\end{tikzcd}\]

Note that as a fault complex describes a graph state, it does not have the ability to describe morphisms, i.e. MBQC-based or circuit-based protocols with input or output qubits.

The standard way to foliate a CSS code $C_\bullet$ from the fault complex perspective is to take $(\CR \otimes C)_\bullet$, where $\CR_\bullet : \CR_1 \rightarrow \CR_0$ has the differential $R$ being either the full-rank parity-check matrix for the repetition code or its dual. In the former, the code is initialised in the $\ket{0}$ state and measured out in the $Z$ basis; in the latter, the code is initialised in the $\ket{+}$ state and measured out in the $X$ basis. Hence the code `idles' for $l$ rounds, where $l$ is the blocklength of the repetition code.

If we do this, the fault complex has the explicit form:
\begin{equation}\begin{tikzcd}
    & \CR_1 \otimes C_1 \arrow[r]\arrow[ddr] & \CR_1 \otimes C_0 \arrow[dr] & \\
    \CR_1 \otimes C_2 \arrow[ur] \arrow[dr] & & & \CR_0 \otimes C_0 \\
    & \CR_0 \otimes C_2 \arrow[r]& \CR_0 \otimes C_1 \arrow[ur] &
\end{tikzcd}\label{eq:tensor_prod_fc}\end{equation}
In the circuit model we have, in order,
\begin{itemize}
    \item $\CR_0\otimes C_0$ corresponds to $X$ detectors.
    \item $\CR_0\otimes C_1$ corresponds to locations for $Z$ Pauli error faults.
    \item $\CR_1 \otimes C_0$ corresponds to locations for $X$ check error faults.
    \item $\CR_0\otimes C_2$ corresponds to locations for $Z$ check error faults.
    \item $\CR_1 \otimes C_1$ corresponds to locations for $X$ Pauli error faults.
    \item $\CR_1\otimes C_2$ corresponds to $Z$ detectors.
\end{itemize}
Note that $\CR_1\otimes C_1$ and $\CR_0\otimes C_1$ correspond to the same data qubits but at different points in time; in the circuit model this is purely convention, and any given data qubit can still undergo either type of error, $Z$ or $X$.

It is easy to compute the homologies using the K{\"u}nneth formula. Then elements of $H_1(F)$ correspond to equivalence classes of logical $\overline{Z}$ faults in spacetime, composed of $Z$ Pauli errors and $X$ check errors, and elements of $H^2(F)$ to logical $\overline{X}$ faults, composed of $X$ Pauli errors and $Z$ check errors.\footnote{There are also the homology groups $H_1(F)$ and $H^4(F)$, corresponding to classes of detector triggers which are impossible.}

For completeness, we have,
\begin{itemize}
    \item $H_1(\CR\otimes C) = H_1(\CR)\otimes H_0(C) \oplus H_0(\CR)\otimes H_1(C)$.
    \item $H^2(\CR\otimes C) = H^1(\CR)\otimes H^1(C) \oplus H^0(\CR)\otimes H^2(C)$.
\end{itemize}
When $\CR_\bullet$ is the repetition code, we have \begin{align*}&H_1(\CR\otimes C) = H_1(\CR)\otimes H_0(C) = \{\underline{0}, \underline{1}\}\otimes C_0/\im H_X\\
&H^2(\CR\otimes C) = H^1(\CR)\otimes H^1(C) = \{\underline{0}, (1,0,0,\cdots,0)\}\otimes H^1(C),
\end{align*}
where $\underline{0}$ and $\underline{1}$ are the all-zero and all-one vectors, and $(1, 0,0, \cdots, 0)$ is a weight 1 vector representing the nonzero equivalence class in $H^1(\CR)$; any vector with odd weight is in the same class. When $\CR_\bullet$ is dual to the repetition code we instead have
\begin{align*}&H_1(\CR\otimes C) = H_0(\CR)\otimes H_1(C) = \{\underline{0}, (1,0,0,\cdots,0)\}\otimes H_1(C)\\
&H^2(\CR\otimes C) = H^0(\CR)\otimes H^2(C) = \{\underline{0}, \underline{1}\}\otimes H^2(C).
\end{align*}

\begin{figure}
\[\input{Figures-Tikz/unit_cell_fc.tikz}\]\caption{A fault complex for a unit cell of the 2D surface code, starting and terminating at $X$-type boundaries. Time flows from bottom to top.}\label{fig:unit_cell}\end{figure}
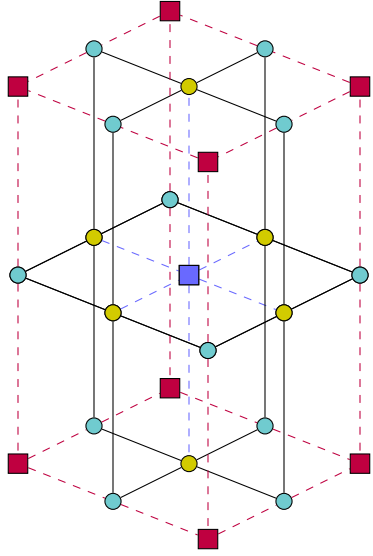

In Fig.~\ref{fig:unit_cell} we show the fault complex for a small surface code as a Tanner graph, where circles are fault locations and squares are detectors, and red, teal, yellow and blue indicate basis elements of $F_0, F_1, F_2,F_3$ respectively.

Fault complexes are a useful algebraic framework for reasoning about fault distance of CSS codes in the phenomenological model: each location is a point in spacetime which can experience a fault, which is also a basis vector in a space (either $F_1$ or $F^2$) and so weights of faults can be related directly to weights of vectors in the fault complex.

We do not want to presume that the code is initialised and measured out in a given basis, because we want to prove that a given protocol has sufficient fault-distance no matter what other (assumed fault-tolerant) operations the logical circuit contains. One approach to this is to replace the parity-check matrix of $\CR_\bullet$ with the parity-check matrix of a different repetition code, that of the cycle graph. This corresponds to `taking the trace' of the circuit operation being performed, and ensures that errors cannot vanish on the initial and final boundaries. However, this leads to other, serious issues: a timelike fault which runs from the start to the finish of the spacetime volume is now a spacetime stabiliser, and there are other logical faults which we encounter that are equivalent to this fault, and so they will be unaccounted for.

Instead, we modify the spacetime volume by adding checks to the start and end, which constitute entry points for timelike errors to enter and exit the volume: depending on whether $\CR_\bullet$ is a full-rank repetition code or its dual, corresponding to initialising in the $\ket{\overline{0}}$ or $\ket{\overline{+}}$ state respectively, the new checks are $Z$ checks or $X$ checks.

\begin{figure}
\[\input{Figures-Tikz/unit_cell_newchecks_fc.tikz}\]\caption{A fault complex for a unit cell of the 2D surface code, starting and terminating at $X$-type boundaries, but with new $X$ checks to allow $\overline{Z}$ timelike errors to enter and exit the volume.}\label{fig:unit_cell_newchecks}\end{figure}
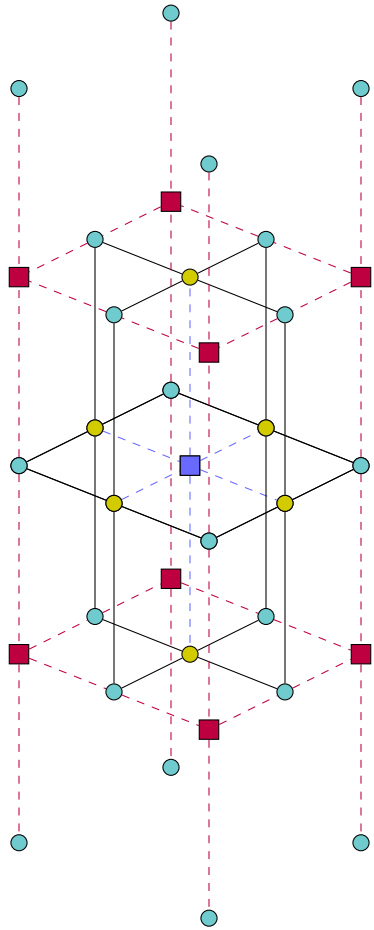

We accomplish this using mapping cones. Explicitly, when $\CR_\bullet$ is the dual of the repetition code, so $\CR_\bullet = \F_2^{l-1} \rightarrow \F_2^l$ for some $l$ determining the number of rounds in the spacetime volume, the code is initialised in $\ket{+}$ and measured out in the $X$-basis; we then introduce new checks for timelike fault entrypoints by taking the mapping cone of the chain map $g_\bullet$,
\[\begin{tikzcd}
    0 \arrow[r]\arrow[d] & 0 \arrow[r]\arrow[d] & 0 \arrow[r]\arrow[d] & C_0\oplus C_0 \arrow[d, "g_0"]\\
    F_3 \arrow[r] & F_2 \arrow[r] & F_1 \arrow[r] & F_0
\end{tikzcd}\]
where $g_0$ maps the first copy of $C_0$ into the top layer of $X$-detectors in the fault complex, and the second copy of $C_0$ into the bottom layer of $X$-detectors. Call the resultant fault complex $F' = \cone(g_\bullet)$. The unit cell from Fig.~\ref{fig:unit_cell}, once modified in this manner, is shown in Fig.~\ref{fig:unit_cell_newchecks}.

The consequent classes of logical faults in $\cone(g_\bullet)$ can be computed using the Snake Lemma, see Eq.~\ref{eq:exact_dims}. We have 
\begin{align*}
    |H_1(F')| &= |H_1(\cone(g))| = |H_1(F)| + |C_0 \oplus C_0| - |\im H_0(g)| \\
    &= |H_1(F)| + 2|C_0| - |\im H_0(g)|.
\end{align*}
By the K{\"u}nneth formula, $H_0(F) = H_0(\CR)\otimes H_0(C) = \{\underline{0}, (1,0,0,\cdots,0)\} \otimes H_0(C)$, so $|H_0(F)| = |H_0(C)|$. By inspection, every element of one layer of $X$-detectors at the bottom of the fault complex is in the image of $g_\bullet$; those at the top are in the same equivalence class. Hence $|\im H_0(g)| = |H_0(F)| = |H_0(C)|$, so 
\[|H_1(F')| = |H_1(\cone(g))| = |H_1(F)| + 2|C_0| - |H_0(C)|.\]

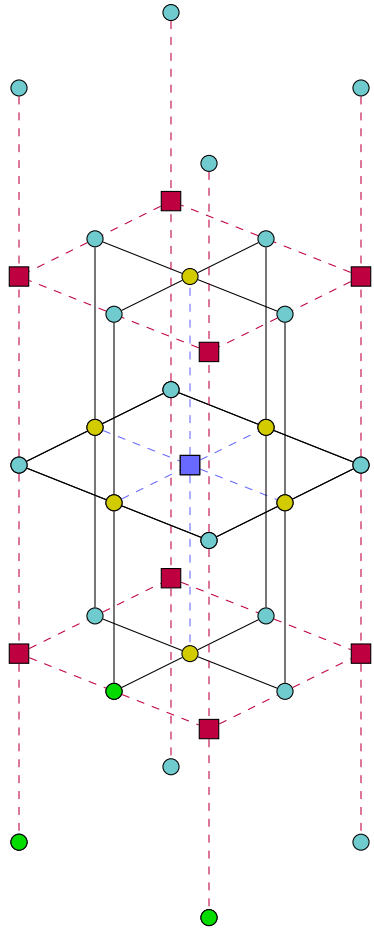
\begin{figure}
\[\input{Figures-Tikz/unit_cell_newchecks_fc_redundant_fault.tikz}\]\caption{The fault complex from Fig.~\ref{fig:unit_cell_newchecks}, where a $\overline{Z}$ timelike fault enters the fault complex and is immediately annihilated by a spacelike fault. The highlighted fault locations are shown in green.}\label{fig:unit_cell_newchecks_redundant_fault}\end{figure}

\begin{figure}
\[\input{Figures-Tikz/unit_cell_newchecks_fc_timelike_fault.tikz}\]\caption{The fault complex from Fig.~\ref{fig:unit_cell_newchecks}, where a $\overline{Z}$ timelike fault extends from one timelike boundary to the other. The highlighted fault locations are shown in green.}\label{fig:unit_cell_newchecks_timelike_fault}\end{figure}
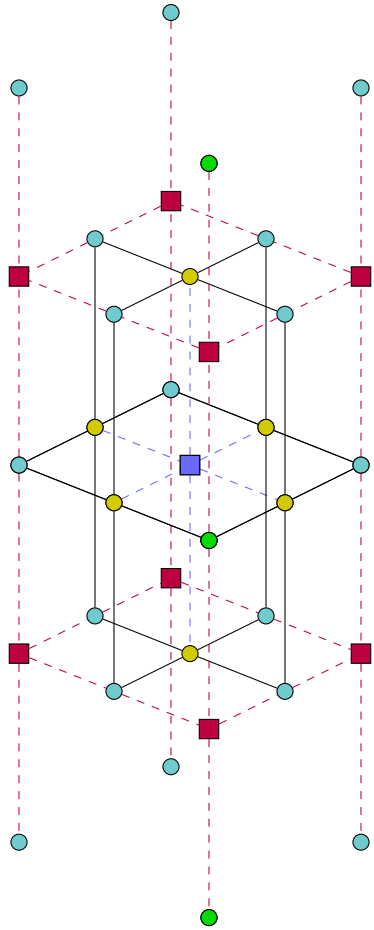

We therefore add $2m_X - r_X$ new basis elements to $H_1(F)$, where $m_X = |C_0|$ is the number of $X$ checks in $C_\bullet$ and $r_X = |H_0(C)|$ is the number of redundant $X$ checks in $C_\bullet$. All new elements of $H_1(F')$ correspond to logical faults which enter from a timelike boundary on the top or bottom;
$m_X-r_X$ of these can be immediately annihilated by spacelike errors on the boundary, or by spacelike errors on further layers. We illustrate this in Fig.~\ref{fig:unit_cell_newchecks_redundant_fault}, where incoming $X$ check errors turn on $X$ detectors, which can then be immediately switched off again by $Z$ data qubit Pauli errors. The remaining $m_X$ are strings of $X$ check failures running from the bottom to the top boundary, illustrated in Fig.~\ref{fig:unit_cell_newchecks_timelike_fault}. In both cases, all new logical faults must have timelike support on the new checks at the boundary, and these cannot be cleaned from the boundary into the bulk.

We can similarly calculate that
\begin{align*}
    |H_0(F')| &= |H_0(F)| + 0 - |\im H_0(g)| - 0 \\
    &= |H_0(F)| - |H_0(F)| = 0.
\end{align*}

For the dual version, where $\CR_\bullet = \F_2^l \rightarrow \F_2^{l-1}$, we instead perform the mapping cone on the cochain complex, and add $2m_Z - r_Z$ new elements to $H^2(F)$. Additionally, $H^3(F)$ becomes 0.

In summary, when the initialisation and measurement is in the $X$ basis the spacetime volume for the idling operation now contains:
\begin{itemize}
    \item All $\overline{Z}$ timelike logical faults, extending through the volume.
    \item Those $\overline{X}$ timelike logical faults with check errors in $H^2(C)$ which are not equivalent to spacelike faults.
    \item All $\overline{Z}$ spacelike logical faults.
    \item No $\overline{X}$ spacelike logical faults.
\end{itemize}
and when it is in the $Z$ basis the volume is the same but with the $\overline{X}$ and $\overline{Z}$ logical fault characterisation inverted. Therefore if we use both of these fault complexes then, combined, we account for all possible logical faults, as the code (and fault complex) is CSS.

We wish to describe complicated surgery protocols, rather than the idling state. Now, a certain fault complex for the idling state is sufficient to study certain stability experiments, but insufficient for the description of surgery protocols involving many logical measurements. For this the fault complex must describe the actual code deformation taking place. This can also be accomplished using mapping cones, as we shall demonstrate presently.

\subsection{Full block reading}

We are now ready to prove the fault-distance of full block reading.
In this case, we start with taking $E_\bullet = \F_2^c \otimes C_\bullet$ as our initial CSS code in the idling state, so the initial fault complex is $F_\bullet = (\CR \otimes E)_\bullet$. We then consider the two cases, one where the idling volume starts and ends with $Z$ basis measurements, and the other with $X$ basis measurements, including the entrypoints for timelike errors, so that we acquire two different fault complexes. For convenience we label them both $F'_\bullet$, and the difference will be made clear by dividing into the case for $X$-type boundaries and $Z$-type boundaries.

In each case, we are applying mapping cones to the idling volume. To perform a full block reading in the $Z$ basis, measuring $\overline{Z}$ logical operators, the mapping cone is on the chain complex, and in the $X$ basis the mapping cone is on the cochain complex.

In the $Z$ basis, the mapping cone on the fault complex is taken from the chain map $f_\bullet$ with components:
\[\begin{tikzcd}
    0 \arrow[r]\arrow[d, "f_3"] & C_2 \arrow[r]\arrow[d, "f_2"] & C_1 \arrow[d, "f_1"]\arrow[r] & C_0 \arrow[d, "f_0"] \\
    F'_3 \arrow[r] & F'_2 \arrow[r] & F'_1 \arrow[r] & F'_0
\end{tikzcd}\]
where the components map the auxiliary code $C_\bullet$ to the codeblocks in $E_\bullet$ to be measured, at a primal layer. In other words, in the case where we start and finish in the $X$ basis we now have the fault complex
\[\begin{tikzcd}
    & & \textnode{$E_0$ \\ (Bottom $X$ check faults)} \arrow[dddr] & \\
    & & \textnode{$E_0$ \\ (Top $X$ check faults)} \arrow[ddr] & \\
    & \textnode{$\CR_1 \otimes E_1$ \\ (Old $X$ Pauli faults)} \arrow[r]\arrow[ddr] & \textnode{$\CR_1 \otimes E_0$ \\ (Old $X$ check faults)} \arrow[dr] & \\
    \textnode{$\CR_1 \otimes E_2$ \\ (Old $Z$ detectors)} \arrow[ur] \arrow[dr] & & & \textnode{$\CR_0 \otimes E_0$ \\ (Old $X$ detectors)} \\
    & \textnode{$\CR_0 \otimes E_2$ \\ (Old $Z$ check faults)} \arrow[r]& \textnode{$\CR_0 \otimes E_1$ \\ (Old $Z$ Pauli faults)} \arrow[ur] & \\
    \textnode{$C_2$ \\ (New $Z$ detectors)} \arrow[ur] \arrow[r] & \textnode{$C_1$ \\ (New $Z$ check faults)} \arrow[ur] \arrow[r] & \textnode{$C_0$ \\ (New $Z$ Pauli faults)} \arrow[uur] & 
\end{tikzcd}\]
where the copies of $E_0$ at the top of the diagram are to add timelike entrypoints to the spacetime volume, and the $C_\bullet$ at the bottom performs the full block reading.

We can now flip the boundary conditions, so that $\CR_\bullet = \F_2^l \rightarrow \F_2^{l-1}$, and flip the additional timelike entrypoints accordingly, but still perform the same full block reading, to acquire the fault complex
\[\begin{tikzcd}
    & \textnode{$E_2$ \\ (Bottom $Z$ check faults)} & & \\
    & \textnode{$E_2$ \\ (Top $Z$ check faults)} & & \\
    & \textnode{$\CR_1 \otimes E_1$ \\ (Old $X$ Pauli faults)}  \arrow[r]\arrow[ddr] & \textnode{$\CR_1 \otimes E_0$ \\ (Old $X$ check faults)}  \arrow[dr] & \\
    \textnode{$\CR_1 \otimes E_2$ \\ (Old $Z$ detectors)} \arrow[ur]\arrow[uur]\arrow[uuur] \arrow[dr] & & & \textnode{$\CR_0 \otimes E_0$ \\ (Old $X$ detectors)} \\
    & \textnode{$\CR_0 \otimes E_2$ \\ (Old $Z$ check faults)} \arrow[r]& \textnode{$\CR_0 \otimes E_1$ \\ (Old $Z$ Pauli faults)} \arrow[ur] & \\
    \textnode{$C_2$ \\ (New $Z$ detectors)} \arrow[ur] \arrow[r] & \textnode{$C_1$ \\ (New $Z$ check faults)} \arrow[ur] \arrow[r] & \textnode{$C_0$ \\ (New $Z$ Pauli faults)} \arrow[uur] & 
\end{tikzcd}\]

Due to the change in boundary conditions, each of these fault complexes has different logical faults, and we must account for them both. 

There are some important consequences of using the fault complex formalism rather than the phenomenological circuit-based model. It dictates that each fault location in the fault complex is a qubit initialised in the $\ket{+}$ state and then measured out in the $X$ basis, with the entangling $CZ$ gates performing space- and time-like Hadamards to exchange error types between layers. As a consequence, unlike in the circuit-based model, there are not separate timesteps on which the new ancilla data qubits are initialised in $\ket{+}$ and measured out in the $X$ basis, as this is included in the fault complex layers. So, in the circuit model, one would initialise the new data qubits in $\ket{+}$, then measure all checks for 1 round, then measure out the new data qubits in the $X$ basis, which would generally be considered 3 timesteps: one each for the `merge', measure, and `split' steps.
In the fault complex picture, this all happens on 1 timestep. This is why there is only one `copy' of $C_\bullet$ used for the auxiliary system in the mapping cone.
    
Additionally, the layers of $Z$ and $X$ checks are alternating, rather than occurring on the same timestep as in the circuit-based model. When a surgery protocol only uses 1 round of $Z$ measurements, as in full block reading, the fault complex does not introduce $X$ checks connections to the new data qubits at all. There are also no new fault locations for $X$ Pauli data qubit errors on the auxiliary system, as they would either act on data qubits in $\ket{+}$ or before the qubits are measured out in the $X$ basis.

\begin{remark}[Multiple measurement rounds]
In Section~\ref{app:pbr_low_distance} below we see cases in which the number of measurement rounds required is greater than 1, in which case there are more than one `copy' of $C_\bullet$ employed, and there are deformed $X$ checks and new fault locations for $X$ Pauli data qubit errors.
\end{remark}

\begin{remark}[Cycle detectors]
For all the full block reading procedures we do not require any cycle checks, or detectors related to those checks. In Thm.~\ref{thm:full_block_amortised} we have imposed a condition that no measured logical Paulis should be products of any others, i.e. the proof requires that the set of measured logical Paulis is minimal for the group it generates. This condition can be removed if one uses detectors for the cycles in the hypergraph: these are not cycle checks, merely detectors inferred from the initial state preparation and final measurement of the new data qubits, so the code will remain LDPC, but the detector matrix may not unless the cycle basis is sparse.
\end{remark}

\FDfullblock*
\proof
We treat the case of $X$-type boundary conditions first.

\textbf{$X$-type boundary conditions.}

By the Snake Lemma, we can compute the following.
\begin{align*}
        |H_1(\cone(f))| &= |H_1(F')| - |\im H_1(f)| + |H_0(C)| - |\im H_0(f)|\\ 
        &= |H_1(F')| - |H_1(C)| + |H_0(C)| - 0 \\
        &= |H_1(F')| - |H_1(C)| + |H_0(C)|.
\end{align*}

Applying the same argument to $H_2(\cone (f))$, we have:
\begin{itemize}
    \item $|H_1(\cone(f))| = |H_1(F')| - |H_1(C)| + |H_0(C)|$.
    \item $|H^2(\cone(f))| = |H^2(F')| - |H^2(C)|$.
\end{itemize}
We first discuss the negative contributions to the classes of logical faults.

The negative contribution to $\overline{Z}$ logical faults (faults that commute with all $X$ detectors) appears because $|H_1(C)|$ logical measurements have occurred in the $Z$ basis, and so there are $|H_1(C)|$ spacelike $\overline{Z}$ errors which no longer affect the logical state.

The negative contribution to $\overline{X}$ logical faults (faults that commute with all $Z$ detectors) appears because the full block reading introduces $Z$ detectors between codeblocks, which require timelike $Z$ check errors to now be in $\ker H_Z$ of the auxiliary code in order to be undetected.

As the space distance of the code and deformed code are each at least $d$, and there are no new equivalence classes of logical faults, we cannot clean any logical fault with these boundary conditions to have weight below $d$.

\textbf{$Z$-type boundary conditions.}
In this case the equivalence classes of logical faults can be computed for $H_1(\cone (f))$ as
\begin{align*}
        |H_1(\cone(f))| &= |H_1(F')| - |\im H_1(f)| + |H_0(C)| - |\im H_0(f)|\\ 
        &= |H_1(F')| - 0 + |H_0(C)| - 0 \\
        &= |H_1(F')| + |H_0(C)|.
\end{align*}
Applying the same argument to $H_2(\cone(f))$ we have:
\begin{itemize}
    \item $|H_1(\cone(f))| = |H_1(F')| + |H_0(C)|$.
    \item $|H^2(\cone(f))| = |H^2(F')| + |H^1(C)| - |H^2(C)|$.
\end{itemize}

where $H_1(F')$ and $H^2(F')$ are equivalence classes already present before the full block reading, and $H_0(C)$ and $H^1(C)$ are new. Again we have a negative contribution of $|H^2(C)|$ to the classes of $Z$ check faults.

Note that some of the contributions to equivalence classes appear in both boundary conditions.

For the new $\overline{X}$ logical faults, it is straightforward that $H^1(C)$ are equivalence classes of undetectable $Z$ check failures on the auxiliary code, which will flip the logical measurement outcomes of the full block reading.

For the new $\overline{Z}$ logical faults, the equivalence classes given by $H_0(C)$ are spacetime logical faults which begin as $Z$ Pauli errors on the auxiliary code, and then move into the original spacetime volume and become timelike errors to then exit from the start or end boundaries. Because the $Z$ Pauli errors are not in $\im(H_X)$, they cannot be cleaned into the original spacetime volume and are genuinely new classes of logical faults. They are also not products of space and time logical faults.

To prove the fault distance of a single full block reading, where there are at least $d$ rounds of measurement before and after the block reading but only $1$ round used for the block reading itself, it is sufficient to show that every element of these equivalence classes, and their combinations, must have weight at least $d$.

Any timelike logical fault in $H_1(F')$ or $H^2(F')$ must originate at the boundaries, and therefore must have at least weight $d$ to reach, and therefore affect, the protocol.  They cannot be cleaned away from the boundary by spacetime stabilisers. Timelike errors which originate at the boundary must extend for $d$ rounds to affect the computation.

Logical faults from the positive contribution of $H_0(C)$ must exit through one of the boundaries of the spacetime volume, and hence must have weights at least $d$.

Spacelike logical faults in $H_1(F')$ and $H^2(F')$ must have weight at least $d$ because the original code and the deformed code during block reading both have distance at least $d$. Adding spacetime stabilisers to these spacelike logical faults can only increase the weights, because whatever qubit faults are removed by cleaning are merely moved to another round in the volume due to the structure of the repetition code $\CR_\bullet$.

Lastly, logical faults in $H^1(C)$, which are new check errors on the auxiliary system, must have weight at least $d$ as timelike errors. Applying spacetime stabilisers to clean these into the original code is possible, but the corresponding spacetime fault must act on at least $d$ data qubits due to the 1-to-1 structure of the full block reading, corresponding to a spacelike logical fault, followed by some timelike errors, and then the same spacelike logical fault, and so all such logical faults must have weight at least $d$.

\endproof

Evidently a dualised proof would apply were the full block reading to be applied in the $X$ basis instead.

\begin{remark}
    As full block reading is only applied for 1 round of measurements, these new classes of spacetime faults corresponding to $H_0(C)$ can be eliminated by adding an extra $C_{-1}$ term to $C_\bullet$, which adds detectors reconstructed from the initialisation and  measurements of the ancilla system in the $X$-basis. Adding an extra $C_{-1}$ term cannot make the deformed code non-LDPC, but can make the detectors high weight. As we shall presently show, these new classes of spacetime faults are not problematic anyway.
\end{remark}

We can then consider the case where multiple full block readings are performed in a row, but with only $\CO(1)$ rounds between them, and still leaving $\CO(d)$ rounds of measurement before and after the procedure.

In this case all the previous calculations apply, by applying the mapping cone to the chain or cochain complex versions of the fault complex. However, the analysis of logical faults requires two more ingredients. First, in the case where only one block reading is performed in the spacetime volume, the elements of $H_0(C)$ must exit through the boundaries of the volume. When there are multiple block readings performed in close proximity, however, these elements can connect and form noncontractible closed curves within the spacetime volume, with low fault weight. 
These noncontractible closed curves appear only when there is a full block reading whose logical measurements are a product of others performed in the same spacetime volume.

Second, we must consider the compacted code of the spacetime volume, see Definition~\ref{def:compacted_code}.

\begin{lemma}
    Let $\Xi$ be a set of $\eta$ full block reading procedures  of $Z$-type such that no block reading procedure in $\Xi$ is a product of any others.
    Then the compacted code $\mathtt{CC}_\bullet$ has distance $d$.
\end{lemma}
\proof
This is identical to the proof of code distance of the code $(C\otimes R)^\bullet$, where $R^\bullet$ is a full-rank matrix, see Lemma~\ref{lem:fullblock_distance}.
\endproof

\algoft*
\proof
Say that the $\eta$ block readings being performed are each defined by a bitstring of length $c$, and the set of bitstrings is divided into sets of size $\eta_Z$ and $\eta_X$ for $Z$ and $X$ type block readings. Each bitstring contains a 1 in a location if that codeblock is involved in the block reading specified by the bitstring, and a 0 otherwise. Whether the block readings are performed simultaneously or spread across multiple rounds, we can describe the block readings being performed by a pair of pattern matrices $\mathbb{P}_Z$ and $\mathbb{P}_X$ over $\F_2$, with $c$ columns and $\eta_Z$ and $\eta_X$ rows respectively. Note that, unlike when performing them simultaneously, these matrices do not need to form a pattern complex.

Assume that there is no block reading performed in the spacetime volume whose logical measurements are products of others. Then $\mathbb{P}_Z$ and $\mathbb{P}_X$ are each full-rank matrices.

Any logical fault composed of $Z$ Pauli errors in $H_0(C)$ on a $Z$-type block reading must extend to timelike $X$ check faults on all blocks in that row of $\mathbb{P}_Z$. To prevent extending all the way to the boundaries of the volume, such timelike faults must reach $Z$ Pauli errors in $H_0(C)$ on a different row; these must subsequently extend onto all blocks in that row of $\mathbb{P}_Z$. Hence in order to form a noncontractible closed curve inside the volume the pattern matrix $\mathbb{P}_Z$ must not be full rank, and so there are some block readings which are products of previous ones.

Dualising the argument gives the same result for $X$ Pauli errors in $H^2(C)$ on an $X$-type block reading.

All other timelike logical faults follow the same arguments as before: they must either extend to a boundary or have distance $d$ because of the new metachecks if they are timelike.

For spacelike errors, we must be more careful, as spacelike errors in the different deformed codes during the spacetime volume can be cleaned to have overlapping spacelike support. In particular, 
one can take products of spacelike logical faults in each deformed code, or the undeformed code. Before applying spacetime stabilisers, such a product is a set of disjoint spacelike logical faults separated by some number of timesteps, which in the worst case is constant.

Upon multiplying by spacetime stabilisers, such spacelike logical faults can have support that is cleaned to the same timestep and hence cancel, introducing some check errors between the timesteps. Say that we have a set of $\vartheta$ spacelike $\overline{Z}$ logical faults $\Lambda_1, \Lambda_2, \cdots, \Lambda_\vartheta$ at different logical timesteps, acting on different logical qubits. These could each be separated by a constant number of timesteps, and induce a constant number of check errors when partially cleaned by a spacetime stabiliser $S$. Thus, if the spacelike component of the product $S\Lambda_1\Lambda_2\cdots\Lambda_\vartheta$ has weight below $d$ the fault-distance may fall below $d$.

We now consider the compacted code $\mathtt{CC}_\bullet$ of the spacetime volume. Each logical operator $\Lambda_i$ in each deformed code is guaranteed to be a logical operator of the $\mathtt{CC}_\bullet$. Let $\Lambda_i'$ be the logical operator $\Lambda_i$ viewed in $\mathtt{CC}_\bullet$. We can hence consider the product $|\prod_i \Lambda_i'|$, which is the minimal weight of the spacelike component of $S\Lambda_1\Lambda_2\cdots\Lambda_\vartheta$. As $\mathtt{CC}_\bullet$ has distance $d$, $|\prod_i \Lambda_i'| \geq d$ for any choice of logical operators $\Lambda_i$ in any deformed codes. Then, $|\prod_i S\Lambda_i| \geq d$.

For spacelike $\overline{X}$ logical operators, the argument is similar. The $X$-distance of the compacted code $\mathtt{CC}_\bullet$ is at least $d$, and so cleaning $\overline{X}$ logical operators to have overlapping spacelike components can never reduce the fault weight below $d$.
\endproof

For full block reading of CSS-type, we require an extension of the previous lemma concerning the compacted code.

\begin{lemma}\label{lem:CSS_compacted}
    If the hypergraph surgeries are CSS-type full block readings such that no logical measurement is the product of any other then the compacted code $\mathtt{CC}_\bullet$ has distance $d$.
\end{lemma}
\proof
This is identical to the proof of code distance of the code $(C\otimes R)^\bullet$, where each pattern matrix in the complex $R^\bullet$ is a full-rank matrix, see Lemma~\ref{lem:fullblock_distance}.
\endproof

\CSSfbreading*

\proof
In this case we again take iterative mapping cones in the same way.

In the worst case, where there is short timelike length between full block readings, the cones are applied to only two layers in the fault complex: $Z$-type measurements to a layer with $X$ detectors (corresponding to a basis element of $\CR_0$) and $X$-type measurements to a close layer with $Z$ detectors (a basis element of $\CR_1$). By virtue of the mapping cone construction, the fault complex commutes as a chain complex.

We then use identical arguments to the proof of $Z$-type full block reading, where this time the constant number of layers between measurements is reduced to 0. As the proof of fault-distance is independent of this number of rounds, and the compacted code has distance $d$, there are no spacelike logical faults which can be cleaned to have fault weight with spacelike component lower than $d$.
\endproof

\subsection{Partial block reading with high-distance subcodes}\label{app:pbr_high_distance}

\FDSubcode*
\proof
We can rerun the arguments for Prop.~\ref{prop:fault_distance_fullblock} but where $A_\bullet \neq C_\bullet$. $A_\bullet$ is a subcode of $C_\bullet$ or can be a subcode thickened up to $d$ times (in the space direction). The thickening does not affect the homology of $A_\bullet$, or the maps on homology $H_i(f)$ into $C_\bullet$.

In partial block reading with high-distance subcodes, unlike full block reading, we include additional $X$-detectors given by the initialisation and measurement of the new qubits in the $X$-basis. These give us detectors corresponding to a basis of cycles in the hypergraph; however, there is no guarantee that this basis is sparse. This does not affect the LDPC property of the code, but does affect the size of the detectors when they are used for decoding. In other words, the deformed code is guaranteed to be LDPC, but the fault complex is not.

Now, for a $Z$-type partial block reading the fault complex is given by $\cone(f_\bullet)$, where $f_\bullet$ has components
\[\begin{tikzcd}
    0 \arrow[r]\arrow[d, "0"] & A_2 \arrow[r]\arrow[d, "f_2"] & A_1 \arrow[d, "f_1"]\arrow[r] & A_0 \arrow[d, "f_0"] \arrow[r] & A_{-1} \arrow[d, "0"] \\
    F'_3 \arrow[r] & F'_2 \arrow[r] & F'_1 \arrow[r] & F'_0 \arrow[r] & 0
\end{tikzcd}\]
with $\im(f_\bullet)$ having support on a single primal layer, across potentially multiple codeblocks, and $A_{-1}$ is a vector space with differential such that $H_0(A) = 0$ to fix the cycles.

For the $X$-type boundary conditions we then have
\begin{align*}
|H_1(\cone(f))| &= |H_1(F')| - |\im H_1(f)| + |H_0(A)| - |\im H_0(f)| \\
&= |H_1(F')| - |H_1(A)| + 0 - 0 \\
&= |H_1(F')| - |H_1(A)|.
\end{align*}
And
\begin{align*}
|H_2(\cone(f))| &= |H_2(F')| - |\im H_2(f)| + |H_1(A)| - |\im H_1(f)| \\
&= |H_2(F')| - |H_2(A)| + |H_1(A)| - |H_1(A)| \\
&= |H_2(F')| - |H_2(A)| \\
&= |H^2(\cone(f)| = |H^2(F')| - |H^2(A)|.
\end{align*}

\begin{itemize}
    \item $|H_1(\cone(f))| = |H_1(F')| - |H_1(A)|$.
    \item $|H^2(\cone(f))| = |H^2(F')| - |H^2(A)|$.
\end{itemize}
For the $Z$-type boundary conditions, the calculation is the same except $|\im H_1(f)| = 0$, so:
\begin{itemize}
    \item $|H_1(\cone(f))| = |H_1(F')|$,
    \item $|H^2(\cone(f))| = |H^2(F')| + |H^1(A)| - |H^2(A)|$.
\end{itemize}

Hence we can categorise the logical faults in an identical way to previously and find that they must all either extend from a boundary of the spacetime volume, or have spacelike weight at least $d$, or have timelike weight at least $d$, and that spacetime stabilisers cannot clean these faults to have weight lower than $d$ by virtue of the repetition code structure of $\CR_\bullet$.

Because we added the $X$-detectors, which are present to allow the space distance of the deformed code to be at least $d$, we also eliminate the spacetime faults which were present for full-block reading and corresponded to contributions of $H_0(C)$. Now, $H_0(A) = 0$ so there are no such contributions, so we need not be concerned about noncontractible loops in the spacetime volume.
\endproof

% I fixed the countings I think, they are all negative contribution (that are already counted in opposite basis complex) so should be fine.
% Can I help you with proof merging? or should I just leave it to you? Awesome! Let me still read then. 

% awesome :) thanks

% I think you can take a well-deserved break, I'll merge stuff in

\FDmanysubcode*
\proof
All logical faults must either extend from a boundary or have distance $d$ because of the new metachecks if they are timelike.

For the spacelike faults, we can apply the same argument as for full block reading. Each logical operator $\Lambda_i$ in each deformed code is guaranteed to be a logical operator of the compacted code $\mathtt{CC}_\bullet$. Let $\Lambda_i'$ be the logical operator $\Lambda_i$ viewed in $\mathtt{CC}_\bullet$. We can hence consider the product $|\prod_i \Lambda_i'|$, which is the minimal  weight of the spacelike component of $S\Lambda_1\Lambda_2\cdots\Lambda_\vartheta$. As $\mathtt{CC}_\bullet$ has distance $d$, $|\prod_i \Lambda_i'| \geq d$ for any choice of logical operators $\Lambda_i$ in any deformed codes. Then, $|\prod_i S\Lambda_i| \geq d$.

For spacelike $\overline{X}$ logical operators, the argument is similar but more straightforward. The $X$-distance of the compacted code $\mathtt{CC}_\bullet$ must always be at least $d$ by Lemma~\ref{lem:X_dist_preserved}, and so cleaning $\overline{X}$ logical operators to have overlapping spacelike components can never reduce the fault weight below $d$.
\endproof

\subsection{Partial block reading with low-distance subcodes}\label{app:pbr_low_distance}

Consider a partial block reading, where the subcode $A_\bullet$ for the partial block reading has distance less than $d$. For all our descriptions here we assume the block reading is measuring in the $Z$ basis, and then one can dualise to acquire the corresponding construction to measure in the $X$ basis.

We also presume that the auxiliary hypergraph for a $Z$-type partial block reading has $X$-checks to gauge-fix cycles, and the same for $X$-type partial block readings with $Z$ checks. These are genuine checks, and not just detectors inferred from the initial and final measurements of the new qubits. In terms of the initial auxiliary system, this still appears as adding an $A_{-1}$ term to $A_\bullet$ such that $H_0(A) = 0$,
\[A_\bullet = \begin{tikzcd}A_2 \arrow[r] & A_1 \arrow[r] & A_0 \arrow[r] & A_{-1}\end{tikzcd}\]
but this is not yet the complex which we will perform the mapping cone with.

As before, we ignore the thickening of $A_\bullet$ in space, as one can compute that this leaves the homology and categorisation of logical fault equivalence classes unaffected, and is necessary only to preserve the spatial distance of the deformed code.

However, as the distance of $A_\bullet$ is now less than $d$, we must thicken $A_\bullet$ in time to make a fault complex $K_\bullet$ corresponding to running the measurement procedure for more than 1 round. Let $d_A$ be the distance of $A_\bullet$ and let $d_A = \frac{1}{\alpha}d$ for $d$ the code distance of $C_\bullet$. 

The fault complex $K_\bullet$ corresponds to taking $(\CP(\alpha)\otimes A)_\bullet$, where
\[\CP(\alpha)_\bullet = \begin{tikzcd}
    \CP(\alpha)_1 \arrow[r] & \CP(\alpha)_0
\end{tikzcd} = \begin{tikzcd}
    \F_2^{\alpha-1} \arrow[r] & \F_2^{\alpha}
\end{tikzcd}\]
is dual to the full-rank repetition code with
length $\alpha$, and has the differential matrix being the incidence matrix of the path graph with $\alpha$ vertices. Explicitly $K_\bullet$ is the chain complex:
\[\begin{tikzcd}
    & \CP(\alpha)_1 \otimes A_1 \arrow[r]\arrow[ddr] & \CP(\alpha)_1 \otimes A_0 \arrow[r]\arrow[ddr] & \CP(\alpha)_1 \otimes A_{-1} \arrow[dr] & \\
    \CP(\alpha)_1 \otimes A_2 \arrow[ur] \arrow[dr] & & & & \CP(\alpha)_0 \otimes A_{-1} \\
    & \CP(\alpha)_0 \otimes A_2 \arrow[r]& \CP(\alpha)_0 \otimes A_1 \arrow[r] & \CP(\alpha)_0 \otimes A_{0} \arrow[ur] &
\end{tikzcd}\]
Observe that $A_\bullet = K_\bullet$ when $\alpha = 1$, recovering the case where we measure for only one round.

Now, for a $Z$-type partial block reading the fault complex is given by $\cone(f_\bullet)$, where $f_\bullet$ has the form:
\[
\begin{tikzcd}
    & \CP(\alpha)_1 \otimes A_1 \arrow[r]\arrow[ddr] & \CP(\alpha)_1 \otimes A_0 \arrow[r]\arrow[ddr] & \CP(\alpha)_1 \otimes A_{-1} \arrow[dr] & \\
    \CP(\alpha)_1 \otimes A_2 \arrow[ur] \arrow[dr] \arrow[dd, "0"] & & & & \CP(\alpha)_0 \otimes A_{-1} \arrow[dd, "0"] \\
    & \CP(\alpha)_0 \otimes A_2 \arrow[r]\arrow[d, "f_2"]& \CP(\alpha)_0 \otimes A_1 \arrow[r]\arrow[d, "f_1"] & \CP(\alpha)_0 \otimes A_{0} \arrow[ur] \arrow[d, "f_0"]& \\
    F'_3 \arrow[r] & F'_2 \arrow[r] & F'_1 \arrow[r] & F'_0 \arrow[r] & 0
\end{tikzcd}
\]
There are also maps from the top terms in the diagram -- $\CP(\alpha)_1 \otimes A_1$, $\CP(\alpha)_1 \otimes A_0$ and $\CP(\alpha)_1 \otimes A_{-1}$ -- but these are zero maps, so we abuse notation slightly by labelling the maps from the bottom layer $f_2, f_1, f_0$. As before, the only nonzero maps in $f_\bullet$ are into primal layers.

The mapping cone then yields the fault complex $\cone(f_\bullet)$ for the measurement,
\[
\begin{tikzcd}
    K_2 \arrow[r] \arrow[dr] & K_1 \arrow[r]\arrow[dr] & K_0 \arrow[r] \arrow[dr] & K_{-1} \\
    F'_3 \arrow[r] & F'_2 \arrow[r] & F'_1 \arrow[r] & F'_0
\end{tikzcd}
\]
where the $K_3$ term, equal to $\CP(\alpha)_1\otimes A_2$, is omitted as it has no physical meaning, being a ``detector on detectors''.

Before continuing further, let us make some orienting remarks regarding the terms and maps in this mapping cone. Below we expand out to the full fault complex $\cone(f_\bullet)$ for the case where the spacetime volume is initialised and measured out in the $X$ basis. As before, the $\CP(\alpha)_1\otimes A_2$ term is omitted. 
The diagram clearly commutes, by inheriting the coherent monic of the subcode $A_\bullet$ into the original codeblocks $E_\bullet$.

\[
\begin{tikzcd}
    \textnode{$\CP(\alpha)_1 \otimes A_1$ \\ (New $Z$ detectors)} \arrow[r]\arrow[ddr] & \textnode{$\CP(\alpha)_1 \otimes A_0$ \\ (New $X$ Pauli faults)} \arrow[r]\arrow[ddr] & \textnode{$\CP(\alpha)_1 \otimes A_{-1}$ \\ (New $X$ check faults)} \arrow[dr] & \\
    & & & \textnode{$\CP(\alpha)_0 \otimes A_{-1}$ \\ (New $X$ detectors)} \\
    \textnode{$\CP(\alpha)_0 \otimes A_2$ \\ (New $Z$ detectors)} \arrow[r]\arrow[dr, "f_2"]& \textnode{$\CP(\alpha)_0 \otimes A_1$ \\ (New $Z$ check faults)} \arrow[r]\arrow[dr, "f_1"] & \textnode{$\CP(\alpha)_0 \otimes A_0$ \\ (New $Z$  Pauli faults)} \arrow[ur] \arrow[ddr, "f_0"]& \\
    & \textnode{$\CR_0 \otimes E_2$ \\ (Old $Z$ check faults)} \arrow[r] & \textnode{$\CR_0 \otimes E_1$ \\ (Old $Z$ Pauli faults)} \arrow[dr] & \\
    \textnode{$\CR_1 \otimes E_2$ \\ (Old $Z$ detectors)} \arrow[ur] \arrow[dr] & & & \textnode{$\CR_0 \otimes E_0$ \\ (Old $X$ detectors)} \\
    & \textnode{$\CR_1 \otimes E_1$ \\ (Old $X$ Pauli faults)} \arrow[uur]\arrow[r]& \textnode{$\CR_1 \otimes E_0$ \\ (Old $X$ check faults)} \arrow[ur] & \\
    & & \textnode{$E_0$ \\ (Bottom $X$ checks)} \arrow[uur] & \\
    & & \textnode{$E_0$ \\ (Top $X$ checks)} \arrow[uuur] & \\
\end{tikzcd}
\]
\begin{itemize}
    \item All terms in $(\CR \otimes E)_\bullet$ are those from the fault complex of the idling operation.
    \item The new $E_0$ terms at the bottom are to add timelike entrypoints to the volume.
    \item $\CP(\alpha)_0\otimes A_{-1}$ is a set of $X$ detectors which arise from cycle checks throughout the surgery procedure.
    \item $\CP(\alpha)_0\otimes A_0$ is a set of fault locations for new qubits which can undergo $Z$ Pauli errors.
    \item $\CP(\alpha)_1\otimes A_{-1}$ is a set of new $X$ checks corresponding to cycles in the hypergraph, which can undergo measurement errors.
    \item $\CP(\alpha)_0\otimes A_1$ is a set of new $Z$ checks, corresponding to vertices in the hypergraph, which can undergo measurement errors.
    \item $\CP(\alpha)_1\otimes A_0$ is a set of fault locations for new qubits which can undergo $X$ Pauli errors. Note that when the block reading was performed for only 1 round these fault locations were not present because of the immediately adjacent initialisations in $\ket{+}$ and measurements in the $X$-basis.
    \item $\CP(\alpha)_0\otimes A_2$ is a set of $Z$ detectors which are metachecks due to the block reading.
    \item $\CP(\alpha)_1\otimes A_1$ is a set of $Z$ detectors due to the vertex checks throughout the surgery procedure.
\end{itemize}

To acquire the opposite spacetime volume, where it is initialised and measured in the $Z$ basis, the $\CR_\bullet$ is dualised, and $F'$ has extra $E_2$ components, rather than $E_0$. This is identical to the full block reading case.

We can hence rerun a similar proof as for the case of full block-reading, and acquire the homology spaces from the Snake Lemma. We can then prove the following.

\FDSubcodeAlpha*

\proof

\textbf{$X$-type boundary conditions.}

By the Snake Lemma, we can compute the following.
\begin{itemize}
    \item $|H_1(\cone(f))| = |H_1(F')| - |H_1(A)|$.
    \item $|H^2(\cone(f))| = |H^2(F')| - |H^2(A)|$.
\end{itemize}
In detail, using the K{\"u}nneth formula and the fact that $|H_1(\CP(\alpha))= 0|$ and $|H_0(\CP(\alpha))| = 1$,
\begin{align*}
    |H_1(\cone(f))| &= |H_1(F')| + |H_0(K)| - |\im H_1(f)| - |\im H_0(f)| \\
    &= |H_1(F')| + |H_0(A)| - |\im H_1(f)| - |\im H_0(f)| \\
    &= |H_1(F')| + 0 - |H_1(A)| - 0\\
    &= |H_1(F')| - |H_1(A)|.
\end{align*}
The other computation works similarly.

\textbf{$Z$-type boundary conditions.}
In this case the equivalence classes of logical faults can be computed as,
\begin{itemize}
    \item $|H_1(\cone(f))| = |H_1(F')|$.
    \item $|H^2(\cone(f))| = |H^2(F')| + |H^1(A)|- |H^2(A)|$.
\end{itemize}

We thus have a similar categorisation of all the equivalence classes of logical faults to those in previous proofs: there are spacelike errors which must always have weight at least $d$ by virtue of the deformed code having distance at least $d$, and there are timelike errors that extend from a boundary and so have weight at least $d$. However, there are also the timelike errors which correspond to the $|H^1(A)|$ term.

As check errors in the auxiliary code, these timelike logical faults must always have weight at least $d$ if they extend for $\lceil \alpha \rceil$ rounds, because they must have weight at least $\frac{1}{\alpha}d$ in each round. Should fewer than $\lceil \alpha \rceil$ rounds be used then there always exists a logical fault with weight lower than $d$, proving the first part of the proposition.

For the second part, we must be careful about the cleaning with spacetime stabilisers. Given such a timelike logical fault with weight at least $d$ in the auxiliary code, at least $d_A$ on each layer, cleaning it entirely into the bulk can leave only $2d_A$ qubit Pauli $X$ errors, one set of size $d_A$ just before the surgery and one set of size $d_A$ just after, and a set of timelike errors connecting them. We must therefore ensure that this set of timelike errors has sufficent size to yield a high fault-distance. It is distributed over $\alpha$ layers, so we ensure that each layer of check errors contains at least $\nu =\frac{(d-2v)}{\alpha}$ faults, where $v$ is the size of the set of spacelike errors at the bottom (or equivalently the top), and $v \geq d_A$. When $v \geq d/2$, $\nu < 0$ so we are done; this is always the case if $\alpha \leq 2$, i.e. $d_A \geq d/2$. The rest of the proof therefore assumes $d_A \leq v < d/2$. Note that $\nu < \frac{d}{\alpha} = d_A$.

Because the minimal size of the bottom set of Pauli errors is $d_A$, there are at least $d_A$ check errors in the next round if $|H_Zv| \geq d_A$, where $v$ is the qubit fault vector for the bottom set of Pauli errors. Now, assume that the $H_Z$ parity matrix of the original code has soundness $\rho_{\CC} \geq \frac{n}{m}$ as a classical code. This is guaranteed if the original code is quantum locally testable with soundness $\rho \geq \frac{2n}{m}$ by Lemma~\ref{lem:fact_17}.

We now show that $d(v, \CC) \geq d_A$, where $\CC = \ker H_Z$. First, if $|v+x|$ was lower than $d_A$ for any $x \in \im H_X^\intercal$ then the distance of the subcode would be lower than $d_A$, as the subcode has connections to all $X$ checks incident to $v$, so multiplying $v$ by $x$ restricted to $A_1$ would result in a nontrivial logical operator in $A_\bullet$ with weight lower than $d_A$. So we must now consider $|v+x|$ for any $x \in \ker H_Z \backslash \im H_X^\intercal$. As $v < d/2$, the maximal overlap of $v$ with any other $\overline{X}$ logical in the original code is $v$. Hence $|v+x| \geq |v|$ $\forall x \in \ker H_Z \backslash \im H_X^\intercal$, and $v \geq d_A$.

As a consequence of the soundness being $\rho_{\CC} \geq \frac{n}{m}$, $|H_Zv| \geq \frac{m}{n}\rho_{\CC} d(v, \CC) \geq d(v,\CC) \geq d_A$, so each layer between the top and the bottom Pauli $X$ errors must have at least $d_A$ check errors. 

By the same argument, applying cleaning to only a portion of the checks in the auxiliary system will distribute the faults such that the portion in the original code and the portion on checks in the auxiliary system at each timestep must sum to $d_A$, and so the fault distance is at least $d$.
\endproof

When performing many partial block readings close together in the same spacetime volume, we strengthen the condition to require soundness $\rho \geq \frac{2n}{m}$ regardless of whether $\alpha$ is above or below 2.

\FDSubcodeTogether*

\proof

By taking sequential mapping cones on the initial fault complex for the idling operation, one finds that the equivalence classes of logical faults are all either:
\begin{itemize}
    \item Timelike faults extending between boundaries.
    \item Spacelike faults.
    \item Timelike faults which act on the auxiliary systems, and cannot be cleaned to have weight below $d$ in the original code.
\end{itemize}

Timelike logical faults in the auxiliary system and between boundaries must still have weight at least $d$.

Cleaning spacelike logical faults $\Lambda_1, \Lambda_2, \cdots, \Lambda_\vartheta$ to have some shared spacelike support must induce at least a proportional weight in check faults, as the initial code has sufficient soundness, so the product $S\Lambda_1\Lambda_2\cdots\Lambda_\vartheta$ cannot fall below weight $d$. Note that the condition on the distance of the compacted code $\mathtt{CC}_\bullet$ is unnecessary in this case.

Therefore the fault-distance of the procedure is at least $d$.
\endproof

\subsection{Fast hypergraph surgery}

\GenHyperSurg*

\proof
This proof proceeds in a largely similar manner to those of Appendix~\ref{app:pbr_high_distance}. For a single hypergraph surgery with an ancillary system $A_\bullet$ we construct the same fault complex
\[\begin{tikzcd}
    0 \arrow[r]\arrow[d, "0"] & A_2 \arrow[r]\arrow[d, "f_2"] & A_1 \arrow[d, "f_1"]\arrow[r] & A_0 \arrow[d, "f_0"] \arrow[r] & A_{-1} \arrow[d, "0"] \\
    F'_3 \arrow[r] & F'_2 \arrow[r] & F'_1 \arrow[r] & F'_0 \arrow[r] & 0
\end{tikzcd}\]
but where $A_\bullet$ is now an arbitrary code such that $f_\bullet$ is non-overlapping on vertices and $H_0(A) = 0$.

\textbf{$X$-type boundary conditions.}
\begin{align*}
    |H_1(\cone(f))| &= |H_1(F')| - |\im H_1(f)| + |H_0(A)| - |\im H_0(f)| \\
    &= |H_1(F')| - |\im H_1(f)|,
\end{align*}
and 
\begin{align*}
    |H_2(\cone(f))| &= |H_2(F')| - |\im H_2(f)| + |H_1(A)| - |\im H_1(f)| \\
    &= |H_2(F')| - |\im H_2(f)| + |H_1(A)| - |H_1(A)| \\
    &= |H_2(F')| - |\im H_2(f)|.
\end{align*}

In summary we have:
\begin{itemize}
    \item $|H_1(\cone(f))| = |H_1(F')| - |\im H_1(f)|$.
    \item $|H^2(\cone(f))| = |H^2(F')| - |\im H^2(f)|$.
\end{itemize}

\textbf{$Z$-type boundary conditions.}
Similar calculations apply but $|\im H_2(f)| = |\im H_1(f)| = 0$.

\begin{itemize}
    \item $|H_1(\cone(f))| = |H_1(F')|$,
    \item $|H^2(\cone(f))| = |H^2(F')| + |H^1(A)| - |\im H_2(f)|$.
\end{itemize}

We can then apply such mapping cones sequentially, and so all logical faults are categorised into equivalence classes with the following representatives:
\begin{itemize}
    \item Logical faults on checks in the auxiliary system.
    \item Logical faults on data qubits in the original or deformed codes.
    \item Logical faults on checks extending through the spacetime volume between boundaries.
\end{itemize}

In order,
\begin{itemize}
    \item Because each auxiliary complex $(A_\bullet)_1, (A_\bullet)_2, (A_\bullet)_3, ..., (A_\bullet)_\eta$ has 1-cosystolic distance $d$, the  weight of a logical fault on checks in the auxiliary system must be at least $d$.
    \item Because the maps between the auxiliary systems and original codes are all non-overlapping on vertices, the weight of such a logical fault when cleaned into the original code must be at least $d$ to flip $d$ checks in the auxiliary system.
    \item Because the compacted code has distance at least $d$, any spacelike fault must have weight at least $d$. This accounts for the fact that  spacelike logical operators can cleaned by spacetime stabilisers to cancel their shared spacelike support.
    \item All other timelike faults must extend from a timelike boundary, so to affect the protocol must have weight at least $d$.
\end{itemize}
\endproof

\begin{remark}
    If the chain maps are overlapping on vertices (i.e. they do not satisfy Definition~\ref{def:inj_on_verts}) then the argument breaks down, as a timelike logical fault in the auxiliary system could be cleaned to a smaller logical fault in the original code, and so additional syndrome rounds are required to increase the weights of these faults. In the extreme case, where multiple auxiliary systems measure a single logical representative, this coincides with Theorem~\ref{thm:proof_connectivity}, which shows that such a measurement procedure will always require $\Omega(d)$ rounds to preserve the fault distance. 
\end{remark}

\section{Non-CSS measurements by full block reading}\label{app:nonCSS_block}

In this appendix we provide a more detailed explanation of how block reading extends to more general logical Pauli measurements on codes that possess transversal gates. 

\subsection{$Y$ measurements}

We can attach an ancillary system of the following form to any CSS code $C$:
\[\input{Figures-Tikz/Y_measure.tikz}\]
where the data qubits $Q$ are in the original code, and the r.h.s. scalable Tanner graph is the ancillary system. The checks must always commute by inspection of the diagram.

However, the properties of this combined code are complicated. Evidently, none of the $\overline{Z}$ or $\overline{X}$ logicals are measured out, as there are no new $Z$ or $X$ checks.

\begin{definition}
Let $C$ have a basis of representatives such that $\supp(\overline{Y}_i) = \supp(\overline{X}_i)=\supp(\overline{Z}_i)$, $\forall i \in [k]$. Then we say $C$ has matching representatives.
\end{definition}

Let $C$ be self-dual, i.e. $H_Z = H_X$. If $C$ has matching representatives then $\overline{Y}_i$ is measured out. As $\overline{Y}_i \in \ker H_Z = \ker H_X$, cleaning by the new $Y$ checks $P_Y$ in bijection with $\supp(\overline{Y}_i)$ gives $P_Y \overline{Y}_i = 0$ on both the new sets of data qubits. If the code does not have matching representatives then the cleaning argument does not work, and so the ancillary system may not measure the logical qubits.

Note that to have $\supp(\overline{X}_i) = \supp(\overline{Z}_i)$, $|\supp(\overline{Y}_i)| = 1 \mod 2$, i.e. it must have odd support, in order for $\overline{X}_i$ to anticommute with $\overline{Z}_i$.

If $C$ has matching representatives, all logical qubits will be measured out in the $Y$ basis. There are no new $\overline{Z}$ logicals, as any $\overline{Z}$ logical on the top right new data qubits must be in $\ker(H_Z^\intercal)$ so be products of old $Z$ checks, and any $Z$ operators on the bottom right new data qubits won't commute with the old $X$ checks. The inverse is true for new $\overline{X}$ checks.
\begin{example}
    2D colour codes are self-dual and have $\overline{Y}$ logicals composed of $\overline{X}$ and $\overline{Z}$ logicals on the same support \cite{landahl2011fault}.
\end{example}

\begin{remark}
If $C$ is self-dual but there exists a product of Pauli $Y$ operators on data qubits, which is a composition of $\overline{Z}_i$ and $\overline{X}_j$ with $\supp(\overline{Z}_i) = \supp(\overline{X}_j)$, but $i \neq j$ so the $Y$ operator is actually a representative of $\overline{Z}_i\otimes\overline{X}_j$, then the $Y$ ancillary system will still send this $Y$ operator to zero when cleaned, but the measurement is a $\overline{Z}_i\otimes\overline{X}_j$ measurement instead.
\end{remark}

For the measurement protocol, assume $C$ is self-dual with matching representatives.
First initialise new data qubits in the top(bottom) right set in the $\ket{0}$($\ket{+}$) states respectively. Then measure all stabilisers. To commute with all $\overline{Y}$ logicals being measured, each new data qubit must be connected to an even number of new checks in bijection with each $\overline{Y}$ logical, and so $X$($Z$) errors on top right(bottom right) qubits will not affect the logical measurement outcome.

As the code is self-dual, products of $Z$ and $X$ checks with the same support are mapped to a subset of new $Y$ checks, such that any nontrivial $\overline{Y}$ measurement error must be in $\ker(H_Z)\backslash\im(H_X^\intercal) = \ker(H_X)\backslash\im(H_Z^\intercal)$, and hence the logical measurement has some protection against measurement faults.

\begin{lemma}
    Let $C$ be a self-dual CSS code with matching representatives. Then $C$ admits a transversal $S$ gate, acting as $S^{\otimes k}$ on the logical space up to a Pauli correction.
\end{lemma}
\proof
From \cite[Prop.~9]{vuillot2022quantum}, $C$ admits a transversal $S$ gate up to a Pauli correction if:
\begin{itemize}
    \item $u\cdot v = 0 \mod 2$ for all $X$ stabilisers $u$, $v$.
    \item For any $\overline{X}$ logical operator $u$ and $\overline{X}$ stabiliser $v$, $u\cdot v = 0 \mod 2$.
\end{itemize}

As $C$ is self-dual the first condition is satisfied. From the matching representatives property, any $\overline{X}_i$ logical operator $u$ has the same support as its matching $\overline{Z}_i$ operator, and $\overline{Z}_i$ must commute with all stabilisers hence $u\cdot v = 0 \mod 2$.
\endproof

Therefore the procedure is not very useful, because the codeblock can have its measurement basis permuted by transversal $S$ and $H$ and so $\overline{Y}$ measurements are unnecessary. The same argument applies to performing block reading of multiple codeblocks in products containing $\overline{Y}$ terms. We do not know if partial block reading for non-CSS measurements, where the subcode is a stabiliser code but not CSS, can allow for fast measurements which cannot be performed using transversal gates to permute bases.

\subsection{$\overline{X}\otimes\overline{Z}$ measurements}
Given a CSS code $C$, let $D$ be the dual of that code, i.e. $H_X^D = H_Z^C$ and $H_Z^D = H_X^C$. Introduce the following ancillary system:
\[\input{Figures-Tikz/H_measure1.tikz}\]
where $C$ is on the l.h.s. and $D$ on the r.h.s. The top right and middle checks are now of mixed-type, checking stabilisers with both $X$ and $Z$ terms.
One can check by symplectic multiplication that the above Tanner graph commutes.
We now also have meta-checks which are of `mixed-type', using both $X$ and $Z$ checks, indicated by the pink diamond. This is a $\overline{Z}\otimes\overline{Z}$ measurement by full block reading, but conjugated by single-qubit Cliffords.

However, because the codeblocks $C$ and $D$ are dual, this could equally be performed by using a modified version of homomorphic measurements but with transversal CX and CZ gates \cite{breuckmann2024foldtransversal}.

\section{Proofs concerning modular expansion}\label{app:mod_exp_proofs}

In this appendix we provide technical proofs related to properties of modular expansion used in the main text.

\DistPreservation*
\proof
We show that for every nontrivial $\overline{\Lambda}_Z$ logical operator in $Q \leftrightarrow \CH$ with some support in $\bigcup_i V_i$, the weight of $\overline{\Lambda}_Z$ cannot drop below $d$ by multiplying with new $Z$ checks (vertices) in $\CH$.

Let $v$ be an arbitrary subset of $\CV$, and define the stabiliser $Z(G^\intercal v)$ to be the product of all checks in $v$. We must show that $|\overline{\Lambda}_Z\circ Z(G^\intercal v)|\geq d$.

First, $G^\intercal v \geq \min(d, |\kappa_\lambda| - |v \cap \kappa_\lambda| + |v \cap \CU \backslash \kappa_\lambda|\ :\ \lambda \in 2^I)$ by the definition of modular expansion. If \[\min(d, |\kappa_\lambda| - |v \cap \kappa_\lambda| + |v \cap \CU \backslash \kappa_\lambda|\ :\ \lambda \in 2^I) = d\] then we are immediately done, as no matter which data qubits in $Q$ are turned off by $Z(G^\intercal v)$, there are at least $d$ new data qubits with a $Z$ Pauli in the ancillary hypergraph. The other case is more difficult.

\begin{figure}
    \centering
    \includegraphics[scale=0.7]{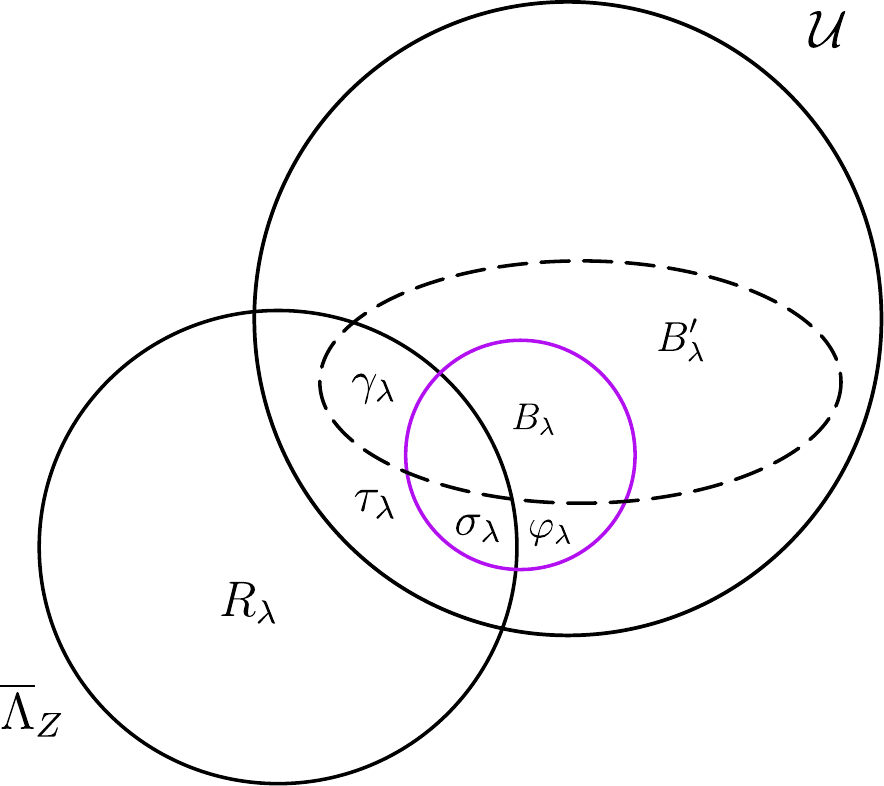}
    \caption{Venn diagram indicating the overlapping supports of a logical operator $\overline{\Lambda}_Z$ and sets of checks $\kappa_\lambda$ (magenta) and $v \cap \CU$ (dashed).}
    \label{fig:supports_venn}
\end{figure}

If $|G^\intercal v| \geq |\kappa_\lambda| - |v \cap \kappa_\lambda| + |v \cap \CU \backslash \kappa_\lambda|$ for some $\lambda \in 2^I$ then we can construct the diagram in Fig.~\ref{fig:supports_venn}.

The set in the dashed region is $v \cap \CU$, the top right circle is $\CU$, the bottom left circle is $\overline{\Lambda}_Z$ and the magenta circle is $\kappa_\lambda$. We compute by inspection of the diagram that 
\[|\overline{\Lambda}_Z\circ Z(G^\intercal v)| \geq |\tau_\lambda| +|\sigma_\lambda| + |B_\lambda| + |B'_\lambda| + |R_\lambda| + |\kappa_\lambda| - |v \cap \kappa_\lambda| + |v\cap \CU \backslash \kappa_\lambda|.\]

Next,
\[|\kappa_\lambda| - |v \cap \kappa_\lambda| + |v\cap \CU \backslash \kappa_\lambda| = |\varphi_\lambda| + |\sigma_\lambda| + |B'_\lambda| + |\gamma_\lambda|\]
and
\[|\overline{\Lambda}_Z \circ \overline{Z}_\lambda| = |R_\lambda| + |\tau_\lambda| + |\gamma_\lambda| + |B_\lambda| + |\varphi_\lambda| \geq d\]
where $\overline{Z}_\lambda = Z(f^{-1}\kappa_\lambda)$ acts as $Z$ on the preimage of $\kappa_\lambda$ in the original code. $\overline{Z}_\lambda$ is a logical operator, but not necessarily a nontrivial one. As $\overline{\Lambda}_Z$ is a nontrivial logical operator of the deformed code,
\[|\overline{\Lambda}_Z \circ \overline{Z}_\lambda| \geq d.\]

Plugging these together we find that
\[|\overline{\Lambda}_Z\circ Z(G^\intercal v)| \geq 2(|\sigma_\lambda| + |B'_\lambda|) + d \geq d.\]

As the cycles are gauge-fixed and no $\overline{Z}$ logical can be reduced below weight $d$, the distance of $Q \leftrightarrow \CH$ is at least $d$.
\endproof

\Thickening*
\proof
Let $G : \F_2\CE \rightarrow \F_2\CV$ be the incidence matrix of $\CH$, $R_L$ the incidence matrix of $\CJ_L$, and $G_L$ the incidence matrix of $\CH_L$. Then
\[G_L = \begin{pmatrix}
     G \otimes I_L & I_n \otimes R_L
\end{pmatrix}\]
where $n = |\CV|$.

We divide up the vertices of $\CH_L$ into sets $\CV^\ell$ for $\ell \in [L]$, corresponding to each level of the thickened graph. Similarly, for a given level $\ell$ we have $V_i^\ell$ and $\CU^\ell = \bigcup_i V_i^\ell$. 

Letting $v^\ell \in \CV^\ell$ be a set of vertices in level $\ell$, we have the vector $v = (v^1\ v^2\ \cdots\ v^L)^\intercal \in \bigcup_\ell\CV^\ell$ seen as a set of vertices in $\CH_L$. Associated to this set we also have $u^\ell = v^\ell \cap \CU^\ell = v\cap \CU^\ell$.

Now, let $r = \arg\min_j \min(t, |\kappa^j_\lambda| - |u^j \cap \kappa^j_\lambda| + |u^j \backslash \kappa^j_\lambda|\ :\ \lambda \in 2^I)$. Using the triangle inequality and the fact that $L\CM_t(\CH) \geq 1$,

\begin{align*}
|G_L^\intercal v| &= |G_L^\intercal (v^1\ v^2\ \cdots\ v^L)^\intercal| = \sum_{j=2}^L |v^{j-1}+v^j| + \sum_{j=1}^L|G^\intercal v^j| \\
&\geq |v^r + v^\ell| + L\CM_t(\CH)\min(t, |\kappa^r_\lambda| - |u^r \cap \kappa^r_\lambda| + |u^r \backslash \kappa^r_\lambda|\ :\ \lambda \in 2^I) \\
& \geq |u^r + u^\ell| + \min(t, |\kappa^r_\lambda| - |u^r \cap \kappa^r_\lambda| + |u^r \backslash \kappa^r_\lambda|\ :\ \lambda \in 2^I).
\end{align*}
Now consider the different cases.

If $\min(\cdot) = t$ then $|u^r + u^\ell| +t \geq t$ and we are done.

If $\min(\cdot) = |\kappa^r_\lambda| - |u^r \cap \kappa^r_\lambda| + |u^r \backslash \kappa^r_\lambda|$ for some $\lambda\in 2^I$ then
\begin{align*}
    & |u^r + u^\ell| + |\kappa^r_\lambda| - |u^r \cap \kappa^r_\lambda| + |u^r \backslash \kappa^r_\lambda| \\
    &= |u^r\cap \kappa^r_\lambda + u^\ell \cap \kappa^r_\lambda| + |u^r\backslash \kappa^r_\lambda + u^\ell \backslash \kappa_\lambda^\ell| + |\kappa_\lambda^\ell| - |u^r\cap \kappa^r_\lambda| + |u^r\backslash \kappa^r_\lambda|\\
    &\geq |\kappa_\lambda^\ell| - |u^\ell\cap \kappa_\lambda^\ell| + |u^\ell\backslash \kappa_\lambda^\ell|,
\end{align*}
making use of the triangle inequality for the last line.

Therefore $|G^\intercal_L v|\geq \min(t,|\kappa_\lambda^\ell|-|v\cap \kappa_\lambda^\ell| + |v\cap \CU^\ell\backslash \kappa_\lambda^\ell| : \lambda \in 2^I )$ for any set of vertices $v$ in $\CH_L$, so $\CH_L$ has modular expansion $\CM_t(\CH_L) \geq 1$.
\endproof

\section{Circuit-equivalence of surgery and homomorphic measurement}\label{app:proof_circuit_equiv}

In this appendix we derive a circuit equivalence between generalized surgery measurement and homomorphic measurement. 
The following proofs use the so-called ``scalable'' ZX-calculus from Ref.~\cite{carette2019szx}. This calculus has previously been used to describe LDPC surgeries in Ref.~\cite{poirson2025engineering}. We assume familiarity with the ZX-calculus for this section; see Ref.~\cite{vandewetering2020zx} for a comprehensive introduction.

In short, the scalable ZX-calculus (SZX-calculus) has thick wires indexed by a positive integer $n$, denoting a register of $n$ qubits. Green and red spiders then act on the $n$ qubits, with spider phases now represented by a vector $\underline{v}$ over $\C$ of length $n$, containing phases acting on each qubit. As in the conventional ZX-calculus, an all-zeros vector is omitted and the green or red spider left unlabelled. We also omit the label denoting the size of the qubit register.

The SZX-calculus has other generators than green and red spiders. For example, we also have the matrix arrows
\begin{equation}\label{eq:right_and_left_arrows}
    \input{Figures-Tikz/rep0.tikz}  \ \ \text{and} \ \ \input{Figures-Tikz/rep0b.tikz} 
\end{equation}
where $A$ and $B$ are matrices over $\F_2$. These generators can be defined by their action on the Hilbert space. Suppose $A \in \F_2^{n\times m}$ and $B \in \F_2^{p\times q}$. For a qubit state in the computational basis $\ket{x}$ with $x \in \F_2^m$, the arrow pointing to the right in Eq.~\ref{eq:right_and_left_arrows} corresponds to the linear map $R_A : \ket{x} \mapsto \ket{Ax}$. Meanwhile, the arrow pointing to the left corresponds to the linear map $H^{\tens p} R_{B^T}H^{\tens q}=H^{\tens p}\left(\ket{x} \mapsto \ket{B^Tx}\right)H^{\tens q}$.

Stabiliser measurement on CSS codes can be neatly described using the SZX-calculus. Given a CSS code with $n$ qubits and the parity-check matrices $H_X$, $H_Z$, the measurements can be described by
\[\input{Figures-Tikz/stab_measure_SZX.tikz}\]
where $s_Z\pi$ is the vector of phases representing syndrome outcomes $s_Z = \{0, 1\}^{m_Z}$, and the same for $s_X\pi$. In particular, when the input state is in the codespace, $s_Z = s_X = \underline{0}$ and the phase vectors can be omitted. The same applies for the case where the the initial state is not in the codespace but the measurements project it into the codespace.

Commutation of stabilisers implies that
\[\input{Figures-Tikz/stab_measure_SZX_empty.tikz} \quad = \quad \input{Figures-Tikz/stab_measure_SZX2_empty.tikz}\]
and furthermore,
\[\input{Figures-Tikz/stab_measure_collide.tikz} \quad = \quad \input{Figures-Tikz/stab_measure_collide2.tikz}\]
with a similar rewrite for $X$-checks. Obviously such rewrites do not generally preserve fault-tolerance of the circuit: we are taking multiple rounds of syndrome measurement and collapsing them into one round, which will generally reduce the fault-distance of the spacetime code.

Along with the merge and copy rules of the ZX-calculus we can now prove that surgery by an auxiliary system and homomorphic measurement using transversal gates are equivalent as circuits.

\SZXHomToSurgery*
\proof
We describe an $Z$-basis logical measurement; the same result for $X$ measurements can be easily acquired similarly. The rewrites required are as follows.
\[\input{Figures-Tikz/SZX_hom_to_surgery1.tikz}\]
The first SZX-diagram on the top left is depicting a homomorphic measurement. The original code is shown on the bottom wire, with parity-check matrices $H_Z$, $H_X$. An ancilla code is initialised in $\ket{0}^{\otimes n'}$ then its stabiliser checks are measured. Then a set of transversal CNOTs defined by a chain map $f_\bullet$ are applied, with the controls on the initial code and the targets on the ancilla code. Then the ancilla code is measured out in the $Z$ basis, with the vector $\alpha\pi$ where $\alpha \in \F_2^{n'}$ determines the logical measurement outcome.

Next, the $\ket{0}$ initialisations are commuted through the $Z$-check measurements, and then the merge rule is used to extend the check qubits of the $X$-check measurements to instead become data qubits of the auxiliary system, initialised in $\ket{+}^{m_X}$ and measured in the $X$-basis. We now see that the data qubits of the homomorphic measurement have become check outcomes of $Z$ measurements, which include both the new data qubits of the auxiliary system and data qubits of the original code. The outcomes $\alpha\pi$ of these $Z$ measurements determine the logical measurement outcome.

Last, we add $X$ checks which gauge-fix the cycles (there are cases, such as in full-block reading and the CKBB scheme \cite{cohen2022low}, where these are not necessary), with the check matrix $N$. We also deform the $X$ checks of the original code to include qubits of the auxiliary system, such that these checks commute with the $Z$ checks that determine the logical measurement outcome. The deformation of these checks is determined by the chain map $f_\bullet$.
\endproof

We can then do the opposite rewriting to convert surgeries to homomorphic measurements.
\SZXSurgeryToHom*
\proof
\[\input{Figures-Tikz/SZX_surgery_to_hom1.tikz}\]
\endproof

Thus we see that the vertices $\mathcal{V}$ of the hypergraph in surgery, which are $Z$-checks, are rewritten to the data qubits in homomorphic measurement; the edges $\mathcal{E}$ of the hypergraph, which are data qubits, are rewritten to be $X$-check outcomes. The cycles vanish from the circuit entirely, which is because they become $X$-type meta-checks on the ancilla code. These relabellings correspond precisely to the degree-shifting of a mapping cone to recover the original chain map $f_\bullet$.

\section{Fast measurement requires multiple representatives}\label{app:proof_connectivity}

In this appendix we provide a proof of our results lower bounding the spacetime volume of a general procedure to measure logical representatives in a quantum code.

\AuxiliaryFaultBound*

\proof
The measurement scheme begins at time $t_i$ and ends at time $t_o = t_i + T$, and measures all the logical operator representatives contained in $S$. Data qubit errors occur at integer timesteps and measurement errors occur at half-integer timesteps. To measure this operator by an auxiliary system, entangling gates are applied from each qubit in $S$ to some other system $\textbf{Aux}$ of checks and data qubits, such that the measurement outcomes of checks in $\textbf{Aux}$ infer the logical measurement outcomes.
The entangling gates are applied while measuring checks in $\textbf{Aux}$ for $T$ rounds before ceasing the application of entangling gates. 

In order to receive an incorrect logical measurement outcome without detecting an error it is sufficient to flip any of those logical operators for the duration of the logical measurement scheme, so long as there are no logical measurement outcomes between representatives which are `mismatched', meaning that measurements of different representatives of the same logical operator yield different outcomes. Thus if any representative in $S$ is flipped, all representatives in the same equivalence class in $S$ must be flipped. Observe that as none of the correct logical measurement outcomes of $S$ are known in advance (which would occur if $Q$ was initialised in a particular known state with deterministic logical measurement outcomes) there are now no detectors which can detect the logical error.

To flip some operators in the round at time $t_i$, apply a Pauli operator fault $f$ to $Q$ with weight $w$ satisfying the conditions specified in the Proposition.

This set of faults is detectable by checks in the original code at timesteps $t_i+\frac{1}{2}, t_i+\frac{3}{2},\cdots$, so in order to conceal the fault we add $s$ check errors at each timestep for the duration of the logical measurement, where $s$ is the number of checks which $f$ flips. In total there are $Ts$ of these check faults.

Lastly, apply an identical Pauli operator fault $f$ after the logical measurement procedure concludes, at time $t_o+1$. This is a logical fault of the measurement procedure with weight $2w + Ts$.

\endproof

Were the code $Q$ left in the idling operation instead, the above logical fault would be a spacetime stabiliser of the code, and so this logical fault weight can be vastly lower than the distance of $Q$ depending on $S$, the time $T$ and the syndrome weights $s$ of operators. Note also that this is just an upper bound on the fault-distance: if the auxiliary system is poorly constructed, for example by connecting every qubit in $S$ to a single new check qubit, the fault-distance may be substantially lower due to check errors on the auxiliary system.

\ConnectivityRequired*

We prove this by demonstrating that otherwise there exists a weight $|f|$ logical fault, where $|f| \in o(d)$.

\proof
Assume w.l.o.g. that the measurement being performed is on a $\overline{Z}$ logical operator, as any Pauli measurement is locally equivalent to this measurement by conjugation of single-qubit Cliffords.

If the logical operator representative has support $v$, the measurement being performed is of the operator $\bigotimes\limits_{i \in v}Z_i$. To measure this operator by an ancillary system, entangling gates are applied from each qubit $i$ to some other system $\textbf{Aux}$ of checks and data qubits, such that the measurement outcomes of checks in $\textbf{Aux}$ infer the logical measurement outcome of $\bigotimes\limits_{i \in v}Z_i$.

The measurement scheme starts at time $t_i$ and concludes at time $t_o$, with $o(d)$ rounds between. Data qubit errors occur at integer timesteps and measurement errors occur at half-integer timesteps.
Assume that the entangling gates are applied while measuring checks in $\textbf{Aux}$ for $o(d)$ rounds before ceasing the application of entangling gates. Regardless of the system of checks and other structure of $\textbf{Aux}$, there exists a low weight fault which flips the logical measurement outcome.

Consider an arbitrary qubit $j \in v$. At any time $t \in [t_i, t_o]$ the application of a Pauli error $X_j^t$, followed by measurement errors on all incident $Y$ and $Z$ checks at time $t+\frac{1}{2}$, and then a Pauli error $X_j^{t+1}$, is a local spacetime stabiliser of the original code, and therefore does not violate any detectors. However, at time $t+\frac{1}{2}$ the outcome of measuring $\bigotimes\limits_{i \in v}Z_i$ is flipped, as $X_j$ anticommutes with $Z_j$. This is then corrected by the subsequent Pauli error $X_j^{t+1}$, and so at time $t+\frac{3}{2}$ the measurement of $\bigotimes\limits_{i \in v}Z_i$ would return the correct outcome again.

Hence a logical fault $f$ on the logical measurement can be constructed by applying a Pauli error $X_j^{t_i}$ and measurement errors on all incident $X$ and $Y$ checks on qubit $j$ from time $t_i+\frac{1}{2}$ to time $t_o+\frac{1}{2}$, and a final Pauli error $X_j^{t_o+1}$. This is a spacetime stabiliser of the original code, as it is composed of local spacetime stabilisers at times $t=t_i, t_i+1,\cdots,t_o$, hence the fault does not violate any detectors. Each round of measurement of $\bigotimes\limits_{i \in v}Z_i$ will be flipped by the original $X_j^{t_i}$ error, and the outcome of $\bigotimes\limits_{i \in v}Z_i$ is then flipped back after the measurement procedure concludes. The fault $f$ is then $X_j^{t_o+1}X_j^{t_i}\bigcup_{t_i \leq t \leq t_o} M_j^{t+\frac{1}{2}}$, where $M_j^{t+\frac{1}{2}}$ is a measurement fault on each $Y$ or $Z$ check incident to qubit $j$ at time $t+\frac{1}{2}$.

Because the code is LDPC, the fault at each round has weight bounded above by a constant, where the precise weight depends on the number of $Y$ and $Z$ checks incident to qubit $j$, but this cannot increase the weight asymptotically. Because the logical measurement is performed for $o(d)$ rounds, the weight $|f| \in o(d)$.

Note that if the correct logical measurement outcome is known in advance, there is an additional detector at each round of $\bigotimes\limits_{i \in v}Z_i$ and so the above argument does not hold. Similarly, the reason the proof no longer applies when measuring multiple logical operator representatives is because there are additional detectors which appear because the measurements of each representative of the same logical operator must agree with each other to be correct; equivalently, these additional detectors correspond to sets of measurements whose correct outcomes are known in advance, as they are stabilisers of the original code.
\endproof

Both of the above proofs are insensitive to the structure of $\textbf{Aux}$, which could be a measurement hypergraph in surgery, an ancilla block in homomomorphic measurement or any other logical measurement gadget.

%% file: Figures-Tikz/unit_cell_fc.tikz
\begin{tikzpicture}
	\begin{pgfonlayer}{nodelayer}
		\node [style={X_detector}] (0) at (0.5, -7) {};
		\node [style={X_detector}] (1) at (-4.5, -5) {};
		\node [style={Z_detector}] (7) at (0, 0) {};
		\node [style={X_detector}] (8) at (4.5, -5) {};
		\node [style={X_detector}] (9) at (-0.5, -3) {};
		\node [style={Z_fault}] (10) at (-2, -6) {};
		\node [style={Z_fault}] (11) at (2.5, -6) {};
		\node [style={Z_fault}] (12) at (2, -4) {};
		\node [style={Z_fault}] (13) at (-2.5, -4) {};
		\node [style={X_fault}] (14) at (0, -5) {};
		\node [style={X_fault}] (15) at (-2, -1) {};
		\node [style={X_fault}] (16) at (-2.5, 1) {};
		\node [style={X_fault}] (17) at (2.5, -1) {};
		\node [style={X_fault}] (18) at (2, 1) {};
		\node [style={Z_fault}] (19) at (0.5, -2) {};
		\node [style={Z_fault}] (20) at (4.5, 0) {};
		\node [style={Z_fault}] (21) at (-0.5, 2) {};
		\node [style={Z_fault}] (22) at (-4.5, 0) {};
		\node [style={X_detector}] (23) at (-0.5, 7) {};
		\node [style={X_detector}] (24) at (4.5, 5) {};
		\node [style={Z_detector}] (25) at (0, 0) {};
		\node [style={X_detector}] (26) at (-4.5, 5) {};
		\node [style={X_detector}] (27) at (0.5, 3) {};
		\node [style={Z_fault}] (28) at (2, 6) {};
		\node [style={Z_fault}] (29) at (-2.5, 6) {};
		\node [style={Z_fault}] (30) at (-2, 4) {};
		\node [style={Z_fault}] (31) at (2.5, 4) {};
		\node [style={X_fault}] (32) at (0, 5) {};
		\node [style={X_fault}] (33) at (2, 1) {};
		\node [style={X_fault}] (34) at (2.5, -1) {};
		\node [style={X_fault}] (35) at (-2.5, 1) {};
		\node [style={X_fault}] (36) at (-2, -1) {};
		\node [style={Z_fault}] (37) at (-0.5, 2) {};
		\node [style={Z_fault}] (38) at (-4.5, 0) {};
		\node [style={Z_fault}] (39) at (0.5, -2) {};
		\node [style={Z_fault}] (40) at (4.5, 0) {};
	\end{pgfonlayer}
	\begin{pgfonlayer}{edgelayer}
		\draw [style={X_edge}] (1) to (13);
		\draw [style={X_edge}] (13) to (9);
		\draw [style={X_edge}] (9) to (12);
		\draw [style={X_edge}] (12) to (8);
		\draw [style={X_edge}] (11) to (8);
		\draw [style={X_edge}] (0) to (11);
		\draw [style={X_edge}] (1) to (10);
		\draw [style={X_edge}] (10) to (0);
		\draw (10) to (14);
		\draw (14) to (13);
		\draw (14) to (11);
		\draw (14) to (12);
		\draw [style={Z_edge}] (7) to (14);
		\draw [style={Z_edge}] (15) to (7);
		\draw [style={Z_edge}] (7) to (17);
		\draw [style={Z_edge}] (7) to (18);
		\draw [style={Z_edge}] (16) to (7);
		\draw (22) to (16);
		\draw (16) to (21);
		\draw (21) to (18);
		\draw (18) to (20);
		\draw (20) to (19);
		\draw (19) to (22);
		\draw (16) to (13);
		\draw (12) to (18);
		\draw (17) to (11);
		\draw (15) to (10);
		\draw [style={X_edge}] (19) to (0);
		\draw [style={X_edge}] (20) to (8);
		\draw [style={X_edge}] (22) to (1);
		\draw [style={X_edge}] (21) to (9);
		\draw [style={X_edge}] (24) to (31);
		\draw [style={X_edge}] (31) to (27);
		\draw [style={X_edge}] (27) to (30);
		\draw [style={X_edge}] (30) to (26);
		\draw [style={X_edge}] (29) to (26);
		\draw [style={X_edge}] (23) to (29);
		\draw [style={X_edge}] (24) to (28);
		\draw [style={X_edge}] (28) to (23);
		\draw (28) to (32);
		\draw (32) to (31);
		\draw (32) to (29);
		\draw (32) to (30);
		\draw [style={Z_edge}] (25) to (32);
		\draw (40) to (34);
		\draw (34) to (39);
		\draw (39) to (36);
		\draw (36) to (38);
		\draw (38) to (37);
		\draw (37) to (40);
		\draw (34) to (31);
		\draw (30) to (36);
		\draw (35) to (29);
		\draw (33) to (28);
		\draw [style={X_edge}] (37) to (23);
		\draw [style={X_edge}] (38) to (26);
		\draw [style={X_edge}] (40) to (24);
		\draw [style={X_edge}] (39) to (27);
	\end{pgfonlayer}
\end{tikzpicture}

%% file: Figures-Tikz/unit_cell_newchecks_fc.tikz
\begin{tikzpicture}
	\begin{pgfonlayer}{nodelayer}
		\node [style={X_detector}] (0) at (0.5, -7) {};
		\node [style={X_detector}] (1) at (-4.5, -5) {};
		\node [style={Z_detector}] (2) at (0, 0) {};
		\node [style={X_detector}] (3) at (4.5, -5) {};
		\node [style={X_detector}] (4) at (-0.5, -3) {};
		\node [style={Z_fault}] (5) at (-2, -6) {};
		\node [style={Z_fault}] (6) at (2.5, -6) {};
		\node [style={Z_fault}] (7) at (2, -4) {};
		\node [style={Z_fault}] (8) at (-2.5, -4) {};
		\node [style={X_fault}] (9) at (0, -5) {};
		\node [style={X_fault}] (10) at (-2, -1) {};
		\node [style={X_fault}] (11) at (-2.5, 1) {};
		\node [style={X_fault}] (12) at (2.5, -1) {};
		\node [style={X_fault}] (13) at (2, 1) {};
		\node [style={Z_fault}] (14) at (0.5, -2) {};
		\node [style={Z_fault}] (15) at (4.5, 0) {};
		\node [style={Z_fault}] (16) at (-0.5, 2) {};
		\node [style={Z_fault}] (17) at (-4.5, 0) {};
		\node [style={X_detector}] (18) at (-0.5, 7) {};
		\node [style={X_detector}] (19) at (4.5, 5) {};
		\node [style={Z_detector}] (20) at (0, 0) {};
		\node [style={X_detector}] (21) at (-4.5, 5) {};
		\node [style={X_detector}] (22) at (0.5, 3) {};
		\node [style={Z_fault}] (23) at (2, 6) {};
		\node [style={Z_fault}] (24) at (-2.5, 6) {};
		\node [style={Z_fault}] (25) at (-2, 4) {};
		\node [style={Z_fault}] (26) at (2.5, 4) {};
		\node [style={X_fault}] (27) at (0, 5) {};
		\node [style={X_fault}] (28) at (2, 1) {};
		\node [style={X_fault}] (29) at (2.5, -1) {};
		\node [style={X_fault}] (30) at (-2.5, 1) {};
		\node [style={X_fault}] (31) at (-2, -1) {};
		\node [style={Z_fault}] (32) at (-0.5, 2) {};
		\node [style={Z_fault}] (33) at (-4.5, 0) {};
		\node [style={Z_fault}] (34) at (0.5, -2) {};
		\node [style={Z_fault}] (35) at (4.5, 0) {};
		\node [style={Z_fault}] (36) at (0.5, -12) {};
		\node [style={Z_fault}] (37) at (4.5, -10) {};
		\node [style={Z_fault}] (38) at (-4.5, -10) {};
		\node [style={Z_fault}] (39) at (-0.5, -8) {};
		\node [style={Z_fault}] (40) at (0.5, 8) {};
		\node [style={Z_fault}] (41) at (4.5, 10) {};
		\node [style={Z_fault}] (42) at (-4.5, 10) {};
		\node [style={Z_fault}] (43) at (-0.5, 12) {};
	\end{pgfonlayer}
	\begin{pgfonlayer}{edgelayer}
		\draw [style={X_edge}] (1) to (8);
		\draw [style={X_edge}] (8) to (4);
		\draw [style={X_edge}] (4) to (7);
		\draw [style={X_edge}] (7) to (3);
		\draw [style={X_edge}] (6) to (3);
		\draw [style={X_edge}] (0) to (6);
		\draw [style={X_edge}] (1) to (5);
		\draw [style={X_edge}] (5) to (0);
		\draw (5) to (9);
		\draw (9) to (8);
		\draw (9) to (6);
		\draw (9) to (7);
		\draw [style={Z_edge}] (2) to (9);
		\draw [style={Z_edge}] (10) to (2);
		\draw [style={Z_edge}] (2) to (12);
		\draw [style={Z_edge}] (2) to (13);
		\draw [style={Z_edge}] (11) to (2);
		\draw (17) to (11);
		\draw (11) to (16);
		\draw (16) to (13);
		\draw (13) to (15);
		\draw (15) to (14);
		\draw (14) to (17);
		\draw (11) to (8);
		\draw (7) to (13);
		\draw (12) to (6);
		\draw (10) to (5);
		\draw [style={X_edge}] (14) to (0);
		\draw [style={X_edge}] (15) to (3);
		\draw [style={X_edge}] (17) to (1);
		\draw [style={X_edge}] (16) to (4);
		\draw [style={X_edge}] (19) to (26);
		\draw [style={X_edge}] (26) to (22);
		\draw [style={X_edge}] (22) to (25);
		\draw [style={X_edge}] (25) to (21);
		\draw [style={X_edge}] (24) to (21);
		\draw [style={X_edge}] (18) to (24);
		\draw [style={X_edge}] (19) to (23);
		\draw [style={X_edge}] (23) to (18);
		\draw (23) to (27);
		\draw (27) to (26);
		\draw (27) to (24);
		\draw (27) to (25);
		\draw [style={Z_edge}] (20) to (27);
		\draw (35) to (29);
		\draw (29) to (34);
		\draw (34) to (31);
		\draw (31) to (33);
		\draw (33) to (32);
		\draw (32) to (35);
		\draw (29) to (26);
		\draw (25) to (31);
		\draw (30) to (24);
		\draw (28) to (23);
		\draw [style={X_edge}] (32) to (18);
		\draw [style={X_edge}] (33) to (21);
		\draw [style={X_edge}] (35) to (19);
		\draw [style={X_edge}] (34) to (22);
		\draw [style={X_edge}] (1) to (38);
		\draw [style={X_edge}] (0) to (36);
		\draw [style={X_edge}] (3) to (37);
		\draw [style={X_edge}] (4) to (39);
		\draw [style={X_edge}] (42) to (21);
		\draw [style={X_edge}] (43) to (18);
		\draw [style={X_edge}] (40) to (22);
		\draw [style={X_edge}] (41) to (19);
	\end{pgfonlayer}
\end{tikzpicture}

%% file: Figures-Tikz/unit_cell_newchecks_fc_redundant_fault.tikz
\begin{tikzpicture}
	\begin{pgfonlayer}{nodelayer}
		\node [style={X_detector}] (0) at (0.5, -7) {};
		\node [style={X_detector}] (1) at (-4.5, -5) {};
		\node [style={Z_detector}] (2) at (0, 0) {};
		\node [style={X_detector}] (3) at (4.5, -5) {};
		\node [style={X_detector}] (4) at (-0.5, -3) {};
		\node [style={Z_fault}] (5) at (-2, -6) {};
		\node [style={Z_fault}] (6) at (2.5, -6) {};
		\node [style={Z_fault}] (7) at (2, -4) {};
		\node [style={Z_fault}] (8) at (-2.5, -4) {};
		\node [style={X_fault}] (9) at (0, -5) {};
		\node [style={X_fault}] (10) at (-2, -1) {};
		\node [style={X_fault}] (11) at (-2.5, 1) {};
		\node [style={X_fault}] (12) at (2.5, -1) {};
		\node [style={X_fault}] (13) at (2, 1) {};
		\node [style={Z_fault}] (14) at (0.5, -2) {};
		\node [style={Z_fault}] (15) at (4.5, 0) {};
		\node [style={Z_fault}] (16) at (-0.5, 2) {};
		\node [style={Z_fault}] (17) at (-4.5, 0) {};
		\node [style={X_detector}] (18) at (-0.5, 7) {};
		\node [style={X_detector}] (19) at (4.5, 5) {};
		\node [style={Z_detector}] (20) at (0, 0) {};
		\node [style={X_detector}] (21) at (-4.5, 5) {};
		\node [style={X_detector}] (22) at (0.5, 3) {};
		\node [style={Z_fault}] (23) at (2, 6) {};
		\node [style={Z_fault}] (24) at (-2.5, 6) {};
		\node [style={Z_fault}] (25) at (-2, 4) {};
		\node [style={Z_fault}] (26) at (2.5, 4) {};
		\node [style={X_fault}] (27) at (0, 5) {};
		\node [style={X_fault}] (28) at (2, 1) {};
		\node [style={X_fault}] (29) at (2.5, -1) {};
		\node [style={X_fault}] (30) at (-2.5, 1) {};
		\node [style={X_fault}] (31) at (-2, -1) {};
		\node [style={Z_fault}] (32) at (-0.5, 2) {};
		\node [style={Z_fault}] (33) at (-4.5, 0) {};
		\node [style={Z_fault}] (34) at (0.5, -2) {};
		\node [style={Z_fault}] (35) at (4.5, 0) {};
		\node [style={Z_fault}] (36) at (0.5, -12) {};
		\node [style={Z_fault}] (37) at (4.5, -10) {};
		\node [style={Z_fault}] (38) at (-4.5, -10) {};
		\node [style={Z_fault}] (39) at (-0.5, -8) {};
		\node [style={Z_fault}] (40) at (0.5, 8) {};
		\node [style={Z_fault}] (41) at (4.5, 10) {};
		\node [style={Z_fault}] (42) at (-4.5, 10) {};
		\node [style={Z_fault}] (43) at (-0.5, 12) {};
		\node [style={green_fault}] (44) at (-4.5, -10) {};
		\node [style={green_fault}] (45) at (-2, -6) {};
		\node [style={green_fault}] (46) at (0.5, -12) {};
	\end{pgfonlayer}
	\begin{pgfonlayer}{edgelayer}
		\draw [style={X_edge}] (1) to (8);
		\draw [style={X_edge}] (8) to (4);
		\draw [style={X_edge}] (4) to (7);
		\draw [style={X_edge}] (7) to (3);
		\draw [style={X_edge}] (6) to (3);
		\draw [style={X_edge}] (0) to (6);
		\draw [style={X_edge}] (1) to (5);
		\draw [style={X_edge}] (5) to (0);
		\draw (5) to (9);
		\draw (9) to (8);
		\draw (9) to (6);
		\draw (9) to (7);
		\draw [style={Z_edge}] (2) to (9);
		\draw [style={Z_edge}] (10) to (2);
		\draw [style={Z_edge}] (2) to (12);
		\draw [style={Z_edge}] (2) to (13);
		\draw [style={Z_edge}] (11) to (2);
		\draw (17) to (11);
		\draw (11) to (16);
		\draw (16) to (13);
		\draw (13) to (15);
		\draw (15) to (14);
		\draw (14) to (17);
		\draw (11) to (8);
		\draw (7) to (13);
		\draw (12) to (6);
		\draw (10) to (5);
		\draw [style={X_edge}] (14) to (0);
		\draw [style={X_edge}] (15) to (3);
		\draw [style={X_edge}] (17) to (1);
		\draw [style={X_edge}] (16) to (4);
		\draw [style={X_edge}] (19) to (26);
		\draw [style={X_edge}] (26) to (22);
		\draw [style={X_edge}] (22) to (25);
		\draw [style={X_edge}] (25) to (21);
		\draw [style={X_edge}] (24) to (21);
		\draw [style={X_edge}] (18) to (24);
		\draw [style={X_edge}] (19) to (23);
		\draw [style={X_edge}] (23) to (18);
		\draw (23) to (27);
		\draw (27) to (26);
		\draw (27) to (24);
		\draw (27) to (25);
		\draw [style={Z_edge}] (20) to (27);
		\draw (35) to (29);
		\draw (29) to (34);
		\draw (34) to (31);
		\draw (31) to (33);
		\draw (33) to (32);
		\draw (32) to (35);
		\draw (29) to (26);
		\draw (25) to (31);
		\draw (30) to (24);
		\draw (28) to (23);
		\draw [style={X_edge}] (32) to (18);
		\draw [style={X_edge}] (33) to (21);
		\draw [style={X_edge}] (35) to (19);
		\draw [style={X_edge}] (34) to (22);
		\draw [style={X_edge}] (1) to (38);
		\draw [style={X_edge}] (0) to (36);
		\draw [style={X_edge}] (3) to (37);
		\draw [style={X_edge}] (4) to (39);
		\draw [style={X_edge}] (42) to (21);
		\draw [style={X_edge}] (43) to (18);
		\draw [style={X_edge}] (40) to (22);
		\draw [style={X_edge}] (41) to (19);
	\end{pgfonlayer}
\end{tikzpicture}

%% file: Figures-Tikz/unit_cell_newchecks_fc_timelike_fault.tikz
\begin{tikzpicture}
	\begin{pgfonlayer}{nodelayer}
		\node [style={X_detector}] (0) at (0.5, -7) {};
		\node [style={X_detector}] (1) at (-4.5, -5) {};
		\node [style={Z_detector}] (2) at (0, 0) {};
		\node [style={X_detector}] (3) at (4.5, -5) {};
		\node [style={X_detector}] (4) at (-0.5, -3) {};
		\node [style={Z_fault}] (5) at (-2, -6) {};
		\node [style={Z_fault}] (6) at (2.5, -6) {};
		\node [style={Z_fault}] (7) at (2, -4) {};
		\node [style={Z_fault}] (8) at (-2.5, -4) {};
		\node [style={X_fault}] (9) at (0, -5) {};
		\node [style={X_fault}] (10) at (-2, -1) {};
		\node [style={X_fault}] (11) at (-2.5, 1) {};
		\node [style={X_fault}] (12) at (2.5, -1) {};
		\node [style={X_fault}] (13) at (2, 1) {};
		\node [style={Z_fault}] (14) at (0.5, -2) {};
		\node [style={Z_fault}] (15) at (4.5, 0) {};
		\node [style={Z_fault}] (16) at (-0.5, 2) {};
		\node [style={Z_fault}] (17) at (-4.5, 0) {};
		\node [style={X_detector}] (18) at (-0.5, 7) {};
		\node [style={X_detector}] (19) at (4.5, 5) {};
		\node [style={Z_detector}] (20) at (0, 0) {};
		\node [style={X_detector}] (21) at (-4.5, 5) {};
		\node [style={X_detector}] (22) at (0.5, 3) {};
		\node [style={Z_fault}] (23) at (2, 6) {};
		\node [style={Z_fault}] (24) at (-2.5, 6) {};
		\node [style={Z_fault}] (25) at (-2, 4) {};
		\node [style={Z_fault}] (26) at (2.5, 4) {};
		\node [style={X_fault}] (27) at (0, 5) {};
		\node [style={X_fault}] (28) at (2, 1) {};
		\node [style={X_fault}] (29) at (2.5, -1) {};
		\node [style={X_fault}] (30) at (-2.5, 1) {};
		\node [style={X_fault}] (31) at (-2, -1) {};
		\node [style={Z_fault}] (32) at (-0.5, 2) {};
		\node [style={Z_fault}] (33) at (-4.5, 0) {};
		\node [style={Z_fault}] (34) at (0.5, -2) {};
		\node [style={Z_fault}] (35) at (4.5, 0) {};
		\node [style={Z_fault}] (36) at (0.5, -12) {};
		\node [style={Z_fault}] (37) at (4.5, -10) {};
		\node [style={Z_fault}] (38) at (-4.5, -10) {};
		\node [style={Z_fault}] (39) at (-0.5, -8) {};
		\node [style={Z_fault}] (40) at (0.5, 8) {};
		\node [style={Z_fault}] (41) at (4.5, 10) {};
		\node [style={Z_fault}] (42) at (-4.5, 10) {};
		\node [style={Z_fault}] (43) at (-0.5, 12) {};
		\node [style={green_fault}] (46) at (0.5, -12) {};
		\node [style={green_fault}] (47) at (0.5, -2) {};
		\node [style={green_fault}] (48) at (0.5, 8) {};
	\end{pgfonlayer}
	\begin{pgfonlayer}{edgelayer}
		\draw [style={X_edge}] (1) to (8);
		\draw [style={X_edge}] (8) to (4);
		\draw [style={X_edge}] (4) to (7);
		\draw [style={X_edge}] (7) to (3);
		\draw [style={X_edge}] (6) to (3);
		\draw [style={X_edge}] (0) to (6);
		\draw [style={X_edge}] (1) to (5);
		\draw [style={X_edge}] (5) to (0);
		\draw (5) to (9);
		\draw (9) to (8);
		\draw (9) to (6);
		\draw (9) to (7);
		\draw [style={Z_edge}] (2) to (9);
		\draw [style={Z_edge}] (10) to (2);
		\draw [style={Z_edge}] (2) to (12);
		\draw [style={Z_edge}] (2) to (13);
		\draw [style={Z_edge}] (11) to (2);
		\draw (17) to (11);
		\draw (11) to (16);
		\draw (16) to (13);
		\draw (13) to (15);
		\draw (15) to (14);
		\draw (14) to (17);
		\draw (11) to (8);
		\draw (7) to (13);
		\draw (12) to (6);
		\draw (10) to (5);
		\draw [style={X_edge}] (14) to (0);
		\draw [style={X_edge}] (15) to (3);
		\draw [style={X_edge}] (17) to (1);
		\draw [style={X_edge}] (16) to (4);
		\draw [style={X_edge}] (19) to (26);
		\draw [style={X_edge}] (26) to (22);
		\draw [style={X_edge}] (22) to (25);
		\draw [style={X_edge}] (25) to (21);
		\draw [style={X_edge}] (24) to (21);
		\draw [style={X_edge}] (18) to (24);
		\draw [style={X_edge}] (19) to (23);
		\draw [style={X_edge}] (23) to (18);
		\draw (23) to (27);
		\draw (27) to (26);
		\draw (27) to (24);
		\draw (27) to (25);
		\draw [style={Z_edge}] (20) to (27);
		\draw (35) to (29);
		\draw (29) to (34);
		\draw (34) to (31);
		\draw (31) to (33);
		\draw (33) to (32);
		\draw (32) to (35);
		\draw (29) to (26);
		\draw (25) to (31);
		\draw (30) to (24);
		\draw (28) to (23);
		\draw [style={X_edge}] (32) to (18);
		\draw [style={X_edge}] (33) to (21);
		\draw [style={X_edge}] (35) to (19);
		\draw [style={X_edge}] (34) to (22);
		\draw [style={X_edge}] (1) to (38);
		\draw [style={X_edge}] (0) to (36);
		\draw [style={X_edge}] (3) to (37);
		\draw [style={X_edge}] (4) to (39);
		\draw [style={X_edge}] (42) to (21);
		\draw [style={X_edge}] (43) to (18);
		\draw [style={X_edge}] (40) to (22);
		\draw [style={X_edge}] (41) to (19);
	\end{pgfonlayer}
\end{tikzpicture}

%% file: Figures-Tikz/Y_measure.tikz
\begin{tikzpicture}
	\begin{pgfonlayer}{nodelayer}
		\node [style={scalable_dot}] (11) at (3.9, -1.925) {};
		\node [style={scalable_dot}] (10) at (-0.075, 2.075) {};
		\node [style={scalable_dot}] (9) at (3.925, 6.075) {};
		\node [style={scalable_box}] (8) at (3.9, 2.1) {};
		\node [style={scalable_box}] (7) at (-0.1, -1.9) {};
		\node [style={scalable_box}] (6) at (-0.1, 6.1) {};
		\node [style={scalable_dot}] (0) at (0, 2) {};
		\node [style={scalable_dot}] (1) at (4, -2) {};
		\node [style={scalable_dot}] (2) at (4, 6) {};
		\node [style={scalable_box}] (3) at (0, -2) {$X$};
		\node [style={scalable_box}] (4) at (4, 2) {$Y$};
		\node [style={scalable_box}] (5) at (0, 6) {$Z$};
		\node [style=none] (12) at (-1.25, 2) {$Q$};
		\node [style=none] (13) at (-1, 4) {$H_Z$};
		\node [style=none] (14) at (-1, 0) {$H_X$};
		\node [style=none] (15) at (2, 6.5) {$I$};
		\node [style=none] (16) at (2, 2.5) {$I$};
		\node [style=none] (17) at (2, -1.5) {$I$};
		\node [style=none] (18) at (5, 4) {$H_Z^\intercal$};
		\node [style=none] (19) at (5, 0) {$H_X^\intercal$};
	\end{pgfonlayer}
	\begin{pgfonlayer}{edgelayer}
		\draw (5) to (0);
		\draw (0) to (3);
		\draw (3) to (1);
		\draw (1) to (4);
		\draw (4) to (0);
		\draw (5) to (2);
		\draw (2) to (4);
	\end{pgfonlayer}
\end{tikzpicture}

%% file: Figures-Tikz/H_measure1.tikz
\begin{tikzpicture}
	\begin{pgfonlayer}{nodelayer}
		\node [style={scalable_meta_mixed}] (31) at (-0.1, -3.925) {};
		\node [style={scalable_box}] (1) at (-4.1, 4.1) {};
		\node [style={scalable_dot}] (2) at (-4.1, 0.075) {};
		\node [style={scalable_dot}] (3) at (-4, 0) {};
		\node [style={scalable_box}] (4) at (-4, 4) {$X$};
		\node [style={scalable_box}] (5) at (-4.1, -3.9) {};
		\node [style={scalable_box}] (6) at (-4, -4) {$Z$};
		\node [style=none] (7) at (-5, 2) {$H_X$};
		\node [style=none] (8) at (-5, -2) {$H_Z$};
		\node [style={scalable_box}] (9) at (-0.1, 0.1) {};
		\node [style={scalable_box}] (10) at (0, 0) {};
		\node [style={scalable_dot}] (11) at (-0.1, 4.075) {};
		\node [style={scalable_dot}] (12) at (0, 4) {};
		\node [style=none] (13) at (1.55, 2) {$[0, H_X^\intercal]$};
		\node [style=none] (14) at (-2, 5) {$I$};
		\node [style=none] (15) at (-2, -1) {$[0, I]$};
		\node [style=none] (17) at (-2, -5) {$I$};
		\node [style=none] (18) at (1, -2.5) {$H_Z$};
		\node [style={scalable_box}] (19) at (3.9, 4.1) {};
		\node [style={scalable_dot}] (20) at (3.9, 0.075) {};
		\node [style={scalable_dot}] (21) at (4, 0) {};
		\node [style={scalable_box}] (22) at (4, 4) {};
		\node [style={scalable_box}] (23) at (3.9, -3.9) {};
		\node [style={scalable_box}] (24) at (4, -4) {$X$};
		\node [style=none] (25) at (5.55, 2) {$[0, H_X]$};
		\node [style=none] (26) at (5.55, -2) {$[H_Z,  0]$};
		\node [style=none] (27) at (2, 5) {$[I,0]$};
		\node [style=none] (28) at (2, -1) {$[I,0]$};
		\node [style=none] (29) at (2, -5) {$I$};
		\node [style={scalable_meta_mixed}] (30) at (0, -4) {};
	\end{pgfonlayer}
	\begin{pgfonlayer}{edgelayer}
		\draw (3) to (6);
		\draw (3) to (4);
		\draw (12) to (10);
		\draw (10) to (3);
		\draw (4) to (12);
		\draw (21) to (24);
		\draw (21) to (22);
		\draw (10) to (21);
		\draw (12) to (22);
		\draw (10) to (30);
		\draw (30) to (24);
		\draw (30) to (6);
	\end{pgfonlayer}
\end{tikzpicture}

%% file: Figures-Tikz/rep0.tikz
\begin{tikzpicture}
	\begin{pgfonlayer}{nodelayer}
		\node [style=none] (0) at (0, 0) {};
		\node [style=rmat] (1) at (1, 0) {};
		\node [style=none] (2) at (2, 0) {};
		\node [style=none] (3) at (1, 0.7) {\footnotesize$A$};
	\end{pgfonlayer}
	\begin{pgfonlayer}{edgelayer}
		\draw [style=very thick] (0.center) to (1);
		\draw [style=very thick] (1) to (2.center);
	\end{pgfonlayer}
\end{tikzpicture}

%% file: Figures-Tikz/rep0b.tikz
\begin{tikzpicture}
	\begin{pgfonlayer}{nodelayer}
		\node [style=none] (0) at (0, 0) {};
		\node [style=lmat] (1) at (1, 0) {};
		\node [style=none] (2) at (2, 0) {};
		\node [style=none] (3) at (1, 0.7) {\footnotesize$B$};
	\end{pgfonlayer}
	\begin{pgfonlayer}{edgelayer}
		\draw [style=very thick] (0.center) to (1);
		\draw [style=very thick] (1) to (2.center);
	\end{pgfonlayer}
\end{tikzpicture}

%% file: Figures-Tikz/stab_measure_SZX.tikz
\begin{tikzpicture}
	\begin{pgfonlayer}{nodelayer}
		\node [style=gred] (0) at (2, 0) {};
		\node [style=umat] (1) at (-2, 1.25) {};
		\node [style=dmat] (2) at (2, 1.25) {};
		\node [style=ggreen] (3) at (-2, 0) {};
		\node [style=ggreen] (4) at (2, 2.5) {};
		\node [style=gred] (5) at (-2, 2.5) {};
		\node [style=none] (6) at (-4, 0) {};
		\node [style=none] (7) at (3.25, 1.25) {$H_X^\intercal$};
		\node [style=none] (8) at (-3, 1.25) {$H_Z$};
		\node [style=none] (9) at (4, 0) {};
		\node [style=none] (10) at (-2, 3.5) {$s_Z \pi$};
		\node [style=none] (11) at (2, 3.5) {$s_X \pi$};
	\end{pgfonlayer}
	\begin{pgfonlayer}{edgelayer}
		\draw [style=boldedge] (6.center) to (3);
		\draw [style=boldedge] (3) to (1);
		\draw [style=boldedge] (1) to (5);
		\draw [style=boldedge] (3) to (0);
		\draw [style=boldedge] (0) to (2);
		\draw [style=boldedge] (2) to (4);
		\draw [style=boldedge] (0) to (9.center);
	\end{pgfonlayer}
\end{tikzpicture}

%% file: Figures-Tikz/stab_measure_SZX_empty.tikz
\begin{tikzpicture}
	\begin{pgfonlayer}{nodelayer}
		\node [style=gred] (0) at (2, 0) {};
		\node [style=umat] (1) at (-2, 1.25) {};
		\node [style=dmat] (2) at (2, 1.25) {};
		\node [style=ggreen] (3) at (-2, 0) {};
		\node [style=ggreen] (4) at (2, 2.5) {};
		\node [style=gred] (5) at (-2, 2.5) {};
		\node [style=none] (6) at (-4, 0) {};
		\node [style=none] (7) at (3.25, 1.25) {$H_X^\intercal$};
		\node [style=none] (8) at (-3, 1.25) {$H_Z$};
		\node [style=none] (9) at (4, 0) {};
	\end{pgfonlayer}
	\begin{pgfonlayer}{edgelayer}
		\draw [style=boldedge] (6.center) to (3);
		\draw [style=boldedge] (3) to (1);
		\draw [style=boldedge] (1) to (5);
		\draw [style=boldedge] (3) to (0);
		\draw [style=boldedge] (0) to (2);
		\draw [style=boldedge] (2) to (4);
		\draw [style=boldedge] (0) to (9.center);
	\end{pgfonlayer}
\end{tikzpicture}

%% file: Figures-Tikz/stab_measure_SZX2_empty.tikz
\begin{tikzpicture}
	\begin{pgfonlayer}{nodelayer}
		\node [style=gred] (0) at (-2, 0) {};
		\node [style=umat] (1) at (2, 1.25) {};
		\node [style=dmat] (2) at (-2, 1.25) {};
		\node [style=ggreen] (3) at (2, 0) {};
		\node [style=ggreen] (4) at (-2, 2.5) {};
		\node [style=gred] (5) at (2, 2.5) {};
		\node [style=none] (6) at (-4, 0) {};
		\node [style=none] (7) at (-3.25, 1.25) {$H_X^\intercal$};
		\node [style=none] (8) at (3, 1.25) {$H_Z$};
		\node [style=none] (9) at (4, 0) {};
	\end{pgfonlayer}
	\begin{pgfonlayer}{edgelayer}
		\draw [style=boldedge] (6.center) to (3);
		\draw [style=boldedge] (3) to (1);
		\draw [style=boldedge] (1) to (5);
		\draw [style=boldedge] (3) to (0);
		\draw [style=boldedge] (0) to (2);
		\draw [style=boldedge] (2) to (4);
		\draw [style=boldedge] (0) to (9.center);
	\end{pgfonlayer}
\end{tikzpicture}

%% file: Figures-Tikz/stab_measure_collide.tikz
\begin{tikzpicture}
	\begin{pgfonlayer}{nodelayer}
		\node [style=umat] (1) at (-2, 1.25) {};
		\node [style=ggreen] (3) at (-2, 0) {};
		\node [style=gred] (5) at (-2, 2.5) {};
		\node [style=none] (6) at (-4, 0) {};
		\node [style=none] (8) at (-3, 1.25) {$H_Z$};
		\node [style=none] (9) at (4, 0) {};
		\node [style=umat] (10) at (2, 1.25) {};
		\node [style=ggreen] (11) at (2, 0) {};
		\node [style=gred] (12) at (2, 2.5) {};
		\node [style=none] (13) at (1, 1.25) {$H_Z$};
	\end{pgfonlayer}
	\begin{pgfonlayer}{edgelayer}
		\draw [style=boldedge] (6.center) to (3);
		\draw [style=boldedge] (3) to (1);
		\draw [style=boldedge] (1) to (5);
		\draw [style=boldedge] (11) to (10);
		\draw [style=boldedge] (10) to (12);
		\draw [style=boldedge] (3) to (9.center);
	\end{pgfonlayer}
\end{tikzpicture}

%% file: Figures-Tikz/stab_measure_collide2.tikz
\begin{tikzpicture}
	\begin{pgfonlayer}{nodelayer}
		\node [style=umat] (0) at (0, 1.25) {};
		\node [style=ggreen] (1) at (0, 0) {};
		\node [style=gred] (2) at (0, 2.5) {};
		\node [style=none] (3) at (-2, 0) {};
		\node [style=none] (4) at (-1, 1.25) {$H_Z$};
		\node [style=none] (5) at (2, 0) {};
	\end{pgfonlayer}
	\begin{pgfonlayer}{edgelayer}
		\draw [style=boldedge] (3.center) to (1);
		\draw [style=boldedge] (1) to (0);
		\draw [style=boldedge] (0) to (2);
		\draw [style=boldedge] (1) to (5.center);
	\end{pgfonlayer}
\end{tikzpicture}

%% file: Figures-Tikz/SZX_hom_to_surgery1.tikz
\begin{tikzpicture}
	\begin{pgfonlayer}{nodelayer}
		\node [style=gred] (2) at (-12.5, 5.5) {};
		\node [style=gred] (3) at (-6.5, 5.5) {};
		\node [style=umat] (4) at (-10.5, 6.75) {};
		\node [style=dmat] (5) at (-6.5, 6.75) {};
		\node [style=ggreen] (6) at (-10.5, 5.5) {};
		\node [style=ggreen] (7) at (-6.5, 8) {};
		\node [style=gred] (8) at (-10.5, 8) {};
		\node [style=gred] (9) at (-2.5, 5.5) {};
		\node [style=gred] (10) at (-0.5, 5.5) {};
		\node [style=umat] (11) at (-2.5, 3.5) {};
		\node [style=ggreen] (12) at (-2.5, 1.5) {};
		\node [style=gred] (13) at (-6.5, 1.5) {};
		\node [style=umat] (14) at (-10.5, 2.75) {};
		\node [style=dmat] (15) at (-6.5, 2.75) {};
		\node [style=ggreen] (16) at (-10.5, 1.5) {};
		\node [style=ggreen] (17) at (-6.5, 4) {};
		\node [style=gred] (18) at (-10.5, 4) {};
		\node [style=none] (19) at (-12.5, 1.5) {};
		\node [style=none] (20) at (-0.5, 1.5) {};
		\node [style=none] (21) at (-3.5, 3.5) {$f_1^\intercal$};
		\node [style=none] (22) at (-0.5, 6.25) {$\alpha\pi$};
		\node [style=none] (23) at (-11.5, 6.75) {$H_Z'$};
		\node [style=none] (24) at (-5, 6.75) {$(H_X')^\intercal$};
		\node [style=none] (25) at (-5.25, 2.75) {$H_X^\intercal$};
		\node [style=none] (26) at (-11.5, 2.75) {$H_Z$};
		\node [style=none] (27) at (1, 4) {$=$};
		\node [style=gred] (29) at (8.5, 5.5) {};
		\node [style=dmat] (31) at (8.5, 6.75) {};
		\node [style=ggreen] (33) at (8.5, 8) {};
		\node [style=gred] (35) at (12.5, 5.5) {};
		\node [style=umat] (37) at (12.5, 3.5) {};
		\node [style=ggreen] (38) at (12.5, 1.5) {};
		\node [style=gred] (39) at (8.5, 1.5) {};
		\node [style=umat] (40) at (4.5, 2.75) {};
		\node [style=dmat] (41) at (8.5, 2.75) {};
		\node [style=ggreen] (42) at (4.5, 1.5) {};
		\node [style=ggreen] (43) at (8.5, 4) {};
		\node [style=gred] (44) at (4.5, 4) {};
		\node [style=none] (45) at (2.5, 1.5) {};
		\node [style=none] (46) at (14.5, 1.5) {};
		\node [style=none] (47) at (11.5, 3.5) {$f_1^\intercal$};
		\node [style=none] (48) at (12.5, 6.25) {$\alpha\pi$};
		\node [style=none] (50) at (10, 6.75) {$(H_X')^\intercal$};
		\node [style=none] (51) at (9.75, 2.75) {$H_X^\intercal$};
		\node [style=none] (52) at (3.5, 2.75) {$H_Z$};
		\node [style=none] (53) at (-14, -7) {$=$};
		\node [style=dmat] (55) at (-2.5, -5.5) {};
		\node [style=ggreen] (56) at (-2.5, -4) {};
		\node [style=gred] (57) at (-2.5, -7) {};
		\node [style=umat] (58) at (-2.5, -8.5) {};
		\node [style=ggreen] (59) at (-2.5, -10) {};
		\node [style=gred] (60) at (-6.5, -10) {};
		\node [style=umat] (61) at (-10.5, -8.75) {};
		\node [style=dmat] (62) at (-6.5, -8.75) {};
		\node [style=ggreen] (63) at (-10.5, -10) {};
		\node [style=ggreen] (64) at (-6.5, -7.5) {};
		\node [style=gred] (65) at (-10.5, -7.5) {};
		\node [style=none] (66) at (-12.5, -10) {};
		\node [style=none] (67) at (-0.5, -10) {};
		\node [style=none] (68) at (-3.5, -8.5) {$f_1^\intercal$};
		\node [style=none] (69) at (-1.5, -7) {$\alpha\pi$};
		\node [style=none] (70) at (-4, -5.5) {$(H_X')^\intercal$};
		\node [style=none] (71) at (-5.25, -8.75) {$H_X^\intercal$};
		\node [style=none] (72) at (-11.5, -8.75) {$H_Z$};
		\node [style=ggreen] (73) at (-12.5, -4) {};
		\node [style=ggreen] (74) at (-0.5, -4) {};
		\node [style=none] (75) at (1, -7) {$=$};
		\node [style=dmat] (76) at (12.5, -5.5) {};
		\node [style=ggreen] (77) at (12.5, -4) {};
		\node [style=gred] (78) at (12.5, -7) {};
		\node [style=umat] (79) at (12.5, -8.5) {};
		\node [style=ggreen] (80) at (12.5, -10) {};
		\node [style=gred] (81) at (8.5, -10) {};
		\node [style=umat] (82) at (4.5, -8.75) {};
		\node [style=dmat] (83) at (8.5, -8.5) {};
		\node [style=ggreen] (84) at (4.5, -10) {};
		\node [style=ggreen] (85) at (8.5, -7) {};
		\node [style=gred] (86) at (4.5, -7.5) {};
		\node [style=none] (87) at (2.5, -10) {};
		\node [style=none] (88) at (14.5, -10) {};
		\node [style=none] (89) at (11.5, -8.5) {$f_1^\intercal$};
		\node [style=none] (90) at (13.5, -7) {$\alpha\pi$};
		\node [style=none] (91) at (11, -5.5) {$(H_X')^\intercal$};
		\node [style=none] (92) at (7.25, -8.5) {$H_X^\intercal$};
		\node [style=none] (93) at (3.5, -8.75) {$H_Z$};
		\node [style=ggreen] (94) at (2.5, -4) {};
		\node [style=ggreen] (95) at (14.5, -4) {};
		\node [style=gred] (96) at (8.5, -4) {};
		\node [style=umat] (97) at (8.5, -5.5) {};
		\node [style=none] (98) at (7.5, -5.5) {$f_2^\intercal$};
		\node [style=ggreen] (99) at (4.5, -1.5) {};
		\node [style=gred] (101) at (4.5, -4) {};
		\node [style=none] (102) at (3.5, -2.75) {$N^\intercal$};
		\node [style=dmat] (103) at (4.5, -2.75) {};
	\end{pgfonlayer}
	\begin{pgfonlayer}{edgelayer}
		\draw [style=boldedge] (19.center) to (16);
		\draw [style=boldedge] (16) to (14);
		\draw [style=boldedge] (14) to (18);
		\draw [style=boldedge] (16) to (13);
		\draw [style=boldedge] (13) to (15);
		\draw [style=boldedge] (15) to (17);
		\draw [style=boldedge] (13) to (12);
		\draw [style=boldedge] (12) to (11);
		\draw [style=boldedge] (11) to (9);
		\draw [style=boldedge] (9) to (3);
		\draw [style=boldedge] (3) to (5);
		\draw [style=boldedge] (5) to (7);
		\draw [style=boldedge] (3) to (6);
		\draw [style=boldedge] (6) to (2);
		\draw [style=boldedge] (6) to (4);
		\draw [style=boldedge] (4) to (8);
		\draw [style=boldedge] (9) to (10);
		\draw [style=boldedge] (12) to (20.center);
		\draw [style=boldedge] (45.center) to (42);
		\draw [style=boldedge] (42) to (40);
		\draw [style=boldedge] (40) to (44);
		\draw [style=boldedge] (42) to (39);
		\draw [style=boldedge] (39) to (41);
		\draw [style=boldedge] (41) to (43);
		\draw [style=boldedge] (39) to (38);
		\draw [style=boldedge] (38) to (37);
		\draw [style=boldedge] (37) to (35);
		\draw [style=boldedge] (35) to (29);
		\draw [style=boldedge] (29) to (31);
		\draw [style=boldedge] (31) to (33);
		\draw [style=boldedge] (38) to (46.center);
		\draw [style=boldedge] (66.center) to (63);
		\draw [style=boldedge] (63) to (61);
		\draw [style=boldedge] (61) to (65);
		\draw [style=boldedge] (63) to (60);
		\draw [style=boldedge] (60) to (62);
		\draw [style=boldedge] (62) to (64);
		\draw [style=boldedge] (60) to (59);
		\draw [style=boldedge] (59) to (58);
		\draw [style=boldedge] (58) to (57);
		\draw [style=boldedge] (55) to (56);
		\draw [style=boldedge] (59) to (67.center);
		\draw [style=boldedge] (55) to (57);
		\draw [style=boldedge] (74) to (56);
		\draw [style=boldedge] (73) to (56);
		\draw [style=boldedge] (87.center) to (84);
		\draw [style=boldedge] (84) to (82);
		\draw [style=boldedge] (82) to (86);
		\draw [style=boldedge] (84) to (81);
		\draw [style=boldedge] (81) to (83);
		\draw [style=boldedge] (83) to (85);
		\draw [style=boldedge] (81) to (80);
		\draw [style=boldedge] (80) to (79);
		\draw [style=boldedge] (79) to (78);
		\draw [style=boldedge] (76) to (77);
		\draw [style=boldedge] (80) to (88.center);
		\draw [style=boldedge] (76) to (78);
		\draw [style=boldedge] (95) to (77);
		\draw [style=boldedge] (94) to (77);
		\draw [style=boldedge] (96) to (97);
		\draw [style=boldedge] (97) to (85);
		\draw [style=boldedge] (99) to (103);
		\draw [style=boldedge] (103) to (101);
	\end{pgfonlayer}
\end{tikzpicture}

%% file: Figures-Tikz/SZX_surgery_to_hom1.tikz
\begin{tikzpicture}
	\begin{pgfonlayer}{nodelayer}
		\node [style=dmat] (0) at (-2.5, 6.5) {};
		\node [style=ggreen] (1) at (-2.5, 8) {};
		\node [style=gred] (2) at (-2.5, 5) {};
		\node [style=umat] (3) at (-2.5, 3.5) {};
		\node [style=ggreen] (4) at (-2.5, 2) {};
		\node [style=gred] (5) at (-6.5, 2) {};
		\node [style=umat] (6) at (-10.5, 3.25) {};
		\node [style=dmat] (7) at (-6.5, 3.5) {};
		\node [style=ggreen] (8) at (-10.5, 2) {};
		\node [style=ggreen] (9) at (-6.5, 5) {};
		\node [style=gred] (10) at (-10.5, 4.5) {};
		\node [style=none] (11) at (-12.5, 2) {};
		\node [style=none] (12) at (-0.5, 2) {};
		\node [style=none] (13) at (-3.5, 3.5) {$F^\intercal$};
		\node [style=none] (14) at (-1.5, 5) {$\alpha\pi$};
		\node [style=none] (15) at (-3.5, 6.5) {$\CG^\intercal$};
		\node [style=none] (16) at (-7.75, 3.5) {$H_X^\intercal$};
		\node [style=none] (17) at (-11.5, 3.25) {$H_Z$};
		\node [style=ggreen] (18) at (-12.5, 8) {};
		\node [style=ggreen] (19) at (-0.5, 8) {};
		\node [style=gred] (20) at (-6.5, 8) {};
		\node [style=umat] (21) at (-6.5, 6.5) {};
		\node [style=none] (22) at (-7.5, 6.5) {$M^\intercal$};
		\node [style=none] (23) at (1, 5) {$=$};
		\node [style=none] (24) at (-3.25, 5) {$\CV$};
		\node [style=none] (25) at (-12.5, 8.75) {$\CE$};
		\node [style=dmat] (30) at (12.5, 6.5) {};
		\node [style=ggreen] (31) at (12.5, 8) {};
		\node [style=gred] (32) at (12.5, 5) {};
		\node [style=umat] (33) at (12.5, 3.5) {};
		\node [style=ggreen] (34) at (12.5, 2) {};
		\node [style=gred] (35) at (8.5, 2) {};
		\node [style=umat] (36) at (4.5, 3.25) {};
		\node [style=dmat] (37) at (8.5, 3.5) {};
		\node [style=ggreen] (38) at (4.5, 2) {};
		\node [style=ggreen] (39) at (8.5, 5) {};
		\node [style=gred] (40) at (4.5, 4.5) {};
		\node [style=none] (41) at (2.5, 2) {};
		\node [style=none] (42) at (14.5, 2) {};
		\node [style=none] (43) at (11.5, 3.5) {$F^\intercal$};
		\node [style=none] (44) at (13.5, 5) {$\alpha\pi$};
		\node [style=none] (45) at (11.5, 6.5) {$\CG^\intercal$};
		\node [style=none] (46) at (7.25, 3.5) {$H_X^\intercal$};
		\node [style=none] (47) at (3.5, 3.25) {$H_Z$};
		\node [style=gred] (50) at (8.5, 8) {};
		\node [style=umat] (51) at (8.5, 6.5) {};
		\node [style=none] (52) at (7.5, 6.5) {$M^\intercal$};
		\node [style=none] (53) at (11.75, 5) {$\CV$};
		\node [style=ggreen] (54) at (6.5, 8) {};
		\node [style=none] (55) at (6.5, 8.75) {$\CE$};
		\node [style=none] (56) at (-14, -6.5) {$=$};
		\node [style=dmat] (57) at (-7.5, -2.25) {};
		\node [style=ggreen] (58) at (-7.5, -1) {};
		\node [style=gred] (59) at (-2.5, -3.5) {};
		\node [style=umat] (60) at (-2.5, -6.5) {};
		\node [style=ggreen] (61) at (-2.5, -9.5) {};
		\node [style=gred] (62) at (-6.5, -9.5) {};
		\node [style=umat] (63) at (-10.5, -8.25) {};
		\node [style=dmat] (64) at (-6.5, -8.25) {};
		\node [style=ggreen] (65) at (-10.5, -9.5) {};
		\node [style=ggreen] (66) at (-6.5, -7) {};
		\node [style=gred] (67) at (-10.5, -7) {};
		\node [style=none] (68) at (-12.5, -9.5) {};
		\node [style=none] (69) at (-0.5, -9.5) {};
		\node [style=none] (70) at (-3.5, -6.5) {$F^\intercal$};
		\node [style=none] (71) at (-1.5, -3.5) {$\alpha\pi$};
		\node [style=none] (72) at (-8.5, -2.25) {$\CG^\intercal$};
		\node [style=none] (73) at (-7.75, -8) {$H_X^\intercal$};
		\node [style=none] (74) at (-11.5, -8.25) {$H_Z$};
		\node [style=gred] (75) at (-7.5, -3.5) {};
		\node [style=gred] (76) at (-12.5, -3.5) {};
		\node [style=none] (80) at (1, -6.5) {$=$};
		\node [style=dmat] (81) at (7.5, -2.25) {};
		\node [style=ggreen] (82) at (7.5, -1) {};
		\node [style=gred] (83) at (12.5, -3.5) {};
		\node [style=umat] (84) at (12.5, -6.5) {};
		\node [style=ggreen] (85) at (12.5, -9.5) {};
		\node [style=gred] (86) at (8.5, -9.5) {};
		\node [style=umat] (87) at (4.5, -8.25) {};
		\node [style=dmat] (88) at (8.5, -8.25) {};
		\node [style=ggreen] (89) at (4.5, -9.5) {};
		\node [style=ggreen] (90) at (8.5, -7) {};
		\node [style=gred] (91) at (4.5, -7) {};
		\node [style=none] (92) at (2.5, -9.5) {};
		\node [style=none] (93) at (14.5, -9.5) {};
		\node [style=none] (94) at (11.5, -6.5) {$F^\intercal$};
		\node [style=none] (95) at (14.5, -2.75) {$\alpha\pi$};
		\node [style=none] (96) at (6.5, -2.25) {$\CG^\intercal$};
		\node [style=none] (97) at (7.25, -8) {$H_X^\intercal$};
		\node [style=none] (98) at (3.5, -8.25) {$H_Z$};
		\node [style=gred] (99) at (7.5, -3.5) {};
		\node [style=gred] (100) at (2.5, -3.5) {};
		\node [style=gred] (101) at (4.5, -1) {};
		\node [style=ggreen] (102) at (4.5, -3.5) {};
		\node [style=umat] (103) at (4.5, -2.25) {};
		\node [style=gred] (104) at (14.5, -3.5) {};
		\node [style=none] (105) at (3.5, -2.25) {$H_Z'$};
		\node [style=ggreen] (106) at (-10.5, 10.5) {};
		\node [style=gred] (107) at (-10.5, 8) {};
		\node [style=none] (108) at (-11.5, 9.25) {$N^\intercal$};
		\node [style=dmat] (109) at (-10.5, 9.25) {};
		\node [style=none] (110) at (-12.5, -2.75) {$\CV$};
		\node [style=none] (111) at (2.5, -2.75) {$\CV$};
		\node [style=none] (112) at (-7.5, -0.25) {$\CE$};
		\node [style=none] (113) at (7.5, -0.25) {$\CE$};
	\end{pgfonlayer}
	\begin{pgfonlayer}{edgelayer}
		\draw [style=boldedge] (11.center) to (8);
		\draw [style=boldedge] (8) to (6);
		\draw [style=boldedge] (6) to (10);
		\draw [style=boldedge] (8) to (5);
		\draw [style=boldedge] (5) to (7);
		\draw [style=boldedge] (7) to (9);
		\draw [style=boldedge] (5) to (4);
		\draw [style=boldedge] (4) to (3);
		\draw [style=boldedge] (3) to (2);
		\draw [style=boldedge] (0) to (1);
		\draw [style=boldedge] (4) to (12.center);
		\draw [style=boldedge] (0) to (2);
		\draw [style=boldedge] (19) to (1);
		\draw [style=boldedge] (18) to (1);
		\draw [style=boldedge] (20) to (21);
		\draw [style=boldedge] (21) to (9);
		\draw [style=boldedge] (41.center) to (38);
		\draw [style=boldedge] (38) to (36);
		\draw [style=boldedge] (36) to (40);
		\draw [style=boldedge] (38) to (35);
		\draw [style=boldedge] (35) to (37);
		\draw [style=boldedge] (37) to (39);
		\draw [style=boldedge] (35) to (34);
		\draw [style=boldedge] (34) to (33);
		\draw [style=boldedge] (33) to (32);
		\draw [style=boldedge] (30) to (31);
		\draw [style=boldedge] (34) to (42.center);
		\draw [style=boldedge] (30) to (32);
		\draw [style=boldedge] (50) to (51);
		\draw [style=boldedge] (51) to (39);
		\draw [style=boldedge] (68.center) to (65);
		\draw [style=boldedge] (65) to (63);
		\draw [style=boldedge] (63) to (67);
		\draw [style=boldedge] (65) to (62);
		\draw [style=boldedge] (62) to (64);
		\draw [style=boldedge] (64) to (66);
		\draw [style=boldedge] (62) to (61);
		\draw [style=boldedge] (61) to (60);
		\draw [style=boldedge] (60) to (59);
		\draw [style=boldedge] (57) to (58);
		\draw [style=boldedge] (61) to (69.center);
		\draw [style=boldedge] (57) to (75);
		\draw [style=boldedge] (75) to (59);
		\draw [style=boldedge] (92.center) to (89);
		\draw [style=boldedge] (89) to (87);
		\draw [style=boldedge] (87) to (91);
		\draw [style=boldedge] (89) to (86);
		\draw [style=boldedge] (86) to (88);
		\draw [style=boldedge] (88) to (90);
		\draw [style=boldedge] (86) to (85);
		\draw [style=boldedge] (85) to (84);
		\draw [style=boldedge] (84) to (83);
		\draw [style=boldedge] (81) to (82);
		\draw [style=boldedge] (85) to (93.center);
		\draw [style=boldedge] (81) to (99);
		\draw [style=boldedge] (99) to (83);
		\draw [style=boldedge] (76) to (75);
		\draw [style=boldedge] (100) to (102);
		\draw [style=boldedge] (102) to (103);
		\draw [style=boldedge] (103) to (101);
		\draw [style=boldedge] (102) to (99);
		\draw [style=boldedge] (83) to (104);
		\draw [style=boldedge] (54) to (50);
		\draw [style=boldedge] (50) to (31);
		\draw [style=boldedge] (106) to (109);
		\draw [style=boldedge] (109) to (107);
	\end{pgfonlayer}
\end{tikzpicture}

%% file: main.bbl
\begin{thebibliography}{125}
\providecommand{\natexlab}[1]{#1}
\providecommand{\url}[1]{\texttt{#1}}
\expandafter\ifx\csname urlstyle\endcsname\relax
  \providecommand{\doi}[1]{doi: #1}\else
  \providecommand{\doi}{doi: \begingroup \urlstyle{rm}\Url}\fi

\bibitem[Shor(1995)]{shor1995scheme}
Peter~W Shor.
\newblock {Scheme for reducing decoherence in quantum computer memory}.
\newblock \emph{Physical review A}, 52\penalty0 (4):\penalty0 R2493, 1995.

\bibitem[Gottesman(1997)]{gottesman1997stabilizer}
Daniel Gottesman.
\newblock \emph{{Stabilizer codes and quantum error correction}}.
\newblock PhD thesis, California Institute of Technology, 1997.
\newblock URL \url{https://thesis.library.caltech.edu/2900/}.

\bibitem[Kitaev(1997)]{Kitaev1997qec}
A~Yu Kitaev.
\newblock {Quantum computations: algorithms and error correction}.
\newblock \emph{Russian Mathematical Surveys}, 52\penalty0 (6):\penalty0 1191–1249, Dec 1997.
\newblock ISSN 1468-4829.
\newblock \doi{10.1070/rm1997v052n06abeh002155}.
\newblock URL \url{http://dx.doi.org/10.1070/RM1997v052n06ABEH002155}.

\bibitem[Aharonov and Ben-Or(1997)]{aharonov1997fault}
Dorit Aharonov and Michael Ben-Or.
\newblock {Fault-tolerant quantum computation with constant error}.
\newblock In \emph{Proceedings of the twenty-ninth annual ACM symposium on Theory of computing}, pages 176--188, 1997.

\bibitem[Shor(1996)]{shor1996fault}
P.W. Shor.
\newblock {Fault-tolerant quantum computation}.
\newblock In \emph{Proceedings of 37th Conference on Foundations of Computer Science}, pages 56--65, 1996.
\newblock \doi{10.1109/SFCS.1996.548464}.
\newblock URL \url{https://ieeexplore.ieee.org/document/548464}.

\bibitem[Bravyi and Kitaev(1998)]{bravyi1998codes}
S.~B. Bravyi and A.~Yu. Kitaev.
\newblock {Quantum codes on a lattice with boundary}.
\newblock \emph{arXiv:quant-ph/9811052}, 1998.

\bibitem[Kitaev(2003)]{Kitaev2003anyon}
A.Yu. Kitaev.
\newblock {Fault-tolerant quantum computation by anyons}.
\newblock \emph{Annals of Physics}, 303\penalty0 (1):\penalty0 2–30, Jan 2003.
\newblock ISSN 0003-4916.
\newblock \doi{10.1016/s0003-4916(02)00018-0}.
\newblock URL \url{http://dx.doi.org/10.1016/S0003-4916(02)00018-0}.

\bibitem[Dennis et~al.(2002)Dennis, Kitaev, Landahl, and Preskill]{dennis2002memory}
Eric Dennis, Alexei Kitaev, Andrew Landahl, and John Preskill.
\newblock {Topological quantum memory}.
\newblock \emph{Journal of Mathematical Physics}, 43\penalty0 (9):\penalty0 4452–4505, Sep 2002.
\newblock ISSN 1089-7658.
\newblock \doi{10.1063/1.1499754}.
\newblock URL \url{http://dx.doi.org/10.1063/1.1499754}.

\bibitem[Bomb{\'\i}n(2015)]{bombin2015gauge}
H{\'e}ctor Bomb{\'\i}n.
\newblock {Gauge color codes: optimal transversal gates and gauge fixing in topological stabilizer codes}.
\newblock \emph{New Journal of Physics}, 17\penalty0 (8):\penalty0 083002, 2015.

\bibitem[ecz(2022)]{eczoo_topological}
Topological code.
\newblock In Victor~V. Albert and Philippe Faist, editors, \emph{The Error Correction Zoo}. 2022.
\newblock URL \url{https://errorcorrectionzoo.org/c/topological}.

\bibitem[Bravyi et~al.(2010)Bravyi, Poulin, and Terhal]{bravyi2010tradeoffs}
Sergey Bravyi, David Poulin, and Barbara Terhal.
\newblock {Tradeoffs for Reliable Quantum Information Storage in 2D Systems}.
\newblock \emph{Phys. Rev. Lett.}, 104:\penalty0 050503, Feb 2010.
\newblock \doi{10.1103/PhysRevLett.104.050503}.
\newblock URL \url{https://link.aps.org/doi/10.1103/PhysRevLett.104.050503}.

\bibitem[ecz(2025)]{eczoo_qldpc}
Quantum ldpc (qldpc) code.
\newblock In Victor~V. Albert and Philippe Faist, editors, \emph{The Error Correction Zoo}. 2025.
\newblock URL \url{https://errorcorrectionzoo.org/c/qldpc}.

\bibitem[Gidney(2021)]{gidney2021stim}
C.~Gidney.
\newblock {Stim: a fast stabilizer circuit simulator}.
\newblock \emph{{Quantum}}, 5:\penalty0 497, Jul 2021.
\newblock ISSN 2521-327X.
\newblock \doi{10.22331/q-2021-07-06-497}.
\newblock URL \url{https://doi.org/10.22331/q-2021-07-06-497}.

\bibitem[Tillich and Z{\'e}mor(2013)]{tillich2013quantum}
Jean-Pierre Tillich and Gilles Z{\'e}mor.
\newblock {Quantum {LDPC} codes with positive rate and minimum distance proportional to the square root of the blocklength}.
\newblock \emph{IEEE Transactions on Information Theory}, 60\penalty0 (2):\penalty0 1193--1202, 2013.

\bibitem[Kovalev and Pryadko(2013)]{kovalev2013quantum}
Alexey~A Kovalev and Leonid~P Pryadko.
\newblock Quantum kronecker sum-product low-density parity-check codes with finite rate.
\newblock \emph{Physical Review A—Atomic, Molecular, and Optical Physics}, 88\penalty0 (1):\penalty0 012311, 2013.

\bibitem[Panteleev and Kalachev(2021)]{panteleev2021degenerate}
Pavel Panteleev and Gleb Kalachev.
\newblock {Degenerate {Q}uantum {LDPC} {C}odes {W}ith {G}ood {F}inite {L}ength {P}erformance}.
\newblock \emph{{Quantum}}, 5:\penalty0 585, Nov 2021.
\newblock ISSN 2521-327X.
\newblock \doi{10.22331/q-2021-11-22-585}.
\newblock URL \url{https://doi.org/10.22331/q-2021-11-22-585}.

\bibitem[Breuckmann and Eberhardt(2021{\natexlab{a}})]{breuckmann2021balanced}
Nikolas~P. Breuckmann and Jens~N. Eberhardt.
\newblock {Balanced Product Quantum Codes}.
\newblock \emph{IEEE Transactions on Information Theory}, 67\penalty0 (10):\penalty0 6653--6674, 2021{\natexlab{a}}.
\newblock \doi{10.1109/TIT.2021.3097347}.

\bibitem[Lin and Pryadko(2024)]{lin2024quantum}
Hsiang-Ku Lin and Leonid~P Pryadko.
\newblock {Quantum two-block group algebra codes}.
\newblock \emph{Physical Review A}, 109\penalty0 (2):\penalty0 022407, 2024.

\bibitem[Bravyi et~al.(2022{\natexlab{a}})Bravyi, Dial, Gambetta, Gil, and Nazario]{Bravyi2022future}
Sergey Bravyi, Oliver Dial, Jay~M. Gambetta, Darío Gil, and Zaira Nazario.
\newblock {The future of quantum computing with superconducting qubits}.
\newblock \emph{Journal of Applied Physics}, 132\penalty0 (16), Oct 2022{\natexlab{a}}.
\newblock ISSN 1089-7550.
\newblock \doi{10.1063/5.0082975}.
\newblock URL \url{http://dx.doi.org/10.1063/5.0082975}.

\bibitem[Bluvstein et~al.(2024)Bluvstein, Evered, Geim, Li, Zhou, Manovitz, Ebadi, Cain, Kalinowski, Hangleiter, et~al.]{bluvstein2024logical}
Dolev Bluvstein, Simon~J Evered, Alexandra~A Geim, Sophie~H Li, Hengyun Zhou, Tom Manovitz, Sepehr Ebadi, Madelyn Cain, Marcin Kalinowski, Dominik Hangleiter, et~al.
\newblock {Logical quantum processor based on reconfigurable atom arrays}.
\newblock \emph{Nature}, 626\penalty0 (7997):\penalty0 58--65, 2024.
\newblock URL \url{https://www.nature.com/articles/s41586-023-06927-3}.

\bibitem[Team(2025)]{psiquantum2025manufacturable}
PsiQuantum Team.
\newblock A manufacturable platform for photonic quantum computing.
\newblock \emph{Nature}, pages 1--3, 2025.

\bibitem[Xu et~al.(2024{\natexlab{a}})Xu, Bonilla~Ataides, Pattison, Raveendran, Bluvstein, Wurtz, Vasić, Lukin, Jiang, and Zhou]{xu2024constant}
Qian Xu, J.~Pablo Bonilla~Ataides, Christopher~A. Pattison, Nithin Raveendran, Dolev Bluvstein, Jonathan Wurtz, Bane Vasić, Mikhail~D. Lukin, Liang Jiang, and Hengyun Zhou.
\newblock {Constant-overhead fault-tolerant quantum computation with reconfigurable atom arrays}.
\newblock \emph{Nature Physics}, 20\penalty0 (7):\penalty0 1084–1090, Apr 2024{\natexlab{a}}.
\newblock ISSN 1745-2481.
\newblock \doi{10.1038/s41567-024-02479-z}.
\newblock URL \url{http://dx.doi.org/10.1038/s41567-024-02479-z}.

\bibitem[Bravyi et~al.(2024{\natexlab{a}})Bravyi, Cross, Gambetta, Maslov, Rall, and Yoder]{bravyi2024highthreshold}
S.~Bravyi, A.~W. Cross, J.~M. Gambetta, D.~Maslov, P.~Rall, and T.~J. Yoder.
\newblock {High-threshold and low-overhead fault-tolerant quantum memory}.
\newblock \emph{Nature}, 627\penalty0 (8005):\penalty0 778–782, Mar 2024{\natexlab{a}}.
\newblock ISSN 1476-4687.
\newblock \doi{10.1038/s41586-024-07107-7}.
\newblock URL \url{http://dx.doi.org/10.1038/s41586-024-07107-7}.

\bibitem[Breuckmann and Burton(2024)]{breuckmann2024foldtransversal}
Nikolas~P. Breuckmann and Simon Burton.
\newblock {Fold-{T}ransversal {C}lifford {G}ates for {Q}uantum {C}odes}.
\newblock \emph{{Quantum}}, 8:\penalty0 1372, Jun 2024.
\newblock ISSN 2521-327X.
\newblock \doi{10.22331/q-2024-06-13-1372}.
\newblock URL \url{https://doi.org/10.22331/q-2024-06-13-1372}.

\bibitem[Quintavalle et~al.(2023)Quintavalle, Webster, and Vasmer]{quintavalle2023partitioning}
A.~O. Quintavalle, P.~Webster, and M.~Vasmer.
\newblock {Partitioning qubits in hypergraph product codes to implement logical gates}.
\newblock \emph{Quantum}, 7\penalty0 (1153), 2023.

\bibitem[Eberhardt and Steffan(2024)]{eberhardt2024operators}
J.~N. Eberhardt and V.~Steffan.
\newblock {Logical Operators and Fold-Transversal Gates of Bivariate Bicycle Codes}.
\newblock \emph{arXiv:2407.03973v1}, 2024.

\bibitem[Zhu et~al.(2023)Zhu, Sikander, Portnoy, Cross, and Brown]{zhu2023gates}
G.~Zhu, S.~Sikander, E.~Portnoy, A.~W. Cross, and B.~J. Brown.
\newblock {Non-Clifford and parallelizable fault-tolerant logical gates on constant and almost-constant rate homological quantum {LDPC} codes via higher symmetries}.
\newblock \emph{arXiv:2310.16982}, 2023.

\bibitem[Scruby et~al.(2024{\natexlab{a}})Scruby, Pesah, and Webster]{scruby2024quantum}
Thomas~R Scruby, Arthur Pesah, and Mark Webster.
\newblock {Quantum rainbow codes}.
\newblock \emph{arXiv preprint arXiv:2408.13130}, 2024{\natexlab{a}}.

\bibitem[Breuckmann et~al.(2024)Breuckmann, Davydova, Eberhardt, and Tantivasadakarn]{breuckmann2024cups}
Nikolas~P Breuckmann, Margarita Davydova, Jens~N Eberhardt, and Nathanan Tantivasadakarn.
\newblock {Cups and Gates I: Cohomology invariants and logical quantum operations}.
\newblock \emph{arXiv preprint arXiv:2410.16250}, 2024.

\bibitem[Hsin et~al.(2024)Hsin, Kobayashi, and Zhu]{hsin2024classifying}
Po-Shen Hsin, Ryohei Kobayashi, and Guanyu Zhu.
\newblock {Classifying Logical Gates in Quantum Codes via Cohomology Operations and Symmetry}.
\newblock \emph{arXiv preprint arXiv:2411.15848}, 2024.

\bibitem[Lin(2024)]{lin2024transversal}
Ting-Chun Lin.
\newblock {Transversal non-Clifford gates for quantum {LDPC} codes on sheaves}.
\newblock \emph{arXiv preprint arXiv:2410.14631}, 2024.

\bibitem[Golowich and Lin(2024)]{golowich2024quantum}
Louis Golowich and Ting-Chun Lin.
\newblock {Quantum {LDPC} Codes with Transversal Non-Clifford Gates via Products of Algebraic Codes}.
\newblock \emph{arXiv preprint arXiv:2410.14662}, 2024.

\bibitem[Malcolm et~al.(2025)Malcolm, Glaudell, Fuentes, Chandra, Schotte, DeLisle, Haenel, Ebrahimi, Roffe, Quintavalle, et~al.]{malcolm2025computing}
Alexander~J Malcolm, Andrew~N Glaudell, Patricio Fuentes, Daryus Chandra, Alexis Schotte, Colby DeLisle, Rafael Haenel, Amir Ebrahimi, Joschka Roffe, Armanda~O Quintavalle, et~al.
\newblock {Computing Efficiently in QLDPC Codes}.
\newblock \emph{arXiv preprint arXiv:2502.07150}, 2025.

\bibitem[Vuillot and Breuckmann(2022)]{vuillot2022quantum}
Christophe Vuillot and Nikolas~P Breuckmann.
\newblock {Quantum pin codes}.
\newblock \emph{IEEE Transactions on Information Theory}, 68\penalty0 (9):\penalty0 5955--5974, 2022.

\bibitem[Sayginel et~al.(2024)Sayginel, Koutsioumpas, Webster, Rajput, and Browne]{sayginel2024fault}
Hasan Sayginel, Stergios Koutsioumpas, Mark Webster, Abhishek Rajput, and Dan~E Browne.
\newblock {Fault-Tolerant Logical Clifford Gates from Code Automorphisms}.
\newblock \emph{arXiv preprint arXiv:2409.18175}, 2024.
\newblock \doi{10.48550/arxiv.2409.18175}.
\newblock URL \url{https://arxiv.org/abs/2409.18175}.

\bibitem[Berthusen et~al.(2025)Berthusen, Gullans, Hong, Mudassar, and Tan]{berthusen2025automorphism}
Noah Berthusen, Michael~J Gullans, Yifan Hong, Maryam Mudassar, and Shi Jie~Samuel Tan.
\newblock Automorphism gadgets in homological product codes.
\newblock \emph{arXiv preprint arXiv:2508.04794}, 2025.

\bibitem[Gottesman(2013)]{gottesman2013fault}
Daniel Gottesman.
\newblock {Fault-tolerant quantum computation with constant overhead}.
\newblock \emph{arXiv preprint arXiv:1310.2984}, 2013.

\bibitem[Fawzi et~al.(2018)Fawzi, Grospellier, and Leverrier]{fawzi2020constant}
Omar Fawzi, Antoine Grospellier, and Anthony Leverrier.
\newblock {Constant Overhead Quantum Fault-Tolerance with Quantum Expander Codes}.
\newblock In \emph{2018 {IEEE} 59th Annu. Symp. Found. Comput. Scie. (FOCS)}, volume~64, pages 106--114. ACM New York, NY, USA, Oct 2018.
\newblock \doi{10.1109/focs.2018.00076}.
\newblock URL \url{https://doi.org/10.1109%2Ffocs.2018.00076}.

\bibitem[Tamiya et~al.(2024)Tamiya, Koashi, and Yamasaki]{tamiya2024polylog}
Shiro Tamiya, Masato Koashi, and Hayata Yamasaki.
\newblock {Polylog-time-and constant-space-overhead fault-tolerant quantum computation with quantum low-density parity-check codes}.
\newblock \emph{arXiv preprint arXiv:2411.03683}, 2024.

\bibitem[Nguyen and Pattison(2024)]{nguyen2024quantum}
Quynh~T Nguyen and Christopher~A Pattison.
\newblock {Quantum fault tolerance with constant-space and logarithmic-time overheads}.
\newblock \emph{arXiv preprint arXiv:2411.03632}, 2024.

\bibitem[He et~al.(2025{\natexlab{a}})He, Nguyen, and Pattison]{he2025composable}
Zhiyang He, Quynh~T. Nguyen, and Christopher~A. Pattison.
\newblock Composable quantum fault-tolerance, 2025{\natexlab{a}}.
\newblock URL \url{https://arxiv.org/abs/2508.08246}.

\bibitem[Huang et~al.(2023)Huang, Jochym-O'Connor, and Yoder]{huang2023homomorphic}
Shilin Huang, Tomas Jochym-O'Connor, and Theodore~J. Yoder.
\newblock {Homomorphic Logical Measurements}.
\newblock \emph{PRX Quantum}, 4\penalty0 (3):\penalty0 030301, Jul 2023.
\newblock \doi{10.1103/PRXQuantum.4.030301}.
\newblock URL \url{https://link.aps.org/doi/10.1103/PRXQuantum.4.030301}.

\bibitem[Xu et~al.(2024{\natexlab{b}})Xu, Zhou, Zheng, Bluvstein, Ataides, Lukin, and Jiang]{xu2024fast}
Qian Xu, Hengyun Zhou, Guo Zheng, Dolev Bluvstein, J~Ataides, Mikhail~D Lukin, and Liang Jiang.
\newblock {Fast and Parallelizable Logical Computation with Homological Product Codes}.
\newblock \emph{arXiv preprint arXiv:2407.18490}, 2024{\natexlab{b}}.

\bibitem[Bergamaschi and Liu(2024)]{bergamaschi2024fault}
Thiago Bergamaschi and Yunchao Liu.
\newblock On fault tolerant single-shot logical state preparation and robust long-range entanglement.
\newblock \emph{arXiv preprint arXiv:2411.04405}, 2024.
\newblock \doi{10.48550/arxiv.2411.04405}.
\newblock URL \url{https://arxiv.org/abs/2411.04405}.

\bibitem[Hong(2025)]{Hong2025singleshot}
Yifan Hong.
\newblock Single-shot preparation of hypergraph product codes via dimension jump.
\newblock \emph{{Quantum}}, 9:\penalty0 1879, October 2025.
\newblock ISSN 2521-327X.
\newblock \doi{10.22331/q-2025-10-07-1879}.
\newblock URL \url{https://doi.org/10.22331/q-2025-10-07-1879}.

\bibitem[Li et~al.(2025)Li, Preskill, and Xu]{li2025transversal}
Christine Li, John Preskill, and Qian Xu.
\newblock {Transversal dimension jump for product qLDPC codes}.
\newblock \emph{arXiv:2510.07269}, oct 2025.
\newblock URL \url{https://arxiv.org/pdf/2510.07269}.

\bibitem[Bomb{\'\i}n and Martin-Delgado(2009)]{bombin2009quantum}
H{\'e}ctor Bomb{\'\i}n and Miguel~Angel Martin-Delgado.
\newblock {Quantum measurements and gates by code deformation}.
\newblock \emph{Journal of Physics A: Mathematical and Theoretical}, 42\penalty0 (9):\penalty0 095302, 2009.

\bibitem[Breuckmann et~al.(2017)Breuckmann, Vuillot, Campbell, Krishna, and Terhal]{breuckmann2017hyperbolic}
Nikolas~P Breuckmann, Christophe Vuillot, Earl Campbell, Anirudh Krishna, and Barbara~M Terhal.
\newblock {Hyperbolic and semi-hyperbolic surface codes for quantum storage}.
\newblock \emph{Quantum Science and Technology}, 2\penalty0 (3):\penalty0 035007, Aug 2017.
\newblock ISSN 2058-9565.
\newblock \doi{10.1088/2058-9565/aa7d3b}.
\newblock URL \url{http://dx.doi.org/10.1088/2058-9565/aa7d3b}.

\bibitem[Lavasani and Barkeshli(2018)]{Lavasani2018low}
Ali Lavasani and Maissam Barkeshli.
\newblock {Low overhead Clifford gates from joint measurements in surface, color, and hyperbolic codes}.
\newblock \emph{Physical Review A}, 98\penalty0 (5), Nov 2018.
\newblock ISSN 2469-9934.
\newblock \doi{10.1103/physreva.98.052319}.
\newblock URL \url{http://dx.doi.org/10.1103/PhysRevA.98.052319}.

\bibitem[Jochym-O'Connor(2019)]{JochymOConnor2019faulttolerantgates}
Tomas Jochym-O'Connor.
\newblock {Fault-tolerant gates via homological product codes}.
\newblock \emph{{Quantum}}, 3:\penalty0 120, Feb 2019.
\newblock ISSN 2521-327X.
\newblock \doi{10.22331/q-2019-02-04-120}.
\newblock URL \url{https://doi.org/10.22331/q-2019-02-04-120}.

\bibitem[Krishna and Poulin(2021)]{krishna2021fault}
Anirudh Krishna and David Poulin.
\newblock {Fault-tolerant gates on hypergraph product codes}.
\newblock \emph{Physical Review X}, 11\penalty0 (1):\penalty0 011023, 2021.

\bibitem[Cohen et~al.(2022)Cohen, Kim, Bartlett, and Brown]{cohen2022low}
Lawrence~Z. Cohen, Isaac~H. Kim, Stephen~D. Bartlett, and Benjamin~J. Brown.
\newblock {Low-overhead fault-tolerant quantum computing using long-range connectivity}.
\newblock \emph{Science Advances}, 8\penalty0 (20):\penalty0 eabn1717, May 2022.
\newblock ISSN 2375-2548.
\newblock \doi{10.1126/sciadv.abn1717}.
\newblock URL \url{http://dx.doi.org/10.1126/sciadv.abn1717}.

\bibitem[Cowtan and Burton(2024)]{cowtan2024css}
Alexander Cowtan and Simon Burton.
\newblock {{CSS} code surgery as a universal construction}.
\newblock \emph{Quantum}, 8:\penalty0 1344, 2024.

\bibitem[Cowtan(2024)]{cowtan2024ssip}
A.~Cowtan.
\newblock {{SSIP}: automated surgery with quantum {LDPC} codes}.
\newblock \emph{arXiv:2407.09423}, 2024.

\bibitem[Cross et~al.(2024{\natexlab{a}})Cross, He, Rall, and Yoder]{cross2024improved}
Andrew Cross, Zhiyang He, Patrick Rall, and Theodore Yoder.
\newblock {Improved {QLDPC} Surgery: Logical Measurements and Bridging Codes}.
\newblock \emph{arXiv preprint arXiv:2407.18393}, 2024{\natexlab{a}}.

\bibitem[Williamson and Yoder(2024)]{williamson2024low}
Dominic~J Williamson and Theodore~J Yoder.
\newblock {Low-overhead fault-tolerant quantum computation by gauging logical operators}.
\newblock \emph{arXiv preprint arXiv:2410.02213}, 2024.

\bibitem[Swaroop et~al.(2024)Swaroop, Jochym-O'Connor, and Yoder]{swaroop2024universal}
Esha Swaroop, Tomas Jochym-O'Connor, and Theodore~J Yoder.
\newblock {Universal adapters between quantum {LDPC} codes}.
\newblock \emph{arXiv preprint arXiv:2410.03628}, 2024.

\bibitem[Ide et~al.(2024)Ide, Gowda, Nadkarni, and Dauphinais]{ide2024fault}
Benjamin Ide, Manoj~G Gowda, Priya~J Nadkarni, and Guillaume Dauphinais.
\newblock {Fault-tolerant logical measurements via homological measurement}.
\newblock \emph{arXiv preprint arXiv:2410.02753}, 2024.

\bibitem[Zhang and Li(2024)]{zhang2024time}
Guo Zhang and Ying Li.
\newblock {Time-efficient logical operations on quantum {LDPC} codes}.
\newblock \emph{arXiv preprint arXiv:2408.01339}, 2024.

\bibitem[Hillmann et~al.(2024)Hillmann, Dauphinais, Tzitrin, and Vasmer]{hillmann2024single}
Timo Hillmann, Guillaume Dauphinais, Ilan Tzitrin, and Michael Vasmer.
\newblock {Single-shot and measurement-based quantum error correction via fault complexes}.
\newblock \emph{arXiv preprint arXiv:2410.12963}, 2024.

\bibitem[Cowtan et~al.(2025)Cowtan, He, Williamson, and Yoder]{cowtan2025parallel}
Alexander Cowtan, Zhiyang He, Dominic~J. Williamson, and Theodore~J. Yoder.
\newblock {Parallel Logical Measurements via Quantum Code Surgery}.
\newblock \emph{arXiv preprint arXiv:2503.05003}, 2025.

\bibitem[He et~al.(2025{\natexlab{b}})He, Cowtan, Williamson, and Yoder]{he2025extractors}
Zhiyang He, Alexander Cowtan, Dominic~J. Williamson, and Theodore~J. Yoder.
\newblock Extractors: {QLDPC} architectures for efficient pauli-based computation.
\newblock \emph{arXiv preprint arXiv:2503.10390}, 2025{\natexlab{b}}.

\bibitem[Yoder et~al.(2025)Yoder, Schoute, Rall, Pritchett, Gambetta, Cross, Carroll, and Beverland]{yoder2025tour}
Theodore~J. Yoder, Eddie Schoute, Patrick Rall, Emily Pritchett, Jay~M. Gambetta, Andrew~W. Cross, Malcolm Carroll, and Michael~E. Beverland.
\newblock Tour de gross: A modular quantum computer based on bivariate bicycle codes, 2025.
\newblock URL \url{https://arxiv.org/abs/2506.03094}.

\bibitem[Poirson et~al.(2025)Poirson, Roffe, and Booth]{poirson2025engineering}
Clément Poirson, Joschka Roffe, and Robert~I. Booth.
\newblock Engineering {CSS} surgery: compiling any {CNOT} in any code.
\newblock \emph{arXiv preprint arXiv:2505.01370}, 2025.

\bibitem[Zheng et~al.(2025)Zheng, Jiang, and Xu]{zheng2025high}
Guo Zheng, Liang Jiang, and Qian Xu.
\newblock High-rate surgery: towards constant-overhead logical operations.
\newblock \emph{arXiv preprint arXiv:2510.08523}, 2025.

\bibitem[Xu et~al.(2025)Xu, Zhou, Bluvstein, Cain, Kalinowski, Preskill, Lukin, and Maskara]{xu2025batched}
Qian Xu, Hengyun Zhou, Dolev Bluvstein, Madelyn Cain, Marcin Kalinowski, John Preskill, Mikhail~D. Lukin, and Nishad Maskara.
\newblock {Batched high-rate logical operations for quantum LDPC codes}.
\newblock oct 2025.
\newblock URL \url{http://arxiv.org/abs/2510.06159}.

\bibitem[Baspin et~al.(2025)Baspin, Berent, and Cohen]{baspin2025fast}
Nouédyn Baspin, Lucas Berent, and Lawrence~Z. Cohen.
\newblock Fast surgery for quantum ldpc codes, 2025.
\newblock URL \url{https://arxiv.org/abs/2510.04521}.

\bibitem[Tan et~al.(2025)Tan, Hong, Lin, Gullans, and Hsieh]{tan2025single}
Shi Jie~Samuel Tan, Yifan Hong, Ting-Chun Lin, Michael~J. Gullans, and Min-Hsiu Hsieh.
\newblock {Single-Shot Universality in Quantum LDPC Codes via Code-Switching}.
\newblock \emph{arXiv:2510.08552}, oct 2025.
\newblock URL \url{https://arxiv.org/pdf/2510.08552}.

\bibitem[Golowich et~al.(2025)Golowich, Chang, and Zhu]{golowich2025constant}
Louis Golowich, Kathleen Chang, and Guanyu Zhu.
\newblock Constant-overhead addressable gates via single-shot code switching.
\newblock \emph{arXiv preprint arXiv:2510.06760}, 2025.

\bibitem[Vuillot et~al.(2019)Vuillot, Lao, Criger, Almud{\'e}ver, Bertels, and Terhal]{vuillot2019code}
Christophe Vuillot, Lingling Lao, Ben Criger, Carmen~Garc{\'\i}a Almud{\'e}ver, Koen Bertels, and Barbara~M Terhal.
\newblock {Code deformation and lattice surgery are gauge fixing}.
\newblock \emph{New Journal of Physics}, 21\penalty0 (3):\penalty0 033028, 2019.

\bibitem[Hastings(2017)]{hastings2016weight}
Mathew~B. Hastings.
\newblock {Weight reduction for quantum codes}.
\newblock \emph{Quant. Inf. Comput.}, 17\penalty0 (15-16):\penalty0 1307--1334, 2017.
\newblock \doi{10.26421/QIC17.15-16-4}.

\bibitem[Hastings(2021)]{hastings2021weight}
M.~B. Hastings.
\newblock {On Quantum Weight Reduction}.
\newblock \emph{arXiv:2102.10030}, 2021.

\bibitem[Sabo et~al.(2024)Sabo, Gunderman, Ide, Vasmer, and Dauphinais]{sabo2024weight}
Eric Sabo, Lane~G. Gunderman, Benjamin Ide, Michael Vasmer, and Guillaume Dauphinais.
\newblock {Weight-Reduced Stabilizer Codes with Lower Overhead}.
\newblock \emph{PRX Quantum}, 5:\penalty0 040302, Oct 2024.
\newblock \doi{10.1103/PRXQuantum.5.040302}.
\newblock URL \url{https://link.aps.org/doi/10.1103/PRXQuantum.5.040302}.

\bibitem[Yuan(2025)]{Yuan2025unified}
Andrew~C. Yuan.
\newblock {Unified Framework for Quantum Code Embedding}.
\newblock jul 2025.
\newblock URL \url{https://arxiv.org/pdf/2507.05361}.

\bibitem[Bravyi et~al.(2016)Bravyi, Smith, and Smolin]{bravyi2016trading}
S.~Bravyi, G.~Smith, and J.~A. Smolin.
\newblock {Trading Classical and Quantum Computational Resources}.
\newblock \emph{Physical Review X}, 6\penalty0 (2), Jun 2016.
\newblock ISSN 2160-3308.
\newblock \doi{10.1103/physrevx.6.021043}.
\newblock URL \url{http://dx.doi.org/10.1103/PhysRevX.6.021043}.

\bibitem[Litinski(2019)]{litinski2019game}
D.~Litinski.
\newblock {A Game of Surface Codes: Large-Scale Quantum Computing with Lattice Surgery}.
\newblock \emph{Quantum}, 3:\penalty0 128, Mar 2019.
\newblock ISSN 2521-327X.
\newblock \doi{10.22331/q-2019-03-05-128}.
\newblock URL \url{http://dx.doi.org/10.22331/q-2019-03-05-128}.

\bibitem[Bombín(2015)]{Bombin2015single}
Héctor Bombín.
\newblock {Single-Shot Fault-Tolerant Quantum Error Correction}.
\newblock \emph{Physical Review X}, 5\penalty0 (3), Sep 2015.
\newblock ISSN 2160-3308.
\newblock \doi{10.1103/physrevx.5.031043}.
\newblock URL \url{http://dx.doi.org/10.1103/PhysRevX.5.031043}.

\bibitem[Campbell(2019)]{campbell2019theory}
Earl~T Campbell.
\newblock {A theory of single-shot error correction for adversarial noise}.
\newblock \emph{Quantum Science and Technology}, 4\penalty0 (2):\penalty0 025006, Feb 2019.
\newblock \doi{10.1088/2058-9565/aafc8f}.
\newblock URL \url{https://dx.doi.org/10.1088/2058-9565/aafc8f}.

\bibitem[Gu et~al.(2024)Gu, Tang, Caha, Choe, He, and Kubica]{gu2024single}
Shouzhen Gu, Eugene Tang, Libor Caha, Shin~Ho Choe, Zhiyang He, and Aleksander Kubica.
\newblock {Single-shot decoding of good quantum LDPC codes}.
\newblock \emph{Communications in Mathematical Physics}, 405\penalty0 (3):\penalty0 85, 2024.

\bibitem[Kubica and Vasmer(2022)]{Kubica2022single}
Aleksander Kubica and Michael Vasmer.
\newblock {Single-shot quantum error correction with the three-dimensional subsystem toric code}.
\newblock \emph{Nature Communications}, 13\penalty0 (1), Oct 2022.
\newblock ISSN 2041-1723.
\newblock \doi{10.1038/s41467-022-33923-4}.
\newblock URL \url{http://dx.doi.org/10.1038/s41467-022-33923-4}.

\bibitem[Scruby et~al.(2024{\natexlab{b}})Scruby, Hillmann, and Roffe]{scruby2024high}
Thomas~R. Scruby, Timo Hillmann, and Joschka Roffe.
\newblock {High-threshold, low-overhead and single-shot decodable fault-tolerant quantum memory}, 2024{\natexlab{b}}.
\newblock URL \url{https://arxiv.org/abs/2406.14445}.

\bibitem[Quintavalle et~al.(2021)Quintavalle, Vasmer, Roffe, and Campbell]{quintavelle2021single}
Armanda~O. Quintavalle, Michael Vasmer, Joschka Roffe, and Earl~T. Campbell.
\newblock Single-shot error correction of three-dimensional homological product codes.
\newblock \emph{PRX Quantum}, 2:\penalty0 020340, Jun 2021.
\newblock \doi{10.1103/PRXQuantum.2.020340}.
\newblock URL \url{https://link.aps.org/doi/10.1103/PRXQuantum.2.020340}.

\bibitem[Aasen et~al.(2025{\natexlab{a}})Aasen, Hastings, Kliuchnikov, Bello-Rivas, Paetznick, Chao, Reichardt, Zanner, da~Silva, Wang, and Svore]{aasen2025topologically}
David Aasen, Matthew~B. Hastings, Vadym Kliuchnikov, Juan~M. Bello-Rivas, Adam Paetznick, Rui Chao, Ben~W. Reichardt, Matt Zanner, Marcus~P. da~Silva, Zhenghan Wang, and Krysta~M. Svore.
\newblock A topologically fault-tolerant quantum computer with four dimensional geometric codes, 2025{\natexlab{a}}.
\newblock URL \url{https://arxiv.org/abs/2506.15130}.

\bibitem[Zhou et~al.(2024)Zhou, Zhao, Cain, Bluvstein, Duckering, Hu, Wang, Kubica, and Lukin]{zhou2024algorithmic}
Hengyun Zhou, Chen Zhao, Madelyn Cain, Dolev Bluvstein, Casey Duckering, Hong-Ye Hu, Sheng-Tao Wang, Aleksander Kubica, and Mikhail~D. Lukin.
\newblock {Algorithmic Fault Tolerance for Fast Quantum Computing}, 2024.
\newblock URL \url{https://arxiv.org/abs/2406.17653}.

\bibitem[Shutty et~al.()Shutty, Gidney, and Higgott]{shutty2025early}
Noah Shutty, Craig Gidney, and Oscar Higgott.
\newblock Early-stop lattice surgery.
\newblock \emph{To appear}.
\newblock URL \url{https://yale.hosted.panopto.com/Panopto/Pages/Viewer.aspx?id=dbfe6994-f408-46e5-8227-b33001046e13}.

\bibitem[Bravyi et~al.(2024{\natexlab{b}})Bravyi, Cross, Gambetta, Maslov, Rall, and Yoder]{bravyi2024high}
Sergey Bravyi, Andrew~W Cross, Jay~M Gambetta, Dmitri Maslov, Patrick Rall, and Theodore~J Yoder.
\newblock {High-threshold and low-overhead fault-tolerant quantum memory}.
\newblock \emph{Nature}, 627\penalty0 (8005):\penalty0 778--782, 2024{\natexlab{b}}.

\bibitem[Carette et~al.(2019)Carette, Horsman, and Perdrix]{carette2019szx}
Titouan Carette, Dominic Horsman, and Simon Perdrix.
\newblock {SZX-Calculus: Scalable Graphical Quantum Reasoning}.
\newblock In Peter Rossmanith, Pinar Heggernes, and Joost-Pieter Katoen, editors, \emph{44th International Symposium on Mathematical Foundations of Computer Science (MFCS 2019)}, volume 138 of \emph{Leibniz International Proceedings in Informatics (LIPIcs)}, pages 55:1--55:15, Dagstuhl, Germany, 2019. Schloss Dagstuhl -- Leibniz-Zentrum f{\"u}r Informatik.
\newblock ISBN 978-3-95977-117-7.
\newblock \doi{10.4230/LIPIcs.MFCS.2019.55}.
\newblock URL \url{https://drops.dagstuhl.de/entities/document/10.4230/LIPIcs.MFCS.2019.55}.

\bibitem[Breuckmann and Eberhardt(2021{\natexlab{b}})]{breuckmann2021ldpc}
Nikolas~P. Breuckmann and Jens~Niklas Eberhardt.
\newblock {Quantum Low-Density Parity-Check Codes}.
\newblock \emph{PRX Quantum}, 2:\penalty0 040101, Oct 2021{\natexlab{b}}.
\newblock \doi{10.1103/PRXQuantum.2.040101}.
\newblock URL \url{https://link.aps.org/doi/10.1103/PRXQuantum.2.040101}.

\bibitem[Panteleev and Kalachev(2022{\natexlab{a}})]{panteleev2022good}
Pavel Panteleev and Gleb Kalachev.
\newblock {Asymptotically good Quantum and locally testable classical {LDPC} codes}.
\newblock In \emph{Proceedings of the 54th Annual ACM SIGACT Symposium on Theory of Computing}, STOC 2022, page 375–388, New York, NY, USA, 2022{\natexlab{a}}. Association for Computing Machinery.
\newblock ISBN 9781450392648.
\newblock \doi{10.1145/3519935.3520017}.
\newblock URL \url{https://doi.org/10.1145/3519935.3520017}.

\bibitem[Weibel(1994)]{Weib1994}
C.~A. Weibel.
\newblock \emph{An Introduction to Homological Algebra}.
\newblock Cambridge Studies in Advanced Mathematics, Cambridge University Press, 1994.
\newblock doi:10.1017/CBO9781139644136.

\bibitem[Audoux and Couvreur(2019)]{audoux2019ontensor}
Benjamin Audoux and Alain Couvreur.
\newblock On tensor products of {CSS} codes.
\newblock \emph{Ann. Inst. Henri Poincaré Comb. Phys. Interact.}, 6:\penalty0 239--287, 2019.
\newblock \doi{10.4171/AIHPD/71}.

\bibitem[McEwen et~al.(2023)McEwen, Bacon, and Gidney]{McEwen2023relaxing}
Matt McEwen, Dave Bacon, and Craig Gidney.
\newblock Relaxing {H}ardware {R}equirements for {S}urface {C}ode {C}ircuits using {T}ime-dynamics.
\newblock \emph{{Quantum}}, 7:\penalty0 1172, November 2023.
\newblock ISSN 2521-327X.
\newblock \doi{10.22331/q-2023-11-07-1172}.
\newblock URL \url{https://doi.org/10.22331/q-2023-11-07-1172}.

\bibitem[Aharonov and Eldar(2015)]{aharonov2015quantum}
Dorit Aharonov and Lior Eldar.
\newblock {Quantum Locally Testable Codes}.
\newblock \emph{SIAM Journal on Computing}, 44\penalty0 (5):\penalty0 1230--1262, 2015.
\newblock \doi{10.1137/140975498}.

\bibitem[Panteleev and Kalachev(2022{\natexlab{b}})]{panteleev2022asymptotically}
Pavel Panteleev and Gleb Kalachev.
\newblock {Asymptotically good Quantum and locally testable classical {LDPC} codes}.
\newblock In \emph{Proceedings of the 54th Annual ACM SIGACT Symposium on Theory of Computing}, STOC 2022, page 375–388, New York, NY, USA, 2022{\natexlab{b}}. Association for Computing Machinery.
\newblock ISBN 9781450392648.
\newblock \doi{10.1145/3519935.3520017}.
\newblock URL \url{https://doi.org/10.1145/3519935.3520017}.

\bibitem[Leverrier et~al.(2022)Leverrier, Londe, and Z{\'{e}}mor]{leverrier2022towards}
Anthony Leverrier, Vivien Londe, and Gilles Z{\'{e}}mor.
\newblock {Towards local testability for quantum coding}.
\newblock \emph{{Quantum}}, 6:\penalty0 661, Feb 2022.
\newblock ISSN 2521-327X.
\newblock \doi{10.22331/q-2022-02-24-661}.
\newblock URL \url{https://doi.org/10.22331/q-2022-02-24-661}.

\bibitem[Eldar and Harrow(2017)]{eldar2015local}
L.~Eldar and A.~W. Harrow.
\newblock {Local Hamiltonians Whose Ground States Are Hard to Approximate}.
\newblock In \emph{2017 IEEE 58th Annual Symposium on Foundations of Computer Science (FOCS)}, pages 427--438, Los Alamitos, CA, USA, Oct 2017. IEEE Computer Society.
\newblock \doi{10.1109/FOCS.2017.46}.
\newblock URL \url{https://doi.ieeecomputersociety.org/10.1109/FOCS.2017.46}.

\bibitem[Cross et~al.(2024{\natexlab{b}})Cross, He, Natarajan, Szegedy, and Zhu]{Cross2024quantumlocally}
Andrew Cross, Zhiyang He, Anand Natarajan, Mario Szegedy, and Guanyu Zhu.
\newblock {Quantum {L}ocally {T}estable {C}ode with {C}onstant {S}oundness}.
\newblock \emph{{Quantum}}, 8:\penalty0 1501, Oct 2024{\natexlab{b}}.
\newblock ISSN 2521-327X.
\newblock \doi{10.22331/q-2024-10-18-1501}.
\newblock URL \url{https://doi.org/10.22331/q-2024-10-18-1501}.

\bibitem[Wills et~al.(2024)Wills, Lin, and Hsieh]{wills2024tradeoff}
Adam Wills, Ting-Chun Lin, and Min-Hsiu Hsieh.
\newblock Tradeoff constructions for quantum locally testable codes.
\newblock \emph{IEEE Transactions on Information Theory}, 2024.

\bibitem[Dinur et~al.(2024)Dinur, Lin, and Vidick]{dinur2024expansion}
Irit Dinur, Ting-Chun Lin, and Thomas Vidick.
\newblock Expansion of high-dimensional cubical complexes: with application to quantum locally testable codes.
\newblock In \emph{2024 IEEE 65th Annual Symposium on Foundations of Computer Science (FOCS)}, pages 379--385. IEEE, 2024.

\bibitem[Kalachev and Panteleev(2025)]{kalachev2025maximally}
Gleb Kalachev and Pavel Panteleev.
\newblock Maximally extendable product codes are good coboundary expanders.
\newblock \emph{arXiv preprint arXiv:2501.01411}, 2025.

\bibitem[Cowtan and Majid(2023)]{cowtan2022algebraic}
Alexander Cowtan and Shahn Majid.
\newblock Algebraic aspects of boundaries in the kitaev quantum double model.
\newblock \emph{Journal of Mathematical Physics}, 64\penalty0 (10):\penalty0 102203, 10 2023.
\newblock ISSN 0022-2488.
\newblock \doi{10.1063/5.0127285}.
\newblock URL \url{https://doi.org/10.1063/5.0127285}.

\bibitem[Bravyi et~al.(2022{\natexlab{b}})Bravyi, Kim, Kliesch, and Koenig]{Bravyi2022}
Sergey Bravyi, Isaac Kim, Alexander Kliesch, and Robert Koenig.
\newblock {Adaptive constant-depth circuits for manipulating non-abelian anyons}.
\newblock may 2022{\natexlab{b}}.
\newblock URL \url{http://arxiv.org/abs/2205.01933}.

\bibitem[Lyons et~al.(2024)Lyons, Lo, Tantivasadakarn, Vishwanath, and Verresen]{Lyons2024}
Anasuya Lyons, Chiu Fan~Bowen Lo, Nathanan Tantivasadakarn, Ashvin Vishwanath, and Ruben Verresen.
\newblock {Protocols for Creating Anyons and Defects via Gauging}.
\newblock nov 2024.
\newblock URL \url{http://arxiv.org/abs/2411.04181}.

\bibitem[Ren et~al.(2025)Ren, Tantivasadakarn, and Williamson]{Ren2025}
Yuanjie Ren, Nathanan Tantivasadakarn, and Dominic~J. Williamson.
\newblock {Efficient Preparation of Solvable Anyons with Adaptive Quantum Circuits}.
\newblock \emph{Physical Review X}, 15\penalty0 (3), aug 2025.
\newblock \doi{10.1103/b9hf-gx4f}.
\newblock URL \url{http://arxiv.org/abs/2411.04985 http://dx.doi.org/10.1103/b9hf-gx4f}.

\bibitem[Davydova et~al.(2025)Davydova, Bauer, de~la Fuente, Webster, Williamson, and Brown]{davydova2025universal}
Margarita Davydova, Andreas Bauer, Julio C.~Magdalena de~la Fuente, Mark Webster, Dominic~J. Williamson, and Benjamin~J. Brown.
\newblock Universal fault tolerant quantum computation in 2d without getting tied in knots.
\newblock \emph{arXiv preprint arxiv:2503.15751}, 2025.

\bibitem[de~Beaudrap and Horsman(2020)]{deBeaudrap2020zx}
Niel de~Beaudrap and Dominic Horsman.
\newblock The {ZX} calculus is a language for surface code lattice surgery.
\newblock \emph{{Quantum}}, 4:\penalty0 218, January 2020.
\newblock ISSN 2521-327X.
\newblock \doi{10.22331/q-2020-01-09-218}.
\newblock URL \url{https://doi.org/10.22331/q-2020-01-09-218}.

\bibitem[Aasen et~al.(2025{\natexlab{b}})Aasen, Haah, Hastings, and Wang]{aasen2025geometrically}
David Aasen, Jeongwan Haah, Matthew~B. Hastings, and Zhenghan Wang.
\newblock Geometrically enhanced topological quantum codes, 2025{\natexlab{b}}.
\newblock URL \url{https://arxiv.org/abs/2505.10403}.

\bibitem[Beverland et~al.(2024)Beverland, Huang, and Kliuchnikov]{beverland2024fault}
M.~E. Beverland, S.~Huang, and V.~Kliuchnikov.
\newblock {Fault tolerance of stabilizer channels}.
\newblock \emph{arXiv preprint arXiv:2401.12017}, 2024.

\bibitem[Litinski(2025)]{Litinski2025Blocklet}
Daniel Litinski.
\newblock {Blocklet concatenation: Low-overhead fault-tolerant protocols for fusion-based quantum computation}.
\newblock jun 2025.
\newblock URL \url{http://arxiv.org/abs/2506.13619}.

\bibitem[Horsman et~al.(2012)Horsman, Fowler, Devitt, and Meter]{Horsman2012LatticeSurgery}
Dominic Horsman, Austin~G Fowler, Simon Devitt, and Rodney~Van Meter.
\newblock {Surface code quantum computing by lattice surgery}.
\newblock \emph{New Journal of Physics}, 14\penalty0 (12):\penalty0 123011, Dec 2012.
\newblock ISSN 1367-2630.
\newblock \doi{10.1088/1367-2630/14/12/123011}.
\newblock URL \url{http://dx.doi.org/10.1088/1367-2630/14/12/123011}.

\bibitem[Litinski and Oppen(2018)]{Litinski2018latticesurgery}
Daniel Litinski and Felix~von Oppen.
\newblock {Lattice {S}urgery with a {T}wist: {S}implifying {C}lifford {G}ates of {S}urface {C}odes}.
\newblock \emph{{Quantum}}, 2:\penalty0 62, May 2018.
\newblock ISSN 2521-327X.
\newblock \doi{10.22331/q-2018-05-04-62}.
\newblock URL \url{https://doi.org/10.22331/q-2018-05-04-62}.

\bibitem[Wang et~al.(2023)Wang, Lin, and Pryadko]{wang2023abelian}
Renyu Wang, Hsiang-Ku Lin, and Leonid~P Pryadko.
\newblock {Abelian and non-Abelian quantum two-block codes}.
\newblock In \emph{2023 12th International Symposium on Topics in Coding (ISTC)}, pages 1--5. IEEE, 2023.

\bibitem[Panteleev and Kalachev(2022{\natexlab{c}})]{panteleev2021quantum}
Pavel Panteleev and Gleb Kalachev.
\newblock {Quantum {LDPC} Codes With Almost Linear Minimum Distance}.
\newblock \emph{IEEE Transactions on Information Theory}, 68\penalty0 (1):\penalty0 213--229, 2022{\natexlab{c}}.
\newblock \doi{10.1109/TIT.2021.3119384}.

\bibitem[Liang et~al.(2025)Liang, Liu, Song, and Chen]{liang2025generalized}
Zijian Liang, Ke~Liu, Hao Song, and Yu-An Chen.
\newblock Generalized toric codes on twisted tori for quantum error correction, 2025.
\newblock URL \url{https://arxiv.org/abs/2503.03827}.

\bibitem[Coecke and Duncan(2011)]{coecke2011interacting}
Bob Coecke and Ross Duncan.
\newblock Interacting quantum observables: categorical algebra and diagrammatics.
\newblock \emph{New Journal of Physics}, 13\penalty0 (4):\penalty0 043016, apr 2011.
\newblock \doi{10.1088/1367-2630/13/4/043016}.
\newblock URL \url{https://dx.doi.org/10.1088/1367-2630/13/4/043016}.

\bibitem[Rodatz et~al.(2025)Rodatz, Poór, and Kissinger]{rodatz2025fault}
Benjamin Rodatz, Boldizsár Poór, and Aleks Kissinger.
\newblock Fault tolerance by construction, 2025.
\newblock URL \url{https://arxiv.org/abs/2506.17181}.

\bibitem[R{\"{u}}sch et~al.(2025)R{\"{u}}sch, Rodatz, and Kissinger]{Rusch2025Completeness}
Maximilian R{\"{u}}sch, Benjamin Rodatz, and Aleks Kissinger.
\newblock {Completeness for Fault Equivalence of Clifford ZX Diagrams}.
\newblock oct 2025.
\newblock URL \url{https://arxiv.org/pdf/2510.08477}.

\bibitem[Freedman and Hastings(2021)]{freedman2021building}
Michael Freedman and Matthew Hastings.
\newblock {Building manifolds from quantum codes}.
\newblock \emph{Geometric and Functional Analysis}, 31\penalty0 (4):\penalty0 855--894, 2021.

\bibitem[Krishna and Z{\'{e}}mor(2025)]{Krishna2025Tradeoffs}
Anirudh Krishna and Gilles Z{\'{e}}mor.
\newblock {Tradeoffs on the volume of fault-tolerant circuits}.
\newblock \emph{arXiv:2510.03057}, oct 2025.
\newblock URL \url{https://arxiv.org/abs/2510.03057v1}.

\bibitem[Higgott and Breuckmann(2023)]{higgott2023improved}
Oscar Higgott and Nikolas~P. Breuckmann.
\newblock Improved single-shot decoding of higher-dimensional hypergraph-product codes.
\newblock \emph{PRX Quantum}, 4:\penalty0 020332, May 2023.
\newblock \doi{10.1103/PRXQuantum.4.020332}.
\newblock URL \url{https://link.aps.org/doi/10.1103/PRXQuantum.4.020332}.

\bibitem[Gidney(2022)]{gidney2022stability}
Craig Gidney.
\newblock Stability {E}xperiments: {T}he {O}verlooked {D}ual of {M}emory {E}xperiments.
\newblock \emph{{Quantum}}, 6:\penalty0 786, August 2022.
\newblock ISSN 2521-327X.
\newblock \doi{10.22331/q-2022-08-24-786}.
\newblock URL \url{https://doi.org/10.22331/q-2022-08-24-786}.

\bibitem[TkZ()]{TkZ}
{TikZit}.
\newblock https://tikzit.github.io/index.html.

\bibitem[Derks et~al.(2024)Derks, Townsend-Teague, Burchards, and Eisert]{derks2024designing}
Peter-Jan H.~S. Derks, Alex Townsend-Teague, Ansgar~G. Burchards, and Jens Eisert.
\newblock Designing fault-tolerant circuits using detector error models, 2024.
\newblock URL \url{https://arxiv.org/abs/2407.13826}.

\bibitem[Landahl et~al.(2011)Landahl, Anderson, and Rice]{landahl2011fault}
Andrew~J. Landahl, Jonas~T. Anderson, and Patrick~R. Rice.
\newblock Fault-tolerant quantum computing with color codes, 2011.
\newblock URL \url{https://arxiv.org/abs/1108.5738}.
\newblock https://doi.org/10.48550/arXiv.1108.5738.

\bibitem[van~de Wetering(2020)]{vandewetering2020zx}
John van~de Wetering.
\newblock Zx-calculus for the working quantum computer scientist, 2020.
\newblock URL \url{https://arxiv.org/abs/2012.13966}.

\end{thebibliography}
